%% file: AspectsFirstLaw.tex
\documentclass[onecolumn,amsmath,amssymb,nofootinbib,12pt]{article}
\pdfoutput=1 
\usepackage{jheppub} 

\usepackage{bbold}
\usepackage{amsfonts,environ}
\usepackage{amsmath}
\allowdisplaybreaks[4]        
\usepackage{amssymb}
\usepackage{euscript}     
\usepackage{color}  
\usepackage{xcolor}       
\usepackage{tensor}        
\usepackage{amsthm}
\usepackage[latin1]{inputenc}
\usepackage{graphicx}
\usepackage{subfigure}
\usepackage{bbm}
\usepackage{enumitem}
\usepackage[header,title,page,titletoc]{appendix}

\usepackage{tikz}
\usepackage{sansmath}
\usetikzlibrary{shadings,intersections}
\usetikzlibrary{calc,fadings,decorations.pathreplacing}
\usepackage[numbers]{natbib}  
\usepackage{float}

\usepackage{enumitem}

\def\n{{\vec{n}}}
\def\k{{\vec{k}}}
\def\m{{\vec{m}}}
\def\p{{\vec{p}}}
\def\o{{\vec{o}}}

\renewcommand{\(}{\left(}
\renewcommand{\)}{\right)}
\renewcommand{\[}{\left[}
\renewcommand{\]}{\right]}

\newcommand{\cO}{{\cal O}} 
\newcommand{\bra}[1]{{\left\langle{#1}\right\vert}}
\newcommand{\ket}[1]{{\left\vert{#1}\right\rangle}}

\newcommand{\del}{\partial}

\newcommand{\be}{\begin{equation}}
\newcommand{\ee}{\end{equation}}
\newcommand{\bea}{\begin{eqnarray}}
\newcommand{\eea}{\end{eqnarray}}
\newcommand{\beq}{\begin{equation}}
\newcommand{\eeq}{\end{equation}}
\newcommand{\beqa}{\begin{eqnarray}}
\newcommand{\eeqa}{\end{eqnarray}}
\newcommand{\beqar}{\begin{eqnarray*}}
\newcommand{\eeqar}{\end{eqnarray*}}

\newcommand{\eg}{{\it e.g.,}\ }
\newcommand{\ie}{{\it i.e.,}\ }
\newcommand{\reef}[1]{(\ref{#1})}

\newcommand{\mt}[1]{\textrm{\tiny #1}}

\newcommand{\mC}{\mathcal{C}}
\newcommand{\mS}{\mathcal{S}}

\newcommand{\mO}{\mathcal{O}}
\newcommand{\mR}{\mathbb{R}}

\newcommand{\mV}{\mathcal{V}}

\newcommand{\eps}{\epsilon}

\newcommand{\veps}{\varepsilon}
\def\S{\Sigma}
\newcommand{\cv}{{\cal C}_\mt{V}}
\newcommand{\ca}{{\cal C}_\mt{A}}

\newcommand{\cev}[1]{\reflectbox{\ensuremath{\vec{ \reflectbox{\ensuremath{#1}}}}}}
\newcommand{\s}{\sigma}

\newcommand{\GN}{G_\mt{N}}

\newcommand{\vk}{{\vec k}}

\def\d{\delta}
\def\r{\rho}
\def\D{\Delta}
\newcommand{\ww}{\omega}

\newcommand{\VO}{\text{Vol}\,\Omega_{d-1}}

\newcommand\pgfmathsinandcos[3]{%
	\pgfmathsetmacro#1{sin(#3)}%
	\pgfmathsetmacro#2{cos(#3)}%
}
\newcommand\LongitudePlane[3][current plane]{%
	\pgfmathsinandcos\sinEl\cosEl{#2} 
	\pgfmathsinandcos\sint\cost{#3} 
	\tikzset{#1/.style={cm={\cost,\sint*\sinEl,0,\cosEl,(0,0)}}}
}
\newcommand\LatitudePlane[3][current plane]{%
	\pgfmathsinandcos\sinEl\cosEl{#2} 
	\pgfmathsinandcos\sint\cost{#3} 
	\pgfmathsetmacro\yshift{\cosEl*\sint}
	\tikzset{#1/.style={cm={\cost,0,0,\cost*\sinEl,(0,\yshift)}}} %
}
\newcommand\DrawLongitudeCircle[2][1]{
	\LongitudePlane{\angEl}{#2}
	\tikzset{current plane/.prefix style={scale=#1}}
	\pgfmathsetmacro\angVis{atan(sin(#2)*cos(\angEl)/sin(\angEl))} %
	\draw[current plane] (\angVis:1) arc (\angVis:\angVis+180:1);
	\draw[current plane,dashed] (\angVis-180:1) arc (\angVis-180:\angVis:1);
}
\newcommand\DrawLatitudeCircle[2][1]{
	\LatitudePlane{\angEl}{#2}
	\tikzset{current plane/.prefix style={scale=#1}}
	\pgfmathsetmacro\sinVis{sin(#2)/cos(#2)*sin(\angEl)/cos(\angEl)}
	\pgfmathsetmacro\angVis{asin(min(1,max(\sinVis,-1)))}
	\draw[current plane] (\angVis:1) arc (\angVis:-\angVis-180:1);
	\draw[current plane,dashed] (180-\angVis:1) arc (180-\angVis:\angVis:1);
}
\tikzset{%
	>=latex, 
	inner sep=0pt,%
	outer sep=2pt,%
	mark coordinate/.style={inner sep=0pt,outer sep=0pt,minimum size=2pt,
		fill=black,circle}%
}

\NewEnviron{eqs}{%
\begin{equation}\begin{split}
  \BODY
\end{split}\end{equation}
 }

\newcommand{\llangle}{{\langle\! \langle\,}}
\newcommand{\rrangle}{{\,\rangle\! \rangle}}

\newtheorem{theorem}{Theorem}

\preprint{arXiv:2002.05779 [hep-th]}
\title{\boldmath Aspects of The First Law of Complexity}

\author[a,b]{Alice Bernamonti,}
\author[b]{Federico Galli,}
\author[c,d]{Juan Hernandez,}
\author[c]{Robert C. Myers,}
\author[c,d]{Shan-Ming Ruan,}
\author[e]{Joan Sim\'on}

\affiliation[a]{Dipartimento di Fisica e Astronomia, Universit\`a di Firenze, Via G. Sansone 1, 50019 Sesto Fiorentino, Italy}
\affiliation[b]{INFN Sezione di Firenze, Via G. Sansone 1, 50019 Sesto Fiorentino, Italy}
\affiliation[c]{Perimeter Institute for Theoretical Physics,  Waterloo, Ontario N2L 2Y5, Canada}
\affiliation[d]{Department of Physics and Astronomy,
	University of Waterloo, Waterloo, Ontario\\ N2L 3G1, Canada}
\affiliation[e]{School of Mathematics and Maxwell Institute for Mathematical Sciences,\\
University of Edinburgh, Edinburgh EH9 3FD, UK}

\emailAdd{alice.bernamonti@unifi.it}
\emailAdd{federico.galli@fi.infn.it}
\emailAdd{jhernandez@pitp.ca}
\emailAdd{rmyers@pitp.ca}
\emailAdd{sruan@pitp.ca}
\emailAdd{j.simon@ed.ac.uk}

\date{\today}

\abstract{We investigate the first law of complexity proposed in \cite{Bernamonti:2019zyy}, \ie the variation of complexity when the target state is perturbed, in more detail. Based on Nielsen's geometric approach to quantum circuit complexity, we find the variation only depends on the end of the optimal circuit. 
We apply the first law to gain new insights into the quantum circuits and complexity models underlying holographic complexity. In particular, we examine the variation of the holographic complexity for both the complexity=action and complexity=volume conjectures in perturbing the AdS vacuum with coherent state excitations of a free scalar field. 
We also examine the variations of circuit complexity produced by the same excitations for the free scalar field theory in a fixed AdS background. In this case, our work extends the existing treatment of Gaussian coherent states to properly include the time dependence of the complexity variation. We comment on the similarities and differences of the holographic and QFT results.}

\begin{document} 

	\maketitle
	\flushbottom

\section{Introduction}\label{intro}
\input{sections/intro}

\section{First law of complexity} \label{firstL}
\input{sections/firstL}

\subsection{Coherent states to probe the first law}
\label{sec:coherent}
\input{sections/coherent}

\section{Holographic complexity} 
\label{sec:hol-comp}
\input{sections/Bulk}

\input{sections/ComAction} 
\input{sections/ComVolume}

\input{sections/CACVcomparison}

\section{Circuit complexity for QFT}
\label{sec:q-circuit}
\input{sections/circuit}
\section{Discussion} 
    \label{discuss}
\input{sections/discuss}

\section*{Acknowledgments}
It is a pleasure to thank Alex Belin, Pablo Bueno, Pawel Caputa, Horacio Casini, Jos\'e M. Figueroa-O'Farrill, Kevin Grosvenor, Javier Mag\'an, Alex Maloney, Simon Ross, James Sully and Tadashi Takayanagi for useful comments and conversations.  Research at Perimeter Institute is supported in part by the Government of Canada through the Department of Innovation, Science and Economic Development Canada and by the Province of Ontario through the Ministry of Economic Development, Job Creation and Trade. 
AB acknowledges support by the program ``Rita Levi Montalcini" for young researchers and the INFN initiative GAST.
FG has received funding from the European Union's Horizon 2020 research and innovation programme under the Marie Sklodowska-Curie grant agreement No 754496. 
JPH is also supported by the Natural Sciences and Engineering Research Council of Canada through a NSERC PGS-D fellowship. 
RCM was supported in part by research funding from the BMO Financial Group and from the Simons Foundation through the ``It from Qubit'' Collaboration.  RCM was also supported in part by a Discovery Grant from the Natural Sciences and Engineering Research Council of Canada. JS is supported by the Science and Technology Facilities Council (STFC) [grant number ST/L000458/1]. JS would also like to thank the Perimeter Institute for all their support and hospitality during the period January-June 2017 when this project started.
 
\begin{appendix}

\section{No contributions to CA from the caustics} 
\label{app:caustic}
\input{sections/append1}

\section{UV cutoffs and vacuum CA} 
\label{app:ads}
\input{sections/append2}

\section{Globally vs locally minimizing geodesics}\label{sec:Riem}
\input{sections/Riemannian}

\subsection{Smoothness of complexity} \label{globall}
\input{sections/append_smoothness}

\section{Geodesics for simple states}\label{sec:simple}
\input{sections/append3}

\end{appendix}

\bibliographystyle{JHEP}
\bibliography{combined}

\end{document}

%% file: sections/intro.tex
In recent years, quantum information perspectives have produced surprising insights into foundational questions about the AdS/CFT correspondence. One fascinating and new concept that has entered this discussion is quantum circuit complexity, which measures how difficult it is to construct a particular target state from a (simple) reference state by applying a set of (simple) elementary gates, \eg for a review see \cite{johnw,AaronsonRev}.  In considering complexity and holography together, two distinct approaches have emerged. First, new holographic complexity conjectures have drawn our attention to new gravitational observables in the AdS/CFT correspondence \cite{Susskind:2014rva,Stanford:2014jda,Brown:2015bva,Brown:2015lvg,Couch:2016exn}. Second, various approaches have been investigated to understand the complexity of states in quantum field theory, \eg Nielsen's geometric approach  \cite{nielsen2006quantum, nielsen2008, Nielsen:2006}. 

It is believed that the gravitational observables dual to complexity in boundary theory can provide more information about the bulk spacetime than that coming from holographic entanglement entropy \cite{Susskind:2014moa}. Under the heading of holographic complexity, a variety of proposals for the bulk description of the complexity of boundary states have been developed. The most studied of these are the complexity=volume (CV) \cite{Susskind:2014rva,Stanford:2014jda} and the complexity=action (CA) \cite{Brown:2015bva,Brown:2015lvg} conjectures. The CV conjecture states that the complexity is dual to the volume of an extremal codimension-one bulk surface anchored at the time slice $\S$ in the boundary on which the state is defined,
\begin{equation}
\cv(\S) =\ \mathrel{\mathop {\rm
max}_{\scriptscriptstyle{\S=\partial \mathcal{B}}} {}\!\!}\left[\frac{\mathcal{V(B)}}{G_N \, \ell_{ \rm bulk}}\right] \, ,\label{defineCV}
\end{equation}
with $\mathcal B$ corresponding to the bulk surface of interest and $G_N$ denoting Newton's constant in the bulk gravitational theory. Further,  $\ell_{\rm bulk}$ is some additional length scale associated with the bulk geometry, \eg see discussion in \cite{Brown:2015bva,Couch:2018phr}. For simplicity, in the following, we will set $\ell_{\rm bulk} = L$, \ie the curvature radius for the (asymptotic) AdS geometry.  On the other hand, the CA proposal states that the complexity is given by evaluating the gravitational action on a region of spacetime, known as the Wheeler-DeWitt (WDW) patch, which can be regarded as the causal development of a space-like bulk surface 
anchored on the boundary time slice $\S$. The CA proposal then suggests
\beq
\ca(\S) =  \frac{I_\mt{WDW}}{\pi\, \hbar}\,. \label{defineCA}
\eeq
These two conjectures have stimulated a wide variety of recent research efforts investigating the properties and applications of holographic complexity, \eg \cite{Susskind:2014rva,Stanford:2014jda,Susskind:2014jwa, Brown:2015bva,Brown:2015lvg,Susskind:2014moa, Susskind:2015toa,Roberts:2014isa,Lehner:2016vdi,Cai:2016xho, Couch:2016exn,Reynolds:2016rvl,Chapman:2016hwi,Carmi:2016wjl, Moosa:2017yvt,Couch:2017yil,Cai:2017sjv,Brown:2017jil, Carmi:2017jqz,Swingle:2017zcd,Flory:2017ftd,Zhao:2017isy, Abt:2017pmf,Abt:2018ywl,Alishahiha:2018tep,An:2018xhv, Fu:2018kcp,Mahapatra:2018gig,Chapman:2018dem, Chapman:2018lsv,Cano:2018aqi,Barbon:2018mxk, Susskind:2018fmx,Susskind:2018tei,Cooper:2018cmb, Numasawa:2018grg,Brown:2018kvn,Goto:2018iay, Agon:2018zso, Chapman:2018bqj,Flory:2018akz,Flory:2019kah,Braccia:2019xxi,Sato:2019kik,Barbon:2015ria,Barbon:2015soa,Auzzi:2018pbc,Auzzi:2018zdu,Bhattacharya:2019zkb,Ghosh:2019jgd}.

Attempts to understand the complexity of QFT states have mainly centered around Nielsen's geometric approach to evaluating circuit complexity
 \cite{nielsen2006quantum, nielsen2008, Nielsen:2006},\footnote{Of course, we should add that a complementary approach based on the Fubini-Study metric for the space of states was also proposed in \cite{Chapman:2017rqy}.} which we review in more detail in section \ref{sec:intro2}. It was first suggested in \cite{Susskind:2014jwa} that this idea may play a role in defining holographic complexity and this connection was pursued further in \cite{Brown:2016wib,Brown:2017jil}. This approach was first applied to a concrete quantum field theory calculation in \cite{Jeff}, where the authors adapted Nielsen's approach to evaluate the complexity of the vacuum state of a free scalar field theory. These calculations have been extended in a number of interesting ways in the past few years, \eg \cite{Khan:2018rzm,Molina-Vilaplana:2018sfn,Hackl:2018ptj,Alves:2018qfv,Magan:2018nmu,Camargo:2018eof,Chapman:2018hou,cohere,Caceres:2019pgf, Ali:2018fcz,Bhattacharyya:2018bbv, Jiang:2018nzg,Chapman:2019clq,Camargo:2019isp,Doroudiani:2019llj,Bueno:2019ajd,Ali:2019zcj,Ge:2019mjt,Caputa:2018kdj}, but we will be particularly interested in \cite{cohere} where the same techniques were applied to explore the complexity of coherent states in the same QFT.

The first law of complexity was introduced in \cite{Bernamonti:2019zyy} as an attempt to build a concrete bridge between the two discussions, \ie to provide a clear connection between holographic complexity and the quantum circuit constructions for QFT complexity.
The main motivation for the present paper is to further explore this first law, together with providing the technical details necessary to explain the preliminary results presented in \cite{Bernamonti:2019zyy} and the extensions described below.

The first law of complexity computes the difference in complexity between two target states for a fixed reference state and set of gates when the second target state is a small perturbation of the first. In \cite{Bernamonti:2019zyy}, we used Nielsen's geometric approach to circuit complexity to derive the first and second order variations $\delta\mC$ for general (but differentiable) cost functions. Furthermore, \cite{Bernamonti:2019zyy} suggested probing this first law using coherent state excitations in the AdS/CFT correspondence. This is because the complexity variations for these states could be independently evaluated in the boundary theory and in the AdS bulk, hence providing a non-trivial bridge between quantum circuit calculations in QFT using coherent states and holographic calculations in the bulk. As stressed in  \cite{Bernamonti:2019zyy}, the equivalence between the Hilbert spaces in AdS and in the CFT is essential to justify the proposed set-up.

There are several reasons to motivate the relevance of the first law of complexity. First, the continuous formulation of circuit complexity using Nielsen's formalism \cite{nielsen2006quantum, nielsen2008, Nielsen:2006} makes it clear that $\delta\mC$ only depends on the endpoints of the circuit \cite{Bernamonti:2019zyy}. Since our holographic understanding of the reference state and gates is poor, one may make sharper the implications of complexity for holography (or to probe/explore the different conjectures) by focusing on the properties of the target state, which is always assumed to have a good gravitational description in these discussions. We are interested in exploring the possible consequences of this fact 
in holography, where we expect the bulk gravitational solution to give us the information about the final state, and the behaviour of the optimal circuit near the end-point.  Second, the study of variations in observables is always physical. Hence, it is very natural to explore variations of complexity as an example of a potentially new dictionary in holography. From a more technical perspective, these variations could be finite, as it occurs with relative entanglement entropy, making them better defined observables than the complexity $\mC$ itself. Third, from a purely gravitational perspective, the proposals reviewed above define new gauge-invariant observables. Studying their properties under small perturbations is not only natural but could lead to important insights. Indeed, the same considerations in black hole physics lead to the deep connection between gravity, spacetime, thermodynamics and entropy/information \cite{Bekenstein:1973ur,Bardeen:1973gs,Hawking:1974sw,Jacobson:1995ab}. Similarly, the first law of entanglement captures the same information as the  linearized Einstein's equations \cite{Nozaki:2013vta,Lashkari:2013koa,Faulkner:2013ica}.  Finally, from a purely  quantum mechanics perspective, it is an important question whether any notion of complexity can be understood as a resource, in the same sense as energy fluctuations above thermal energy allow to do work in thermodynamics or the existence of correlations in the boundary theory explain the connectivity in the bulk geometry in the AdS/CFT correspondence \cite{VanRaamsdonk:2010pw,Bianchi:2012ev,Maldacena:2013xja}. The first law studied in this work is a balanced equation that any such notion of complexity should satisfy.

The organization of this work is as follows: In section \ref{firstL}, we derive the first law of circuit complexity by considering the variation of complexity between two near target states. The quantum coherent states suggested to probe the first law are reviewed in section \ref{sec:coherent}, where both a boundary and bulk descriptions are provided. In section \ref{sec:hol-comp}, we develop the tools to first, evaluate the complexity=action and complexity=volume variations for the relevant spacetime configurations realizing these coherent state excitations, and second, to analyze and compare the main features of these holographic variations. The tools and evaluation of circuit complexity using quantum field theory in AdS are presented in section \ref{sec:q-circuit}. Finally, we summarize our results and discuss different aspects of the first law of complexity in section \ref{discuss}. Some further technical details on different aspects of this work are presented in appendices \ref{app:caustic}-\ref{sec:simple}.

%% file: sections/firstL.tex
This section derives the first law of quantum circuit complexity. This notion of complexity and Nielsen's geometric approach to its evaluation are first reviewed in section \ref{sec:intro2}. The latter maps the problem of determining the optimal circuit into solving for geodesic trajectories in the space of unitaries that prepare the quantum states.  Within this approach, in section \ref{sec:firstcm} we analyze the general form of complexity variations under small perturbations of the target state and formulate the first law for such variations.  Finally, in section \ref{sec:coherent} we describe the holographic framework describing small-amplitude coherent states that we suggest to probe the first law.

\subsection{Nielsen, geometry and complexity}
\label{sec:intro2}

\begin{figure}[htbp]
		\centering
		\subfigure{\includegraphics[width=5.5in]{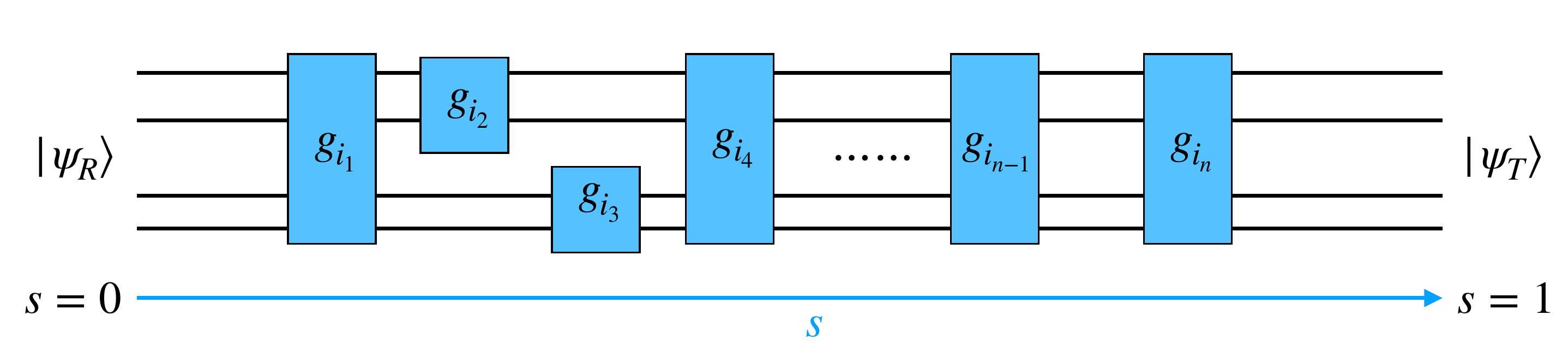}}
		\caption{A general quantum circuit where $\ket{\psi_\mt{T}}$ is prepared beginning with $\ket{\psi_\mt{R}}$ and applying a sequence of elementary unitaries $g_{i}$. We also indicate  the intermediate states  that are produced after every step, \ie
$\ket{\psi_k}=g_{i_\mt{k}}g_{i_\mt{k--1}}\cdots g_{i_\mt{2}}\,g_{i_\mt{1}}\ket{\psi_\mt{R}}$.}\label{general_circuit}
	\end{figure}

In the context of quantum circuit complexity discussions, complexity $\mC$ is defined as the optimal cost to prepare a particular target state $\ket{\psi_\mt{T}}$ starting with a certain reference state $| \psi_\mt{R} \rangle$ by applying a series of elementary gates,
\begin{equation}\label{circuit}
	\ket{\psi_\mt{T}}= U_\mt{TR}\, \ket{\psi_\mt{R}}= g_{i_\mt{n}}\cdots g_{i_\mt{2}}\,g_{i_\mt{1}}\ket{\psi_\mt{R}}\,,
\end{equation} 
as illustrated in figure~\ref{general_circuit}.
That is, given a fixed gate set $\{g_1,\cdots,g_\mt{N}\}$, the complexity $\mC(\ket{\psi_\mt{T}})$ is the \emph{minimum} number of such gates needed to construct the unitary $U_\mt{TR}$ transforming $| \psi_\mt{R} \rangle$ to $\ket{\psi_\mt{T}}$.

Nielsen and collaborators \cite{nielsen2006quantum,nielsen2008,Nielsen:2006} developed a geometric method to identify this optimal circuit. This approach was adopted to evaluate the complexity of quantum field theory states in \cite{Jeff}, and subsequently applied in a variety of different settings, \eg \cite{Chapman:2017rqy,Khan:2018rzm,Molina-Vilaplana:2018sfn,Hackl:2018ptj,Alves:2018qfv,Magan:2018nmu,Camargo:2018eof,Chapman:2018hou,cohere,Caceres:2019pgf, Ali:2018fcz,Bhattacharyya:2018bbv, Jiang:2018nzg,Chapman:2019clq,Camargo:2019isp,Doroudiani:2019llj,Bueno:2019ajd,Ali:2019zcj,Ge:2019mjt,Yang:2017nfn,Caputa:2018kdj,Yang:2018nda}. The idea is to construct a continuum representation of the unitary transformations acting on the states
\begin{equation}
  U(\s) = \cev{\mathcal{P}} \exp \[ -i \int^\s_0\!\!\! d s\, {\cal H}( s)\], \quad \text{where} \quad {\cal H}(s)\equiv \sum_I Y^I(s)\,\mathcal{O}_I \, ,
\label{unitaries}
\end{equation}
where $s$ parametrizes the position (or distance) along the circuit,  while $\cev{\mathcal{P}} $ indicates right-to-left path ordering in interpreting the exponential operator. The instantaneous (path-dependent) Hamiltonian ${\cal H}(s)$ is a linear combination of the Hermitian operators $\mathcal{O}_I$. One might think of these operators as the generators of elementary gates $g_I\sim\exp[-i\varepsilon \mO_I]$ (where $\varepsilon$ would be an infinitesimal parameter) in the corresponding gate set applied in eq.~\eqref{circuit}. The coefficients $Y^I(s)$ in the above expression \eqref{unitaries} are control functions specifying which gates (and how many times they) are being applied at a particular point $s$ along the circuit. 

Eq.~\reef{unitaries} specifies a trajectory $U(\s)$ in the space of unitaries, or equivalently, in the space of states using $\ket{\psi(\s)}=U(\s)\ket{\psi_\mt{R}}$. Assuming $0\le\s\le1$, circuits satisfying eq.~\eqref{circuit} correspond to trajectories satisfying the boundary conditions:
\begin{equation}
  U(\s=0)= \mathbbm{I}\,, \qquad  U(\s=1)= U_\mt{TR}\,.
\label{boundary_condition}
\end{equation}
From this perspective, $Y^I(s)$ is the tangent vector to the trajectories and the instantaneous Hamiltonian can be reconstructed as
\begin{equation}
  {\cal H}(s)=\sum_I Y^I(s)\,\mathcal{O}_I = i \,\partial_s U(s)\,U^{-1}(s)\,.
\label{tangent}
\end{equation}
There are many trajectories or circuits  (\ie an infinite number)  satisfying eq.~\reef{boundary_condition}.

Nielsen's approach to identifying the optimal circuit is to minimize the cost defined as
\begin{equation}
  \mathcal{D}(U(\s))\equiv \int^1_0 ds ~ F \( U(s), Y^I(s)  \),  
\label{costD}
\end{equation}
where $F$ is a local cost function assumed to depend only on the position $U(s)$ and the tangent vector $Y^I(s)$.  While the precise form of the cost function $F$ is not fixed, there are a number of desirable features for  reasonable cost functions \cite{Nielsen:2006}: 1) Smoothness, 2) Positivity, 3) Triangle inequality and 4) Positive homogeneity -- see \cite{Jeff,Bueno:2019ajd} for more recent thorough discussions.\footnote{We note that while \cite{Jeff} suggests dropping the homogeneity property due to holographic considerations, \cite{Bueno:2019ajd} argues that any such measure may not provide a lower bound on quantum circuit complexity and could violate Lloyd's bound \cite{Lloyd_2000}.} Two simple examples of cost functions satisfying these constraints are 
	\begin{equation}\label{function_F}
	F_1(U,Y)=\sum_I \left|Y^I\right|~,\qquad\qquad
	F_2(U,Y)=\Big[\sum_I  \(Y^I\)^2\Big]^{1/2}~.
	\end{equation}
The circuit complexity is then the cost evaluated for the optimal trajectory,\footnote{When working with discrete gates, as in eq.~\reef{circuit}, the target state is prepared within some tolerance $\varepsilon$, \eg $\Vert \,\ket{\psi_\mt{T}}-U_\mt{TR}\ket{\psi_\mt{R}}\Vert^2 \le \varepsilon $. However, with the continuous unitaries \reef{unitaries}, one is always able to prepare the target state exactly with a finite cost, and so our discussion will involve no tolerance. \label{footy77}}
 \ie
\beq 
\mC(\ket{\psi_\mt{T}}) \equiv {\rm Min}\ \mathcal{D}\,. 
\label{compD}
\eeq
With this approach, the task of determining the optimal circuit has been mathematically mapped to the geometric problem of identifying globally minimizing geodesics in a geometry defined by the cost function on the space of unitaries. 

Given this geometrical formulation, it is natural to choose coordinates $x^a$ covering the space of unitary operators $U(x^a)$. Trajectories $x^a(s)$ in this space correspond to unitaries \reef{unitaries} evolving as
\begin{equation}
  U(x(\s)) = \cev{\mathcal{P}} \exp \[ -i \int^\s_0\!\!\! d s\, {\cal H}( x(s))\]  \quad {\rm with}\ \  {\cal H} = \sum_I Y^I(s)\,\mathcal{O}_I\equiv \sum_a \dot{x}^a(s)\,\mathcal{O}_a(x)\,,
\label{new-unitaries}
\end{equation} 
where $\dot{x}^a(s)=\partial_s x^a(s)$ and $\mathcal{O}_a(x)$ are the (position-dependent)  Hermitian operators generating the evolution in the $x^a$ direction, \ie
\begin{equation}
  i\, \frac{\partial U(x)}{\partial x^a} = {\mathcal{O}}_a(x)\ U(x)\,.
\label{PDevolution}
\end{equation}
Each $\mathcal{O}_a(x)$ corresponds to an independent linear combination of the $\cO_I$ appearing in eq.~\reef{unitaries}. The $x$-dependence indicates this linear combination varies from point to point in the space of unitaries.

Using these coordinates, the cost \reef{costD} becomes 
\begin{equation}
  \mathcal{D}= \int^1_0 ds ~ F\! \( x^a(s), \dot{x}^a(s)  \)\,, 
\label{costD2}
\end{equation}
where $F$ is only a function of the coordinates $x^a$ and the velocities $\dot{x}^a$. Given this form, extremizing the cost is analogous to solving for the trajectory of a particle in classical mechanics where $F$ is the Lagrangian (and $s$ the time). Hence the extremal trajectory satisfies the Euler-Lagrange equations
\begin{equation}\label{EOM_EL}
\frac{\partial F}{ \partial x^a}-\frac{\partial\ }{\partial s}\left(\frac{\partial  F}{\partial  \dot{x}^a}\right)=0\,,
\end{equation}
and the boundary conditions 
\beq
  x^a(s=0)=x^a_0\,,\qquad x^a(s=1)=x^a_1\,,
\label{bc0}
\eeq 
are chosen in accord with eq.~\reef{boundary_condition}, \ie
$U(x^a_0)= \mathbbm{ I }$ and $U(x^a_1)= U_\mt{TR}$.
The circuit complexity is then given by evaluating the 
cost on-shell, \ie substituting the extremal trajectory into eq.~\reef{costD2},\footnote{In general, there may be a family of extremal trajectories or unitaries producing the desired transformation \reef{circuit}. In this case, one must still minimize eq.~\reef{costD} over this family to determine the complexity, \eg see \cite{Jeff,Hackl:2018ptj,Chapman:2018hou}.}
\begin{equation}
  \mC\(\ket{\Psi_\mt{T}}\) \equiv \textrm{Min} \int^1_0 ds ~ F\! \( x^a (s), \dot{x}^a (s)  \)\,.
\label{eq:qc-comp}
\end{equation}

Before proceeding, let us comment on the group-theoretic structure that naturally appears in various settings for the evaluation of the complexity of QFT states. To make the latter a tractable problem, one typically chooses a restricted basis of operators $\mO_I$ to construct the unitaries \reef{unitaries}. However, it is natural that this restricted basis should form a closed  algebra, and typically, the $\mO_I$ provide a representation of a Lie algebra $\mathfrak{g}$, \ie $[\mO_I,\mO_J]=i f_{IJ}{}^K \mO_K$. For example, a $\mathrm{GL}(N,\mathbb{R})$ group appears in evaluating the complexity of the  ground state of a free scalar field \cite{Jeff}, and the latter was extended to a $\mathrm{Sp}(2N,\mathbb{R})$ group in examining the corresponding thermofield double state \cite{Chapman:2018hou} -- see also \cite{Hackl:2018ptj}.\footnote{The symmetry closed by the gate generators was used in \cite{Magan:2018nmu} to physically argue for some natural choice of cost functions. This approach was later related to Kirillov's geometric action \cite{kirillov2004lectures} in the context of 2d CFTs and the Virasoro group in \cite{Caputa:2018kdj}. See also \cite{Bueno:2019ajd} for a general discussion on geometric actions and circuit complexity.} In the following, we will find that the affine symplectic group, \ie $\mathbb{R}^{2N} \rtimes \mathrm{Sp}(2N,\mathbb{R})$ plays a central role in evaluating the complexity of the coherent states of interest. The utility of this group-theoretic perspective is that it relegates the physical details of the basis operators $\mO_I$ to the background. Instead, the generators in eq.~\reef{unitaries} are simply elements of the Lie algebra $\mathfrak{g}$, and we can choose the most convenient  representation for the calculations of interest.\footnote{Within this group theoretic framework, we might add that when the cost function does not explicitly depend on the position $U(s)$, as in eq.~\reef{function_F}, the measure becomes right invariant \cite{Brown:2017jil,Susskind:2018pmk,Brown:2019whu}.  This additional symmetry greatly simplifies solving for the corresponding geodesics, \eg see \cite{Jeff,cohere,Hackl:2018ptj}.}

\subsection{First law of circuit complexity} \label{sec:firstcm}

\begin{figure}[t]
		\centering
	\includegraphics[width=4.5in]{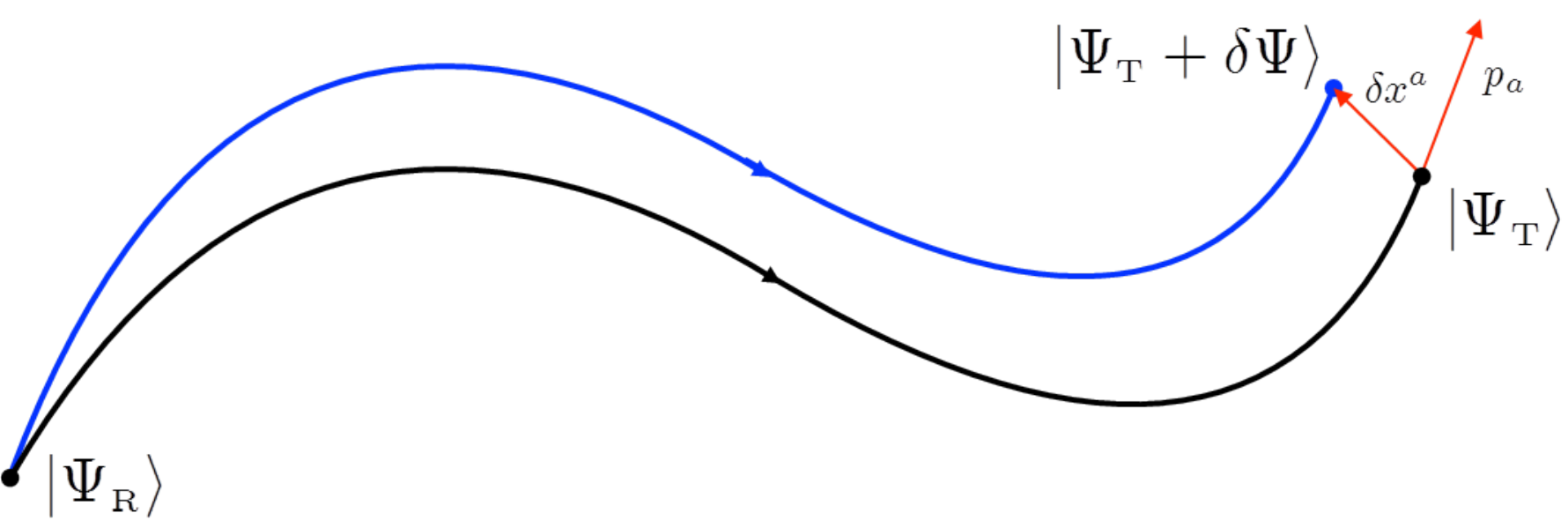}
	\caption{\label{variation} The variation of the Nielsen circuit due to a perturbation $\ket{\Psi_{\mt{T}} +\delta\Psi}$ of the target state $\ket{\Psi_{\mt{T}}}$. }
\end{figure}

Next, we examine the behaviour of the circuit complexity \eqref{eq:qc-comp} under small perturbations. 
Our main focus will be to study the variation in complexity for a fixed reference state $\ket{\Psi_{\mt{R}}}$, when the target state $\ket{\Psi_{\mt{T}}}$ is perturbed to $\ket{\Psi_{\mt{T}} +\delta\Psi}$, 
\begin{equation}
  \delta\mC = \mC\( \ket{\Psi_{\mt{T}} +\delta\Psi} \) - \mC\(\ket{\Psi_\mt{T}}\)\,.
\label{eq:var-comp}
\end{equation}
This variation is illustrated in figure \ref{variation},  as the variation of the corresponding geodesics in the space of states. 

Let us begin by assuming that we have a smooth family -- see  comments on this assumption below -- of geodesic solutions $x^a(s,z)$ satisfying the boundary conditions
\begin{equation}
  x^a(s=0,z)=x^a_0(z)\,,\qquad x^a(s=1,z)=x^a_1(z)\,,
\label{bc1}
\end{equation}
where $z$ parameterises the family. Now for a small variation $\delta z$ around $z=0$, we can write
\begin{equation}\label{assumption}
  x^a(s,z) = x^a(s) +  \delta x^a  \qquad {\rm where}\ \ 
  \delta x^a= v^a(s)\,\delta z\,,
\end{equation}
with $x^a(s)\equiv x^a(s,z=0)$ and $v^a(s)\equiv \partial_z x^a(s,z)|_{z=0}$. The change in the complexity \reef{eq:qc-comp} can then be expressed as
\begin{equation}\label{monkey}
\begin{split}
\delta \mathcal{C}&= \int^1_0 ds ~ \left[F\! \( x^a(s)+v^a(s) \delta z, \dot{x}^a(s) +\dot{v}^a(s) \delta z  \)- F\! \( x^a(s), \dot{x}^a(s)  \)\right] \\
&= \mC'\, \delta z   + \frac12  \mC''\, \delta z^2 +\cdots \,,
\end{split}
\end{equation}
where the first- and second-order coefficients are given by
\begin{equation}\label{monkey2}
\begin{split}
\mC' &=   \[ \frac{\partial F}{\partial \dot{x}^a}v^a \]^{s=1}_{s=0} + \int^1_0 ds ~  \[\frac{\partial F}{\partial x^a}-\frac{\partial\ }{\partial s}\(\frac{\partial  F}{\partial  \dot{x}^a} \)\]v^a \,,\\
\mC'' &=   \int^1_0 ds ~  \[\frac{\partial^2 F}{\partial x^a\partial x^b} v^av^b+2 \frac{\partial^2 F}{\partial x^a\partial  \dot{x}^b} v^{a} \dot{v}^{b} + \frac{\partial^2 F}{\partial \dot{x}^a\partial \dot{x}^b} \dot{v}^a\dot{v}^b\] \,.\\
\end{split} 
\end{equation}
Since $x^a(s)$ is a geodesic solution satisfying the Euler-Lagrange equations \eqref{EOM_EL}, the first order variation $\mC'$ reduces to the boundary term and to leading order, the variation of the complexity \reef{monkey} becomes
\begin{equation}\label{monkey3}
\delta \mC^{(1)}
= p_a\,\delta x^a\big|_{s=1} - p_a\,\delta x^a\big|_{s=0}\,,
\end{equation}
where following the analogy with classical mechanics, we introduced the notation
\begin{equation}
  p_a \equiv \frac{\partial F}{\partial \dot{x}^a} \,.
  \label{monkey4}
\end{equation}
From this classical mechanics perspective, eq.~\reef{monkey3} is a well known result for the variation of the action under perturbations of the boundary conditions. One of the 
interesting features of this result is that $\delta \mC^{(1)}$ only depends on data at the endpoints of the original extremal trajectory, \ie $\delta x^a$ and $p_a$ at $s=0$ and 1.
If we are interested in variations of the complexity where the reference state is kept fixed, as in eq.~\eqref{eq:var-comp}, eq.~\reef{monkey3} reduces to the single boundary term
\begin{equation}
  \delta \mC^{(1)} = p_a\, \delta x^a\big|_{s=1}\,.
\label{first1}
\end{equation}
We refer to eq.~\eqref{first1} as the { first law of complexity}. 

Since the right-hand side of eq.~\reef{first1} involves the inner product of two vectors, it may vanish even though the corresponding vectors are nonvanishing. That is, we may find the variation $\delta x^a$ is orthogonal to the direction that the original circuit is running, as specified by the `momentum' $p_a$. With $\delta\mC^{(1)} =0$, we must examine the second-order variation in eq.~\reef{monkey} to determine the change in the complexity. This will indeed be the case for the coherent state setup we study in this paper. 

Consider the second-order coefficient $\mC''$ in eq.~\reef{monkey2}. Integrating by parts to eliminate the $s$ derivative acting on one of the $v$'s in the last term, and performing a similar integration by parts for one contribution in the $v^a \dot v^b$ term, it reduces to 
\begin{equation}
\begin{split}
\mC'' = &\[    \frac{\partial^2 F}{\partial x^{a}\partial  \dot{x}^{b}} v^a v^b+  \frac{\partial^2 F}{\partial \dot{x}^{a}\partial  \dot{x}^{b}} v^a\dot{v}^b  \]^{s=1}_{s=0}  \\
&+ \int^1_0 ds ~  \[  \frac{\partial^2 F}{\partial x^a\partial x^b}v^b +  \frac{\partial^2 F}{\partial x^a\partial \dot{x}^b}\dot{v}^b - \frac{\partial }{\partial s} \( \frac{\partial^2 F}{\partial \dot x^a \partial  x^b} v^b + \frac{\partial^2 F}{\partial \dot{x}^a\partial \dot{x}^b} \dot{v}^b  \) \] v^a \,.\\
\end{split} 
\end{equation} 
Since these variations are between nearby geodesics, the variation $v^a$ must itself satisfy the perturbed Euler-Lagrange equations.  These require the squared brackets in the integral contribution to $\mC''$ to vanish.\footnote{This is equivalent to 
\begin{equation}\label{Jacobi_equations}
  \[\frac{\partial^2 F}{\partial x^a\partial x^b} -\frac{\partial }{\partial s}\( \frac{\partial^2 F}{\partial \dot{x}^{a}\partial  {x}^{b}} \) \] v^{b}-  \frac{\partial}{\partial s} \(\frac{\partial^2 F}{\partial \dot{x}^a\partial \dot{x}^b} \dot{v}^b\) + 2 \frac{\partial^2 F}{\partial x^{[a}\partial  \dot{x}^{b]}} \dot{v}^{b} = 0\, ,
\end{equation}
which corresponds to a 
generalization of Jacobi's equation $ \frac{d }{d x} \(p(x)\, \frac{d y}{d x} \) -q(x)\, y=0\,.
$}
Hence, the second order variation of complexity $\delta \mC^{(2)}$ with fixed reference state is again determined by a boundary term at $s=1$,
\begin{equation}\label{first2}
\begin{split}
\delta \mC^{(2)} &= \frac{1}{2}  \[    \frac{\partial^2 F}{\partial x^{a}\partial  \dot{x}^{b}}\, \delta x^a\, \delta x^b+  \frac{\partial^2 F}{\partial \dot{x}^{a}\partial  \dot{x}^{b}}\, \delta x^a\, \delta \dot{x}^b  \]  \bigg|_{s=1}\\
&= \frac{1}{2} \delta\!\( \frac{\partial F}{\partial  \dot{x}^a} \)\delta x^{a} \bigg|_{s=1}
= \frac{1}{2} \delta p_a\, \delta x^a\big|_{s=1} \,.
\end{split}
\end{equation}

Combining the first- and second-order variations  in eqs.~\reef{first1} and \reef{first2}, we summarise the first law of complexity as
\begin{equation}
 \delta\mC = p_a\, \delta x^a\big|_{s=1} + \frac{1}{2}\, \delta p_a \,\delta x^a\big|_{s=1} \,,
\label{eq:first-law}
\end{equation}
using the definition of $p_a$ in eq.~\reef{monkey4}. One of the most interesting features of this result is that the variation $\delta\mC$ is entirely determined by data at the final endpoint, \ie at $s=1$.


\noindent{\bf Smoothness of circuit space.} An important assumption at the outset of our derivation of eq.~\reef{eq:first-law} was that the optimal trajectories or circuits form a smooth continuous family as we vary the parameters characterizing the target state. In particular, we are assuming that the optimal circuit preparing the perturbed target state remains close to the original optimal circuit. This assumption typically fails in the original framework introduced in eq.~\reef{circuit} based on using discrete gates.\footnote{Such a complexity model also requires some finite tolerance but this feature is no longer necessary with Nielsen's approach, for the same reason described above -- see also footnote \ref{footy77}.} However, it becomes fairly milder within Nielsen's geometric approach to complexity because the control functions $Y^I(s)$ in eq.~\reef{unitaries} take real values and so effectively we are able to apply arbitrary fractional gates at any point along the circuit. This provides the key difference from the (standard) complexity model with discrete gates, and hence we can expect the optimal circuits themselves form a smooth geometry with Nielsen's approach. We illustrate this distinction with a simple example in appendix \ref{globall}. Let us further add that we certainly find smooth families of optimal circuits in the simple examples studied below. 

Implicitly, our assumption above also maintains that this smooth family of optimal circuits minimizes the cost {\it globally}. That is, solving eq.~\reef{EOM_EL} only provides a solution as the saddle point in the cost, but we assume the solutions $x^a(s,z)$ provide a family of global minima over all possible circuits. As reviewed in appendix~\ref{sec:conjugate}, the absence of conjugate points guarantees the stability of the geodesic, \ie to be {\it locally length minimizing}. In general, the space of states has an interesting topology and our assumption may fail, \ie  the global minimum may shift discontinuously even when considering circuits preparing nearby states, as was emphasized in \cite{Brown:2017jil,Balasubramanian:2019wgd}. However, we will still assume that the family of globally minimizing circuits is continuous in the amplitude of the perturbation.  While one can imagine simple examples where this is not the case (\eg geodesics between `nearly' conjugate points on a sphere -- see appendix \ref{sec:conjugate} for more discussion), our expectation is that this assumption is valid for the coherent states studied below. We note that this was already seen to be the case for similar complexity calculations for coherent states in \cite{cohere}.\footnote{Further, we will see in section \ref{sec:q-circuit} that the amplitude of the expectation values is controlled by the $\mathbb{R}^{2N}$ factor in the $\mathbb{R}^{2N} \rtimes \mathrm{Sp}(2N,\mathbb{R})$ algebra of generators used to prepare the states of interest. The fact that the topology of this factor is trivial would seem to support our assumption.} Of course, it would also be interesting to identify situations (in either QFT or holography) where our assumption does not hold.

%% file: sections/coherent.tex
To embed the quantum circuit complexity discussion in holography, one would require a proper understanding of the reference state $\ket{\Psi_{\mt{R}}}$, the gates $g_i$ and the path $U(\sigma)$ in the space of unitaries or states. However, our knowledge of any of these is very limited. On the other hand, in situations where the conjectures for holographic complexity in eqs.~\reef{defineCV} and \reef{defineCA} are applied, we do have a clear understanding of the target states $\ket{\Psi_{\mt{T}}}$. In particular, these correspond to quantum states in the boundary CFT which are dual to smooth configurations in the bulk gravitational theory in the large-N limit. 

The first law of complexity \eqref{eq:first-law} provides an interesting framework to examine holographic complexity. In particular, eq.~\eqref{eq:first-law} describes the variation of the complexity when the target state is perturbed and the result only depends on data at the endpoint of the quantum circuit. Hence in the holographic context where the target states are well understood, we should have good control of the variations of the target state, and the variations in the holographic complexity may provide insight into identifying the relevant local cost function or to clarify how the action of the gates builds up the spacetime.

To provide an explicit example of exploring holographic complexity using the first law of complexity, we consider Einstein gravity coupled to a negative cosmological constant and a massive free scalar field,
\beq
I_{\mt{bulk}} = \frac1{16\pi\GN} \int d^{d+1} y \sqrt{-g}\, \bigg[{\cal R}+ \frac{d(d-1)}{L^2}-
\frac1{2} \nabla^\mu \Phi \nabla_\mu \Phi -\frac12\,m_\Phi^2\Phi^2 \bigg]\,,
\label{Bact} 
\eeq
as a $(d+1)$-dimensional bulk theory. The latter is dual to a $d$-dimensional boundary CFT, with a scalar operator $\mathcal{O}$ with conformal dimension \cite{Witten:1998qj}
\begin{equation}
  \Delta=\sqrt{m_\Phi^2L^2 +\frac{d^2}{4}}+\frac{d}{2}\,.
  \label{cdimen}
\end{equation}

As our initial target state $\ket{\Psi_{\mt{T}}}$, we consider the AdS$_{d+1}$ vacuum, which in global coordinates, is described by the following metric
\begin{equation}
  ds^2_{\mt{AdS}}= \frac{L^2}{\cos^2 \rho}\( -dt^2+d\rho^2 +\sin^2 \!\rho \ d\Omega_{d-1}^2\)\,,
\label{eq:adsmetric}
\end{equation}
where $L$ denotes the radius of curvature. According to the AdS/CFT correspondence, this bulk configuration (\ie all bulk fields in their vacuum state in the background AdS geometry) is dual the CFT vacuum state, \ie $\ket{\Psi_{\mt{T}}} = \ket{0}$.

As the perturbed target state $\ket{\Psi_{\mt{T}} +\delta\Psi}$, we consider a coherent state where a classical expectation value (with small amplitude) is turned on for a scalar primary CFT operator $\mathcal{\hat O}$ and its descendants. According to the AdS/CFT correspondence, in the  large-N limit, the bulk Hilbert space of a \emph{free} bulk scalar field is equivalent to the CFT Hilbert space. Hence, there is an equivalent description of these excited states involving coherent states  built out of the quantum  scalar field operator $\hat{\Phi}$ in the bulk.\footnote{In the quantum error correction interpretation of the AdS/CFT correspondence, this equivalence is understood to hold in a subspace of the full Hilbert space, known as the \emph{code subspace}. For the excited states in this work, this is the subspace spanned by products of local bulk operators $\hat \Phi(y^\mu)$ acting on the vacuum \cite{Papadodimas:2013jku,Almheiri:2014lwa,Harlow:2018fse}.} Here, the latter then corresponds to turning a classical expectation value for the bulk scalar, and in the regime where the amplitude of the latter is small, we can evaluate the backreaction of the scalar on the spacetime geometry perturbatively. Having determined the backreacted geometry to leading order, we can  evaluate the variation of the holographic complexity for either complexity=volume \reef{defineCV} or complexity=action  \reef{defineCA}.

We would like to stress how the large-N limit of the AdS/CFT correspondence allows us to circumvent the technical difficulty of computing the complexity variation between states in the strongly coupled boundary CFT. Using the isomorphism between Hilbert spaces \cite{ElShowk:2011ag,Fitzpatrick:2011jn,kaplan2013lectures,Terashima:2017gmc,Berenstein:2019tcs} \ie  the vacuum state $|0\rangle$ and the Hilbert space spanned with a set of free field annihilation $\hat{a}_\n$ and creation $\hat{a}^{\!\dagger}_\n$ operators (see below), we can perform both calculations in the bulk, as we will describe in detail in future sections, providing a much more detailed account of our earlier results in \cite{Bernamonti:2019zyy}.

To fulfill the outlined strategy, we review the construction of bulk coherent state excitations in section~\ref{sec:bulk-coh} and their equivalent description, within the code subspace, in terms of generalized free fields in section~\ref{sec:boundary-coh}. We will turn to calculate the variations of the holographic complexity in section \ref{sec:hol-comp}. The actual quantum circuit complexity calculation of the analogous coherent states for a free scalar field propagating in a fixed AdS$_{d+1}$ geometry \reef{eq:adsmetric} is postponed till section \ref{sec:q-circuit}, where we will use  the tools developed for free QFTs and coherent states \cite{Jeff,cohere}.

\subsubsection{Bulk coherent states}
\label{sec:bulk-coh}

Consider a free real massive scalar field $\Phi(y^\mu)$ propagating in the $\text{AdS}_{d+1}$ geometry described by eq.~\eqref{eq:adsmetric}. The scalar part of the bulk action \reef{Bact} can be written as
\begin{equation}
\label{eq:class-action}
  I_{\mt{matter}}= -\frac{1}{32\pi\GN} \int_{\mt{AdS}} d^{d+1} y \sqrt{-g} \(g^{\mu\nu}\nabla_\mu \Phi \nabla_\nu \Phi + m_\Phi^2  \Phi^2\)\,.
\end{equation}
Notice the appearance of the (additional) prefactor $(16\pi\GN)^{-1}$ in the above action, a natural normalisation from the perspective of the gravitational action \reef{Bact}. The latter will simplify the backreaction calculations on the background spacetime and make the scalar field $\Phi$ dimensionless. The action \reef{eq:class-action} yields the Klein-Gordon (KG) classical field equation 
\begin{equation}
  \(-\square + m_\Phi^2\) \Phi= -\frac{1}{\sqrt{-g}} \partial_{\mu} \(\sqrt{-g}g^{\mu\nu}\partial_\nu\Phi \) +m_\Phi^2\Phi=0\,.
\label{eq:KGsol}
\end{equation}
Any classical solution $\Phi_{\mt{cl}}(y)$ of the KG equation \eqref{eq:KGsol} can be expanded
\begin{equation}
  \Phi_{\mt{cl}}(y) = \sum_\n \left(\alpha_\n\,u_\n(y) + \alpha_\n^\ast\,u^\ast_\n(y)\right)
\label{eq:classical}
\end{equation}
in terms of the set of eigenfunctions $u_\n(y^\mu)$ solving eq.~\eqref{eq:KGsol} \cite{Avis:1977yn,Burgess:1984ti,Cotabreveescu:1999em,Fitzpatrick:2011jn,kaplan2013lectures,Terashima:2017gmc,Berenstein:2019tcs}
\begin{equation}\label{eigenwaves}
u_\n(y^\mu) ={N}_{\vec n}\ \sin^\ell\! \rho\,\cos^\Delta\! \rho\ {}_2 F_1\!\[-j,\Delta +j+\ell; \frac{d}{2}+\ell;\sin^2\!\rho\] \,Y_{\ell\vec{m}}^{d-1} \!\(\theta^i \) \,e^{- i \omega_{n} t}\, .
\end{equation}
Here, $Y_{\ell m}^{d-1}$ are spherical harmonics in ($d$--1)-dimensions, and we collectively denote the quantum numbers $\vec n \equiv (j,\ell, \vec m)$. Hence, $\ell$ and $\vec m$ describe the angular mode, whereas $j$ describes a radial one. 
The corresponding spectrum of dimensionless frequencies $\omega_\n$ is given by
\begin{equation}\label{eigenenergy}
  \omega_\n = \Delta +2j+\ell \,,
\end{equation}
where $\Delta$ is the conformal dimension \reef{cdimen} of the dual CFT operator.
The  normalisation constants ${N}_{\n}$ are fixed by the  inner product on a constant time slice $\Sigma_t$ 
\begin{equation}\label{innerp}
\begin{split}
\langle u_\n, u_{\n'} \rangle &=\frac{-i}{16\pi \GN}\int_{\Sigma_t} d^{d} y\, \sqrt{-g}\,g^{tt}  \( u_\n^\ast \overleftrightarrow{\partial_t} u_{\n'}\) = \delta_{\n\n'}\,,\qquad 
\langle u_\n, u^\ast_{\n'} \rangle =0\,.\\
\end{split}
\end{equation}
where $u_\n^\ast \overleftrightarrow{\partial_t} u_{\n'}= u^\ast_\n\, {\partial_t} u_{\n'}
-{\partial_t}u^\ast_\n \, u_{\n'}$. This yields \cite{Cotabreveescu:1999em}\footnote{The overall sign is chosen here to simplify the discussion of the variation of the holographic complexity.}
\begin{equation}\label{normalization}
{N}_{\n}= \(-1\)^j \sqrt{\frac{16 \pi \GN}{L^{d-1}}}\sqrt{ \frac{\Gamma(j+\ell+\frac{d}{2})\Gamma(\Delta+j+\ell)}{j! \left[\Gamma(\ell+\frac{d}{2})\right]^2\Gamma(\Delta+j+1-\frac{d}{2})}}\,. 
\end{equation}

With this normalization, when canonically quantising the scalar field in AdS$_{d+1}$, the scalar field operator $\hat{\Phi}(y^\mu)$ is decomposed into creation $\hat{a}^\dagger_\n$ and annihilation $\hat{a}_\n$ operators
\begin{equation}
  \hat{\Phi}(y^\mu) = \sum_\n  \( \hat{a}_\n \,  u_\n(y^\mu)  +    \hat{a}^{\dagger}_\n\, u^*_\n(y^\mu) \) \,,
\label{scalq}  
\end{equation}
satisfying $[\hat{a}_\n,\hat{a}^\dagger_{\n'}]=\delta_{\n\n'}$. 
These operators generate a basis of states for the Hilbert space in the quantum theory 
\begin{equation}
  \prod_\n \( a_\n^\dagger\)^{r_\n} \ket{0} \,, \quad r_\n \in \mathbb{N}\,.
\label{Fock}
\end{equation}

Consider a coherent state excitation $|\alpha_\n \rangle$, with $\alpha_\n = |\alpha_\n |\,e^{i\theta_\n}$, within this Hilbert space. The latter can be defined as an eigenstate of the annihilation operator
\begin{equation}
  a_\n |\alpha_\n \rangle = \alpha_\n \,|\alpha_\n \rangle\,.
\end{equation}
Alternatively, these states can be constructed by acting with the displacement operator $e^{D(\alpha_\n)}$ on the vacuum, \ie
\begin{equation}
  |\alpha_\n \rangle = e^{D(\alpha_\n)}|0\rangle\quad \quad \text{with} \quad \quad D(\alpha_\n ) = \alpha_\n \,\hat{a}^\dagger_\n - \alpha^\ast_\n \,\hat{a}_\n\,.
\label{eq:Da}
\end{equation}
Since $D^\dagger(\alpha_\n) = -D(\alpha_\n) = D(-\alpha_\n)$, the displacement operator is unitary. Using the Baker-Campbell-Hausdorff formula, it follows
\begin{equation}
  |\alpha_\n \rangle = e^{-|\alpha_\n|^2/2}\,e^{\alpha_\n \hat{a}^\dagger_\n} |0\rangle\,.
\end{equation}

Returning to the quantum field \eqref{scalq}, notice the inner product \reef{innerp} allows to write the annihilation and creation operators as
\begin{equation}
\begin{aligned}
  \hat{a}_\n&= \langle u_\n, \hat\Phi \rangle =\frac{-i}{16\pi \GN}\int_{\Sigma_t} d^d y\, \sqrt{-g}\,g^{tt} \( u_\n^\ast \overleftrightarrow{\partial_t} \hat\Phi\)\,, \\
  \hat{a}_\n^\dagger &= -\langle  u^\ast_\n ,\hat\Phi\rangle =\frac{i}{16\pi \GN}\int_{\Sigma_t} d^d y\, \sqrt{-g}\,g^{tt} \( u_\n \overleftrightarrow{\partial_t} \hat\Phi\) \,.
\label{aadagger}
\end{aligned} 
\end{equation}
Plugging these into \eqref{eq:Da}
\begin{equation}
  D(\alpha_\n) =\frac{i}{16\pi \GN} \int_{\Sigma_t} d^d y \,\sqrt{-g}\,g^{tt} \(\alpha_\n u_\n+ \alpha_\n^\ast u_\n^\ast \) \overleftrightarrow{\partial_t} \hat{\Phi} (y^\mu) \,,
\label{displacement}
\end{equation}
it follows
\begin{equation}
  \left[D(\alpha_\n),\hat{\Phi}(y)\right] = -\left(\alpha_\n\,u_\n(y) + \alpha^\ast_\n\,u^\ast_\n(y)\right)\,.
\end{equation}
This allows one to show \cite{IZ}
\begin{equation}
  \bra{\alpha_\n} \hat{\Phi}(y^\mu) \ket{\alpha_\n}=\left(\alpha_\n\,u_\n+\alpha_\n^\ast\,u^\ast_\n \right)\,.
\end{equation}
Thus the coherent state $\ket{\alpha_\n}$ turns on the $\n$-th mode with classical amplitude $\alpha_\n$. For multi-mode coherent states involving a set $\{ \vec{q} \}$ of modes 
\begin{equation}\label{Def_coherent}
 \ket{\varepsilon \alpha_{\vec{q}}}=  e^{\varepsilon \sum D(\alpha_{\vec{q}})}  \ket{0} \ \quad {\rm with}\ \quad 
 D(\alpha_{\vec{q}})= {\alpha_{\vec{q}} a_{\vec{q}}{}^{\!\dagger}-\alpha^{\ast}_{\vec{q}}a_{\vec{q}}}\,,
\end{equation}
the overall amplitude equals the classical field \eqref{eq:classical}
\begin{equation}
\bra{\varepsilon \alpha_{\vec{q}}} \hat{\Phi}  \ket{\varepsilon \alpha_{\vec{q}}}=\varepsilon
\sum
\!\big(\alpha_{\vec{q}}\,u_{\vec{q}}+\alpha_{\vec{q}}^\ast\,u^\ast_{\vec{q}} \big)\equiv \veps\,\Phi_{\mt{cl}}\,,
\label{class}
\end{equation}
for that specific choice of modes. This is the main  property of coherent states we are interested in exploiting here. Further, note that we have introduced a small parameter $\veps$ (\ie $\veps\ll1$)
to control the overall amplitude of the expectation value \reef{class}. This will become our perturbative parameter in evaluating the gravitational backreaction of the bulk scalar.

\subsubsection{Boundary CFT coherent states}
\label{sec:boundary-coh}

In the large-N limit, there exists a generalised free field CFT operator that captures the same physics just described. Here, we review the construction of this generalised free field operator, following \cite{Fitzpatrick:2011jn}, in order to construct the dual coherent states in the CFT. 

The dual CFT is defined on the cylinder $\mathbb{R}\times \mathbb{S}^{d-1}$ with metric
\begin{equation}
  ds^2_{\mt{CFT}} = -dT^2 + R^2 d\Omega^2_{{d-1}}\,, 
\label{cylinder_metric2}
\end{equation}
where $T= Rt$ is a dimensionful boundary time. One can view this metric as induced on the AdS regulator surface located at
\begin{equation}
  \rho(\epsilon)= \frac{\pi}{2}- \frac{L}{R}\,\epsilon\,,
\end{equation}
in the limit $\epsilon\to 0$, after a proper scaling of the asymptotic AdS metric \eqref{eq:adsmetric}. 

Within this choice, the CFT operator $\hat\cO$ generates a spectrum of states with energies 
\begin{equation}\label{PPP}
  {\Omega}_\n = \frac{\omega_\n }{R} = \frac{\Delta +2j+\ell }{R} \,.
\end{equation}
Using the operator--state correspondence, these are excitations of the vacuum generated by $\hat\cO$ and its descendants 
\begin{equation}
   s^{\mu_1\mu_2\cdots \mu_\ell}_{\ell  m} P_{\mu_1}P_{\mu_2}\cdots P_{\mu_\ell}\ (P^2)^j\, \hat\cO\,,
\label{descend}
\end{equation}
where $P_{\mu}$ are the momentum generators, and $s^{\mu_1\mu_2\cdots \mu_l}_{lm} $ is a symmetric traceless tensor, \eg see \cite{Terashima:2017gmc}. 

The AdS/CFT prescription to construct the generalized free field operator from the bulk scalar field operator $\hat{\Phi}(y^\mu)$ in eq.~\reef{scalq} is \cite{kaplan2013lectures,Fitzpatrick:2011jn} 
\begin{equation}
\begin{aligned}
  \hat \mO (T,\theta^i)  &=  \gamma(d,\Delta)\, \lim_{\rho(\epsilon) \to \frac{\pi}{2}} \frac{ \hat{\Phi}(t,\rho(\epsilon),\theta^i)}{\cos^\Delta\!\rho(\epsilon)} \\
  &=\sum_\n \( \tilde u_\n(T,\theta^i)\ \hat{a}_\n +  \tilde u^{\ast}_\n(T,\theta^i)\ \hat{a}^{\dagger}_\n \) \,,
\end{aligned}
\label{newop}
\end{equation}
where the CFT eigenmodes are given by
\begin{equation}\label{eigenwavess}
\tilde u_\n(T,\theta^i)
=\widetilde{N}_{\n}\  Y_{\ell\vec{m}}^{d-1}\(\theta^i\) \,e^{- i \Omega_{n} T} 
\end{equation}
with normalisation constants $\widetilde{N}_{\n}$ determined by requiring the CFT two-point functions to take the standard form \cite{Fitzpatrick:2011jn} 
\begin{equation}
  \widetilde{N}_{\n}=\sqrt{\frac{2\pi^{d/2}\,\Gamma(\Delta+ j+\ell)\,\Gamma(\Delta +j + 1- \frac{d}{2})}{j!\,\Gamma(\Delta)\Gamma(j+\ell+\frac{d}{2})\,\Gamma(\Delta +1-\frac{d}{2})}}\ .
\label{eq:cftnorm}
\end{equation}
To derive this normalisation we already used the volume of a unit $(d-1)$-sphere equals $\text{Vol}\,\Omega_{d-1}=2\pi^{d/2}/\Gamma(d/2)$. The matching of the bulk normalisation \eqref{normalization} with the CFT normalisation \eqref{eq:cftnorm} requires 
\begin{equation}
  \gamma (d,\Delta) \equiv \sqrt{ \frac{\pi^{(d-2)/2}\, L^{d-1}}{8\,\GN}}\sqrt{\frac{ \Gamma(\Delta + 1 - \frac{d}{2})}{\Gamma(\Delta)}}\,.
\end{equation}

The creation and annihilation operators in \eqref{newop} can be extracted from the boundary operator $\hat\mO$ using 
\begin{equation}\label{arms}
  \hat{a}_\n= \llangle \tilde{u}_\n, \hat  \mO  \rrangle\,, \qquad \hat{a}_\n^\dagger= - \llangle \tilde{u}^\ast_\n, \hat \mO \rrangle \,,
\end{equation}
where we defined the boundary ``inner product'' satisfying 
\begin{equation}\label{inner}
\llangle \tilde{u}_\n, \tilde{u}_{\n'} \rrangle =\frac{i }{ 4\pi R\,  \Omega_n\, \tilde{N}_{\n}^2}\int^{2\pi R}_0\!\! dT \int d^{d-1}\Omega  \ \tilde{u}^\ast_\n\, \overleftrightarrow{\partial_T}  \tilde{u}_{\n'} = \delta_{\n\n'}\,,\qquad 
\llangle \tilde{u}_\n, \tilde{u}^\ast_{\n'} \rrangle=0\,.
\end{equation}
Note the inner product involves an integral over boundary time $T$ because the spatial part of the wavefunctions $\tilde u_\n$ is not sensitive to the (radial) quantum number $j$, \ie the bulk radial quantum number. Hence, to ensure proper orthogonality, one requires such time integration. A more traditional approach would associate the creation operators to the states created by the boundary operator and its descendants in the Euclidean theory,  \eg see \cite{Terashima:2017gmc}.  The present  construction (in particular eq.~\reef{newop}) makes clear that in both the bulk and boundary theories, we are working with the same Hilbert space \reef{Fock}. 

Once the bulk operator is reconstructed using the generalized free field \reef{newop}, the corresponding CFT coherent states \reef{Def_coherent} can be constructed using \eqref{arms} for the boundary theory $\hat{a}_\n$ and $\hat{a}_\n^\dagger$. It follows
\begin{equation}
  \bra{\varepsilon \alpha_{\vec{q}}} \hat \mO (T,\theta^i) \ket{\varepsilon \alpha_{\vec{q}}}=\varepsilon \sum_{\{\vec q\}}\!\big(\alpha_{\vec{q}}\,\tilde u_{\vec{q}}+\alpha_{\vec{q}}^\ast\,\tilde u^\ast_{\vec{q}} \big)
  \equiv \veps\,\mO_{ \mt{cl}}(T,\theta^i)\,,
\label{expect99}
\end{equation}
where
\begin{equation}\label{new99}
\mO_{\mt{cl}}(T,\theta^i)  = \gamma(d,\Delta) \lim_{\rho(\epsilon) \to \frac{\pi}{2}} \frac{ {\Phi}_\mt{cl}(t,\rho(\epsilon),\theta^i )}{\cos^\Delta\! \rho(\epsilon)}\,.
\end{equation}

As a final note, let us add that our description of coherent states is conventional from a QFT perspective. However, the usual discussions of coherent states in the context of the AdS/CFT correspondence focus on the Euclidean path integral preparation of these states by the introduction of sources in the boundary theory, \eg  \cite{BottaCantcheff:2015sav, Marolf:2017kvq,BottaCantcheff:2019apr}.  Ultimately, we are considering the same states as in those constructions.

%% file: sections/Bulk.tex
The main ideas in section~\ref{firstL} were to study the variation in complexity due to a change in the target state and to implement the latter in the AdS/CFT correspondence using coherent states. Here, we evaluate the variation in holographic complexity for both the CA and the CV proposals, in eqs.~\reef{defineCV} and \reef{defineCA}, respectively.

To be more precise, in the large-N limit, we consider Einstein gravity in ($d$+1)-dimensions with a negative cosmological constant coupled to a free massive real scalar field $\Phi$, as described by the bulk action \reef{Bact}. The dual boundary description is given by a $d$-dimensional CFT with a scalar primary operator $\mathcal{O}$, with the conformal dimension given by eq.~\reef{cdimen}. Taking the vacuum as the initial target state, \ie $ |\Psi_{\rm{T}}\rangle =|0\rangle$, the bulk description is
the global AdS metric $g_0$ in eq.~\eqref{eq:adsmetric} with a vanishing scalar field. The (divergent) holographic complexity of $\text{AdS}$ vacuum equals \cite{Chapman:2016hwi}
\begin{equation}
\ca(\S,g_0,0) =  \frac{I[g=g_0, \Phi=0]_\mt{WDW}}{\pi\, \hbar}\,, \quad \quad  \cv(\S,g_0) =\ \mathrel{\mathop {\rm
max}_{\scriptscriptstyle{\S=\partial \mathcal{B}}} {}\!\!}\left[\frac{\mathcal{V(B)}[g=g_0]}{G_N \, L}\right] \,.
\end{equation}
The notation stresses that both $\ca$ and $\cv$ are explicitly functionals of the metric, but $\ca$ also explicitly depends on the scalar field configuration. When turning on a small amplitude scalar field as a perturbation, its backreaction on the geometry induces a second-order perturbation
\begin{equation}
\Phi=\varepsilon\,\Phi_{\mt{cl}}\quad\longrightarrow\quad
  g = g_0 + \varepsilon^2\,\delta g\,.
\end{equation}
The perturbed configuration corresponds to the large-N description of the perturbed target state $|\Psi_{\mt{T}} + \delta\Psi\rangle$ whose holographic complexity equals
\beqa
\ca(\S,g_0 + \varepsilon^2\,\delta g,\varepsilon\,\Phi_{\mt{cl}}) &=&  \frac{I[g=g_0 + \varepsilon^2\,\delta g,\Phi=\varepsilon\,\Phi_{\mt{cl}}]_\mt{WDW}}{\pi\, \hbar}\,,
\nonumber\\ 
\cv(\S,g_0 + \varepsilon^2\,\delta g) &=&\ \mathrel{\mathop {\rm
max}_{\scriptscriptstyle{\S=\partial \mathcal{B}}} {}\!\!}\left[\frac{\mathcal{V(B)}[g=g_0 + \varepsilon^2\,\delta g]}{G_N \, L}\right] \,.
\eeqa
What the first law of complexity quantifies is the variation
\beqa
 \delta \ca(\S) &=& \ca(\S,g_0 + \varepsilon^2\,\delta g, \,\varepsilon\,\Phi_{\mt{cl}}) - \ca(\S,g_0,0)\,, 
\nonumber\\
\delta \cv(\S) &=& \cv(\S,g_0 + \varepsilon^2\,\delta g) - \cv(\S,g_0)\,,
\eeqa
keeping the boundary Cauchy surface fixed and without turning on boundary sources. These are the quantities we compute and discuss in this section.\\

In section~\ref{sec:AdSsetup}, we will introduce the details of the perturbative bulk setup we will consider. Section~\ref{c-action} is devoted to the evaluation and discussion of the variation of CA in this perturbative setup,  while section~\ref{s-volume} deals with CV. A comparison between these two results is performed in section~\ref{sec:comparison}.

\subsection{Bulk AdS setup}
\label{sec:AdSsetup}

The bulk action was given in eq.~\reef{Bact}, and using global coordinates, the AdS$_{d+1}$ vacuum solution, corresponding to $\Phi=0$, was given in \eqref{eq:adsmetric}
\be 
ds^2_{\mt{AdS}}   = \frac{L^2}{\cos^2 \rho}\(- dt^2 + d\rho^2 + \sin^2 \rho\,  d\Omega^2_{d-1} \)\,,
\label{eq:gads}
\ee
where $d\Omega^2_{d-1}$ stands for the metric of the unit ($d$--1)-sphere. Notice that all of the coordinates are dimensionless, measured in units of the AdS radius $L$. Further, $\rho\in [0,\pi/2)$, with $\rho=0$ corresponding to the centre of AdS$_{d+1}$ and $\rho\to\frac{\pi}{2}$ to its conformal boundary. 
More generally, we will denote the (dimensionless) bulk coordinates as $y^\mu=(t,\rho,\theta^i)$, as in eq.~\reef{Bact}.

We are interested in perturbing the vacuum by turning on the scalar field in a coherent state, as in eq.~\eqref{class}, while accounting for its backreaction on the spacetime geometry. For spherically symmetric perturbations $\Phi=\Phi(t,\rho)$, the most general compatible metric ansatz is \cite{Bizon:2011gg,Buchel:2012uh,Buchel:2013uba,Kim:2014ida}
\begin{equation}
  ds^2 = \frac{L^2}{\cos^2 \rho}\( - a(t,\rho) e^{-2b(t,\rho)} dt^2 + \frac{d\rho^2}{a(t,\rho)} + \sin^2 \rho\,  d\Omega^2_{d-1} \)\,.
\label{eq:ansatz}
\end{equation}  
The classical dynamics are governed by the KG scalar equation of motion \reef{eq:KGsol}
\begin{equation}
  \partial_t\left(e^b\,a^{-1}\partial_t \Phi\right) - \frac{1}{\tan^{d-1}\rho}\partial_\rho\left(a\,e^{-b}\tan^{d-1}\rho \partial_\rho\Phi\right) + \frac{m_\Phi^2L^2}{\cos^2\rho}\,e^{-b}\,\Phi = 0\,.
\label{eq:KG}
\end{equation}
and the three nontrivial components of Einstein's equations, which reduce to
\begin{equation}
\begin{split}
  \partial_\rho b &= - \frac{1}{2(d-1)}\sin\rho\cos\rho\left(\frac{e^{2b} }{a^2}\,(\partial_t\Phi)^2 + (\partial_\rho\Phi)^2\right)\,, \\
  \partial_\rho a &= a\,\partial_\rho b + \frac{d-2\cos^2\rho}{\sin\rho\cos\rho}\,(1-a) - \frac{m_\Phi^2L^2}{2(d-1)}\tan\rho\,\Phi^2\, ,\\
\partial_t a &= - \frac{1}{ (d-1)}  a\,  \sin\rho\cos\rho \,  \del_t \Phi \del_\rho \Phi\,.
\end{split}
\end{equation}
These correspond to linear combinations of the  $\rho\rho$, $t\rho$ and $tt$ components of Einstein's equations. Note that as a result of the Bianchi identity, only two of these three equations are independent.

The space of excitations is determined by, first, imposing regularity conditions at the origin $\rho \to 0 $
\begin{equation}
\begin{split}
\Phi(t,\rho)&= \phi(t) + \cO(\rho^2) \, , \\
a(t,\rho)&= 1 + \cO(\rho^2) \, ,\\
b(t,\rho)&= b_0(t) + \cO(\rho^2) \, , 
\end{split}
\label{eq:reg}
\end{equation}
which exclude the existence of horizons. Second, by imposing asymptotically boundary AdS conditions at $\pi/2 -\rho \equiv \eps_\rho \to 0$  
{\begin{equation}
\begin{split}
\Phi(t,\rho)&= \phi_{\Delta}(t)  \eps_\rho^\Delta \, , \\
a(t,\rho)&= 1 - \frac{{M}}{d-1} \eps_\rho^d \, ,  \\
b(t,\rho)&= 0+  \cO(\eps_\rho^{2\Delta}) \,.
\end{split}
\label{eq:asymp}
\end{equation}
Notice that absence of boundary sources was assumed and the AdS/CFT relation $m_\Phi^2L^2  = \Delta (\Delta -d)$ was used \cite{Witten:1998qj}. Further, we used the same residual gauge freedom, as in \cite{Buchel:2012uh}, to set the leading $\cO(1)$ term in $b(t,\rho)$ to zero.  These asymptotic conditions are valid for  $\Delta > d/2$, the range of conformal dimensions that we shall consider in this work.

\subsubsection{Perturbative solutions}

To describe the small amplitude perturbation considered in the quantum circuit discussion, set $\Phi(t,\rho) = \varepsilon\,\Phi_{\mt{cl}}(t,\rho)$, with $\veps$ being the parameter controlling the expansion, as in eq.~\eqref{class}. This induces a small amplitude expansion of the metric perturbations
\begin{equation}
\begin{split}\label{perturb}
a(t,\rho) &= 1+ \varepsilon^2\,a_2(t,\rho)  +\mO( \varepsilon^4) \,, \\
e^{b(t,\rho)}&=1+  \varepsilon^2\,b_2(t,\rho)  +\mO( \varepsilon^4)\,,\\
\end{split}
\end{equation}
which is compatible with the linearised Einstein's equations 
\begin{equation}\label{eq:osc-linear}
\begin{split}
\partial_\rho a_2& + \frac{d-2\cos^2 \rho}{\cos\rho\,\sin \rho}\, a_2 = \partial_\rho b_2 - \frac{ m_\Phi^2L^2}{2 (d-1)} \tan \rho\,   \Phi_{\mt{cl}}^2 \,,\\
\partial_\rho b_2 &=- \frac{1}{2(d-1)}\, \sin \rho\,\cos \rho\ \Big( (\partial_\rho\Phi_{\mt{cl}})^2 + (\partial_t\Phi_{\mt{cl}})^2 \Big)\,, \\
\partial_t a_2 &= - \frac{1}{d-1}\, \sin \rho\,\cos \rho\  \partial_\rho\Phi_{\mt{cl}}\, \partial_t \Phi_{\mt{cl}} \, .
\end{split}
\end{equation}

The scalar perturbation $\Phi_{\mt{cl}}(t,\rho)$ dynamics is controlled by the linearised KG equation, \ie the KG equation in global AdS \eqref{eq:gads} obtained by setting $a(t,\rho)=1$ and $b(t,\rho)=0$ in \eqref{eq:KG}
\be \label{eq:linearisedKG}
  \partial_t^2 \Phi  - \frac{1}{\tan^{d-1}\rho}\partial_\rho\left(\tan^{d-1}\!\rho\ \partial_\rho\Phi\right) + \frac{m_\Phi^2L^2}{\cos^2\rho}\,\Phi = 0\,.
\ee

Time translation invariance of global AdS together with reality of the bulk scalar field allows to describe these perturbations as 
\begin{equation}
  \Phi_{\mt{cl}} (t,\rho) = \sum_j 2|\alpha_j|\,\cos\left(\omega_j t - \theta_j\right)\,e_j(\rho)\,,
\label{eq:s-mode}
\end{equation}
where $\alpha_j = |\alpha_j|\,e^{i\theta_j}$ is the coherent state label and $e_j(\rho)$ are solutions to the Sturm-Liouville problem $\hat{L}[e_j(\rho)]=\omega_j^2 e_j(\rho)$ with operator $\hat{L}$ given by
\begin{equation}
  \hat{L}[e_j(\rho)] = -\frac{1}{\tan^{d-1} \rho}\frac{d}{d\rho}\left[\tan^{d-1}\rho \frac{d}{d\rho}e_j(\rho)\right] + \frac{\Delta(\Delta-d)}{\cos^2\rho}e_j(\rho)\,,
\label{eq:radialEOM}
\end{equation}
and  $\omega_j = 2j + \Delta$. The normalised eigenfunctions are given by 
\begin{equation} \label{eq:eigenfct}
e_j(\rho)  \equiv  A_j \cos^\Delta\rho\,~{}_2 F_1\[-j,\Delta +j, \frac{d}{2};\sin^2 \rho\]
\end{equation}
where
\be
A_j \equiv N_{(j,0,\vec 0)} = \(-1\)^j \sqrt{\frac{16 \pi \GN}{L^{d-1}}}\sqrt{ \frac{\Gamma(j+\frac{d}{2})\Gamma(j+\Delta)}{j! \[ \Gamma(\frac{d}{2})\]^2 \Gamma(j+\Delta+1-\frac{d}{2})}}\,.
\label{eq:Aj}
\ee
Due to the spherical symmetry of our perturbations, these correspond to the s-wave modes, \ie $\ell =\vec m=  0$, in the general discussion \eqref{eigenwaves}.

Taking into account the regularity conditions \eqref{eq:reg} at the origin and  the AdS boundary conditions \eqref{eq:asymp}, the first two equations in \eqref{eq:osc-linear} can be integrated for any $\Phi_{\mt{cl}}(t,\rho)$ yielding
\begin{eqnarray}
 a_2(t,\rho) &=&  - \frac{1}{2(d-1)} \frac{\cos^d \rho}{\sin^{d-2} \rho}\int^\rho_0 dy \tan^{d-1}y  \Big( (\partial_y\Phi_{\mt{cl}})^2 + (\partial_t\Phi_{\mt{cl}})^2    + \frac{m_\Phi^2L^2}{\cos^2 y} \Phi_{\mt{cl}}^2 \Big) \nonumber \\
  &=& -\frac{1}{ d-1}\frac{\cos^d \rho}{\sin^{d-2} \rho}\int^\rho_0 dy \tan^{d-1}y\,T^{\mt{bulk}}_{tt}\, \\
\label{eq:a2}
  b_2(t,\rho) &=&  \frac{1}{2(d-1)}\int^{\pi/2}_\rho dy \sin y\cos y\,  \Big( (\partial_y\Phi_{\mt{cl}})^2 + (\partial_t\Phi_{\mt{cl}})^2 \Big) \nonumber \\
  &= &  \frac{1}{2 (d-1)} \int^{\pi/2}_\rho dy\sin y\cos y\,\left(T^{\mt{bulk}}_{tt} + T^{\mt{bulk}}_{\rho\rho}\right)\, .
  \label{eq:b2} 
\end{eqnarray}
Notice the third equation in \eqref{eq:osc-linear} is satisfied whenever $\Phi_{\mt{cl}}(t,\rho)$ is on-shell. For later convenience,   we have also expressed $a_2$ and $b_2$ in terms of the bulk stress tensor determined by the scalar perturbation  $\Phi_{\mt{cl}}(t,\rho)$ and sourcing the metric perturbations at second order
\begin{equation}
  \veps^2\,T^{\mt{bulk}}_{\mu\nu}  = -\frac{32 \pi \GN}{\sqrt{|g|}}\frac{\delta I_{\mt{matter}} }{\delta g^{\mu\nu}}=  \veps^2\left[\del_\mu\Phi_{\mt{cl}} \del_\nu \Phi_{\mt{cl}}  -  \frac{1}{2} g_{\mu\nu} (\del \Phi_{\mt{cl}}^2  + m^2_\Phi \Phi_{\mt{cl}}^2)\right]\, .
\label{eq:bulk-stensor}
\end{equation}
The bulk energy density $T^{\mt{bulk}}_{tt}$ also sources the conserved gravitational mass of these linearised solutions. Looking at the asymptotic expansion in eq.~\eqref{eq:asymp}, the (dimensionless) mass parameter $M$ is given by
\begin{equation}
  M = \frac{\varepsilon^2}{2}\int_{0}^{\pi/2} \!\!\! d\rho  \tan^{d-1}\!\rho\,\left[(\del_t\Phi_{\mt{cl}})^2 +(\del_\rho \Phi_{\mt{cl}})^2 + \frac{m^2_\Phi L^2}{\cos^2 \rho}\Phi_{\mt{cl}}^2\right] = \veps^2 \int_{0}^{\pi/2} d\rho  \tan^{d-1}\!\rho\  T^{\mt{bulk}}_{tt}\,.
\label{eq:l-mass}
\end{equation}

\subsubsection{Wheeler-DeWitt  patch}
\label{sec:WDW}

The Wheeler-DeWitt patch is a region of spacetime defined as the domain of dependence of a bulk spatial slice anchored on a Cauchy surface at the boundary $\Sigma$, \ie typically, constant time slice. Since the complexity=action proposal \reef{defineCA} for holographic complexity involves evaluating the action functional on-shell over the WDW patch, the geometry of the latter is described here. This is done for global AdS $(\text{WDW})$ and for its second-order spherically symmetric perturbations $(\delta\text{WDW})$  given by
\begin{equation}
\begin{split} \label{eq:metricsplit}
ds^2 &=  \(  g_{0,\mu\nu}    + \delta g_{\mu\nu} \) \, dy^\mu dy^\nu \\
&= \frac{L^2}{\cos^2 \rho}\bigg[ - \left(1+\varepsilon^2 (a_2-2b_2)\right) dt^2 + \left(1-\varepsilon^2a_2\right) d\rho^2 + \sin^2\rho\, d\Omega_{d-1}^2\bigg] \, . 
\end{split}
\end{equation}

By definition, the WDW patch is bounded by a null hypersurface. Given the spherical symmetry of the geometry \eqref{eq:metricsplit}, the latter is generated by radial null geodesics emanating from the boundary surface $\Sigma$ and intersecting at the origin $\rho=0$ in a caustic. We shall distinguish between the null boundaries for global AdS $(\partial\text{WDW})$ and for the second-order perturbations $(\partial\delta\text{WDW})$.

Let us denote the boundary time picking the Cauchy surface $\Sigma$ by $t_\Sigma$. The past and future boundaries of the WDW patch originating at $\r = \pi/2$ at time $t_\Sigma$ can be described by (see figure~\ref{fig:WDW})
\begin{equation} 
  t_\pm(\rho) = t_{0\pm}(\rho) + \delta t_{\pm}(\rho)\,.
\end{equation}  
$t_{0\pm}(\rho)$ describes the undeformed past and future boundary of the WDW patch in global AdS, whereas $\d t_{\pm}(\rho)$ describes its deformation due to the perturbation \eqref{eq:metricsplit}.  Both functions are determined solving order by order the null condition
\be
- \left( 1+\varepsilon^2 (a_2-2b_2)\right) dt^2 + \left(1-\varepsilon^2a_2\right) d\rho^2 =0 \, .
\ee
This yields
\begin{equation}\label{eq:nullsurf}
\begin{split}
 t_{0\pm}(\rho) &= t_\Sigma \pm \(\frac{\pi}{2} - \rho\) \,, \\
\d t_{\pm}(\rho)&=\mp \,  \veps^2 \int^{\pi/2}_\rho  \Big(a_2(t_{0\pm}(y),y)-b_2(t_{0\pm}(y),y) \Big) dy \,.
\end{split}
\end{equation}
\begin{figure}[ht]
	\centering
	\includegraphics[width=0.35\textwidth]{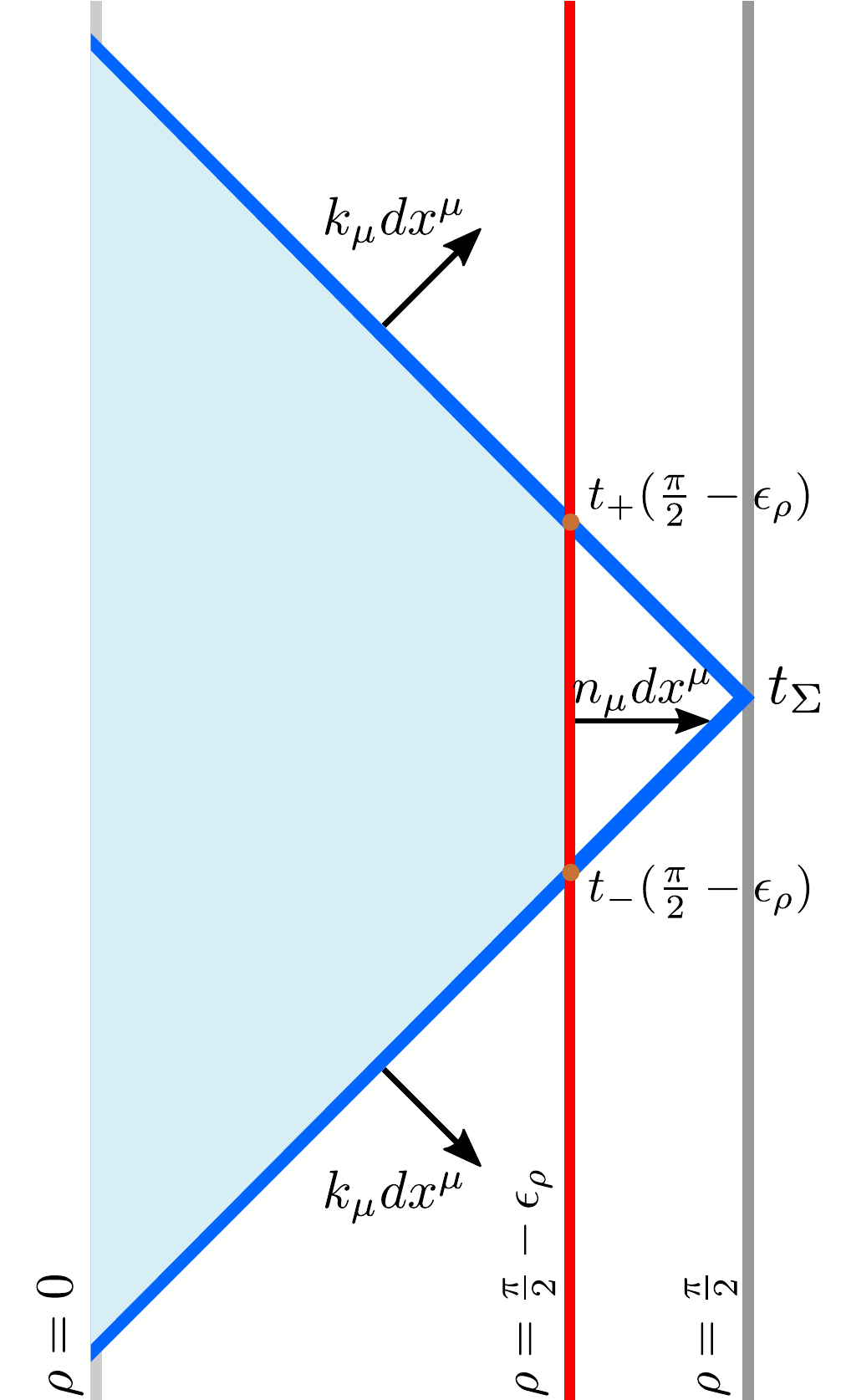}	
	\caption{Representation of the   WDW patch. The WDW patch is  bounded by the future and past null surfaces  $t_\pm(\rho)$ (thick blue lines) joining at $t_\Sigma$ on the AdS conformal boundary (grey line).  $k_{\mu} dx^\mu$ is the outward directed normal one-form to the null WDW boundary. The regulated asymptotic AdS boundary (red line) cuts the WDW patch at $\rho =\pi/2 - \eps_\rho$, and has outward directed normal $n_\mu dx^\mu$. The   $\rho =\pi/2 - \eps_\rho$ regulator surface and null hypersurfaces intersect at the null joint codimension-2 surfaces at   $t_\pm(\pi/2 -\eps_\rho)$. } \label{fig:WDW}
\end{figure}

In order to evaluate the divergent action functional on the WDW patch, one needs to introduce an infinitesimal cutoff $\eps_\rho$ at the AdS boundary $\rho=\pi/2-\eps_\rho$. As  depicted in  figure~\ref{fig:WDW}, this   procedure gives rise to a timelike boundary for the WDW patch, the portion of the AdS regulator surface where time runs from $t_-(\pi/2 -\eps_\rho)$ to $t_+(\pi/2 -\eps_\rho)$.\footnote{An alternative procedure would be to anchor the WDW patch directly to the AdS regulator surface. This was considered, \eg in \cite{Carmi:2016wjl}, where it was shown that for CA the two choices lead to the same structure of UV divergences.} This regulator surface and the null boundaries of the WDW patch intersect at the {\it null joints}, codimension-2 surfaces of constant $t=t_\pm(\pi/2 -\eps_\rho) $ and $\rho=\pi/2-\eps_\rho$ (see figure~\ref{fig:WDW}).

To sum up, the boundary of the WDW patch is made of the future and past null surfaces \eqref{eq:nullsurf} together with the portion described above of the AdS regulator surface at constant $\rho=\pi/2-\eps_\rho$ and the null joints where these meet. In what follows, we introduce some geometric quantities characterizing this boundary.

We define the outward-pointing normal one-form and the corresponding null normal vector to the null WDW boundaries to be
\begin{equation}
\begin{split}
  k_\mu\, dx^\mu  &\equiv (k_{0,\mu} +\delta k_\mu) dx^\mu =  L\(\pm dt  + d\rho - \varepsilon^2   \(a_2 - b_2 \) d\rho\) \, , \\
  k^\mu\, \del_\mu & \equiv (k_0^\mu +\delta k^\mu)\del_\mu =    \frac{\cos^2\rho}{L}  \[  \mp \del_t + \del_\rho   +   \varepsilon^2    \( \pm   \( a_2 - 2 b_2  \) \del_t  +  b_2  \del_\rho \) \] \, .
\end{split}
\label{dk}
\end{equation}
The upper (lower) sign corresponds to the future (past) boundary of the WDW patch. For later convenience, we distinguished between the global AdS null normal vector  $k_0^\mu$ and its $\mathcal{O}(\varepsilon^2)$  perturbation $\delta k^\mu$. 

We can define a null coordinate $s$ parameterizing the null translations along the WDW boundaries through $\del_s \equiv k^{\mu}\del_{\mu}$. Hence, the null hypersurface bounding the WDW patch can be conveniently parameterized by the ($d$--1)-dimensional unit sphere in  \eqref{eq:metricsplit} and the null coordinate $s$. The induced metric $\gamma$ on this null surface coincides with the angular part of the metric \eqref{eq:metricsplit} and has no perturbative corrections. Namely 
\be \label{dsdWDW}
ds^{2}_{\del \mt{WDW}} = L^2\, \tan^{2} \!\rho \ d\Omega_{d-1}^{2}
\ee
with its determinant being\footnote{Given the spherical symmetry of our setup and to avoid clutter we are not explicitly including the angular part of the metric in the determinant here and everywhere else in what follows. In other words, we are implicitly picking coordinates for the unit $S^{d-1}$ such that the metric determinant associated to $d\Omega_{d-1}^2$  equals 1. We will denote the corresponding  integration as $\int d\Omega_{d-1} = \VO=2\pi^{d/2}/\Gamma(d/2)$.}
\be
\gamma = L^{2(d-1)}\,\tan^{2(d-1)}\!\rho\, .
\ee
Notice the parameter $s$ is affine only at leading order in the perturbative expansion. This can be seen from explicitly evaluating
\be
k^{\mu} \nabla_{\mu} k_{\nu} = \kappa\,  k_{\nu}\, 
\ee
which shows that  $\kappa$ vanishes only at leading order    
\begin{equation}
  \kappa =  \delta\kappa = \pm  \varepsilon^2 \frac{\cos^2\rho}{L}\del_t (a_2-b_2)\, . 
\label{kappadef}
\end{equation}

Similarly, for the AdS regulator surface, the outward directed normal one-form and vector read
\begin{eqs}\label{dn}
n_{\mu}dx^{\mu} &\equiv \( n_{0,\mu} + \delta n_{\mu} \)dx^{\mu} = \frac{L}{\cos \rho}\(1- \frac{\veps^2 }{2}a_2 \)  d\rho   \,  \\
n^{\mu} \del_{\mu} &\equiv \( n_0^{~\mu} + \delta n^{\mu} \)\del_{\mu}  = \frac{\cos \rho}{L}\(1+ \frac{\veps^2 }{2}a_2 \)  \del_\rho  \,.
\end{eqs}
The induced metric on the AdS regulator surface equals $h_{\mu \nu}= g_{\mu\nu}  - n_\mu n_\nu$. With an analogous notation as for the other geometric quantities, we will distinguish between the AdS, $h_{0,\mu\nu}$, and the perturbed part, $\delta h_{\mu\nu}$, of the metric $h_{\mu\nu}$.

Finally, the codimension-2 null joint surfaces have induced metric $\sigma$. It reduces to the angular part of the metric  \eqref{eq:metricsplit}. Thus,  $\sigma$ coincides with $\gamma$ and has no perturbative corrections in $\varepsilon$.

%% file: sections/ComAction.tex
\subsection{Complexity=Action}
\label{c-action}

The complexity=action conjecture \cite{Brown:2015bva,Brown:2015lvg} suggests the complexity of a boundary state on the time slice $\S$ can be calculated holographically as the gravitational action evaluated on the Wheeler-DeWitt patch, \ie
\begin{equation}
\label{defineCAx}
\ca(\S) =  \frac{I_\mt{WDW}}{\pi}\,. 
\end{equation}

The evaluation of the holographic complexity \eqref{defineCAx} in the purely gravitational sector requires the addition of boundary contributions to the effective action to have a well defined variational principle due to the boundaries of the WDW patch \cite{Lehner:2016vdi}. Following the conventions adopted in \cite{Chapman:2018dem}, the  action including these gravitational boundary terms reads
\begin{equation}
\begin{aligned}
I &= I_{\mt{bulk}} + I_{\mt{GHY}} + I_{\mt{jt}} + I_{\kappa} + I_{\mt{ct}}  \\
&=  \frac1{16\pi\GN} \int d^{d+1} y \sqrt{|g|} \bigg[{\cal R}+ \frac{d(d-1)}{L^2} -
\frac{1}{2}  g^{\mu\nu}\nabla_\mu \Phi \nabla_\nu \Phi  -\frac{1}{2}m_{\Phi}^2\Phi^2\bigg] \\
&\,+  \frac{1}{8\pi \GN} \int_{\mt{regulator}}\!\!\!\!\!   d^{d}x\,\sqrt{|h|}\,K + \frac{1}{8\pi \GN} \int_{\rm joints}\!\!\!\!\!  d\Omega_{d-1} \,\sqrt{\sigma}\,a_{\mt{jt}} \\
&\,+ \frac{1}{8\pi \GN} \int_{\del{\rm WDW}}\!\!\!\!\!  ds\,d\Omega_{d-1}\,\sqrt{\gamma}\,\kappa  + \frac{1}{8\pi \GN} \int_{\del{\rm WDW}}\!\!\!\!\! ds \,d\Omega_{d-1}\,\sqrt{\gamma}\, \Theta \log (\ell_{\rm ct} \Theta) 
\,. 
\end{aligned}
\label{eq:faction}
\end{equation}
The bulk action \eqref{Bact}
\begin{equation}
  I_{\mt{bulk}} = I_{\mt{EH}} + I_{\mt{matter}}\,,
\end{equation}
splits into $I_{\mt{EH}}$, the Einstein-Hilbert action with a negative cosmological constant, and $I_{\mt{matter}}$, describing the coupling of the real massive scalar field to gravity, as isolated in eq.~\reef{eq:class-action}. These match the bulk physics reviewed in section \ref{sec:AdSsetup}. The remaining terms are surface terms evaluated on the different pieces of the boundary of the WDW patch:  $I_{\mt{GHY}}$ is the usual Gibbons-Hawking-York term \cite{PhysRevLett.28.1082, PhysRevD.15.2752} defined on the AdS boundary regulator surface, $I_{\kappa}$ and $I_{\mt{ct}}$ involve integration over the null boundaries of the WDW patch, whereas $I_{\mt{jt}}$ is the null joint term evaluated where the null boundaries of the WDW patch intersect the AdS boundary regulator surface \cite{Lehner:2016vdi}.  
Notice that,  as for  vacuum AdS  solutions \cite{Chapman:2016hwi}, there  is no additional contribution associated to the caustics at the tips of the WDW patch (see appendix \ref{app:caustic}).\\

Due to the presence of $I_{\mt{matter}}$, the first question to ask is whether the matter sector of the effective action also requires the addition of boundary contributions to preserve the well definiteness of the variational principle. To analyse this, compute the variation
\begin{equation}
\begin{split} 
\delta I_{\mt{matter}}   &=  \frac{1}{16 \pi\GN }  \int d^{d+1}y \sqrt{|g|}   \d \Phi \(    \Box \Phi - m_\Phi^2\,\Phi   \)  -  \frac{1}{16 \pi\GN }  \int d^{d}y \sqrt{|h|} \,\delta \Phi\, n^\mu \del_{\mu}\Phi \Big|_{\rho=\rho_\eps}  \\
&- \frac{1}{16 \pi\GN }  \int_{\partial\mt{WDW}} ds\,\sqrt{\gamma}\,\delta\Phi\,\partial_s\Phi  \,.
\end{split}
\label{eq:mvariation}
\end{equation}
The first term is the Klein-Gordon equation of motion and vanishes on-shell. The second and third terms correspond to boundary contributions at the AdS boundary regulator surface and the null boundary of the WDW patch, respectively. 

The second term is the standard one considered in AdS/CFT. In the range of conformal dimensions $\Delta >d/2$, the asymptotic expansion for the bulk scalar field (\eg \cite{Klebanov:1999tb} )
\be
\Phi  = \eps_{\r}^{d-\Delta} \phi_{d-\Delta} + \dots +  \eps_{\r}^{\Delta} \phi_{\Delta} +O( \eps_{\r}^{\Delta+2})
\ee
gives a boundary term contribution proportional to
\begin{equation}
\begin{split}
-  \int d^{d}x \sqrt{|h|} \delta \Phi n^\mu \del_{\mu}\Phi &=  (d -\Delta ) \eps_{\r}^{d-2\Delta} \phi_{d-\Delta} \delta  \phi_{d-\Delta}  +\dots \\
&+ \[\Delta \delta \phi_{d-\Delta}  \phi_{ \Delta}   + (d-\D)\delta \phi_{\Delta} \phi_{d-\Delta} \] + \Delta\phi_\Delta\delta\phi_\Delta\,\epsilon_\rho^{2\Delta+1-d}
\end{split} 
\end{equation}
where the omitted terms are intermediate powers and functionals of the mode $\phi_{d-\Delta}$ only. Imposing Dirichlet boundary conditions with vanishing leading mode, \ie $ \delta \phi_{d-\Delta} =\phi_{d-\Delta}=0$, this boundary term vanishes when removing the cutoff.\footnote{This analysis must be reconsidered in the range $\frac{d}2-1\le\Delta\le \frac{d}2$, where the alternate quantization scheme calls for additional boundary terms, \eg see \cite{Klebanov:1999tb,Casini:2016rwj}. \label{footy33}}

Regarding the third term, we proceed as in the gravitational sector \cite{Lehner:2016vdi}. Hence, we assume Dirichlet boundary conditions along the null boundary of the WDW patch so that $\delta \Phi = 0$ in this term, \ie we do not impose any additional boundary conditions for the bulk scalar field along the null boundary.\footnote{One may question the consistency of this boundary condition with the one considered on the AdS boundary regulator surface at the intersection of the latter with the null boundary. That is, one may ask if an additional joint term is required at the intersection of these two surfaces, but our calculations suggest that such a boundary term is not needed.}

The discussion above indicates the existence of a good variational principle for the bulk scalar field when $\Delta >d/2$ {\it without} the addition of any further boundary contributions. This extends the argument in \cite{Lehner:2016vdi} to the full effective action \eqref{eq:faction} in this range of conformal dimensions. 

This result allows us to compute the variation of the holographic complexity $\delta \ca(\S)$ using eq.~\eqref{defineCAx} to second order in the bulk scalar field amplitude $\veps$. To organise our discussion, we split $\delta \ca(\S)$ into the three types of contributions that in principle appear
\begin{equation}
\delta\ca(\Sigma) = \frac{1}{\pi}\left(
\delta I_{\mt{WDW}}+ \delta I_{\delta\mt{WDW}} +  \delta I_{\delta \mt{cutoff}}\right)\,.
\label{varC}
\end{equation} 
$ \delta I_{\mt{WDW}}$ is the variation due to the change in the background fields within the original WDW patch,  $\delta I_{\delta\text{WDW}}$ is the variation due to the change in the shape of the WDW patch and $\delta I_{\delta \mt{cutoff}}$ is the variation due to the change of the radial location of the AdS boundary regulator surface.

A detailed description of the contribution from each of the terms in \eqref{eq:faction} to $\delta I_{\mt{WDW}}$ and $\delta I_{\delta\mt{WDW}}$ appears in the next section. We also show that in the present case $\delta I_{\delta \mt{cutoff}}$ actually vanishes. Readers not interested in the details of their evaluation can skip to section \ref{sec:CA-result}, where the net result is summarized.

\subsubsection{Action variation evaluation}\label{sec:actionevaluation}

In this section, we start by showing that the variation of the location of the radial cutoff has no impact on the variation of the action. 
We then compute the contributions to $\delta I_{\mt{WDW}}$ and $\delta I_{\delta\mt{WDW}}$ originating from the different terms in  \eqref{eq:faction}.

\paragraph{Variation of the cutoff $ \delta I_{\delta\mt{cutoff}}$.} Before computing $\delta I_{\mt{WDW}}$ and $\delta I_{\delta\mt{WDW}}$, we show the contribution $\delta I_{\delta \mt{cutoff}}$ vanishes, to second order in the amplitude $\veps$, whenever the conformal dimension $\Delta > d/2$.

 The origin of $\delta I_{\delta \mt{cutoff}}$ is the usual procedure to fix the cutoff by going to the Fefferman-Graham coordinates \cite{fefferman1985elie,fefferman2007ambient} (see \cite{Emparan:1999pm,deHaro:2000vlm,Skenderis:2002wp} for standard holographic renormalisation applications). In appendix \ref{app:ads}, we show the global AdS $(\eps_\rho)$ and the perturbed solution $(\eps_{\mt{pert}})$ cutoffs differ by an order $\cO(\veps^2)$ term  
\be
\eps_{\mt {pert}} = \eps_\rho  \( 1+ \frac{1}{2} \varepsilon^2  a_2(t, \pi/2-\eps_\rho )\)\, .
 \label{eq:cutoffmatched}
\ee  
Since this difference is already second order, to compute $\delta I_{\delta \mt{cutoff}}$ reduces to evaluating \eqref{eq:faction} for global AdS integrating up to $\eps_{\mt{pert}}$ (see appendix \ref{app:ads} for details)
\begin{equation}
 I_{\mt{vac}} = \frac{\VO L^{d-1} }{8\pi \GN}\eps_{\mt {pert}}^{1-d} \( 2(d-1) - \frac{1}{d-1} + \log \frac{  \ell_{\mt{ct}}(d-1)}{L }  +\dots  \) \,
\end{equation}
where dots indicate subleading terms in the cutoff expansion. Using \eqref{eq:cutoffmatched}, this term results in an extra contribution to $\delta \ca(\S)$, which reads 
\begin{equation}\label{eq:deltaIcutoff}
\delta I_{\delta\mt{cutoff}} = \frac{\varepsilon^2 \VO L^{d-1} }{16\pi \GN}\eps_\rho^{1-d} a_2(t, \pi/2-\eps_\rho ) \( 2(d-1)^2 -1 + (d-1)\log \frac{  \ell_{\mt{ct}}(d-1)}{L }  +\dots  \) \, . 
 \end{equation}
However, given the asymptotic boundary conditions \eqref{eq:asymp}, it follows $a_2 \sim \eps^d_\rho$. Hence, $\delta I_{\delta\mt{cutoff}}$ vanishes linearly in the cutoff $\eps_\rho$. The corrections to $\delta I_{\mt{WDW}}$ and $\delta I_{\delta\mt{WDW}}$ due to \eqref{eq:cutoffmatched} are higher order in the $\veps$ perturbative expansion we are considering. Hence, in what follows, we will simply identify both cutoffs.\\

\paragraph{Gravitational bulk term.}
\label{c-bulk}
To evaluate the contributions to $\delta I_{\mt{WDW}}$ and $\delta I_{\delta\mt{WDW}}$  we start with the variation of the Einstein-Hilbert action coupled to a cosmological constant term
\begin{equation}
  I_{\mt{EH}} = \frac1{16\pi\GN} \int d^{d+1} y \sqrt{|g|} \bigg[{\cal R}+ \frac{d(d-1)}{L^2}\bigg]\,.
 \label{eq:gbulk}
\end{equation}
Following the general discussion, its second order variation splits into two contributions
\begin{equation} \label{eq:dgbulk}
  \delta  I_{\mt{EH}}  = \delta I_{\mt{EH,\,WDW}} + \delta I_{\mt{EH},\,\delta\mt{WDW}}\,.
\end{equation}
$\delta I_{\mt{EH,\,WDW}}$ comes from the second order variation of the action evaluated on the undeformed WDW patch. Since the variation of the action is computed around a solution to the equations of motion, this term reduces to a total derivative
\begin{equation}
  \delta I_{\mt{EH,\,WDW}} = \frac{1}{16\pi \GN}\int_{\mt{WDW}}   \!\!\!\!\!d^{d+1}y\,\sqrt{|g_0|}\, \nabla_\sigma\left(g_0^{\sigma\nu}\nabla^\mu\delta g_{\mu\nu} - \nabla^\sigma\delta g^\mu_\mu  \right)\,.
\end{equation}
Notice that all covariant derivatives are vacuum AdS derivatives. 
Using Stokes' theorem, $\delta I_{\mt{EH,\,WDW}}$ is localized on the boundary of the (regulated) WDW patch
\begin{equation}
\begin{split}
  \delta I_{\mt{EH,\,WDW}} &= \frac{1}{16\pi \GN}\int_{\del\mt{WDW}} \!\!\!\!ds\, d\Omega_{d-1}\,\sqrt{\gamma}\, k_{0,\sigma}\left(g_0^{\sigma\nu}\nabla^\mu\delta g_{\mu\nu} - \nabla^\sigma\delta g^\mu{}_\mu  \right)\,  \\
 &~+ \frac{1}{16\pi \GN}  \int_{t_{0-}(\rho)}^{t_{0+}(\rho) \!\!\!} dt \,  d\Omega_{d-1}  \sqrt{|h_0|} n_{0,\sigma}\left(g_0^{\sigma\nu}\nabla^\mu\delta g_{\mu\nu} - \nabla^\sigma\delta g^\mu{}_\mu  \right)\, \Bigg|_{\rho  =\frac{\pi}{2}- \eps_\rho} \\
 &\equiv \delta I_{\mt{EH,\,null}} + \delta I_{\mt{EH,\,reg }}\,
\end{split}
\end{equation}
This boundary term splits into two contributions (see figure~\ref{fig:WDW}):  the first is evaluated on the null hypersurface $\del$WDW up to the regulator surface. This has induced metric determinant $\gamma = L^{2(d-1)}\,\tan^{2(d-1)}\!\rho\,$ and normal one-form $k_{\sigma}$, as in \eqref{dk}. The second, is evaluated on the time-like regulator surface $\rho =\frac{\pi}{2} -  \epsilon_\rho$ with induced (unperturbed) metric determinant $ |h_0| = L^{2d}\,\tan^{2(d-1)}\!\rho /\cos^2\! \rho $ and normal $n_{\sigma}$ as in \eqref{dn}.\footnote{Apart from the restricted range of  integration, the latter is the same contribution that appears in  the variation of the gravitational action and gives rise to the GHY term when posing a well defined variational principle for the action with Dirichlet boundary conditions in AdS. That is, this term is completely cancelled by an opposite contribution coming from the variation of the GHY term.  An analogous cancellation would clearly occur in our case. However,  given that, as we will discuss, in our case this kind of contributions vanish linearly in the cutoff $\eps_\rho$  and because of the presence of additional terms, this type of cancellation will not be explicitly included in what follows.} 

Substituting the explicit expressions, using integration by parts in some of the terms and taking into account the metric perturbation  regularity conditions at the origin \eqref{eq:reg} and  fall-offs at the AdS boundary \eqref{eq:asymp}, yields for the null surface contribution
\begin{eqs} \label{eq:bulkvarwdw}
   \delta I_{\mt{EH,\,null}} =&  \frac{\veps^2  }{8 \pi \GN  L} \int_{\del\mt{WDW}} \!\!\!\!ds \, d\Omega_{d-1} \,  \,\sqrt{\gamma} \Big[ \mp   \cos^2 \rho \,   \del_t (a_2-b_2) - (d-1)  \cot \rho  \, b_2   
   \\
   & - \sin \rho \cos \rho   \(a_2 -b_2 \)  \Big]   -\frac{\veps^2  }{16 \pi \GN }\int_{\rm joints}\!\!d\Omega_{d-1} \, \sqrt{\gamma}  \( a_2- 2b_2\) 
\end{eqs}
where, as before, the upper (lower) sign refers to the upper (lower) part of the WDW patch boundary. 
The last term arises from integrating by parts, and it is evaluated at the location of the joints between the original WDW boundary and the regulator surface.  
 
Similarly the integral along the regulator surface gives
\begin{eqs}
  \delta I_{\mt{EH,\,reg }} & =     -\frac{\veps^2}{16\pi \GN L}    \int_{t_{0-}(\rho)}^{t_{0+}( \rho)}  \!\! dt \,   d\Omega_{d-1} \ \sqrt{|h_0|} \  \bigg[ \frac{d- \cos^2\rho}{\sin \rho} a_2   \\ 
  &~~~~~~~~~~~~~~~~~~~~~~~~  + \sin \rho(a_2-  2  b_2 ) + \cos\rho \, \del_{\rho} \(a_2 - 2 b_2\) \bigg]  \Bigg|_{\rho  =\frac{\pi}{2}- \eps_\rho}\, . 
\end{eqs} 

The second contribution to  $\delta I_{\mt{EH}}$ in eq.~\eqref{eq:dgbulk} arises from the background AdS action evaluated over the geometric variation of the WDW patch described by eq.~\eqref{eq:nullsurf}:
\begin{eqs}
 \delta I_{\mt{EH},\,\delta\mt{WDW}} &=  \frac{1}{16\pi \GN}\int_{\delta\mt{WDW}} \!\!\!\!\!
  d^{d+1}y\,\sqrt{|g_0|}\bigg[{\cal R}_0+ \frac{d(d-1)}{L^2}\bigg] \\
  &= - \frac{ d \,  }{8\pi \GN L^2} \int d\rho  \, d\Omega_{d-1}    \sqrt{- g_0} \,  \( \delta t_{+} (\rho) - \delta t_{-}(\rho) \) \,.
\end{eqs}
In writing the second line we made explicit use of the vacuum AdS$_{d+1}$  value of ${\cal R}_0 = -d(d+1)/ L^2$. Using the integral expression  \eqref{eq:nullsurf}  for  $\delta t_{\pm} (\rho)$ and rearranging the order of integration, this contribution can also be recast in the form of an integral over the boundary of the undeformed WDW
\begin{equation} \label{eq:bulkvardwdw}
\begin{split}
  \delta I_{\mt{EH},\,\delta\mt{WDW}}   =   \frac{\veps^2}{8\pi \GN L}   \int_{\del \mt{WDW}} \!\!\!\!\! ds\,d\Omega_{d-1}\,\sqrt{\gamma}\,  \sin \rho \cos \rho  \(a_2 - b_2 \) \, .
\end{split}
\end{equation}
Since this cancels one of the terms in eq.~\eqref{eq:bulkvarwdw}, the complete variation $\delta I_{\mt{EH}}$ equals
\begin{eqs}\label{eq:bulkvar}
   \delta I_{\mt{EH}} &=  \frac{\veps^2}{8 \pi \GN L } \int_{\del\mt{WDW}} \!\!\!\! ds\,d\Omega_{d-1}  \sqrt{\gamma} \, \Big[ \mp   \cos^2 \rho \,   \del_t (a_2-b_2) - (d-1)  \cot \rho  \, b_2  \Big]\,  \\
 &  -\frac{\veps^2}{16\pi \GN L}    \int_{t_{0-}(\rho)}^{t_{0+}( \rho)} dt \,   d\Omega_{d-1} \sqrt{|h_0|} \[ \frac{d- \cos^2\rho}{\sin \rho} a_2 + \sin \rho(a_2-  2  b_2) + \cos\rho \, \del_{\rho} \(a_2 - 2 b_2\) \]  \Bigg|_{\rho  =\frac{\pi}{2}- \eps_\rho}\\
& -\frac{\veps^2  }{16 \pi \GN  } \int_{\rm joints} d\Omega_{d-1}  \sqrt{\gamma}  \( a_2-2 b_2\)  \\
\end{eqs}
For conformal dimensions $\Delta > d/2$, both the second and third line contributions vanish when removing the cutoff $\eps_\rho$ due to the asymptotic boundary conditions  \eqref{eq:asymp}. More concretely, the vanishing of the unit sphere integral in the joint term follows from expanding the integrand for $\eps_\rho\to 0$. Since, $\sqrt\gamma \sim \eps_\rho^{1-d}$, $a_2 \sim \eps_\rho^d$ and $b_2 \sim \eps_\rho^{2\Delta}$, the conclusion follows for $2\Delta > d$. The AdS regulator surface term has a constant contribution when expanding near the AdS boundary, but the integration along the time direction between $t_{0+}(\pi/2 - \eps_\rho)$ and $t_{0-}(\pi/2 - \eps_\rho)$   (see eq. \eqref{eq:nullsurf}) yields an overall linear dependence in the cutoff for small  $\eps_\rho$. Thus, in the limit where the regulator surface is removed,  $\delta I_{\mt{EH}}$ in eq.~\eqref{eq:bulkvar} reduces to
\be \label{eq:bulkvarfin}
  \delta I_{\mt{EH}} =  \frac{\veps^2}{8 \pi \GN L } \int_{\del\mt{WDW}} \!\!\!\! ds\,d\Omega_{d-1}  \sqrt{\gamma} \, \Big[ \mp   \cos^2 \rho \,   \del_t (a_2-b_2) - (d-1)  \cot \rho  \, b_2  \Big]\, . \\
\ee

\paragraph{GHY term.}
\label{c-ghy}

The GHY term in eq.~\eqref{eq:faction}
\begin{equation}
   I_{\mt{GHY}} =  \frac{1}{8\pi \GN} \int_{\mt{regulator}}\!\!\!\!\!  \!\!\!\!  d^{d}x\,\sqrt{|h|}\,K\, 
\label{eq:GHY}
\end{equation}
involves the integral of the trace, $K =h^{\mu\nu}K_{\mu\nu}$, of the extrinsic curvature $K_{\mu\nu} = h_{~\mu}^\sigma h_{~\nu}^\rho\,\nabla_\sigma n_\rho$ of the  asymptotic regulator surface $\rho = \pi/2 -\eps_\rho$, where the WDW patch gets cut off \cite{Carmi:2016wjl}. Here $ n_\rho$ is the outward directed normal to the regulator surface -- see eq.~\eqref{dn}. 

Following the general discussion around eq.~\eqref{varC}, the second order variation $\delta I_{\mt{GHY}}$ involves two contributions
\begin{equation}
  \delta I_{\mt{GHY}} = \delta I_{\mt{GHY} ,\,\mt{WDW}}  + \delta I_{\mt{GHY} ,\,\delta\mt{WDW}}\,.
\end{equation}
$\delta I_{\mt{GHY} ,\,\mt{WDW}}$ comes from integrating the second order variation of $\sqrt{|h|}\,K$ along the segment of  the AdS regulator surface intersecting the original WDW patch (see figure~\ref{fig:WDW})
\begin{equation}
\begin{split}
\delta I_{\mt{GHY} ,\,\mt{WDW}}  &=  \frac{1}{8\pi \GN}  \int_{t_{0-}(\rho)}^{t_{0+}(\rho) \!\!\!} dt \,  d\Omega_{d-1}  \sqrt{|h_0|}    \Big[ \frac{1}{2}K_0 \, h_0^{\mu\nu}\delta h _{\mu\nu}  + \delta K\Big] \Bigg|_{\rho  =\frac{\pi}{2}- \eps_\rho}  \\
&= \frac{1}{16\pi \GN}  \int_{t_{0-}(\rho)}^{t_{0+}(\rho) \!\!\!} dt \,  d\Omega_{d-1}  \sqrt{|h_0|} \Big[K_0 \, h_0^{\mu\nu}\delta h _{\mu\nu} - K_0^{\mu\nu}\delta g_{\mu\nu}  \\
&\hspace{6cm}- n_0^{\nu}  \left(\nabla^\mu\delta g_{\mu\nu}   - \nabla_\nu\delta g^\mu{}_\mu  \right) \Big] \Bigg|_{\rho  =\frac{\pi}{2}- \eps_\rho}  \, . 
\end{split}
\end{equation}
Following the notation used so far,  $K_0$ indicates the AdS value of the extrinsic curvature and $\delta K$ its second order variation. In writing  the  second expression  we have used  (see \eg \cite{usefulformulas})
\begin{equation}
  \delta K = -\frac{1}{2} K_0^{\mu\nu}\delta g_{\mu\nu} - \frac{1}{2} n_{0,\sigma}  \left(g_0^{\sigma\nu}\nabla^\mu\delta g_{\mu\nu} - \nabla^\sigma\delta g^\mu{}_\mu  \right) + \mathcal{D}_{\mu} c^{\mu}  \, 
\end{equation}
and the fact that $c_{\mu} = - \frac{1}{2} h^{\,~~\sigma}_{0,\mu} n_0^{\nu} \delta g_{\sigma\nu}$ identically vanishes (here $\mathcal{D}_{\mu}$ is the covariant derivative on the regulator surface compatible with the induced metric).
 
$\delta I_{\mt{GHY} ,\,\delta\mt{WDW}}$ involves the background value $\sqrt{|h_0|}\,K_0$ evaluated over the intersection between the deformation of the WDW patch   and the regulator surface
\be
\delta I_{\mt{GHY} ,\,\delta\mt{WDW}}   = \frac{1}{8\pi \GN}   \int d\Omega_{d-1}      \sqrt{|h_0|}  K_0 \Big( \delta t_{+}(\rho)  -  \delta t_{-}(\rho) \Big)   \Bigg|_{\rho  =\frac{\pi}{2}- \eps_\rho}   \, .
\ee 
Explicit calculation gives rise to 
\begin{eqs} \label{eq:GHYWDW}
\delta I_{\mt{GHY} ,\,\mt{WDW}}  =     \frac{\veps^2}{8\pi \GN L}    \int_{t_{0-}(\rho)}^{t_{0+}( \rho)} dt \,   d\Omega_{d-1} \sqrt{|h_0|} \( \frac{d-  \cos^2\rho}{\sin \rho} (a_2 - b_2) + \frac{\cos\rho}{2} \del_{\rho} \(a_2 - 2b_2\) \)  \Bigg|_{\rho  =\frac{\pi}{2}- \eps_\rho} 
\end{eqs}
and 
\begin{eqs}\label{eq:GHYdWDW}
\delta I_{\mt{GHY} ,\,\delta\mt{WDW}}   = -  \frac{\veps^2}{8\pi \GN L }   \int d\Omega_{d-1}       \sqrt{|h_0|}  \, \frac{d - \cos^2\rho}{\sin \rho}   \int^{\pi/2}_{\rho} dr \Big(a_2 - b_2 \Big)_{t=t_{\pm}(r)} \Bigg|_{\rho  =\frac{\pi}{2}- \eps_\rho}  \,.
 \end{eqs}
 
Both integrals vanish linearly in $\eps_\rho$ when using the asymptotic boundary conditions \eqref{eq:asymp} in the range $\Delta > d/2$. More precisely, the integrand in \eqref{eq:GHYWDW} has a finite term but time integration gives rise to $t_{0+}(\rho)-t_{0-}(\rho)$ which is linear in $\eps_\rho$ according to \eqref{eq:nullsurf}. Regarding \eqref{eq:GHYdWDW}, the radial $(r)$ integral scales as $\eps_\rho^{d+1}$, whereas $\sqrt{|h_0|}  \, \frac{d - \cos^2\rho}{\sin \rho} \sim \eps_\rho^{-d}$, giving an overall linear scaling. Hence $\delta I_{\mt{GHY}}$ does not contribute to the variation of the full action.

\paragraph{Joint terms.}
\label{c-joint}

\begin{figure}[htbp]
	\centering 	\subfigure{\includegraphics[width=0.3 \textwidth]{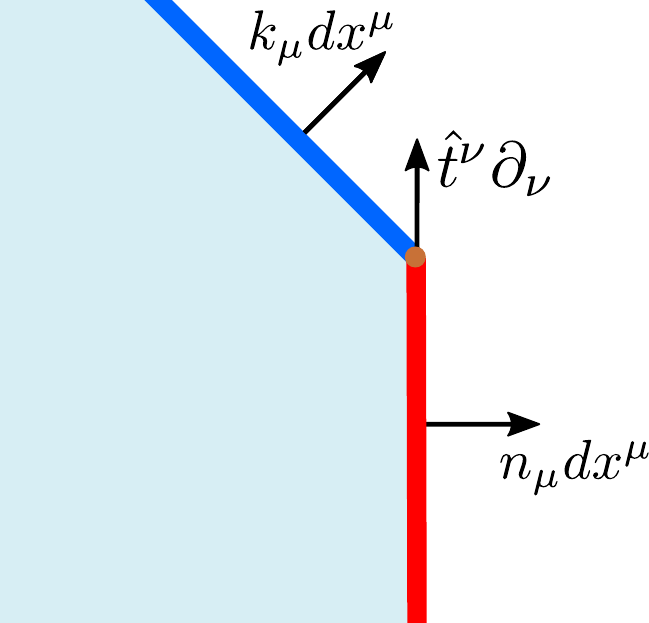}}
	\caption{ The joint between the null surface and the time-like regulator surface. $k_\mu dx^\mu$  and $n_\mu dx^\mu$  are the outward directed normal one forms. 
	$\hat t^{\mu} \del_{\mu}$ is a unit vector in the tangent space to the boundary time-like surface and  outward directed with respect to the boundary of this surface.} \label{fig:jointsign}
\end{figure}

The boundary term in \eqref{eq:faction} evaluated at the joint between the null WDW patch boundary and the timelike regulator surface equals
\be
  I_{\mt{jt}} =  \frac{1}{8\pi \GN} \int_{\rm joints}\!\!\!\!\! d\Omega_{d-1} \,\sqrt{\sigma}\,a_{\mt{jt}} \, .
\label{eq:jt}
\ee 
$\sqrt{\sigma}$ stands for the induced measure at the joint, which in the present case coincides with $\sqrt{\gamma}$.  The quantity $a_{\mt{jt}}$ is defined in terms of the outward directed normal to the WDW boundary, $k_\mu dx^\mu$ in \eqref{dk} and the outward directed normal to the regulator surface, $n_\mu dx^\mu$ in \eqref{dn}, as
\begin{equation}
a_{\mt{jt}} = \varsigma \log | k_{\mu} n^{\mu}|  
\, .
\label{eq:ajt}
\end{equation}
$\varsigma$ is a sign defined in \cite{Lehner:2016vdi} (see also \cite{Carmi:2016wjl}) in terms of the outward directed normal one-forms and of the auxiliary vector $\hat t^{\mu} \del_{\mu}$ tangent to the time-like surface and outward directed from its boundary (see figure~\ref{fig:jointsign}):
\be
\varsigma \equiv - ~\text{sign}\(  k_{\mu} n^{\mu}\)  \text{sign} \(k_{\nu}\hat t^{\nu}\) \, .
\ee

As mentioned earlier, given the spherical symmetry of our ansatz, the metric $\gamma$ is unchanged by the perturbation. Hence, the entire variation $\delta I_{\mt{jt}}$ comes from the variation of \eqref{eq:ajt}.\footnote{The variation due to the explicit change in the shape of the WDW boundary, \ie the shift of the joint location along the time direction following from \eqref{eq:nullsurf}, is irrelevant here because  the background value of $a_{\mt{jt}}$ is time translational invariant.} Using \eqref{dk} and \eqref{dn}, this equals
\be
\delta a_{\mt{jt}}  =  - \frac{k_{0,\mu}\delta n^{\mu}  + \delta k_{\mu} n_0^{\mu} }{ k_{0,\mu}n_0^{\mu} }   =  \frac{ \varepsilon^2}{2} \( a_2-2b_2\) \, 
\ee
and $\varsigma=-1$ for both the past and future WDW-regulator joint.

Integrating this at the location of the joint formed by the original WDW patch with the regulator surface, one obtains the joint term variation
 \be \label{eq:deltaIjt}
\d I_{\mt{jt}} =   \frac{\veps^2}{16\pi \GN  }  \int_{\rm joints} d\Omega_{d-1}   \sqrt{\gamma} \( a_2 - 2 b_2 \) \, . 
\ee
Comparing with \eqref{eq:bulkvar}  we see that this exactly cancels  with the joint term arising in $\delta I _{ \mt{bulk} }$. Nonetheless,  this term is also vanishing by itself when the regulator surface is removed.

\paragraph{$\kappa$ term.}
\label{c-kappa}

This term involves the integral of the parameter $\kappa$ quantifying by how much $s$ fails to be an affine parameter along the null boundary of the WDW patch 
\begin{equation}
  I_{\kappa} = \frac{1}{8\pi \GN} \int_{\del{\rm WDW}}\!\!\!\!\!  ds\,d\Omega_{d-1}\,\sqrt{\gamma}\,\kappa\,.
\label{eq:k}
\end{equation}
As computed in  \eqref{kappadef}, in our choice of parametrization, $\kappa = \delta \kappa$ vanishes at order $\veps^0$ but is non-vanishing at second order in $\veps$. Hence, the variation of $I_{\kappa}$ equals the integral of $\kappa$ over the boundary of the original WDW patch
 \begin{eqs}
 \delta I_{\kappa} &= \frac{1}{8\pi \GN} \int_{\del{\rm WDW}}\!\!\!\!\!  ds\,d\Omega_{d-1}\,\sqrt{\gamma}\,\delta \kappa  \\
 &= \pm \frac{\veps^2}{8\pi \GN L} \int_{\del{\rm WDW}}\!\!\!\!\!  ds\,d\Omega_{d-1}\,\sqrt{\gamma}\, \cos^2\rho \,  \del_t \,(a_2-b_2) \,.
\label{eq:deltaIk}
\end{eqs}
As always the $\pm$ sign is associated to the contribution integrated along the the future and past part of the null boundary of the WDW patch,  respectively.

\paragraph{Counterterm.}
\label{c-counterterm}

The remaining term in  \eqref{eq:faction} 
\begin{equation}
 I_{\mt{ct}} = \frac{1}{8\pi \GN} \int_{{\del\mt{WDW}}}\!\!\!\!\! ds \,d\Omega_{d-1}\,\sqrt{\gamma}\, \Theta \log (\ell_{\rm ct} \Theta)\,.
\label{eq:ct0}
\end{equation}
was introduced in \cite{Lehner:2016vdi} to ensure that the action is invariant under reparametrisations of the null boundary. It depends on an arbitrary scale $\ell_{\mt ct}$ and the expansion scalar of null generators $\Theta =  \partial_s \log \sqrt{\gamma}$. 

The variation of this term is slightly more subtle than the previous ones. 
Indicating  with $\Theta_0+\delta\Theta$ the background value and the variation  of the expansion scalar,  we shall consider the variation
\beqa
  \delta I_{\mt{ct}} &=&   \frac{1}{8\pi \GN}\bigg[ \int_{\partial\delta\mt{WDW}}\!\!\!\! ds\,d\Omega_{d-1}\,\sqrt{\gamma}\, (\Theta_0+\delta\Theta)\log \ell_{\mt{ct}}(\Theta_0+\delta\Theta) 
  \nonumber\\
  &&\qquad\qquad\qquad- \int_{\partial\mt{WDW}}\!\!\!\! ds\,d\Omega_{d-1}\,\sqrt{\gamma}\,\Theta_0\log \ell_{\mt{ct}}\Theta_0\,\bigg]\,.
\label{dCT}
\eeqa
to order $\veps^2$. $\Theta_0$  is the background value of the  expansion and spherical symmetry guarantees the perturbed expansion $\delta\Theta$ is only due to the change in the tangent vectors \reef{dk} along the null boundaries 
\begin{equation}
  \delta \Theta = \delta k^\mu\,\partial_\mu \log \sqrt{\gamma}\,.
\end{equation}

To order $\mathcal{O}(\veps^2)$, all the terms involving $\delta \Theta$ are integrated over the WDW setting to zero its deformation, \ie these give the part of the variation  $\delta I_{\mt{ct}}$ integrated over the original WDW
\be
  \delta I_{\mt{ct} , \mt{WDW} } =  \frac{1}{8\pi \GN}  \int_{\partial \mt{WDW}}\!\!\!\! ds\,d\Omega_{d-1}\,\sqrt{\gamma}\, \( \delta\Theta \log \ell_{\mt{ct}}\Theta_0 + \delta \Theta\)\, .
\ee
The part of the variation arising from integrating over the deformed WDW boundary is instead given by
\beqa
   \delta I_{\mt{ct} , \delta  \mt{WDW} } &=& \frac{1}{8\pi \GN} \bigg[ \int_{\partial\delta\mt{WDW}}\!\!\!\!\!\! ds\,d\Omega_{d-1}\,\sqrt{\gamma} \Theta_0  \log\ell_{\mt{ct}}\Theta_0 -   \int_{\partial\mt{WDW}}\!\!\!\!\!\! ds\,d\Omega_{d-1}\,\sqrt{\gamma} \Theta_0  \log\ell_{\mt{ct}}\Theta_0\,\bigg]  
   \nonumber\\
   &=& - \frac{1}{8\pi \GN}  \int_{\partial\mt{WDW}}\!\!\!\! ds\,d\Omega_{d-1}\sqrt{\gamma} \left(\delta k^\mu\partial_\mu\log\sqrt{\gamma}\right)\log \ell_{\mt{ct}}\Theta_0\, .  \label{eq:Ictdwdw}
\eeqa
 
A  direct way of understanding  how the second line arises  is to translate the integrals into radial integrals by making use of the relation between the two parametrizations encoded in tangent vector expression in eq.~$\eqref{dk}$. That is, by noticing that $d\rho = k_0^{\rho} \, ds$ for  the  original WDW patch, and   $d\rho = \(k_0^{\rho} +\delta k_0^{\rho} \) ds$ for the deformed WDW patch. The second line of eq.~\eqref{eq:Ictdwdw} is obtained when using the explicit expression for $\Theta_0$.

Therefore, the complete variation reduces to 
\begin{eqs} \label{eq:ct}
   \delta I_{\mt{ct}}& =   \delta I_{\mt{ct},\mt{ WDW} } + \delta I_{\mt{ct} , \,  \delta  \mt{WDW} } \\
   &= \frac{1}{8\pi \GN}  \int_{\partial\mt{WDW}} \!\!\!\! ds\,d\Omega_{d-1}\,\sqrt{\gamma}\,\delta k^\mu\partial_\mu \log \sqrt{\gamma} \\
    &  = \frac{\veps^2}{8\pi \GN L}  \int_{\partial\mt{WDW}} \!\!\!\! ds\,d\Omega_{d-1}\,\sqrt{\gamma}\, (d-1)  \cot\! \rho\ b_2\,,
\end{eqs}
with all the dependence on the arbitrary scale $\ell_{\mt{ct}}$ dropping out of this final expression.

\paragraph{Matter term.}
\label{c-matter}
The remaining contribution to evaluate is the variation in the matter part of the bulk action, given in eq.~\reef{eq:class-action}.
Since the perturbation $\Phi(y)=\varepsilon\Phi_{\mt{cl}}(t,\rho)$ is on top of the vacuum solution $\Phi(y)=0$, the variation of the matter action equals the on-shell matter action of the perturbation. Using the equations of motion, $\delta I_{\mt{matter}}$ will always yield a total derivative
\begin{equation}
  \delta I_{\mt{matter}}  = -\frac{\varepsilon^2}{32\pi\GN} \int_{\mt{WDW}} d^{d+1}y\,\sqrt{|g_0 |}\,\nabla_\mu\left(g_0^{\mu\nu} \Phi_{\mt{cl}}\nabla_\nu \Phi_{\mt{cl}}\right) \,.
\end{equation}
Using Stokes' theorem, as for the EH term,  this variation splits into two boundary contributions
\begin{equation}
\begin{split}
  \delta I_{\mt{matter}} =& -\frac{\varepsilon^2 }{32\pi\GN} \int_{\partial\mt{WDW}}\!\!\!ds\, d\Omega_{d-1}\, \sqrt{\gamma}\,k_{0,\sigma} \, g_0^{\sigma\nu} \Phi_{\mt{cl}}\partial_\nu \Phi_{\mt{cl}}   \\
  & -\frac{\varepsilon^2 }{32\pi\GN}  \int^{t_{0+}(\rho)}_{t_{0-}(\rho)}dt\, d\Omega_{d-1} \, \sqrt{|h_0|}\,n_{0,\sigma}  \, g_0^{\sigma\nu} \Phi_{\mt{cl}}\partial_\nu \Phi_{\mt{cl}}  \Big|_{ \rho = \frac{\pi}{2}-\eps_\rho}\,, 
\label{eq:tbulk}
\end{split}  
\end{equation}
one along the null WDW boundary and a second one along the regulator surface near the AdS boundary. Given the asymptotic fall-off of the scalar field $\Phi_{\mt{cl}} \sim \eps_\rho^\Delta$, with $\Delta > d/2$, the term localized along the regulator surface  vanishes when $\eps_\rho \to 0 $. In fact, rewriting the relevant part of the corresponding integral in a more explicit fashion, we find 
\be
 \int^{t_{0+}(\rho)}_{t_{0-}(\rho)}dt\, \, \sqrt{|h_0|}\,n_{0,\sigma} g_0^{\sigma\nu} \Phi_{\mt{cl}}\partial_\nu \Phi_{\mt{cl}}  \Big|_{ \rho = \frac{\pi}{2}-\eps_\rho} = \frac{L^{d-1} }{2} \int^{t_{0+}(\rho)}_{t_{0-}(\rho)}dt \tan^{d-1}\! \rho \,\del_\rho \Phi_{\mt{cl}}^2  \Big|_{ \rho = \frac{\pi}{2}-\eps_\rho}  \sim \eps_{\rho}^{2\Delta - d}\, .
\ee 
It then follows that 
\be
  \delta I_{\mt{matter}} =  -\frac{\varepsilon^2 }{64\pi\GN}  \int_{\partial\mt{WDW}}\!\!\! \!\!\! ds\, d\Omega_{d-1}\, \sqrt{\gamma} \partial_s \Phi^2_{\mt{cl}}  \, .
\label{eq:tbulk1}
\ee

\subsubsection{Action variation results}
\label{sec:CA-result}

Let us add up the individual variations discussed in the previous subsection. A priori, the full action variation could get contributions from all terms, but we showed 
\begin{equation}
  \delta I_{\mt{jt}}, \delta I_{\mt{GHY}}\to 0 \quad \quad \text{when} \quad \quad  \epsilon_\rho \to 0
\end{equation}
for $\Delta > d/2$.  Hence, the full action variation equals
\begin{equation}
  \delta I = \delta I_{\mt{EH}} + \delta I_\kappa + \delta I_{\mt{ct}} + \delta I_{\mt{matter}}\,,
\end{equation}
where all contributions are null boundary integrals over $\partial\text{WDW}$, \ie the null boundary of the undeformed WDW patch in global AdS.

Interestingly, the sum of the (finite) contributions from the gravitational sector
\begin{equation}
  \delta I_{\mt{EH}} + \delta I_\kappa + \delta I_{\mt{ct}} = 0\,,
\end{equation}
vanishes when using the explicit expressions in eqs.~\eqref{eq:bulkvarfin}, \eqref{eq:deltaIk} and \eqref{eq:ct}.  Thus, there is no net contribution to the action variation coming from the gravitational sector of the action, and the full action variation equals the matter variation in eq.~\eqref{eq:tbulk1}
\begin{equation}
  \delta I = \delta I_{\mt{matter}} =  -\frac{\varepsilon^2\,\text{Vol}\,\Omega_{d-1}\,L^{d-1}}{64\pi\GN} \int_{\partial\mt{WDW}}\!\!\!\!\!\! ds \tan^{d-1}\!\rho\ \partial_s \Phi_{\mt{cl}}^2 \,,
\label{eq:matter-var}
\end{equation}
where the integral over the ($d$--1)-sphere was performed. 

Further, integrating  eq.~\eqref{eq:matter-var} by parts, the variation can be written as
\begin{eqs}
  \delta I  =& \frac{\varepsilon^2\, (d-1)\text{Vol}\,\Omega_{d-1}\,L^{d-2}}{64\pi\GN}  \int_{\partial\mt{WDW}} ds\,\tan^{d-2}\!\rho\ \Phi_{\mt{cl}}^2 \,. 
\end{eqs}
This result already ignores all possible boundary contributions since they vanish for $\Delta > d/2$.  This is manifest for $s=0$ $(\rho=0)$ given the finiteness of the scalar field $\Phi_{{\mt{cl}}}(t,\rho=0)$ and it also holds at $s\to \infty$ $(\rho\to \frac{\pi}{2})$ due to the asymptotic behaviour of the scalar field $\Phi_{{\mt{cl}}} \sim \eps^\D_\rho$.  

The final variation is  {\it finite}. Thus, removing the regulator and writing the null integral in terms of  the radial variable $\rho$, the variation reduces to
\begin{equation}
   \delta I =  \frac{\varepsilon^2\,(d-1)\,\text{Vol}\,\Omega_{d-1}\,L^{d-1}}{64\pi\GN} \int^{\pi/2}_{0} d\rho\,\frac{\tan^{d-2}\rho}{\cos^2\rho}\,\big[ \Phi_{\mt{cl}}^2(t_{0+}(\rho),\rho) + \Phi_{\mt{cl}}^2(t_{0-}(\rho),\rho) \big]
\label{eq:final-CA}   
\end{equation}
where  the two terms account for the integration along the future and past boundaries of the WDW patch, respectively.

\paragraph{Analytic results for $\delta \mC_{\mt A}$.} We wish to evaluate \eqref{eq:final-CA} for a general linear superposition of (spherically symmetric) modes as in eq.~\eqref{eq:s-mode}, \ie
\begin{equation}
  \Phi_{\mt{cl}} (t,\rho) = \sum_j 2|\alpha_j|\,\cos\left(\omega_j t - \theta_j\right)\,e_j(\rho)\,,
\end{equation}
where the normalized eigenfunctions $e_j(\rho)$ are given in eq.~\eqref{eq:eigenfct}. Since $j$ is a positive integer, the hypergeometric function in \eqref{eq:eigenfct} reduces to a polynomial 
\be
{}_2F_1\left[-j,\D+j,\frac d 2;\sin^2\rho\right]  = \sum_{n=0}^{j} (-1 )^n \binom{j}{n} \frac{(\Delta + j)_n}{(d/2)_n}\, \sin^{2n} \!\rho \,,
\label{eq:hyper-id}
\ee
with $(b)_n \equiv \Gamma(b+n)/\Gamma(b)$. This allows us to write the action variation \eqref{eq:final-CA} and thus the variation $\delta \mC_{\mt A}$ for a linear superposition of coherent states \eqref{eq:s-mode} in the form
\begin{equation}
\delta \mC_{\mt A}=\frac{\delta I}{\pi} = \frac{\varepsilon^2}{\pi^2} \sum_{j,k} |\alpha_j\alpha_k| \left(\cos \(\omega_j t_\Sigma - \theta_j\)\cos \(\omega_k t_\Sigma - \theta_k\) \mC^\mt{A}_{j,k} 
+ \sin\(\omega_j t_\Sigma - \theta_j\)\sin\(\omega_k t_\Sigma - \theta_k\) \mS^\mt{A}_{j,k} \right)\,,
\label{eq:CAvar-final}
\end{equation}
with amplitudes $\mC^\mt{A}_{j,k}$ and $\mS^\mt{A}_{j,k}$ from each pair of frequencies $(\omega_j,\,\omega_k)$  defined as
\begin{eqs}
  \mC^\mt{A}_{j,k}  &= \frac{(d-1)\,\text{Vol}\,\Omega_{d-1}\,L^{d-1}}{8 \GN}  A_jA_k \sum_{n=0}^j \sum_{m=0}^k (-1)^{n+m}\binom{j}{n}\binom{k}{m} \frac{(\Delta+j)_n}{(d/2)_n} \frac{(\Delta+k)_m}{(d/2)_m} \\
  &\times   \int^{\pi/2}_0 d\rho\,\sin^{d+2(n+m)-2}\rho\,\cos^{2\Delta-d}\rho \cos\( \ww_j (\pi/2 -\rho) \) \cos\( \ww_k(\pi/2 -\rho )\)\,,  \\
   \mS^\mt{A}_{j,k}  &= \frac{(d-1)\,\text{Vol}\,\Omega_{d-1}\,L^{d-1}}{8 \GN} A_jA_k \sum_{n=0}^j \sum_{m=0}^k (-1)^{n+m}\binom{j}{n}\binom{k}{m} \frac{(\Delta+j)_n}{(d/2)_n} \frac{(\Delta+k)_m}{(d/2)_m} \\
  &\times  \int^{\pi/2}_0 d\rho\,\sin^{d+2(n+m)-2}\rho\,\cos^{2\Delta-d}\rho \sin\( \ww_j (\pi/2 -\rho) \) \sin\( \ww_k(\pi/2 -\rho )\)\,. 
\end{eqs}
Notice that $\mC^\mt{A}_{j,k}$ and $\mS^\mt{A}_{j,k}$ are dimensionless numbers where we absorbed all normalization factors except for the coherent state amplitudes $|\alpha_j|$ and $|\alpha_k|$, the $\pi^2$ factor originating from the definition of CA and the scalar action normalization, and the explicit trigonometric functions determining the oscillating behaviour of the variation. 

Using the Euler exponential representation for the cosine and sine functions, both integrals determining $\mC^\mt{A}_{j,k}$ and $\mS^\mt{A}_{j,k}$ can be evaluated in terms of the following building block (\eg see \cite{gradshteyn2007})
\begin{eqs} \label{eq:integralCA}
I_{\gamma}[\alpha, \beta] &\equiv e^{i\frac\pi2\, \beta} \int^{\pi/2}_0 d\rho\, e^{i \alpha\rho}  \sin^{d+2\gamma-2}\rho\,\cos^{2 \Delta -d}\rho  \, = \frac{e^{i\frac\pi2\, \beta}  e^{i\frac{\pi}{2}\( d  -1\)} }{2^{2 \Delta + 2 \gamma -1}} \Gamma\(1 -\Delta - \gamma + \frac{\alpha}{2} \)\\
&\Bigg\{ \frac{  e^{{-i \pi }\( \Delta   - \frac{\alpha}{2}  \)}   \Gamma\( 2 \Delta - d + 1 \) }{\Gamma\( 2 -  d - \gamma  + \Delta +  \frac{\alpha}{2}   \)} \, {}_2F_1\[2 - d  - 2 \gamma , 1 - \Delta - \gamma +  \frac{\alpha}{2}  , 2 -  d - \gamma  + \Delta +  \frac{\alpha}{2} ; -1\]  \\
&+\frac{ e^{i \pi\gamma} \Gamma\(  d + 2 \gamma - 1 \)}{\Gamma\(d   - \Delta + \gamma +  \frac{\alpha}{2}\)} 
\, {}_2F_1\[d -  2 \Delta , 1-  \Delta -\gamma +  \frac{\alpha}{2}  , d   - \Delta + \gamma +  \frac{\alpha}{2} ; -1\] \Bigg\}\, .
\end{eqs}
Notice that $I^\star_\gamma[n,m]=I_\gamma[-n,-m]$. This allows to write $\mC^\mt{A}_{j,k}$ and $\mS^\mt{A}_{j,k}$ as
 \begin{eqs}
  \mC^\mt{A}_{j,k}  &=  \frac{(d-1)\,\text{Vol}\,\Omega_{d-1}\,L^{d-1}}{16 \GN}  A_jA_k  \sum_{n=0}^j \sum_{m=0}^k (-1)^{n+m}\binom{j}{n}\binom{k}{m} \frac{(\Delta+j)_n}{(d/2)_n} \frac{(\Delta+k)_m}{(d/2)_m} \\
  &\qquad\times  \bigg[   I_{n+m}\left[-(\omega_j + \omega_k),  (\omega_j + \omega_k) \right] + I_{n+m} \left[ (\omega_j + \omega_k), -  (\omega_j + \omega_k) \right] \\
 &\qquad~~+ I_{n+m}\left[ - (\omega_j -  \omega_k),  (\omega_j - \omega_k) \right]+ I_{n+m}\left[(\omega_j - \omega_k), -(\omega_j - \omega_k) \right]\bigg]\,,\\
   \mS^\mt{A}_{j,k}  &= -\frac{(d-1)\,\text{Vol}\,\Omega_{d-1}\,L^{d-1}}{16 \GN}  A_jA_k \sum_{n=0}^j \sum_{m=0}^k (-1)^{n+m}\binom{j}{n}\binom{k}{m} \frac{(\Delta+j)_n}{(d/2)_n} \frac{(\Delta+k)_m}{(d/2)_m} \\
  &\qquad\times  \bigg[   I_{n+m}\left[-(\omega_j + \omega_k), (\omega_j + \omega_k) \right] + I_{n+m} \left[ (\omega_j + \omega_k), -  (\omega_j + \omega_k) \right] \\
 &\qquad~~- I_{n+m}\left[ - (\omega_j -  \omega_k),  (\omega_j - \omega_k) \right]- I_{n+m}\left[(\omega_j - \omega_k), -(\omega_j - \omega_k) \right]  \bigg] \,.
 \label{eq:action-coeff}  
 \end{eqs}

It is challenging to provide exact analytic results for general values of the boundary dimension $d$, conformal dimension $\Delta$ and mode frequencies $(\omega_j, \omega_k)$. However, it is possible to do so for a \emph{fixed} pair $(d,\Delta)$. Below, we consider $d=3$, to compare with our earlier results in \cite{Bernamonti:2019zyy}, arbitrary frequencies and different specific conformal dimensions $\Delta$ corresponding to marginal, irrelevant and relevant dual operators, respectively.\\

\noindent {\bf Marginal operator.} Consider $\Delta=d=3$ for arbitrary frequencies. This corresponds to the massless scalar field discussed in  \cite{Bernamonti:2019zyy}, though in the latter, we fixed $t_\Sigma=0$ and real amplitudes $\alpha_j$ (\ie $\theta_j=0$), in which case the variation of the holographic complexity \reef{eq:CAvar-final} reduces to
\beq
\delta \mC_{\mt A}=\frac{\varepsilon^2}{\pi^2} \sum_{j,k}  \mC^\mt{A}_{j,k}\ \alpha_j\,\alpha_k  \, .
\label{eq:CAvarX}
\eeq
Here we extend our results by allowing for general boundary times $t_\Sigma$ and arbitrary phases $\theta_j$ for the coherent state amplitudes. Hence, the action variation $\delta \mC_{\mt A}$ generically depends on $\mC^\mt{A}_{j,k}$, as well as $\mS^\mt{A}_{j,k}$. Performing explicitly the finite sums with Mathematica, we find
\begin{equation}
\begin{aligned}
  \mC^\mt{A}_{j,k} &= \sqrt{\frac{(j+\frac32)(k+\frac32)}{\left(j+1\right) \left(j+2\right)\left(k+1\right) \left(k+2\right)}} \\
&\times\, \bigg( H_{j+\frac{1}{2}}+H_{j+\frac{3}{2}} +H_{k+\frac{1}{2}}+H_{k+\frac{3}{2}}  
-H_{j+k+\frac{5}{2}}-H_{j-k-\frac{1}{2}}-2+4\log 2\bigg)\,, \\
  \mS^\mt{A}_{j,k} &= \frac{1}{\sqrt{(j+1) (k+1)}} \( -H_{j-k-\frac{1}{2}}-H_{j+k+\frac{3}{2}}+2 H_{j+\frac{1}{2}}+2 H_{k+\frac{1}{2}}+4 \log 2 \) \,,
\end{aligned}
\label{33}
\end{equation}
in terms of harmonic numbers $H_n$. When $n$ is a positive integer, these are defined by 
\begin{equation}
  H_n\equiv \sum_{k=1}^n \frac{1}{k} = \int_{0}^1 \frac{1-x^n}{1-x}dx\,.
\end{equation}
The latter expression allows an analytic continuation to arbitrary real and complex numbers $\alpha$ that is related to the Gamma function by
\begin{equation}
  H_{\alpha} = \gamma + \frac{d \log \Gamma(\alpha+1)}{d\alpha} \,,
\label{eq:har-num}
\end{equation}
where $\gamma$ is the Euler-Mascheroni constant. In particular, $H_{-\frac{1}{2}}=-\log 4$. 

Since harmonic numbers have an asymptotic expansion
\begin{equation}
  H_n = \log n + \gamma + \frac{1}{2n} - \sum_{k=1}^\infty \frac{B_{2k}}{2k\,n^{2k}}\,, \quad n\gg 1
\end{equation}
where $B_{2k}$ are the Bernouilli numbers, our analytic results allow us to analyse the mathematical behaviour of the action variation when one of the frequencies $\omega_j = 3 + 2j$ is large, \ie $j\gg 1$. Consider a perturbation \eqref{eq:s-mode} with a \emph{single} mode at large $j$. There are then only diagonal contributions to $\delta \mC_{\mt A}$, with amplitudes approximated by
\begin{equation}
  \mC^\mt{A}_{j,j} \sim 3 \frac{\log j}{j} + \mathcal{O}(j^{-1})\,, \quad \quad \mS^\mt{A}_{j,j} \sim \frac{\log j}{j} + \mathcal{O}(j^{-1})\,.
\end{equation}
We learn both coefficients at leading order are suppressed with the same functional dependence on the frequency, but different coefficients of order one. Subleading contributions also differ by order one coefficients.

Let us now consider a linear combination of two modes with frequencies $\omega_j$ and $\omega_k$ with large $j\gg 1$. There are two natural cases to consider:
\begin{enumerate}[label=(\alph*)] 
\item If $k\sim\mathcal{O}(1)$, the diagonal and off-diagonal amplitudes behave like
\begin{equation}
\begin{aligned}
  &\mC^\mt{A}_{j,j} \sim 3 \frac{\log j}{j} + \mathcal{O}(j^{-1})\,, \quad &\mC^\mt{A}_{k,k} &\sim \mathcal{O}(1)\,, \quad
  \mC^\mt{A}_{j,k} \sim \frac{g(k)}{j^{1/2}} + \mathcal{O}(j^{-3/2}) \\
  &\mS^\mt{A}_{j,j} \sim \frac{\log j}{j} + \mathcal{O}(j^{-1})\,, \quad &\mS^\mt{A}_{k,k} &\sim \mathcal{O}(1)\,, \quad
  \mS^\mt{A}_{j,k} \sim \frac{h(k)}{j^{3/2}} + \mathcal{O}(j^{-5/2})
\end{aligned}
\end{equation}
with finite functions $g(k)$ and $h(k)$. For example, for $\Delta=d=3$ from eq.~\reef{33}, we find: 
\beq
\small
h(k) = \sqrt{\frac{8(k+1) (k+2)}{2 k+3}}\,,
\quad
g(k)=\sqrt{\frac{2 k+3}{2(k+1)(k+2)}} \( H_{k+\frac{1}{2}} + H_{k+\frac{3}{2}}-2+\log 16 \)\,.
\eeq
\item If $k=j + \delta j$ with $|\delta j|\ll j$, the diagonal and off-diagonal amplitudes behave like
\begin{equation}
\begin{aligned}
  \mC^\mt{A}_{j,j} &\sim \mC^\mt{A}_{k,k} \sim  \mC^\mt{A}_{j,k} \sim 3 \frac{\log j}{j} + \mathcal{O}(j^{-1})\,, \\
  \mS^\mt{A}_{j,j} &\sim \mS^\mt{A}_{k,k} \sim \mS^\mt{A}_{j,k} \sim \frac{\log j}{j} + \mathcal{O}(j^{-1})\,,
\end{aligned}
\end{equation}
where the subleading corrections are $\delta j$ dependent.
\end{enumerate}
Since these statements hold for any pair $(\omega_j, \omega_k)$, we reach the following conclusions. The dominant contribution to the action variation $\delta \mC_{\mt A}$ always comes from the low frequency modes. In particular when both frequencies are of order one, the action variation will typically have off-diagonal terms that are expected to be of the same order of magnitude as the diagonal ones, and both are expected to be of order one. In the large frequency sector, all amplitudes are suppressed by $\log j/j$. The off-diagonal amplitudes between large and small frequency sectors are rationally suppressed. In particular, we observe that for $\Delta=d=3$, $\mS^\mt{A}_{j,k} \sim \mC^\mt{A}_{j,k}/j \sim j^{-3/2}$. \\

\noindent {\bf Irrelevant operator.} Let us keep the boundary dimension $d=3$ fixed, but consider a positive mass perturbation with $\Delta = 4$. Using Mathematica, we find
\begin{equation}
\begin{aligned}
  \mC^\mt{A}_{j,k} &=\frac{1}{4} \sqrt{\frac{(2 j+3) (2 j+5) (2 k+3) (2 k+5)}{(j+1) (j+2) (j+3) (k+1) (k+2) (k+3)}} \\
 &\times\, \bigg(  -\frac{24 (j+2) (k+2)}{(2 j+3) (2 j+5) (2 k+3) (2 k+5)} +H_{\frac{7}{2}+j+k} -H_{j-k-\frac{1}{2}}     \bigg) 
 \label{440}\\
  \mS^\mt{A}_{j,k} &= \sqrt{\frac{( j+\frac{3}{2}) ( j+\frac{5}{2}) (k+\frac{3}{2}) ( k+\frac{5}{2})}{(j+1) (j+2) (j+3) (k+1) (k+2) (k+3)}} \bigg(    -H_{j-k-\frac{1}{2}}-H_{j+k+\frac{7}{2}}+2 H_{j+\frac{1}{2}} \\
& +2 H_{k+\frac{1}{2}}+4 \log (2)-\frac{ \left( 6 j^2 k^2+20 j^2 k+20 j k^2+14 j^2+14 k^2+64 j k+41 j+41 k+21\right)}{( j+\frac{3}{2}) (j+\frac{5}{2}) (k+\frac{3}{2}) (k+\frac{5}{2})} \bigg)
\end{aligned}
\end{equation}
Consider a pair of frequencies $(\omega_j,\omega_k)$, with $j\gg 1$. Depending on the value of $k$, we find the following asymptotic behaviours
\begin{enumerate}[label=(\alph*)] 
\item If $k\sim\mathcal{O}(1)$, the diagonal and off-diagonal amplitudes behave like
\begin{equation}
\begin{aligned}
  \mC^\mt{A}_{j,j} \sim 4 \frac{\log j}{j} + \mathcal{O}(j^{-1})\,, \quad \mC^\mt{A}_{k,k} &\sim \mathcal{O}(1)\,, \quad
  \mC^+_{j,k} \sim \frac{g(k)}{j^{3/2}} + \mathcal{O}(j^{-5/2}) \\
   \mS^\mt{A}_{j,j} \sim 3\frac{\log j}{j} + \mathcal{O}(j^{-1})\,, \quad \mS^\mt{A}_{k,k} &\sim \mathcal{O}(1)\,, \quad
  \mS^\mt{A}_{j,k} \sim \frac{h(k)}{j^{1/2}} + \mathcal{O}(j^{-3/2})
\end{aligned}
\end{equation}
for finite functions $g(k)$ and $h(k)$.

\item If $k=j + \delta j$ with $|\delta j|\ll j$, the diagonal and off-diagonal amplitudes behave like
\begin{equation}
\begin{aligned}
  \mC^\mt{A}_{j,j} &\sim \mC^\mt{A}_{k,k} \sim  \mC^\mt{A}_{j,k} \sim 4 \frac{\log j}{j} + \mathcal{O}(j^{-1})\,, \\
   \mS^\mt{A}_{j,j} &\sim \mS^\mt{A}_{k,k} \sim \mS^\mt{A}_{j,k} \sim 3\frac{\log j}{j} + \mathcal{O}(j^{-1})\,,
\end{aligned}
\end{equation}
where the subleading corrections are $\delta j$ dependent.
\end{enumerate}
The conclusions are similar to the ones for the marginal case, \ie $\Delta=3$. The dominant contribution to the action variation $\delta \mC_{\mt A}$ comes from the low frequency modes. The amplitudes in the large frequency sector are suppressed by $\log j/j$, whereas the off-diagonal amplitudes between large and small frequency sectors are rationally suppressed. Contrary to the $d =\Delta=3$ case, for $d=3$ and $\Delta=4$, we find $\mC^\mt{A}_{j,k} \sim \mS^\mt{A}_{j,k}/j \sim j^{-3/2}$, \ie $\mC^\mt{A}_{j,k}$ is smaller than $\mS^\mt{A}_{j,k}$ for $\Delta=4$.\\

\noindent {\bf Relevant operator.} Finally, consider $\Delta=2$ in $d=3$. Using Mathematica, we find
\begin{equation}
\begin{aligned}
  \mC^\mt{A}_{j,k} &= \sqrt{\frac{1}{(1+j)(1+k)}} \(H_{j+k+\frac{3}{2}}-H_{j-k-\frac{1}{2}} \)\,, \\
  \mS^\mt{A}_{j,k} &=\frac{1}{\sqrt{(j+1) (k+1)}} \( -H_{j-k-\frac{1}{2}}-H_{j+k+\frac{3}{2}}+2 H_{j+\frac{1}{2}}+2 H_{k+\frac{1}{2}}+4 \log 2 \) \,.
\end{aligned}  
\end{equation}
Consider a pair of frequencies $\omega_j$ and $\omega_k$, with $j\gg 1$. Depending on the value of $k$, we find the following asymptotic behaviours
\begin{enumerate}[label=(\alph*)] 
\item If $k\sim\mathcal{O}(1)$, the diagonal and off-diagonal amplitudes behave like
\begin{equation}
\begin{aligned}
  \mC^\mt{A}_{j,j} \sim \frac{\log j}{j} + \mathcal{O}(j^{-1})\,, \quad \mC^\mt{A}_{k,k} &\sim \mathcal{O}(1)\,, \quad
  \mC^\mt{A}_{j,k} \sim \frac{2\sqrt{k+1}}{j^{3/2}} + \mathcal{O}(j^{5/2}) \\
   \mS^\mt{A}_{j,j} \sim 3\frac{\log j}{j} + \mathcal{O}(j^{-1})\,, \quad \mS^\mt{A}_{k,k} &\sim \mathcal{O}(1)\,, \quad
  \mS^\mt{A}_{j,k} \sim \frac{h(k)}{j^{1/2}} + \mathcal{O}(j^{3/2})
\end{aligned}
\end{equation}

\item If $k=j + \delta j$ with $|\delta j|\ll j$, the diagonal and off-diagonal amplitudes behave like
\begin{equation}
\begin{aligned}
  \mC^\mt{A}_{j,j} &\sim \mC^\mt{A}_{k,k} \sim  \mC^\mt{A}_{j,k} \sim \frac{\log j}{j} + \mathcal{O}(j^{-1})\,, \\
   \mS^\mt{A}_{j,j} &\sim \mS^\mt{A}_{k,k} \sim \mS^\mt{A}_{j,k} \sim 3\frac{\log j}{j} + \mathcal{O}(j^{-1})\,.
\end{aligned}
\end{equation}

\end{enumerate}
The conclusions are analogous to those for the previous case. The dominant contribution to the action variation $\delta \mC_{\mt A}$ comes from the low frequency modes. The amplitudes in the large frequency sector are suppressed by $\log j/j$, whereas the off-diagonal amplitudes between large and small frequency sectors are rationally suppressed. Contrary to $d=\Delta=3$, but as for $d=3$ and $\Delta=4$, we find $\mC^\mt{A}_{j,k} \sim \mS^\mt{A}_{j,k}/j \sim j^{-3/2}$.

\begin{figure}[htbp]
 	\centering
 	\subfigure{\includegraphics[width=0.60 \textwidth]{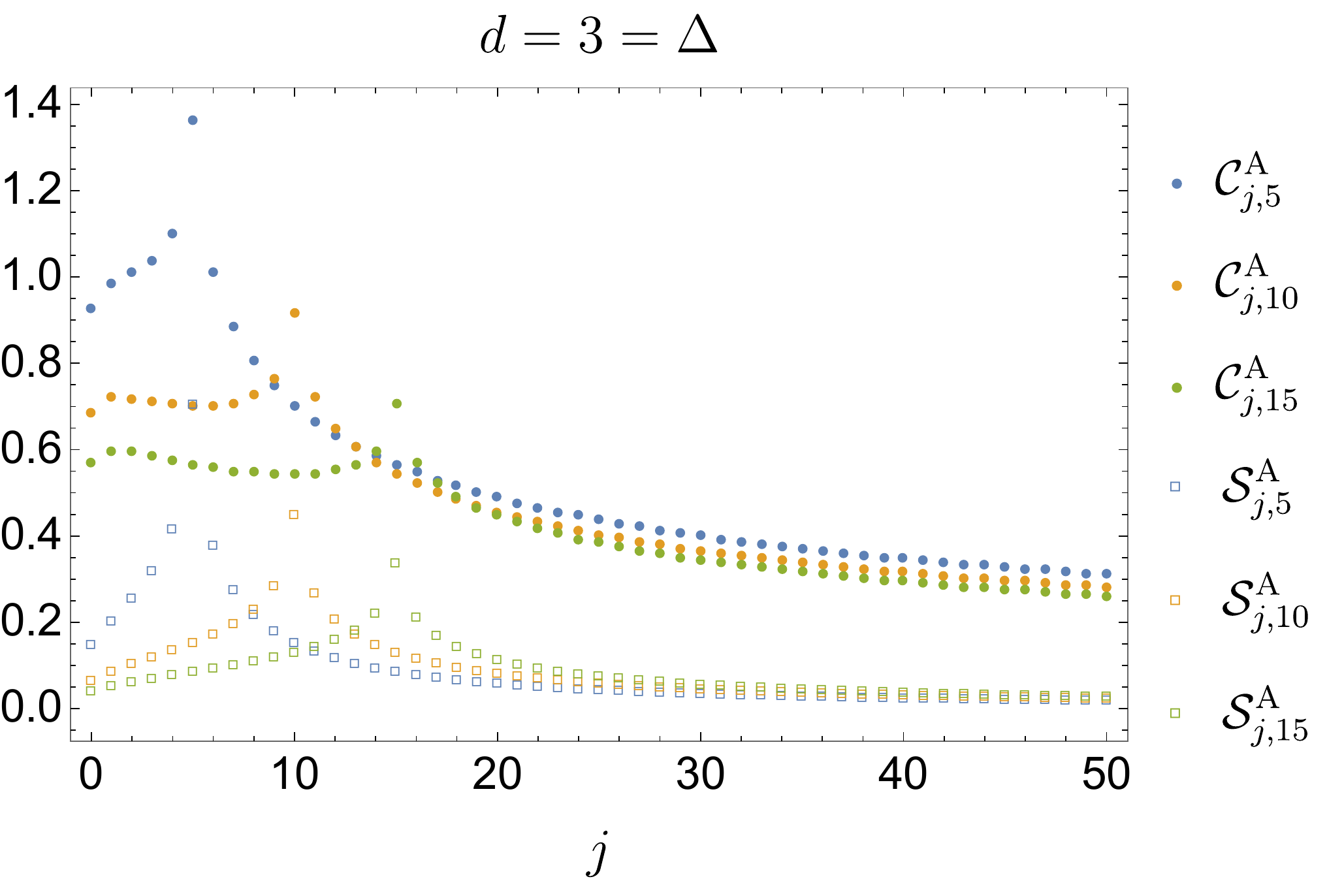}}
 	\caption{Explicit results for $\mC^\mt{A}_{j,k}$ and $\mS^\mt{A}_{j,k}$ with fixed $k$ and $d=3=\Delta$.\label{Cplusminus} }
 \end{figure}

 \paragraph{Numerical results for $\delta \mC_{\mt A}$.} Our analysis thus far focused on $d=3$, $\Delta \sim \mathcal{O}(1)$ and generic frequency. Here we numerically explore whether our findings are generic.
All our plots and discussion below refer to the expressions for $\mC^\mt{A}_{j,k}$ and $\mS^\mt{A}_{j,k}$ in eq.~\eqref{eq:action-coeff}, using eq.~\eqref{eq:integralCA}. We start this discussion with figure~\ref{Cplusminus} plotting  $\mC^\mt{A}_{j,k}$ and $\mS^\mt{A}_{j,k}$ for fixed $d=\Delta=3$, and for different values of $k$ as a function of $j$. All curves decay at large $j$, with $k$ dependent amplitude. Notice amplitudes increase for smaller values of $j$, reaching a maximum when $j\sim k$, matching our discussion derived from analytic considerations.

\begin{figure}[htbp]
 	\centering
 	\subfigure{\includegraphics[width=0.49 \textwidth]{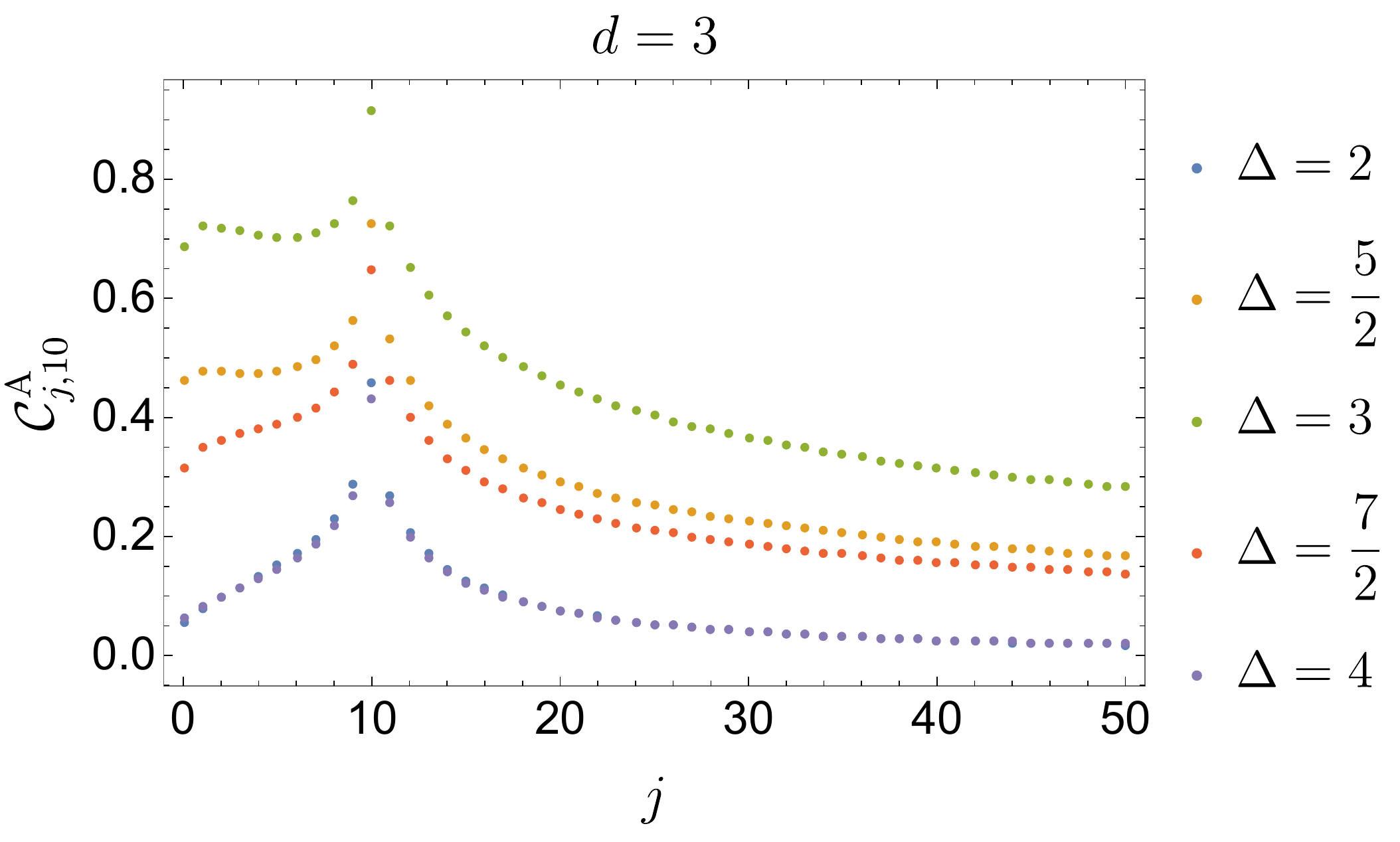}}
 	\subfigure{\includegraphics[width=0.49 \textwidth]{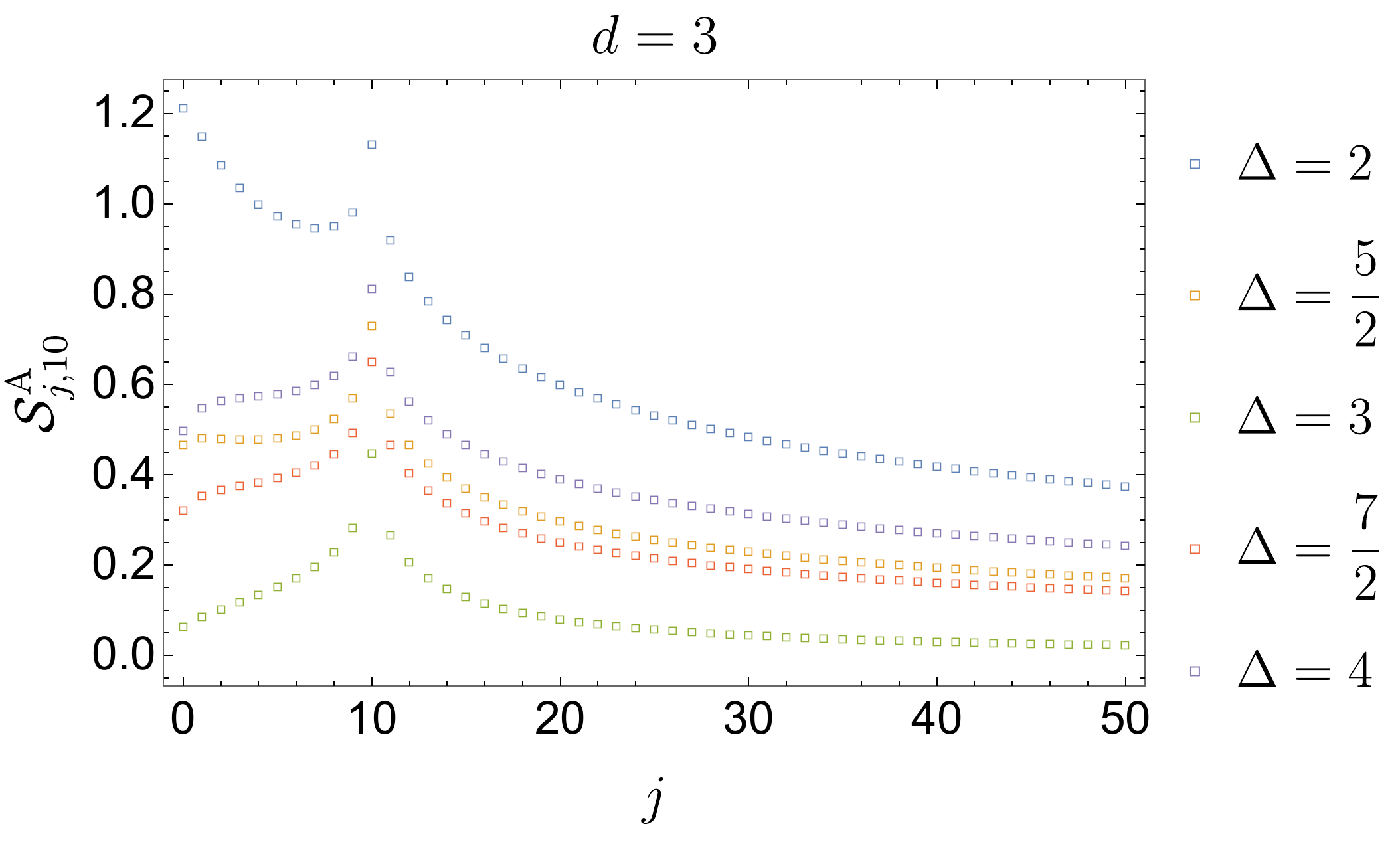}}
 	\caption{The dependence of $\mC^\mt{A}_{j,k}$ (\emph{Left}) and $\mS^\mt{A}_{j,k}$ (\emph{Right}) on $\Delta$ for $d=3$, $k=10$. For $\mC_{j,k}^\mt{A}$, the massless scalar case with $\Delta=3$ is the top one, whereas it is the smallest for $\mS_{j,k}^\mt{A}$. 
\label{deltaC_Delta}}.
\end{figure}

\begin{figure}[htbp]
	\centering
	\subfigure{\includegraphics[width=0.49 \textwidth]{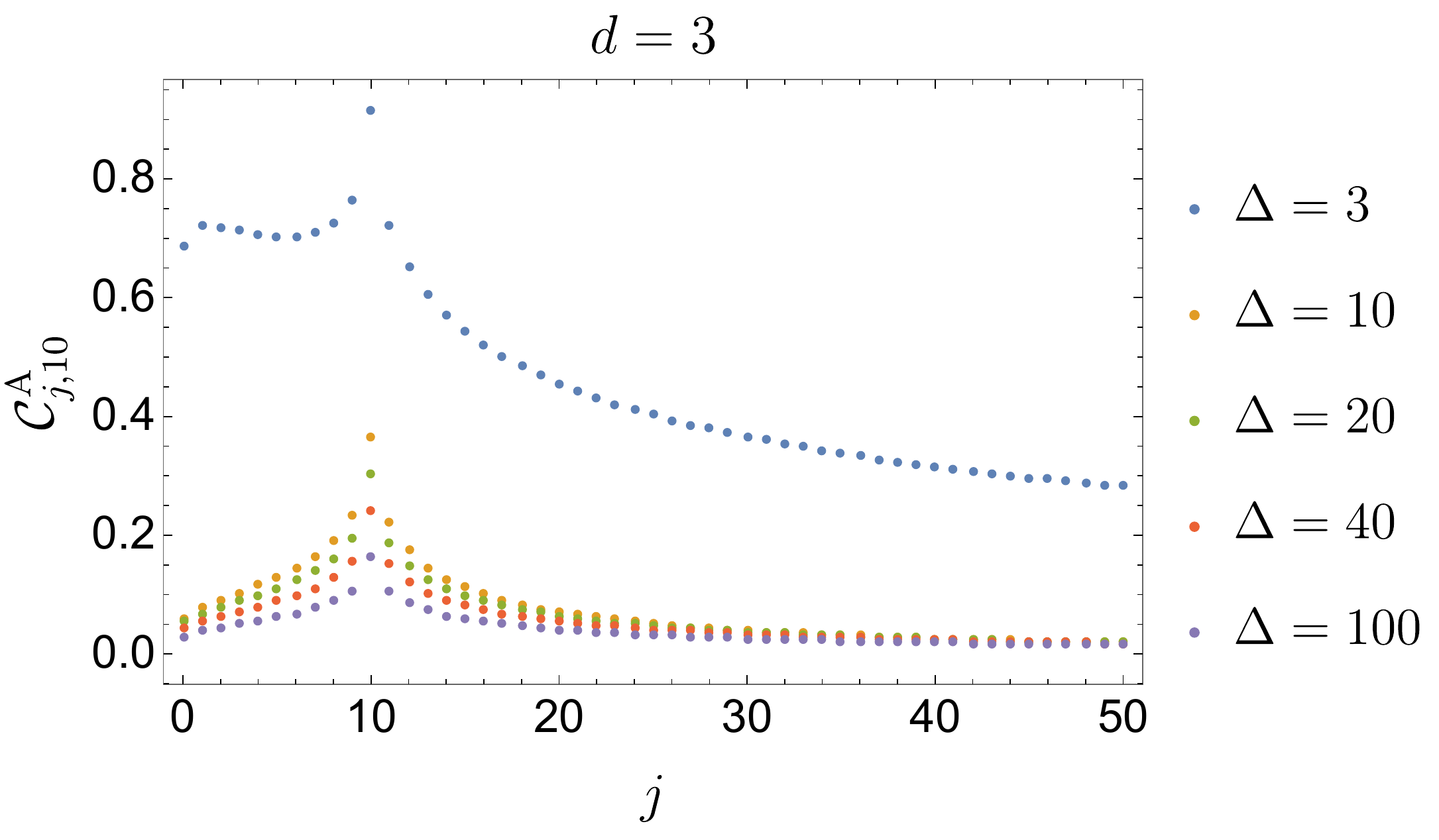}}
	\subfigure{\includegraphics[width=0.49 \textwidth]{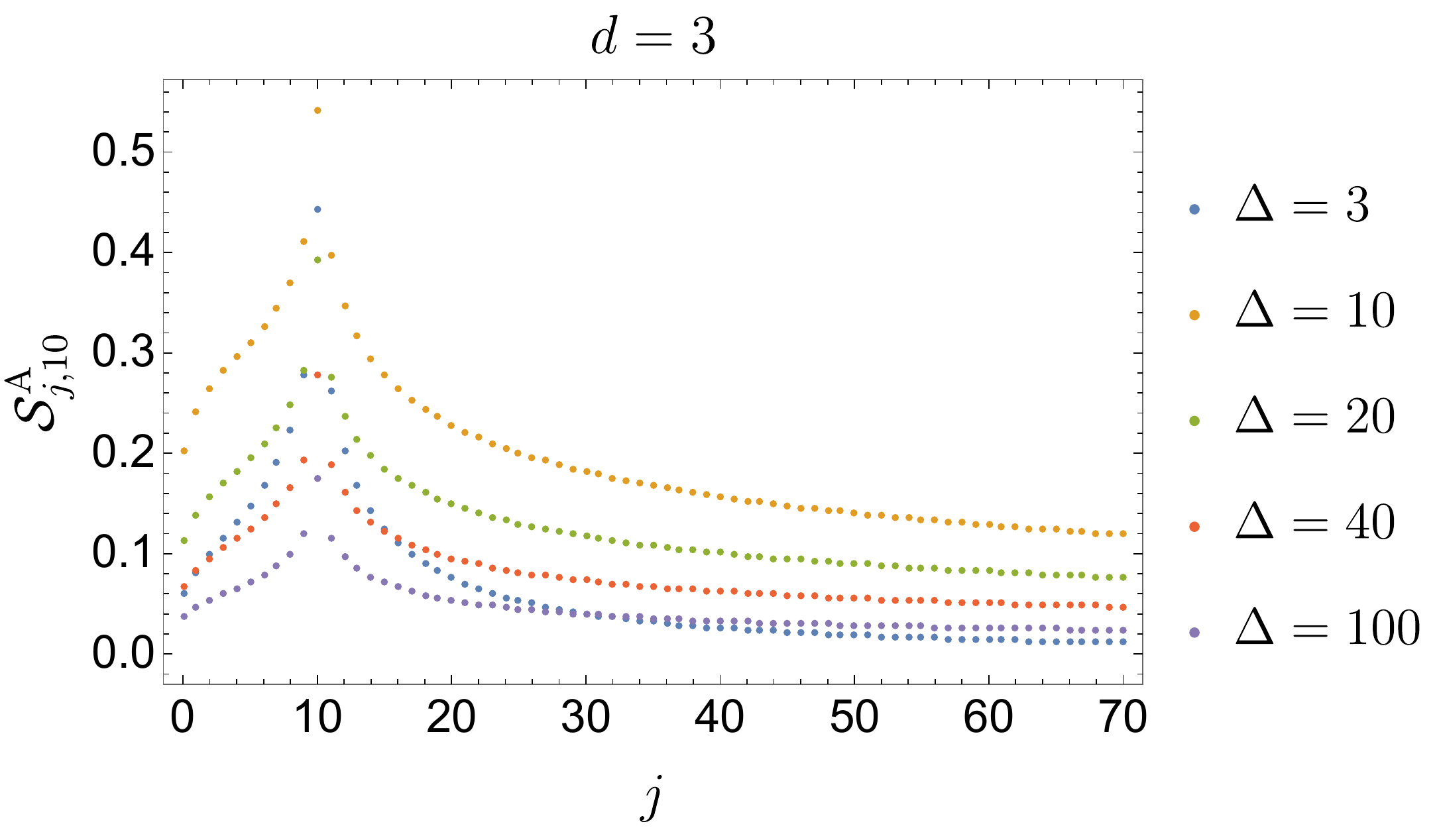}}
	\caption{The dependence of $\mC^\mt{A}_{j,k}$ (\emph{Left}) and $\mS^\mt{A}_{j,k}$ (\emph{Right}) on large $\Delta$ for $d=3$, $k=10$. Notice the crossing of some of the curves in the right panel is a generic feature. \label{deltaC_Delta2}}
\end{figure}

Next, in figure~\ref{deltaC_Delta}, we keep $d=3$, fix $k=10$ and study the dependence on $\Delta$ as a function of $j$. We observe the decay is $\Delta$ dependent, for $\Delta\sim\mathcal{O}(1)$, but the peaks at $j\sim k$ remain. The same peaks persist at large $\Delta$, as can be seen in figure~\ref{deltaC_Delta2}, but whereas the amplitudes $\mC_{j,k}^\mt{A}$ have a universal decay, \ie independent of $\Delta$, at large $j$, the amplitudes $\mS_{j,k}^\mt{A}$ are still $\Delta$ dependent in this regime. 

The existence of these peaks can be understood using, as an example, our particular analytic result for $\mC^\mt{A}_{j,k}|_{d=3,\Delta=2}$. Harmonic numbers satisfy
\begin{equation}
  H_\alpha = H_{\alpha-1} + \frac{1}{\alpha}\,.
\end{equation}
This allows us to prove
\begin{equation}
\begin{aligned}
(\mC^\mt{A}_{j+1,k}-\mC^\mt{A}_{j,k})|_{d=3,\Delta=2} &= \mC^\mt{A}_{j,k}|_{d=3,\Delta=2}\left(\sqrt{\frac{j+1}{j+2}}-1\right)- \sqrt{\frac{k+1}{j+2}}\,\frac{2}{\(j+k+\frac 5 2\)\(j-k+\frac 1 2\)}\,.
\end{aligned}
\label{eq:action-explain}
\end{equation}
This is an exact result showing $\mC^\mt{A}_{j,k}|_{d=3,\Delta=2}$ is monotonically decreasing at large $j$. Since the right hand side coefficient multiplying $\mC^\mt{A}_{j,k}|_{d=3,\Delta=2}$ is negative and approaches zero quickly, we can estimate the point where the monotonically increasing behaviour changes into monotonically decreasing as the value of $j$ where the second term flips sign. This is achieved at $j=k - 1/2$. Hence, in this particular case, we can easily, and quite accurately, explain the existence and location for these peaks at $j\sim k$, even if both are $\mathcal{O}(1)$.\\
 

%% file: sections/ComVolume.tex
\subsection{Complexity=Volume}
\label{s-volume}

The complexity=volume conjecture \reef{defineCV} suggests the complexity of a quantum state defined on a boundary time slice $\Sigma$ equals the volume of an extremal codimension-one bulk hypersurface ${\cal B}$ meeting the asymptotic boundary on $\Sigma$, \ie\footnote{Recall that for simplicity, we substitute the AdS radius $L$ for the scale $\ell_\mt{bulk}$ appearing in eq.~\reef{defineCV}.}
\begin{equation}
  {\cal C}_{\text{V}}(\Sigma) =  \max \limits_{\Sigma=\partial{\cal B}} \left[\frac{{\cal V}({\cal B})}{G_N\,L}\right] \,.
\label{CV}
\end{equation}
To determine these codimension-one hypersurfaces, one extremises the volume functional 
\begin{equation}
  {\cal V}({\cal B}) = \int d^{d}\sigma\,\sqrt{\text{det}\,{\cal G}}\,,
\end{equation}
where $\sigma^a$ are the $d$-dimensional coordinates on the surface and ${\cal G}$ is its induced metric from the bulk metric $g$, \ie in components, ${\cal G}_{a b} = g_{\mu\nu}(X)\partial_a X^\mu(\sigma)\partial_b X^\nu(\sigma)$. 

Given a background metric $g_0$ with extremal surface $X^\mu_0(\sigma)$, we are interested in computing the volume variation due to the variation of the metric $g_0+\delta g$. The contributions to the volume variation can be organised as for the action variation
\be \label{eq:volvar}
  \delta   {\cal V}({\cal B})  = \delta   {\cal V}_{{\cal B}_0} +  \delta{\cal V}_{\delta X} +   \delta  {\cal V}_{\delta \mt{cutoff}} \, . 
\ee
The first two terms are variations produced by the deformation of the background metric $g_0+\delta g$ and the deformation of the extremal volume surface $X^\mu_0(\sigma) + \delta X^\mu(\sigma)$, respectively, while $\delta {\cal V}_{\delta \mt{cutoff}}$ is due to the correction of the AdS boundary cutoff.

Before evaluating $\delta {\cal V}_{{\cal B}_0} + \delta {\cal V}_{\delta X}$, let us analyse whether $ \delta{\cal V}_{\delta \mt{cutoff}}$ contributes for the coherent state perturbations we consider. We follow the same strategy as in section \ref{sec:actionevaluation}. The structure of divergences coming from the volume of vacuum global AdS is
\begin{equation}
  {\cal V}({\cal B}_0) \sim \eps^{-d} \sum^{d}_{k=\mt{odd}} v_k \eps^k  + v_{\log}\log\eps+ \dots  \, .
\end{equation}
with the $\log$ term   only present for odd $d$, \ie an odd number of boundary dimensions, and the dots indicate terms that are vanishing as the cutoff is removed.
On the other hand, as discussed in section \ref{sec:actionevaluation}, the perturbative corrections to the cut-off start at $\mathcal{O}(\eps^d)$. Hence, we conclude $\delta {\cal V}_{\delta \mt{cutoff}}$ vanishes as the AdS regulator is removed, \ie 
$\delta {\cal V}_{\delta \mt{cutoff}}\sim \eps$.

Let us now discuss $\delta {\cal V}_{{\cal B}_0} +  \delta {\cal V}_{\delta X}$. Working at linear order in the variation, using the identity
\begin{equation}
  \sqrt{\text{det}\,{\cal G}} = \sqrt{\text{det}\,{\cal G}_0} + \frac{1}{2}\sqrt{\text{det}\,{\cal G}_0}\,{\cal G}_0^{ab} \delta{\cal G}_{ab}\,,
\end{equation}
and the variation of the induced metric
\begin{equation}
  \delta{\cal G}_{ab} = \delta g_{\mu\nu} \partial_a X_0^\mu \partial_b X_0^\nu + 2g_{\mu\nu}^0 \partial_a \delta X^\mu\partial_b X_0^\nu + \delta X^{\rho}\del_{\rho} g_{\mu\nu}^0\partial_a X_0^\mu \partial_b X_0^\nu\,,
\end{equation}
one can write the variation of the volume as
\begin{equation} \label{eq:volvar1}
\begin{split}
  \delta {\cal V}({\cal B}) &= \frac{1}{2} \int d^{d}\sigma\,\sqrt{\text{det}\,{\cal G}_0}\,{\cal G}_0^{ab} \delta g_{\mu\nu} \partial_a X_0^\mu \partial_b X_0^\nu \\
  &+  \frac{1}{2} \int d^{d}\sigma\,\sqrt{\text{det}\,{\cal G}_0}\,{\cal G}_0^{ab}\left(2g_{\mu\nu}^0 \partial_a\delta X^\mu\partial_b X_0^\nu + \delta X^{\rho}\del_{\rho} g_{ab}^0\partial_a X_0^\mu \partial_b X_0^\nu \right)
  \end{split}
  \end{equation}
The first term corresponds to $\delta {\cal V}_{{\cal B}_0} $,  the change in volume of the undeformed surface $X_0^\nu$ due to the deformation of the background. The terms in the second line equal $\delta{\cal V}_{\delta X}$, the contribution due to the deformation of the extremal volume surface. Upon integrating by parts in the first term, the remaining bulk integral in $\delta {\cal V}_{\delta X}$ is proportional to the background equations of motion and thus vanishes. However, the integration by parts produces the following boundary term  
\begin{equation}
\int_{\partial{\cal B}} \sqrt{\text{det}\,{\cal G}_0} \, n_{0,a} \, {\cal G}_0^{ab} g_{\mu\nu}^0 \delta X^\mu\partial_b X_0^\nu\,,
\end{equation}
with $n_{0,a}$ denoting the normal to $\partial{\cal B}$.  While this term may in general be non-zero, in the spherically symmetric setup described in section~\ref{sec:AdSsetup}, it vanishes since it is proportional to $\delta X^\rho$, which can be set to zero by a gauge choice.

Thus, the volume variation reduces to
\begin{equation}
  \delta {\cal V}({\cal B}) =  \delta {\cal V}_{{\cal B}_0}  = \frac{1}{2} \int d^{d}\sigma\,\sqrt{\text{det}\,{\cal G}_0}\,{\cal G}_0^{ab} \delta g_{\mu\nu} \partial_a X_0^\mu \partial_b X_0^\nu\, .
\label{eq:vol-var}
\end{equation}
As we shall see, this is finite and thus we henceforth remove the AdS boundary regulator. 

Since constant time slices are extremal surfaces in the AdS geometry \reef{eq:gads}, \ie $X_0^0=t_\Sigma$ and $X_0^a=\sigma^a$, the volume variation \eqref{eq:vol-var} reduces to
\begin{equation}
  \delta {\cal V}({\cal B}) = \frac{1}{2} \int d^{d}x\, \sqrt{\text{det}\,{\cal G}_0} \,g_0^{ij}\delta g_{ij}\,,
\label{eq:svol-var}
\end{equation}
where the integral was rewritten as a bulk space integral, \ie $i,j$ stand for spacelike directions in global AdS. Restricting to the spherically symmetric perturbations in eq.~\eqref{eq:ansatz} and working at the linearised level \eqref{perturb}, the volume variation yields
\begin{equation}
\begin{split}
   \delta {\cal V}({\cal B}) &= \frac{1}{2}\,L^{d-2}\, \text{Vol}\,\Omega_{d-1}\int^{\pi/2}_0 d\rho\,\tan^{d-1}\!\rho \ \cos \rho\, \delta g_{\rho\rho} \\
   &=  -\frac{\varepsilon^2}{2}\,L^{d}\, \text{Vol}\,\Omega_{d-1}\int^{\pi/2}_0 d\rho\,\frac{\tan^{d-1}\!\rho}{\cos \rho}\ a_2(t_\Sigma,\rho)\,.
\end{split}
\end{equation}

Using eq.~\eqref{eq:a2}, the volume variation can be written as sourced by the matter stress tensor as 
\begin{equation}
\begin{aligned}
\delta {\cal V} ({\cal B})&= \frac{ \varepsilon^2 \, L^{d}}{2(d-1)}\, \text{Vol}\,\Omega_{d-1}\int^{\pi/2}_0 d\rho \sin \rho \int_0^\rho dy \tan^{d-1} y \, T_{tt}^{\mt{bulk}}(t_\Sigma,y)\\
&= \frac{\varepsilon^2}{2(d-1)} \int_{t_\Sigma} d\rho\, d\Omega_{d-1} \sqrt{|h|}\,\cos^2\rho\ T^{\mt{bulk}}_{tt}(t_\Sigma,\rho)\,.
\end{aligned}
\label{eq:our-vol-var}
\end{equation}
To produce the final expression, we exchanged the order of the integrals in the first line, performed the $\rho$ integral, substituted $y\to\rho$, and rewrote the resulting integral in terms of the induced metric on the extremal surface of global AdS at $t =t_\Sigma$, \ie
\be
\sqrt{|h|} = L^{d}  \, \frac{\sin^{d-1}\!\rho}{\cos^d \rho}\,\sqrt{h_\Omega}\,,
\ee
where $\sqrt{h_\Omega}$ is the angular measure on the unit ($d$--1)-sphere. Inserting the stress tensor \eqref{eq:bulk-stensor}, we find
\begin{equation}\label{eq:fvol-var}
  \delta {\cal V} ({\cal B})= \frac{\varepsilon^2\,L^d\, \text{Vol}\,\Omega_{d-1}}{4(d-1)}\,\int^{\pi/2}_0 dy\,\tan^{d-1}\!y\,\cos y\left[(\partial_t\Phi_{\mt{cl}})^2 + (\partial_\rho\Phi_{\mt{cl}})^2 + \frac{m_\Phi^2L^2}{\cos^2\rho}\Phi_{\mt{cl}}^2\right]\,.
\end{equation}
Notice that this expression is close to that given for the mass in eq.~\reef{eq:l-mass} but contains an extra factor of $\cos y$. As a result that in contrast to the mass, one finds that the above expression is {\it not} a positive definite quantity. In particular,  for a relevant operator $(\Delta < d)$, $m^2$ is negative  and for the low frequency modes, the derivative terms may not be large enough to compensate for this negative contribution to the integral.

\subsubsection{Volume variation evaluation}

Integrating by parts and using the equation of motion \eqref{eq:linearisedKG}, the volume variation \eqref{eq:fvol-var} becomes
\begin{equation}
\begin{aligned}
   \delta {\cal V}({\cal B}) =&\frac{\veps^2}{4(d-1)}\,L^d\,\text{Vol}\,\Omega_{d-1} \Bigg\{ \int^{\pi/2}_0 dy\,\tan^{d-1}y\left[\cos y\left((\partial_t\Phi_{\mt{cl}})^2-\Phi_{\mt{cl}}\partial^2_t\Phi_{\mt{cl}}\right) + \sin y\,\Phi_{\mt{cl}}\partial_y\Phi_{\mt{cl}}\right] \\
   &+\left. \tan^{d-1}y\cos y\, \Phi_{\mt{cl}}\partial_y\Phi_{\mt{cl}}\right|^{\pi/2}_0 \Bigg\}\,. 
\label{eq:deltaVintermediate}
\end{aligned}
\end{equation}
For the range of conformal dimensions considered in this work, $\Delta > d/2$, the boundary term cancels for any choice of frequencies. Hence, this contribution is ignored in the following. 

We evaluate eq.~\eqref{eq:deltaVintermediate} for the superposition of modes \eqref{eq:s-mode}
\begin{equation}
  \Phi_{\mt{cl}} (t,\rho) = \sum_j 2|\alpha_j|\,\cos\left(\omega_j t - \theta_j\right)\,e_j(\rho)\,.
\end{equation}
Here, we define 
\begin{equation}
\begin{aligned}
  (\partial_t\Phi_{\mt{cl}})^2-\Phi_{\mt{cl}}\partial^2_t\Phi_{\mt{cl}} =& \sum_{j,k} T_{jk}(t) \,e_j\,e_k \\
   \Phi_{\mt{cl}}\partial_y\Phi_{\mt{cl}} = & \sum_{j,k} Y_{jk} (t)  \frac{d(e_j e_k)}{dy}\,,
\end{aligned}
\end{equation}
where all the time dependence was kept in
\begin{equation}
\begin{aligned} 
  T_{jk}(t) &= 2 |\alpha_j\alpha_k| \(\omega_j^2+ \omega_k^2\) \cos\(\omega_j t-\theta_j \) \cos\(\omega_k t-\theta_k  \)  \\
  & + 4 |\alpha_j\alpha_k|\, \omega_j \omega_k \sin\(\omega_j t-\theta_j  \) \sin\(\omega_k t-\theta_k \) \,,\\
 Y_{jk}(t)&= 2 |\alpha_j \alpha_k| \cos\(\omega_j t-\theta_j  \) \cos\(\omega_k t-\theta_k \)\,.
\end{aligned}
\label{eq:kin-pot}
\end{equation}

The normalised eigenfunctions $e_j(y)$ are given in eq.~\eqref{eq:eigenfct}, and eq.~\eqref{eq:hyper-id} still holds since $j$ is a positive integer. Consider first the time derivatives in eq.~\eqref{eq:deltaVintermediate}. Since
\begin{equation*}
  e_j(y)e_k(y) = A_jA_k \cos^{2\Delta}y \sum_{n=0}^j \sum_{m=0}^k (-1)^{n+m}\binom{j}{n}\binom{k}{m} \frac{(\Delta+j)_n}{(d/2)_n} \frac{(\Delta+k)_m}{(d/2)_m} \sin^{2(n+m)} y\,,
\end{equation*}
the radial integral yields
\begin{equation}
  \int^{\pi/2}_0 dy\,\sin^{d+2(n+m)-1}y\cos^{2\Delta-d+2}y =\frac{\Gamma\left(\Delta-\frac{d-3}{2}\right)\Gamma(m+n+d/2)}{2\Gamma(m+n+\Delta+3/2)}\,,
\end{equation}
where we used the identity 
 \begin{equation}
 \int^{\pi/2}_0 dy\,\sin^a y\cos^b y = \frac{1}{2} \frac{\Gamma\left(\frac{b+1}{2}\right)\Gamma\left(\frac{a+1}{2}\right)}{\Gamma\left(\frac{a+b}{2} + 1\right)}\,.
\label{eq:vol-int-id}
\end{equation}

Consider the term involving $\Phi\partial_y \Phi$ in eq.~\eqref{eq:deltaVintermediate}. Integrating by parts yields
\begin{equation}
\begin{aligned}
  \int^{\pi/2}_0 dy\,\tan^{d-1}y\sin y\,\frac{d(e_je_k)}{dy} &= \left.\tan^{d-1}y\sin y\,e_je_k\right|^{\pi/2}_0 \\
  &-\int^{\pi/2}_0 dy\,\tan^{d-1}y\cos y\,e_je_k\left(1 + \frac{d-1}{\cos^2y}\right)\,.
\end{aligned}
\end{equation}
Once more, the boundary contribution vanishes since $\Delta > d/2$. The remaining radial integral equals
\begin{align}
  \int^{\pi/2}_0 dy\,\sin^{d+2(n+m)-1}y& \cos^{2\Delta-d+2}y \(1+\frac{d-1}{\cos^2 y}\) = \nonumber \\
  &(d\D +(m+n)(d-1))\frac{\Gamma\left(\Delta-\frac{d-1}{2}\right)\Gamma(m+n+d/2)}{2\Gamma(m+n+\Delta+3/2)}
\end{align}
using eq.~\eqref{eq:vol-int-id}. Altogether, the volume variation \eqref{eq:deltaVintermediate} can be written as
\beqa
  \delta {\cal V}({\cal B}) &=& \veps^2\,L^d\,\text{Vol}\Omega_{d-1} \ \sum_{j,k} |\alpha_j||\alpha_k|\big[\cos(\omega_j t_\Sigma-\theta_j)\,\cos(\omega_k t_\Sigma-\theta_k)\,C^\mt{V}_{j,k}
  \nonumber\\
  &&\qquad\qquad\qquad\qquad + \sin(\omega_j t_\Sigma-\theta_j)\,\sin(\omega_k t_\Sigma-\theta_k)\,S^\mt{V}_{j,k}\big]\,,
\label{eq:finalvol-var}
\eeqa
where the time dependence was parameterised as in \eqref{eq:CAvar-final} for $\delta\ca(\Sigma)$, to facilitate the comparison, and the coefficients $   C^\mt{V}_{j,k},  S^\mt{V}_{j,k}, $ are given by
\beqa
 C^\mt{V}_{j,k} &\equiv& \frac{\Gamma\left(\Delta-\frac{d-1}{2}\right)}{4(d-1)}\,A_j\,A_k\,\sum_{n=0}^j \sum_{m=0}^k (-1)^{m+n}\binom{j}{n}\binom{k}{m} \frac{\Gamma(m+n+d/2)}{\Gamma(m+n+\Delta+3/2)} \frac{(\Delta+j)_n}{(d/2)_n} \frac{(\Delta+k)_m}{(d/2)_m} \nonumber \\
  && \qquad\times\ \  \left[\left(\Delta - \frac{d-1}{2}\right)(\omega_j^2 + \omega_k^2) - \left(d\Delta + (m+n)(d-1)\right)\right]\,, \\
  S^\mt{V}_{j,k} &\equiv& \frac{\Gamma\left(\Delta-\frac{d-1}{2}\right)}{4(d-1)}\,A_j\,A_k\,\sum_{n=0}^j \sum_{m=0}^k (-1)^{m+n}\binom{j}{n}\binom{k}{m} \frac{\Gamma(m+n+d/2)}{\Gamma(m+n+\Delta+3/2)} \frac{(\Delta+j)_n}{(d/2)_n} \frac{(\Delta+k)_m}{(d/2)_m} \nonumber \\
  && \qquad\times\ \  \left[2\Delta - (d-1)\right] \omega_j\omega_k\,.\nonumber
\eeqa
The explicit quadratic dependence in the frequencies allows to easily keep track of the source of these contributions when comparing to eq.~\eqref{eq:kin-pot}, where $T_{jk}(t_\Sigma)$ and $Y_{jk}(t_\Sigma)$ account for the kinetic and potential energy contributions to the volume variation. Taking the CV conjecture \reef{CV}, we can rewrite the variation of holographic complexity as 
\begin{equation}\label{howse}
\delta \mC_{\mt V} = \frac{\veps^2}{\pi^2} 
 \sum_{j,k} |\alpha_j\,\alpha_k|\left[\cos(\omega_j t_\Sigma-\theta_j)\,\cos(\omega_k t_\Sigma-\theta_k)\,\mathcal{C}^\mt{V}_{j,k} + \sin(\omega_j t_\Sigma-\theta_j)\,\sin(\omega_k t_\Sigma-\theta_k)\,\mathcal{S}^\mt{V}_{j,k}\right]\,,
\end{equation}
by redefining the dimensionless parameters 
\begin{equation}
  \mathcal{C}^\mt{V}_{j,k}=\frac{\pi^2L^{d-1}\,{\rm Vol}\Omega_{d-1}}{\GN }C^\mt{V}_{j,k} \quad \text{and} \quad  
  \mathcal{S}^\mt{V}_{j,k}=\frac{\pi^2L^{d-1}\,{\rm Vol}\Omega_{d-1}}{\GN }S^\mt{V}_{j,k}\,.
\end{equation} 
With this new normalization, the coefficients $\mathcal{C}^\mt{V}_{j,k}$ and $\mathcal{S}^\mt{V}_{j,k}$ are purely numerical quantities, and eq.~\reef{howse} for $\delta \mC_{\mt V}(\Sigma)$ is readily compared with eq.~\eqref{eq:CAvar-final} for $\delta\ca(\Sigma)$.

\subsubsection{Volume variation results}

To start our analysis of the variations $\delta \mC_{\mt V}$ in eq.~\eqref{howse}, we consider a coherent state $\ket{\alpha_j}$ with a \emph{single} mode excited. There is a single diagonal contribution to the sums in eq.~\eqref{howse} that we shall denote as $\delta \mC_{{\mt V}|j,j}$
\begin{equation}\label{howse2}
  \delta \mC_{{\mt V}|j,j} = \frac{\veps^2}{\pi^2}\,|\alpha_j|^2 \left[\cos^2(\omega_j t_\Sigma-\theta_j)\,\mC^\mt{V}_{j,j} + \sin^2(\omega_j t_\Sigma-\theta_j)\,\mS^\mt{V}_{j,j}\right]\,.
\end{equation}

Since this expression is proportional to eq.~\eqref{eq:fvol-var}, it is positive definite if $\Delta \geq d$, \ie when $m^2L^2 \geq 0$. However, our analysis also applies in the range $\frac{d}{2}\le \Delta<d$, and hence for $\Delta = \frac{d}{2} + \delta$ with $\delta \in (0,d/2)$, eq.~\reef{howse2} could be negative. It is natural to examine this issue for the smallest frequency $\omega_0=\Delta$, \ie $j=0$.\footnote{In fact, one finds $\delta \mC_{\mt V|j,j} < 0$ is only possible for $j=0$. Of course, the sign of off-diagonal terms is not fixed and depends on $\theta_{j}-\theta_k$ and $t_{\Sigma}$.}  In this case,
\begin{equation}
\begin{aligned}
  \mC^\mt{V}_{0,0} &= \frac{4\pi^3{\rm Vol}\Omega_{d-1}}{(d-1)}\frac{\Gamma\left(\Delta-\frac{d}{2}+\frac{1}{2} \right)\Gamma(\Delta)}{\Gamma\left(\Delta-\frac{d}{2}+1 \right)\Gamma(\Delta+\frac{3}{2}) }\,\left[2\Delta^2\left(\Delta - \frac{d-1}{2}\right)-d\Delta\right]\,, \\
  \mS^\mt{V}_{0,0} &= \frac{4\pi^3{\rm Vol}\Omega_{d-1}}{(d-1)}\frac{\Gamma\left(\Delta-\frac{d}{2}+\frac{1}{2} \right)\Gamma(\Delta)}{\Gamma\left(\Delta-\frac{d}{2}+1 \right)\Gamma(\Delta+\frac{3}{2}) }\,\left[2\Delta - (d-1)\right]\,\Delta^2\,.
\end{aligned}
\end{equation}
The last factor $2\Delta - (d-1) = 2\delta +1$ in $\mS^\mt{V}_{0,0}$ is always positive. Thus, the complexity variation may only be negative when $\mC^\mt{V}_{0,0}$ is. Since its zeroes satisfy
\begin{equation}
  \delta_\pm = \frac{d+1}{4}\left(-1 \pm \sqrt{1+ \frac{4d}{(d+1)^2}}\right)\,,
\end{equation}
and $\delta_- < 0$ $\forall\,\,d$, which is non-physical, we conclude that for $\delta \in (0,\delta_+)$, the variation of the holographic complexity can be negative. Notice that for large $d$, we may Taylor expand the square root, and then only keeping the first term, we find
\begin{equation}
  \delta_+ \approx \frac{1}{2} < \frac{d}{2}
\end{equation}
Exact evaluation of $\delta_+$ for $d\geq 2$ confirms $\delta_+ < \frac{d}{2}$ $\forall\,\,d$. Hence, $\mC^{\mt V}_{0,0}$ is negative in any dimension. Time dependence makes $\delta \mC_{{\mt V}|0,0}$ oscillate, from the maximum attained whenever $\Delta t_\Sigma - \theta_0 = \frac{(2k+1)\pi}{2}$, where it is positive, to the minimum $\Delta t_\Sigma - \theta_0 = k\pi$, where it is negative. 

The negativity of $\delta {\cal C}_{V|0,0}$ only happens for $\delta < \frac{d}{2}$. For $\Delta \gg 1$, which corresponds to a large frequency limit, one can approximate the Gamma functions using Stirling's formula 
\begin{equation}
  \frac{\Gamma\left(\Delta - (d-1)/2\right)}{\Gamma(\Delta+3/2)} \sim \frac{1}{\Delta^{1+d/2}}\,, \quad \quad \frac{\Gamma(\Delta)}{\Gamma\left(\Delta-\frac{d}{2}+1 \right)} \sim\Delta^{d/2-1}\,,
\end{equation}
yielding
\begin{equation}
  \mC^\mt{V}_{0,0} \simeq  \mS^\mt{V}_{0,0} \simeq  \frac{8\pi^3{\rm Vol}\Omega_{d-1} }{d-1}\,\Delta\,.
\end{equation}
Hence, $\delta {\cal C}_{V|0,0}$ grows linearly in the large frequency limit when $j=0$ $(\Delta \gg 1)$.\\ 

\paragraph{Analytic results for $\delta \mC_{\mt V}$.} It is possible to provide analytic formulas for $\delta \mC_{{\mt V}|j,j} $ for a specific choice of the pair $(\Delta,d)$. For example, when $\Delta=d=3$, to compare with our earlier results in \cite{Bernamonti:2019zyy}, one has
\begin{equation}\label{Tjj_term}
  T_{jj}(t_\Sigma) = 4|\alpha_j|^2  \omega_j^2 \,, \quad Y_{jj}(t_\Sigma)= 2 |\alpha_j|^2 \cos^2\(\omega_j t_\Sigma-\theta_j  \)\,,
\end{equation}
with $\omega_j = 3 + 2j$. Notice $T_{jj}$ is time independent and quadratic in the frequency. The overall volume variation equals
\begin{equation}\label{deltaCV_d3}
\begin{split}
\delta \mC_{{\mt V}|j,j} ({\cal B}) &=4\pi \varepsilon^2|\alpha_j|^2 \( \mV_{T}-\mV_{Y} \)\,,\\
\mV_{T}&=\frac{ 4(j+1) (2 j+3)}{j+2}  \left( \frac{1}{4 j+7}+\frac{7}{4 j+5}+2+\frac{\gamma +\log (4)}{2(j+1)^2}+\frac{\psi ^{(0)}\left(2 j+\frac{5}{2}\right)}{2(j+1)^2}\right)\,, \\
\mV_{Y}&= \frac{2(2 j+3)\cos^2 \(\omega_j t_\Sigma-\theta_j\) }{(j+1) (j+2)} \left( H_{2 j+\frac{3}{2}}-\frac{8 (j+1)^2}{(4j+5)(4j+7)}+ \log (4)\right)  \,.\\
\end{split}
\end{equation}
where $H_\alpha$ are the harmonic numbers in eq.~\eqref{eq:har-num} and $\psi^{(0)}(z) = \Gamma'(z)/\Gamma(z)$ is the logarithmic derivative of the gamma function $\Gamma(z)$.  

\begin{figure}[htbp]
	\centering
	\subfigure{\includegraphics[width=0.45 \textwidth]{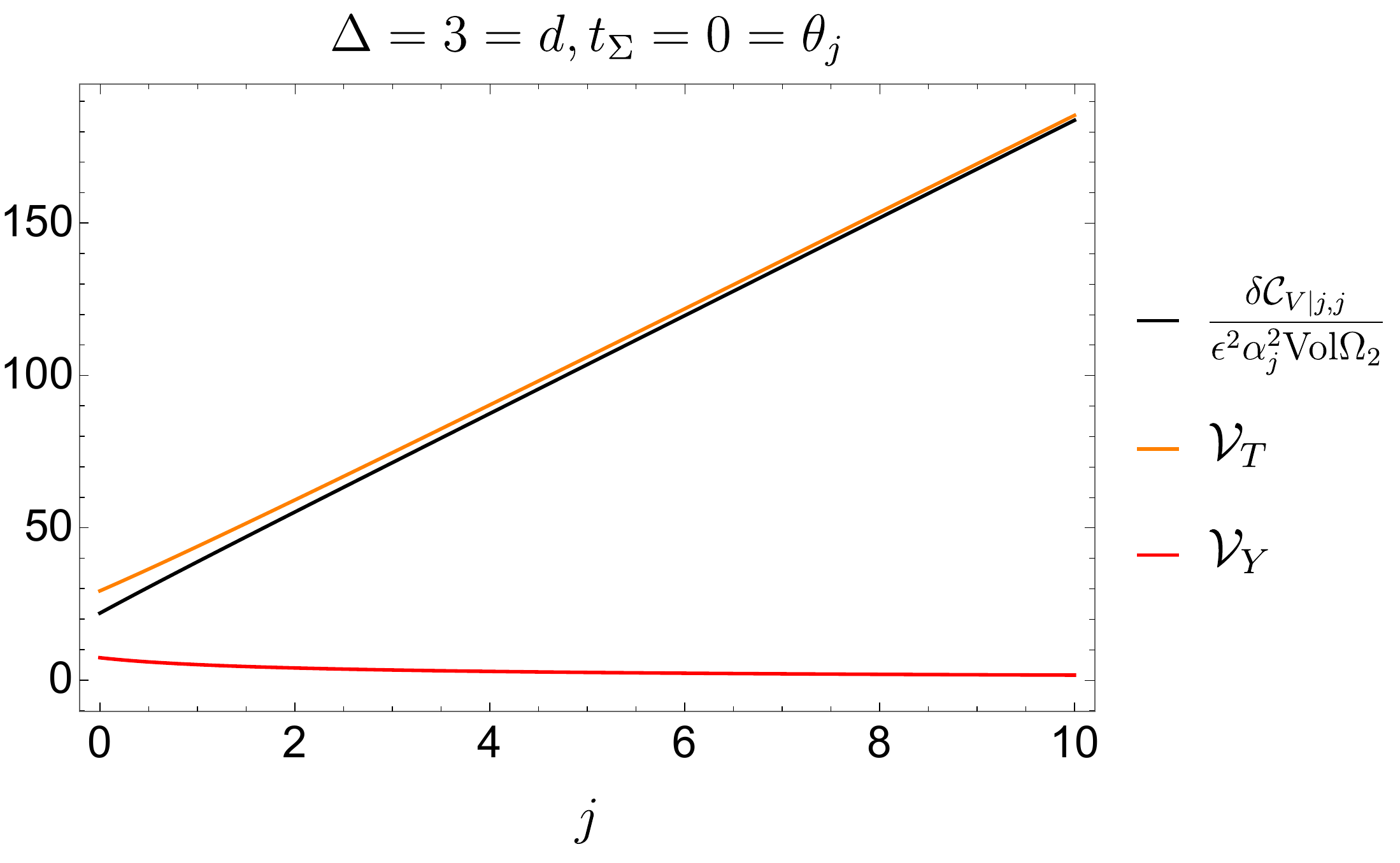}}
	\subfigure{\includegraphics[width=0.51 \textwidth]{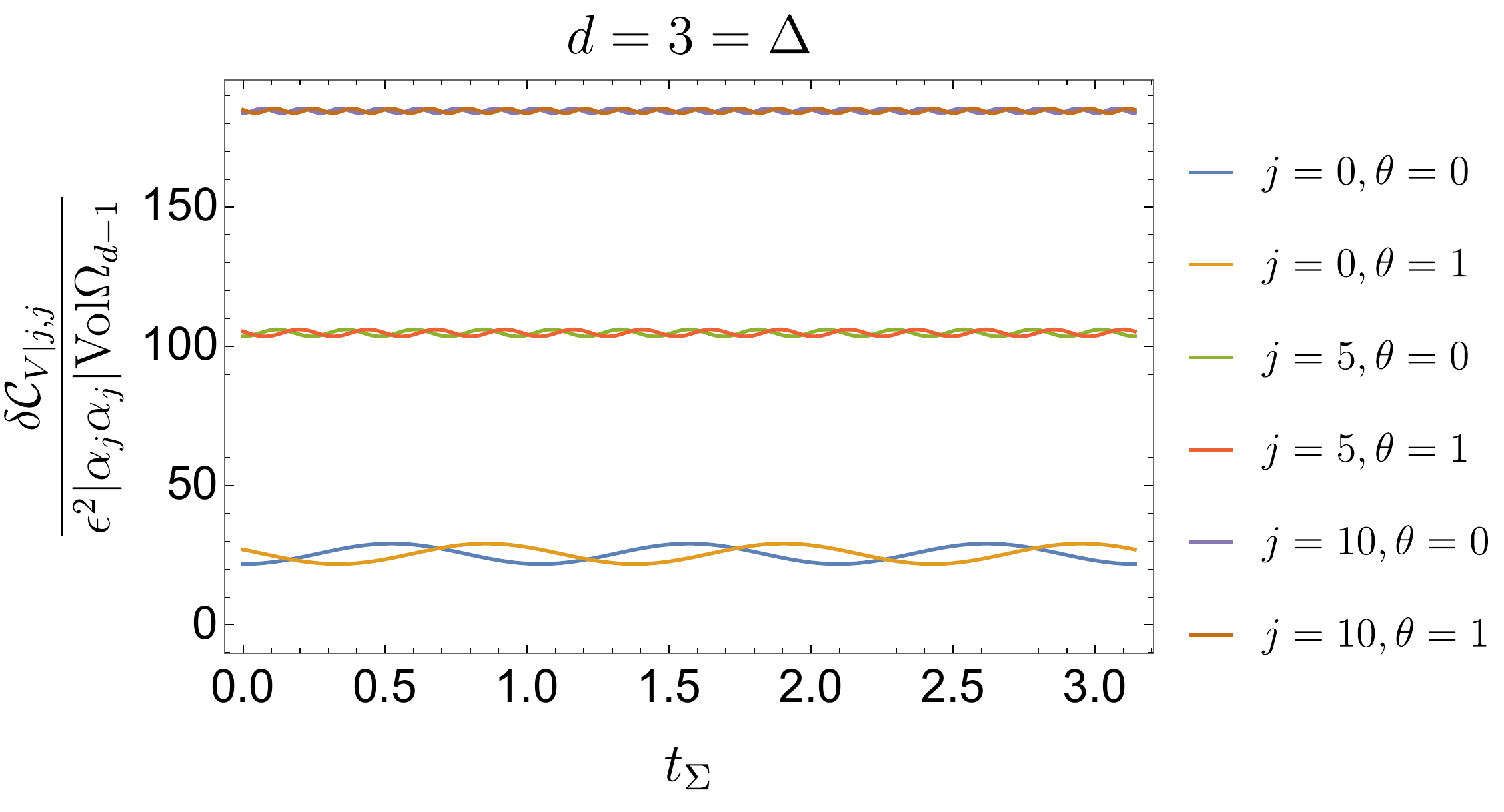}}
	\caption{\emph{Left}: Volume variation for one mode with $\theta_j=t_\Sigma =0$ and its dependence on the frequency $\omega_j=2j+\Delta$. \emph{Right}: Time dependence of $\delta \mC_{\mt V}$ for a single mode with different frequencies and $\theta_j$. Even if $j$ is an integer, for simplicity we plot the smooth function derived from the analytical expression \eqref{deltaCV_d3}. Here we have set $d= \Delta =3$ in both panels. \label{deltaCV_d=3}}
\end{figure}

In figure \ref{deltaCV_d=3}, this specific $\delta {\cal V}_{j,j}({\cal B})$ is plotted for $t_\Sigma=\theta_j=0$. Even though the contribution of the time-dependent factor $\mV_{Y}$ is maximal at this point, its boundedness makes this term subleading as soon as $j$ grows. Indeed, the large frequency limit of eq.~\eqref{deltaCV_d3} for any $t_\Sigma$ and $\theta_j$ yields
\begin{equation}
  \delta \mC_{{\mt V}|j,j} ({\cal B})\sim 16\pi \,\varepsilon^2|\alpha_j|^2 \( 2\omega_j + \frac{ \log j}{j}\sin^2 \(\omega_j t_\Sigma-\theta_j\)+ \mathcal{O}\(\frac{1}{j}\)\) \,.
\label{eq:large-jV}
\end{equation}
The first leading contribution is time independent and originates from $\mV_{T}$ due to the time independence of $T_{jj}$. The time-dependent contribution appears at order $\mathcal{O}(\log j/j)$.  In fact, the exact plots shown in the left panel of figure \ref{deltaCV_d=3} indicate the linear behaviour in $j$ remains a good approximation for small $j$. Numerically, we observe the linear behaviour
\begin{equation}
\delta \mC_{{\mt V}|j,j} ({\cal B})\sim \frac{8 j}{d-1}+\frac{2 \pi  \Delta}{d-1}  \frac{\Gamma (\Delta +1) \Gamma \left(-\frac{d}{2}+\Delta +\frac{3}{2}\right)}{\Gamma \left(\Delta +\frac{3}{2}\right) \Gamma \left(-\frac{d}{2}+\Delta +1\right)}
\end{equation}
is still valid for $j\sim \mathcal{O}(1)$ for a fixed pair $(\Delta,d)$. The right panel of figure \ref{deltaCV_d=3} confirms the subdominant nature of the $\mC_{V,Y}$ contribution because of the small modulation, in agreement with eq.~\eqref{eq:large-jV}.

\begin{figure}[htbp]
 	\centering
 	\subfigure{\includegraphics[width=0.6 \textwidth]{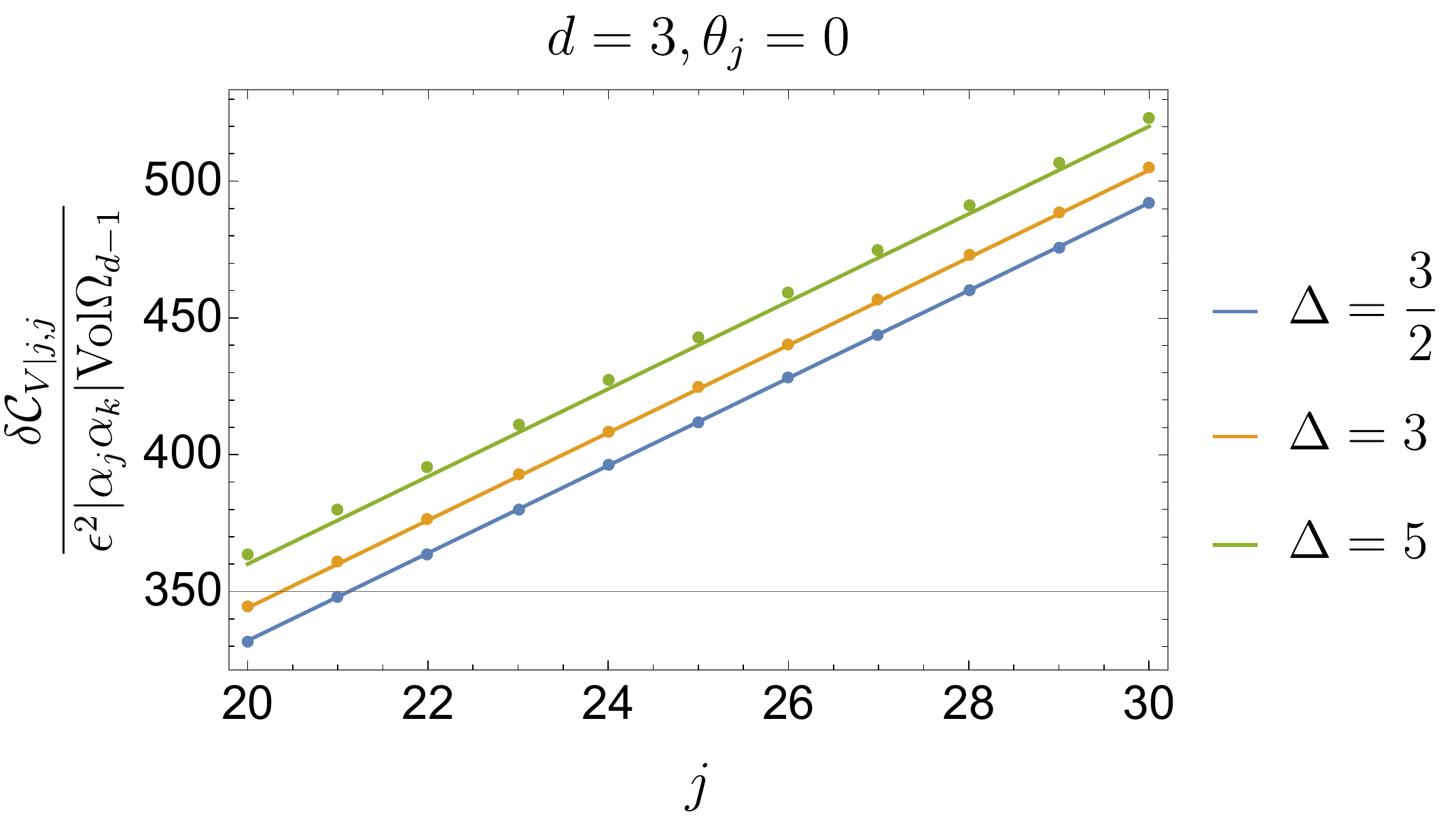}}
 	\caption{
	First two orders of $\delta \mC_{{\mt V}|j,j}$ in the large $j$ expansion for a single mode with $\theta_j=t_{\Sigma}=0$ in $d=3$ (solid line). The dots are obtained by direct evaluation of the finite sum. 
	}
\label{leading_CV_d_3}	
\end{figure}

Given our findings for $\Delta=d=3$, it is natural to analyse the large frequency limit in $\delta \mC_{{\mt V}|j,j}$ for any pair $(d,\Delta)$, with $\Delta > d/2$. This yields
\begin{equation}\label{dletaCVjj}
\delta \mC_{{\mt V}|j,j} ({\cal B})\sim \varepsilon^2|\alpha_j|^2   \,{\rm Vol}\Omega_{d-1}\(\frac{16 \omega_j}{d-1} + \mathcal{O}\(\frac{\log (j)}{j}\)\) \,.
\end{equation} 
This extends our previous large frequency $\omega_j = \Delta + 2j$ results to arbitrary $(\Delta,d)$ and confirms that time dependence appears in subleading contributions. As an example, we compare the analytical form of the complexity variation with the linear approximation in $d=3$ for few $\Delta$ in fig.~\ref{leading_CV_d_3}.

As soon as our perturbations \eqref{eq:s-mode} involve more than one mode, there will be off-diagonal contributions to the volume variation that we shall denote by $\delta \mC_{{\mt V}|j,k}$ with $j\neq k$. Let us follow a similar strategy to the one for diagonal terms and study the simpler case $\delta \mC_{{\mt V}|0,k}$ by focusing on the terms
\begin{equation}
\begin{aligned}
  \mC^\mt{V}_{0,k} &= \frac{\pi^2L^{d-1}\,{\rm Vol}\Omega_{d-1}}{4(d-1)\GN }\,A_0\,A_k\, \frac{\Gamma\left(\Delta - (d-1)/2\right)\Gamma(d/2)}{\Gamma(\Delta+k)}  \\
  &\times \sum_{m=0}^k (-1)^m \binom{k}{m} \frac{\Gamma(\Delta + k+m)}{\Gamma(\Delta + m + 3/2)} \left[\left(\Delta - \frac{(d-1)}{2}\right)(\Delta^2 + \omega_k^2) - (d\Delta + m)(d-1)\right]\,.
\end{aligned}
\label{eq:d0k}
\end{equation}
Let us focus on the large $\Delta\gg 1$ regime. Using the asymptotic expansion for the Gamma function 
\begin{equation}
  \Gamma(z) \sim z^{z-1/2}\,e^{-z}\,\sqrt{2\pi}\left[1 + \frac{1}{12z} + \frac{1}{288z^2} + \mathcal{O}(z^{-3})\right]
\label{eq:gamma-exp}
\end{equation}
for large argument $z$, together with $\Delta \gg k$, we observe
\begin{equation}
\begin{aligned}
  \frac{\Gamma(\Delta + k+m)}{\Gamma(\Delta + m + 3/2)} &\simeq \Delta^{k-3/2}\,e^{-(k-3/2)}\left(1 + \left(k-\frac{3}{2}\right) \frac{k+2m+1/2}{2\Delta} + \right. \\
  & \left. +\frac{(k-3/2)}{24\Delta^2}\left( h_1\,m^2 + h_2\,m + h_3\right) +  \mathcal{O}(\Delta^{-2})\right)
\end{aligned}  
\label{eq:gamma-exp1}
\end{equation}
where $h_i(k)$ with $i=1,2,3$ are computable functions of $k$ with highest power $h_i(k) \sim k^i$. There are two types of corrections in this expansion. One originates from the $z^{z-1/2}$ piece in eq.~\eqref{eq:gamma-exp} and gives rise to an infinite power series of the form $P_s(m)/\Delta^s$ for a polynomial of degree $s$ in the variable $m$ with $k$ dependent coefficients. The second originates from the $z^{-n}$ corrections in eq.~\eqref{eq:gamma-exp}. In our case, these give rise to terms with the same functional dependence as the ones explicitly written, but with a further $\mathcal{O}(\Delta^{-2})$ suppression, at least. This is what the notation in our expansion \eqref{eq:gamma-exp1} tries to capture.

Equipped with eq.~\eqref{eq:gamma-exp1}, we can go back to the summation over $m$ in eq.~\eqref{eq:d0k}. The first two leading contributions correspond to sums of the form:
\begin{equation}
  \sum_{m=0}^k (-1)^m \binom{k}{m} = 0\,,\,\,\, k\neq 0 \quad \quad \text{and} \quad \quad
  \sum_{m=0}^k (-1)^m \binom{k}{m}\,m = (-1)\delta_{k,1}
\label{eq:vol-explain}
\end{equation}
The first identity says that whenever we consider $\delta \mC_{{\mt V}|0,k}$ with $k\neq 0$, the leading terms in eq.~\eqref{eq:d0k} combine to cancel out. The second identity says that, among the subleading contributions, the first one to give a non-vanishing contribution is for $k=1$. One can check the off-diagonal terms with $k=1$ are $\Delta^{-1}$ suppressed with respect to the diagonal ones in the limit $\Delta \gg 1$. Off-diagonal modes with $k\geq 2$ are suppressed by, at least, $\Delta^{-2}$.

We can extend the analytic large $\Delta$ regime analysis for generic $j,k\sim\mathcal{O}(1)$. One can show the dominant contributions to $\mC^\mt{V}_{j,k}$ and $\mS^\mt{V}_{j,k}$ are equal and proportional to
\begin{equation}
  \mC^\mt{V}_{j,k} \sim \mS^\mt{V}_{j,k}  \sim \Delta\,\sum_{n=0}^j (-1)^n \binom{j}{n} \sum_{m=0}^k (-1)^m\binom{k}{m} \frac{\Gamma(m+n+d/2)}{\Gamma(m+d/2)\Gamma(n+d/2)} + \mathcal{O}(\Delta^0)
\end{equation}
This expression is symmetric in the pair $(j,k)$. Without loss of generality, let us consider $j\leq k$. To simplify the mathematical discussion, let us focus on $d=4$. The quotient of Gamma functions equals
\begin{equation}
  \frac{\Gamma(m+n+d/2)}{\Gamma(m+d/2)\Gamma(n+d/2)} = \binom{m+n+1}{m+1}\,\frac{1}{n+1} = \frac{1}{n+1} \frac{(m+n+1)(m+n)\dots (m+2)}{n!}\,.
\end{equation}
As a function of $m$, the expression above is a polynomial of degree $n$. Importantly, it is known from the theory of finite differences that
\begin{equation}
  \sum_{m=0}^k (-1)^m \binom{k}{m}\,P(k-m) = k!\,a_k
\label{eq:finite-difference}  
\end{equation}
where $P(x)$ is a polynomial of degree $k$ and $a_k$ is its k-th coefficient. It follows from these considerations that the dominant contribution to $\mC^\mt{V}_{j,k}$ and $\mS^\mt{V}_{j,k}$ can only occur for $j=k$, since it is only for $n=k$ that the above sum is non-zero and that corresponds to the upper bound on $n$, \ie $j=k$. Since for smaller values of $n$, the sum over $m$ vanishes, we conclude
\begin{equation}
  \mC^\mt{V}_{j,k} \sim \mS^\mt{V}_{j,k}   \sim \frac{\Delta}{j+1}\delta_{j,k} \qquad \text{for}\quad d=4
\end{equation}
Following similar arguments, one can also show
\begin{equation}
\begin{aligned}
  \mC^\mt{V}_{j,k} \sim \mS^\mt{V}_{j,k}    &\sim \Delta\delta_{j,k} \qquad\quad\qquad &\text{for}\quad d=2 \\
  \mC^\mt{V}_{j,k} \sim \mS^\mt{V}_{j,k}   &\sim \frac{\Delta}{(j+1)(j+2)}\delta_{j,k} \ \qquad &\text{for}\quad d=6
\end{aligned}
\end{equation}

\begin{figure}[htbp]
	\centering
	\subfigure{\includegraphics[width=0.49 \textwidth]{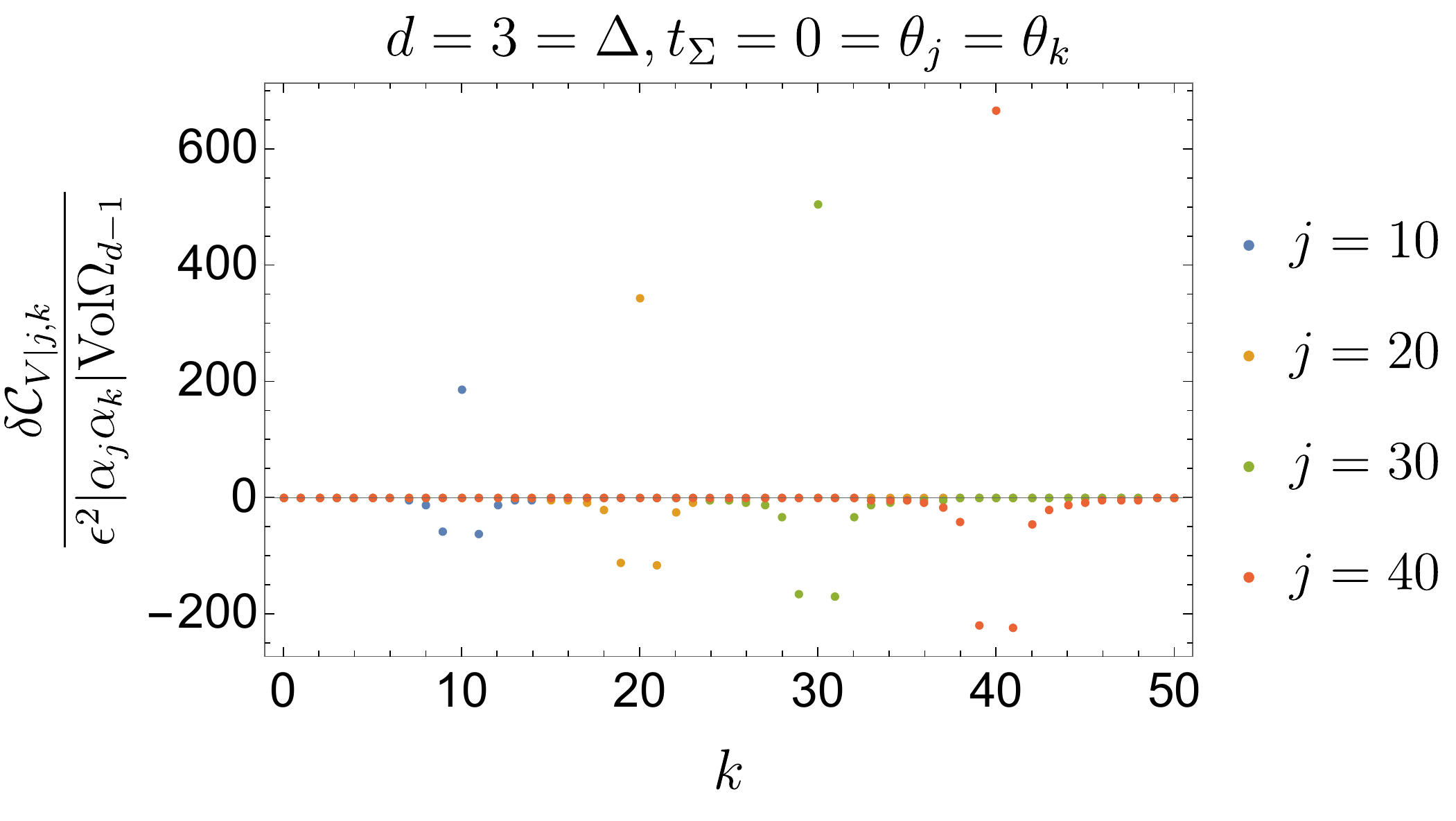}}
	\subfigure{\includegraphics[width=0.49 \textwidth]{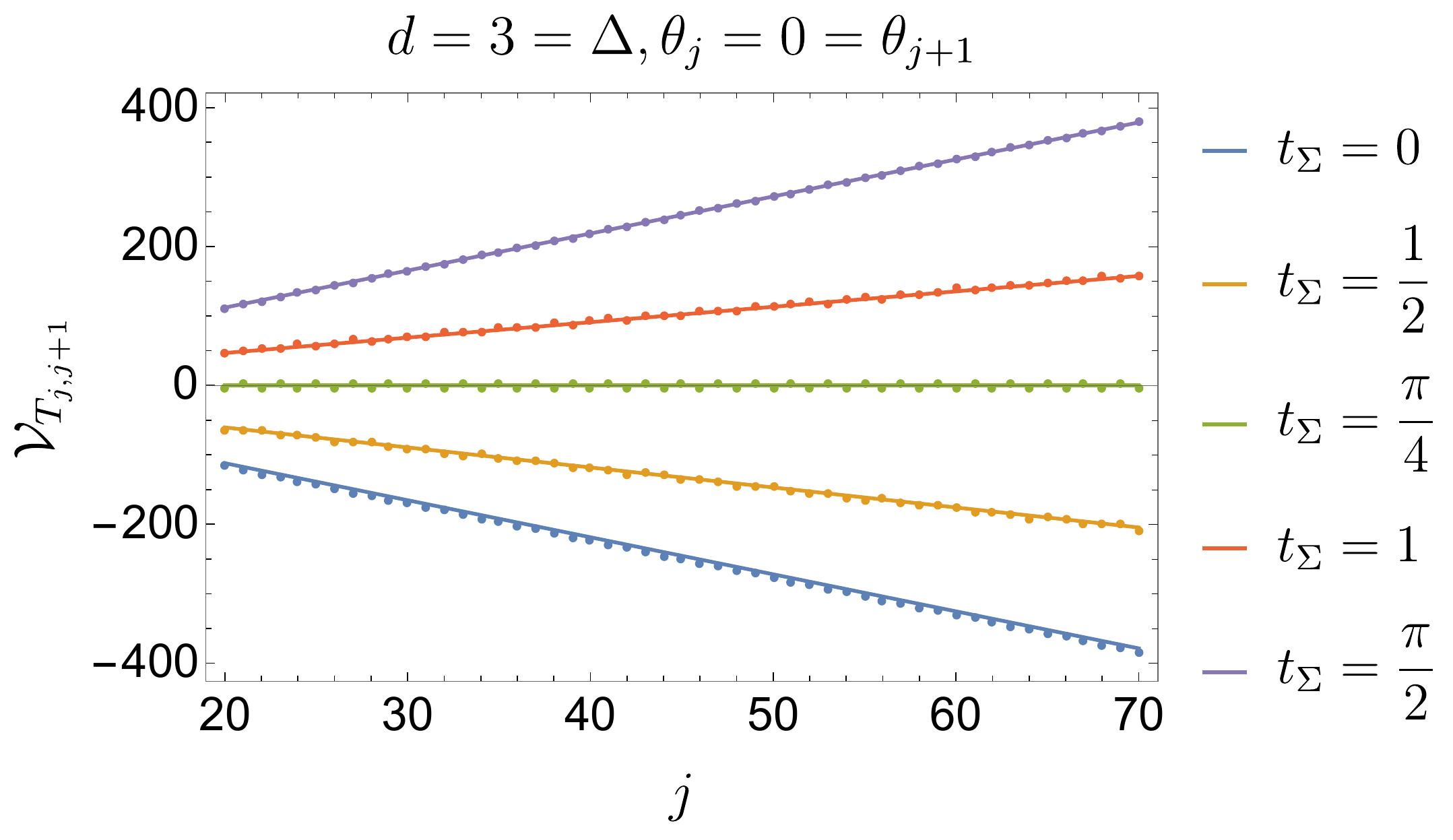}}
	\caption{\emph{Left}: Volume variation as a function of $k$ for various fixed $j$ in $d=3=\Delta$. The value of off-diagonal terms are suppressed by a factor $\frac{1}{|j-k|}$. \emph{Right}: The off-diagonal term $\mV_{T_{j,j+1}}$. The solid lines are the leading order approximation \eqref{eq:off-diag-exp}, while the dots represent the full result, obtained by direct evaluation of the finite sum.  \label{off-dia}}
\end{figure}

Motivated by our analytic large $\Delta$ results, we numerically explored the dominant contributions among the off-diagonal $\delta \mC_{{\mt V}|j,k}$ with $j\neq k$ in the left plot in fig.~\ref{off-dia}. 
These confirm the main contributions are due to $k=j\pm 1$.\footnote{In our discussion of  $\mC^\mt{V}_{0,k}$ the option $k=-1$ was not allowed. This is why we did not discover it in that special case.} This extends our previous claim to a generic choice of off-diagonal modes $(j,k)$. 
Taking the large $j\gg 1$ limit of these dominant off-diagonal contributions, we observe  
\begin{equation}
T_{j,j+1} (t)\sim 4 |\alpha_j\alpha_{j+1}| \omega_j^2 \cos\( \delta \theta +2t_\Sigma\), \qquad \delta \theta = \theta_j -\theta_{j+1} \,.
\end{equation}
This leads to the dominant off-diagonal volume variations
\begin{equation}
\begin{split}
\delta \mC_{{\mt V}|j,j\pm 1}({\cal B}) &=\varepsilon^2|\alpha_{j}\alpha_{j\pm 1}| {\rm Vol}\Omega_{d-1}   \( \mV_{T_{j,j\pm 1}}-\mV_{Y_{j,j\pm 1}} \)\,,\\
\text{with }\qquad \mV_{T_{j,j+1}}&\sim -\frac{4}{3} \left(\frac{4 (\Delta - 1)}{d-1}+\frac{8 j}{d-1}\right)\cos\( \delta \theta +2t_\Sigma\) +\mathcal{O}\(\frac{\log j}{j}\) \\
&\sim -\frac{16 \left(\omega _j- 1\right)}{3 (d-1)}\cos\( \delta \theta +2t_{\Sigma}\)  \,,\\
\qquad \mV_{T_{j,j-1}}&\sim -\frac{16 \left(\omega _j- 3\right)}{3 (d-1)}\cos\( \delta \theta +2t_{\Sigma}\) \,,
\end{split}
\label{eq:off-diag-exp}
\end{equation}
where we only kept the first two leading order contributions (see the right panel in fig.~\ref{off-dia}).\footnote{There is no contradiction between the claim \eqref{eq:off-diag-exp} and the large $\Delta$ behaviour of $\mC^\mt{V}_{j,k}$ and $\mS^\mt{V}_{j,k}$. The latter was computed in the regime where $j,k\sim\mathcal{O}(1)$, whereas the former requires $k=j\pm 1 \gg 1$. Technically, the larger the values of $(j,k)$ are, the more difficult are the sums in $n$ and $m$ appearing in $\mC^\mt{V}_{j,k}$ and $\mS^\mt{V}_{j,k}$.}  Time dependence makes the volume variation oscillate between positive and negative values.

\paragraph{Numerical results for $\delta \mC_{\mt V}$.} To further support the previous analytic considerations and to ease the comparison with the CA discussion in section~\ref{sec:CA-result}, below we present some extra numerical results for $\delta \mC_{\mt V}$.

\begin{figure}[htbp]
	\centering
	\subfigure{\includegraphics[width=0.49 \textwidth]{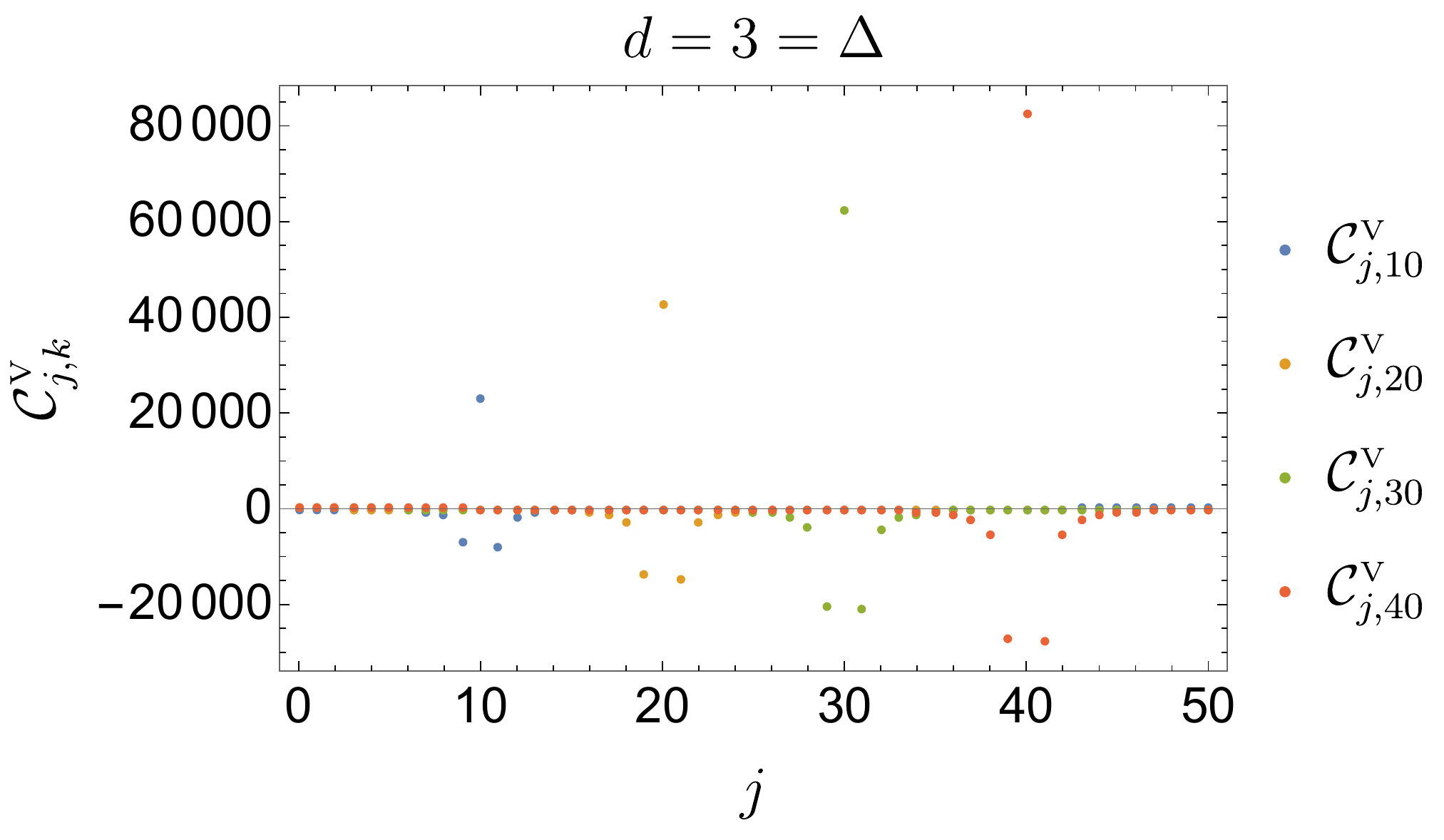}}
	\subfigure{\includegraphics[width=0.49\textwidth]{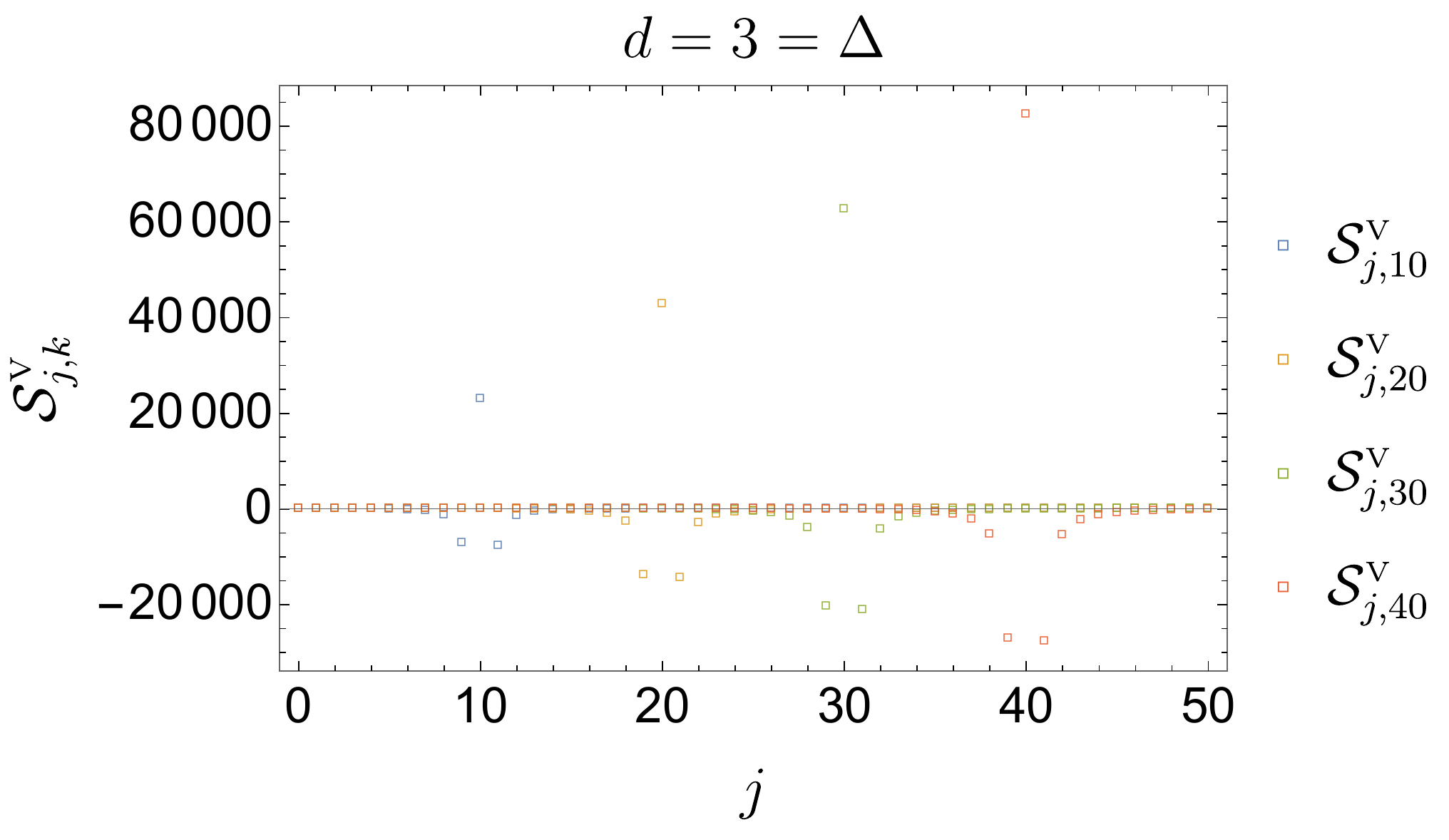}}
	\caption{Different values of $\mC^\mt{V}_{j,k}$ (\emph{Left}) and $\mS^\mt{V}_{j,k}$ (\emph{Right}) as a function of $j$, for various fixed $k$ and $ d =3= \Delta$. Both are clearly dominated by the diagonal terms $j=k$, which show linear growth in $j$ in agreement with \eqref{dletaCVjj}.}
\label{CV_DP}
\end{figure}

First, consider the coefficients $\mC^\mt{V}_{j,k}$ and $\mS^\mt{V}_{j,k}$ controlling $\delta \mC_{\mt V}$.
These are plotted in figure~\ref{CV_DP}  in the case $d=3=\Delta$ for fixed $k$ as a function of $j$, where we can see how fast the amplitudes decay to zero away from $j=k\pm 1$. The dominant contributions are indeed diagonal and they increase linearly in agreement with eq.~\eqref{dletaCVjj}. 

\begin{figure}[htbp]
	\centering
	\subfigure{\includegraphics[width=0.49 \textwidth]{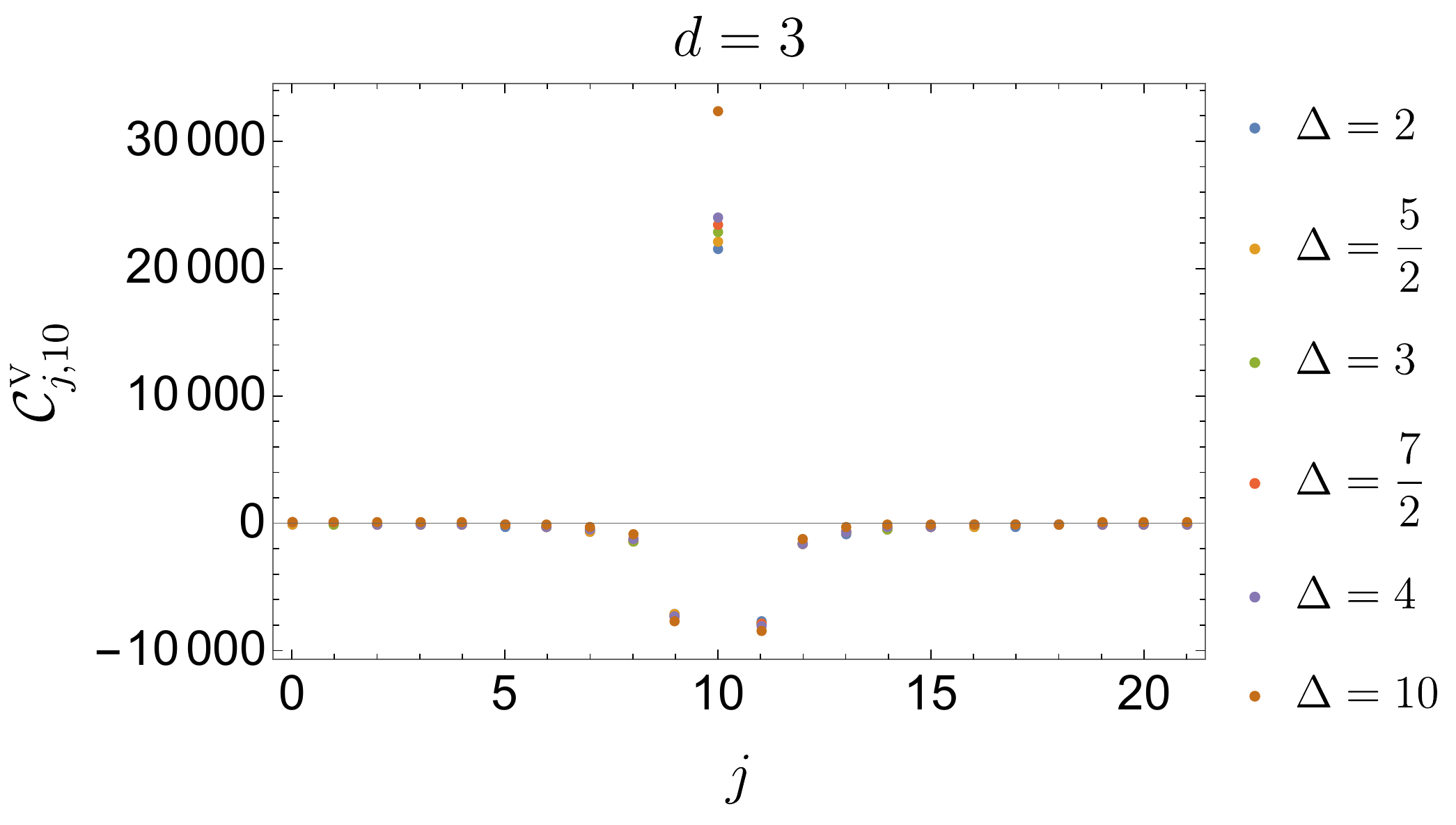}}
	\subfigure{\includegraphics[width=0.49 \textwidth]{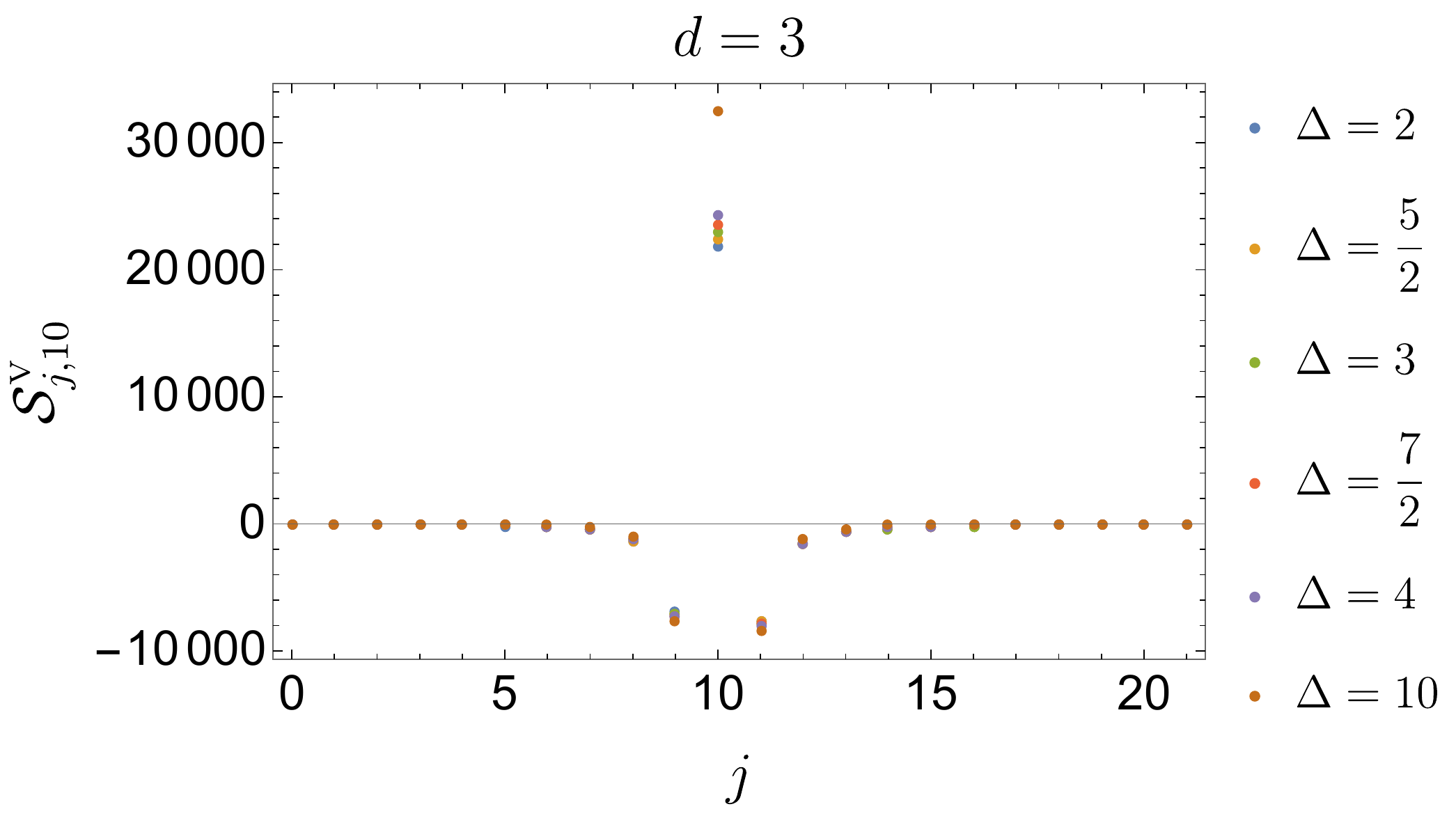}}
	\caption{$\mathcal{C}^\mt{V}_{j,k}$ (\emph{Left}) and $\mathcal{S}^\mt{V}_{j,k}$ (\emph{Right}) as a function of $j$ for various $\Delta$, $d=3$ and $k=10$. Notice the difference in amplitudes with respect to those of $\delta \mC^\mt{A}$, as shown in figure~\ref{deltaC_Delta}.}
 \label{CV_DP02}
\end{figure}

To study the dependence on $\Delta$, we consider the  amplitudes $\mC^\mt{V}_{j,10}$ and $\mS^\mt{V}_{j,10}$ for $d=3$ 
in figure \ref{CV_DP02}. Once more, we observe the dominant contribution stems from $j=k$, with a value that increases linearly in $\Delta$ and a fast decay in the amplitudes whenever $j \neq k\pm 1$, independently of the value of $\Delta$.
 
\subsubsection{Connection with previous work}
\label{sec:connection}

Similar independent discussions have appeared in \cite{Belin:2018bpg,Jacobson:2018ahi} relating the volume variation to an integral of the matter stress tensor perturbation, as in \eqref{eq:our-vol-var}, even though their derivations are different. We show below their results are equivalent to our explicit volume extremisation.

First, in \cite{Belin:2018bpg},\footnote{Let us add here that the authors of \cite{Belin:2018fxe,Belin:2018bpg} suggested a boundary interpretation of holographic complexity using the CV approach. However, this interpretation was in terms of the complexity using the Fubini-Study metric (analogous to the approach introduced in \cite{Chapman:2017rqy}) but relative to the vacuum state. That is, the UV divergent complexity of the AdS vacuum was set to zero with a new renormalization scheme. Hence their suggestion diverges somewhat from the perspective that guides our present work.} the volume (complexity) variation is related to the integral of the matter stress tensor
\be
  2 (d-1)\, \d {\cal V} ({\cal B})= \int_{\tau =0}d^dx \sqrt{|h|} \d T_{\tau \tau}^{\rm scalar}\,,
\label{eq:belin}
\ee
on the spacelike surface defined by setting the timelike WDW coordinate $\tau=0$.  The description of the AdS$_{d+1}$ geometry in \cite{Belin:2018bpg} uses FRW coordinates
\begin{equation}
  ds^2 = L^2\[-d\tau^2 + \cos^2\tau\,d\Sigma_d^2\]\,,
\label{eq:eads}
\end{equation}
where $d\Sigma_d^2$ is a $d$-dimensional Euclidean AdS metric with unit curvature, \ie an Einstein metric satisfying $R_{ab} = -(d-1)h_{ab}$. These coordinates cover a single WDW patch in the AdS$_{d+1}$ background with $-\frac\pi2\le\tau\le\frac\pi2$.

We can match this metric with the global coordinates \reef{eq:gads} on AdS$_{d+1}$ by first rewriting the spatial part in \eqref{eq:eads} as 
\be
  ds^2 = L^2\[-d \tau^2 + \cos^2 \tau \( dy^2+\sinh^2\!y\,d\Omega^2_{d-1}\)\]\,,
\ee
where $d\Omega_{d-1}$ is the line element on a unit ($d$--1)-sphere, followed by the coordinate transformation 
\beqa
\tan t&=&\frac{\tan\tau}{\cosh y}\,,
\nonumber\\
\tan\rho&=&\cos\tau\,\sinh y\,.
\label{eq:3dchange}
\eeqa
Notice the surface at $\tau=0$ corresponds to the constant time slice $t=0$ and satisfies $\left.\frac{\partial \rho}{\partial \tau}\right|_{\tau=0} = 0$. Hence, the stress tensor components restricted to the surface $\tau=0$ transform as
\begin{equation}
  \left.\delta T^{\mt{scalar}}_{\tau\tau}\right|_{\tau =0} = \varepsilon^2\left. \left(\frac{\partial t}{\partial\tau}\right)^2 \,T^{\mt{bulk}}_{tt}\ \right|_{\tau =0} = \varepsilon^2\left.\cos^2\!\rho \ T^{\mt{bulk}}_{tt}\,\right|_{\tau =0}\,.
\end{equation}
Plugging this into eq.~\reef{eq:belin}, our expression for $\delta\mV$ in eq.~\reef{eq:our-vol-var} is recovered.
 

Second, the authors in \cite{Jacobson:2018ahi} use Wald's formalism to derive the first law for causal diamonds, the domain of  causal dependence of a bulk region  $B$. 
As pointed out in  \cite{Jacobson:2018ahi}, taking the bulk region to be a spacelike $(d-1)$-dimensional ball of radius $R$ in AdS,  the infinite  volume limit  $R/L\to \infty$ gives rise  to a first law for WDW patches of $\text{AdS}$
\begin{equation}
  \delta H^{\mt{matter}}_\zeta = -\frac{\kappa}{8\pi G_{{\mt{N}}}} \left[\delta  A - \frac{d-1}{L}\delta {\cal V}\right]\,.
\label{eq:WDWlaw}
\end{equation}
On the left hand side of this expression $\zeta$ is the conformal Killing vector of the (undeformed) causal diamond, \ie the generator of the conformal isometry that preserves the causal diamond, and  $\delta H^{\mt{matter}}_\zeta$ is the matter Hamiltonian associated with the flow generated by $\zeta$.  The right hand side is purely geometrical:  $\kappa$ is the (constant) surface gravity on the conformal Killing horizon associated to $\zeta$, which coincides with the boundary of the WDW patch.  $\delta {\cal V}$ stands for the volume variation of the maximal slice of the causal diamond, 
whereas $\delta A$ is the variation in the area of the edge of the WDW patch.   The latter  corresponds to a variation in the AdS boundary metric at $t=t_\Sigma$, which vanishes for the type of perturbations we consider in this work.  Thus, once more, we are left to relate $\delta {\cal V} $ to an integral of the matter stress tensor. 

Relating the  notation and  conventions of \cite{Jacobson:2018ahi} to the one used in our work,\footnote{In particular, the definition of the stress energy tensor in \cite{Jacobson:2018ahi} differs by a factor $\varepsilon^2/16 \pi G_N$ from ours. Furthermore, they use coordinates where global AdS is
\begin{equation}\label{getrid}
  ds^2 = -(1+r^2/L^2)\,d\hat t^2 + \frac{dr^2}{1+(r^2/L^2)} + r^2\,d\Omega_{d-1}^2\,,
\end{equation}
with the conformal Killing vector $\zeta$, defined as $\mathcal{L}_{\zeta}g_{ab}=2\alpha g_{ab}$,  is given by
\be
\zeta  = \frac{L^2}{R}\[\( 1- \frac{\sqrt{1+ (R/L)^2}}{\sqrt{1+ (r/L)^2}} \cos \frac{\hat t}{L}\) \del_{\hat t} - \frac{r}{L} \sqrt{\( 1+ (R/L)^2 \) \(1+ (r/L)^2\)}\cos \frac{\hat t}{L}\del_{r} \]\,,
\ee 
with $\alpha$ an arbitrary function.} we are led to consider the WDW patch anchored to the boundary time slice $\Sigma$ at $t_\Sigma = 0$. Specialising to our pertubative setup 
\begin{equation}
  \delta H^{\mt{matter}}_\zeta = \frac{\varepsilon^2}{16 \pi G_N} \int_{t= 0} \sqrt{h} \, T^{\mt{bulk}}_{\mu\nu}\, \zeta^\mu \, s^\nu \, ,
\end{equation}
with $s$  the future directed unit normal to the AdS $t_\Sigma=0$ slice, which in the global AdS coordinates \eqref{eq:gads} reads
\be
s^{\mu} \del_\mu= \frac{\cos \rho}{L} \, \del_t \,  .
\ee
 Translating the results of \cite{Jacobson:2018ahi} to the AdS coordinates \eqref{eq:gads}  and taking the large volume limit yields \begin{equation}
\left.\zeta_{\mt{WDW}}\right|_{t_\Sigma=0}  = \cos \rho \, \del_t \qquad \text{with} \qquad  \kappa =1\, 
\end{equation}
 (see \cite{Jacobson:2018ahi} for the details). Substituting the explicit expressions into \eqref{eq:WDWlaw} we thus obtain
\be
\delta  {\cal V}    = \frac{\varepsilon^2}{2 (d-1)} \int_{t_\Sigma=0} \sqrt{h} \cos^2\rho \, T^{bulk}_{tt} \, 
\ee
 which exactly matches \eqref{eq:our-vol-var}.
 
Hence we may conclude that our results will agree with those arising from the analysis of coherent states in both \cite{Belin:2018bpg} and \cite{Jacobson:2018ahi}.

%% file: sections/CACVcomparison.tex
\subsection{Comparing CA and CV results}
\label{sec:comparison}

In eqs.~\eqref{eq:CAvar-final} and \eqref{eq:finalvol-var}, the holographic CA and CV variations  between coherent states of small amplitude $\varepsilon$ in global AdS 
are written as
\begin{equation}\label{deltaCACV}
 \begin{split}
 \delta \mC_{\mt A}& = \frac{\varepsilon^2}{\pi^2}\sum_{j,k} |\alpha_j\alpha_k| \left(\cos \(\omega_j t_\Sigma - \theta_j\)\cos \(\omega_k t_\Sigma - \theta_k\) \mC^{\mt A}_{j,k} 
 + \sin\(\omega_j t_\Sigma - \theta_j\)\sin\(\omega_k t_\Sigma - \theta_k\) \mS^{\mt A}_{j,k} \right)\,,\\
 \delta \mC_{\mt V} &= \frac{\varepsilon^2}{\pi^2} 
 \sum_{j,k} |\alpha_j\alpha_k|\left(\cos(\omega_j t_\Sigma-\theta_j)\,\cos(\omega_k t_\Sigma-\theta_k)\,\mC^{\mt V}_{j,k} + \sin(\omega_j t_\Sigma-\theta_j)\,\sin(\omega_k t_\Sigma-\theta_k)\,\mS^{\mt V}_{j,k}\right)\,.\\
 \end{split}
 \end{equation}
Since the source of the perturbation is the matter scalar field $\Phi_{\mt{cl}}$ in \eqref{eq:s-mode}, both quantities were expected to have the same quadratic dependence on the amplitudes $|\alpha_j|$ of the modes and to have a time dependence through the combination $\omega_j t_\Sigma-\theta_j$, because the coherent states are parameterized by the amplitudes $|\alpha_j|$ and phases $\theta_j$ for each of the modes. This is \emph{not} to say that time dependence is the same effect in both quantities since the two variations depend on the distinct holographic amplitudes
 $\mC^{\mt A}_{j,k},\mS^{\mt A}_{j,k}$ and $\mathcal{C}^{\mt V}_{j,k},\mS^{\mt V}_{j,k}$. Below, we summarize the main similarities and differences between the holographic results presented in sections \ref{c-action} and \ref{s-volume}.
 
Both holographic complexities are second order in the (small) amplitude  of the coherent states, \ie $\delta \mC_{\mt A}\sim \delta \mC_{\mt V} \sim \mathcal{O}(\varepsilon^2)$. This is obvious holographically since the source of the perturbation is the matter field and the latter backreacts on the metric and to the action at second order. In the discussion section, we will comment on the consequences of this fact when comparing with the quantum circuit complexity first law \eqref{eq:first-law}.

It may be natural to expect that the variation of the holographic complexity should always be positive for perturbations around the AdS vacuum, as considered in this work.\footnote{If true, this would imply that the holographic complexity of the vacuum state maybe some kind of minimum for some choices of reference state and complexity measures.  
} However, our calculations show this is \emph{not} the case for $\delta \mC_{\mt V}$, though it does hold for $\delta \mC_{\mt A}$. A window appears where $\delta \mC_{\mt V} <0$ because for relevant operators in the boundary CFT, the corresponding bulk scalar has a tachyonic mass, \ie $m_\Phi^2<0$. 

Regarding the behaviour of the different amplitudes $\mC^{\mt A}_{j,k},\mS^{\mt A}_{j,k}$ and $\mathcal{C}^{\mt V}_{j,k},\mS^{\mt V}_{j,k}$, the main features and comparisons can be summarized as follows:
\begin{itemize}
\item When one of the excited modes satisfies $j\gg 1$, both $\delta \mC_{\mt A}$ and $\delta \mC_{\mt V}$ are dominated by diagonal amplitude contributions, \ie $k=j$. However, whereas $\delta \mC_{\mt A}$ decays as $\frac{\log j}{j}$,  $\delta \mC_{\mt V}\sim \omega_j = \Delta + 2j$ increases linearly. This linear behaviour remains a good approximation for $j=k\sim \mathcal{O}(1)$, whereas $\delta \mC_{\mt A}$ has more structure in this regime and it is generically more sensitive to the conformal dimension $\Delta$ of the boundary perturbation.

\item  When the coherent state perturbation involves more than a single mode, both $\delta \mC_{\mt A}$ and $\delta \mC_{\mt V}$ contain off-diagonal terms. These are subleading when one of these modes satisfies $j\gg 1$. However, these effects subleading decay more slowly with the distance $|j-k|$ in $\delta \mC_{\mt A}$. In fact, the decay in $\delta \mC_{\mt V}$ is so fast that these contributions mainly come from $k=j\pm 1$. This fact remains a good approximation for $j\sim \mathcal{O}(1)$, whereas $\delta \mC_{\mt A}$ shows more structure on the details of the modes.

\item  Although the time dependence shown in eq.~\eqref{deltaCACV} is the same for $\delta \mC_{\mt A}$ and $\delta \mC_{\mt V}$, the amplitudes of the various terms are very different. 
Time dependence is a sub-leading effect in $\delta\mC_{\mt V}$, since the leading contribution $T_{jj}$ is time-independent as indicated in eq.~\eqref{eq:large-jV}. On the other hand, time oscillations are leading effects for  $\delta\mC_{\mt A}$. 
\end{itemize}

%% file: sections/circuit.tex
In section~\ref{sec:coherent}, we proposed to explore the first law of complexity \eqref{eq:first-law} in a set-up, involving the large-N limit of the AdS/CFT correspondence, where both, the holographic complexity conjectures \eqref{defineCV} and \eqref{defineCA}, together with the quantum circuit complexity \eqref{costD}, or \eqref{costD2}, were computable using the free Hilbert space \eqref{Fock}, describing both the generalized free field in the boundary CFT and the dual scalar field $\hat{\Phi}$ propagating in the bulk AdS geometry \eqref{scalq}. Having evaluated and discussed the holographic complexity variations $\delta\mC_\mt{A}$ and $\delta\mC_\mt{V}$ in section \ref{sec:hol-comp}, we now turn to the calculation of the corresponding quantum circuit complexity variation.

As stressed in section~\ref{sec:coherent}, within this free Hilbert space, both the initial target state, \ie the vacuum, and the perturbed target states, \ie the coherent states \eqref{eq:Da}, are Gaussian states. 
This calls to mind the techniques developed in \cite{Jeff,cohere} to evaluate the circuit complexity of Gaussian states in a free quantum field theory. Here we apply these techniques to consider variations in the complexity of a free scalar in a fixed AdS$_{d+1}$ background. These calculations very explicitly reveal the underlying circuits and trajectories, and our expectation is that this exercise will allow us to  develop new insight and intuition for our holographic results of section~\ref{sec:hol-comp} -- see discussion in section \ref{discuss}. While we will summarize the results for the various cost functions studied in the previous works, we will focus on cost functions that exhibit some qualitative similarities to the holographic complexity results, such as the $\kappa=2$ complexity introduced in \cite{Jeff}. 

We develop the free quantum scalar field formalism introduced in section~\ref{sec:bulk-coh} to use the results of \cite{cohere} to evaluate circuit complexity of coherent states. We extend \cite{cohere}  to allow for non-zero momentum Gaussian states, a necessary step to describe the time evolution in circuit complexity, \ie to follow the variations of the complexity as the state evolves in time.  We use the covariance matrix and displacement vector representation of Gaussian states and find a canonical $\mathbb{R}^{2N} \ltimes \mathrm{Sp}(2N,\mathbb{R})$ algebra of gates generated by linear and quadratic field operators, extending the results of~\cite{Jeff,cohere}. 

The $N$ specifying this algebra arises from the UV cutoff. In previous work, the field theory cutoff was implemented with a lattice regularization for which the number of modes $N \sim V/\delta^d$, where $V$ is the spatial volume of the lattice and $\delta$ is the lattice spacing. In the present case of a free scalar in a fixed AdS background, recall from eq.~\reef{eigenwaves} that the modes are labeled by discrete quantum numbers $\vec n = (j,\ell, \vec m)$. Hence we implement the UV cutoff here by only keeping the lowest $N$ energy eigenmodes, \ie we will focus on spherically symmetric configurations and so only consider the modes with $j\le N$ and $\ell=0=\vec m$. Because the quantum number $j$ is related to the number of nodes in the radial profile, this choice effectively introduces a short distance $\delta\sim L/N$ in our analysis.

Calculations similar to previous works \cite{Jeff,cohere} show their results extend in a natural way: 1) the optimal circuits for non-coherent Gaussian states are straight line geodesics in the $\mathbb{R}^N$ subspace of scaling gates, and 2) geodesics for coherent states with perturbatively small amplitudes in $K$ modes remain in $\(\mathbb{H}^2\){}^K \times \mathbb{R}^{N-K}$ subspaces  of scaling and shifting gates.  

\subsection{Quantized scalar field in $\text{AdS}_{d+1}$} \label{sec:quantizedscalar}

Consider the same massive real scalar field $\Phi$ propagating in AdS$_{d+1}$ described in section~\ref{sec:bulk-coh}, with action \eqref{eq:class-action} and eigenfunctions $u_\n(y^\mu)$  in eq.~\eqref{eigenwaves}. The canonical quantum scalar field can be decomposed into creation and annihilation operators as in eq.~\eqref{scalq}
\begin{equation}
  \hat{\Phi}(y^\mu) = \sum_{\n} \left(u_\n(y^\mu)\,\hat{a}_\n + u^*_\n(y^\mu)\,\hat{a}^\dagger_\n\right)\,,
\label{eq:Phi-1}
\end{equation}
acting on the Hilbert space defined at some Cauchy surface $\Sigma_t$. From the classical conjugate momentum 
\begin{equation}
\Pi(y^\mu) \equiv \frac{\delta {\cal L}}{\delta\,\partial_t{\Phi}(y^\mu)} = - {\frac{ \sqrt{-g} }{16\pi G_N}}\,g^{tt}\, \partial_t{\Phi}(y^\mu)\,,
\end{equation}
where we used the fact that the metric AdS metric \reef{eq:adsmetric} is diagonal, one defines the canonically conjugate momentum operator. This can be expanded in the same basis of creation and annihilation operators as 
\begin{equation}
\label{eq:pi to a}
\hat{\Pi}(y^\mu) = i \frac{ \sqrt{-g} }{16\pi G_N} \,g^{tt} \sum_\n \omega_\n \left( u_\n(y^\mu)\,\hat{a}_\n - u^*_\n(y^\mu)\, \hat{a}^\dagger_\n\right)\,,
\end{equation}
satisfying the standard commutation relations 
\begin{equation}
[\hat{\Phi}(x^\mu),\hat{\Pi}(y^\mu)]= i\,\delta^{(d)}(x^\mu,y^\mu)\,,
\end{equation}
where $\delta^{(d)}(x,y)$ is the generalized delta-function on $\Sigma_t$. The expansions~\eqref{eq:Phi-1} and~\eqref{eq:pi to a} can be inverted using the inner product \eqref{innerp} to find 
\begin{equation}
\begin{aligned}
\hat{a}^\dagger_\n =- \int_{\Sigma_t}  d^dy \left(\frac{  \sqrt{-g}}{16\pi G_N}\, \omega_\n \, g^{tt} \, u_\n(y^\mu) \hat{\Phi}(y^\mu) + i u_\n(y^\mu) \hat{\Pi}(y^\mu) \right)\,,\\
\hat{a}_\n = - \int_{\Sigma_t } \, d^dy \left( \frac{ \sqrt{-g} }{16\pi G_N}\, \omega_\n \, g^{tt} \, u^*_\n(y^\mu) \hat{\Phi}(y^\mu)- i u^*_\n(y^\mu) \hat{\Pi}(y^\mu) \right)\,.
\end{aligned}
\end{equation}
The Hamiltonian can be factorized into
\begin{equation}
\hat{H} = \int_{\Sigma_t} \, d^dy  :\!\! \dot{\hat{\Phi}}(y^\mu) \hat{\Pi}(y^\mu)\!\!:\  = \sum_\n \omega_\n \, \hat{a}^\dagger_\n \hat{a}_\n \,.
\end{equation}
We used the normal ordering $: \cdots :$ where creation operators are moved to the left and annihilation operators to the right. 

Just like in flat spacetime, we can find the normal modes in AdS from eqs.~\eqref{eq:Phi-1} and \eqref{eq:pi to a}.
We choose, for simplicity, the Cauchy slice $\Sigma_0$ at constant time $t=0$ and implicitly choose a real basis of spherical harmonics. The general time-dependent case will simply add a position dependent phase between the two terms in the expansion in eq.~\eqref{eq:Phi-1}. With this choice $u_\n^*=u_\n$ and 
\begin{equation}
\hat{\Phi}(y^\mu) = \sum_\n \sqrt{2\omega_\n}\,u_\n(y^\mu)\,\hat{\phi}_\n\,,\qquad \hat{\Pi}(y^\mu) =-\frac{  \sqrt{-g} }{16\pi G_N} g^{tt} \sum_\n \sqrt{2\omega_\n}\,u_\n(y^\mu)\,\hat{\pi}_\n\,,  
\end{equation}
in terms of the normal modes 
\begin{equation}
\label{eq:normal}
\hat{\phi}_\n = \frac{1}{\sqrt{2\omega_\n}} \left(\hat{a}_\n+\hat{a}_{{\n}}^\dagger\right)\,, \qquad \hat{\pi}_\n =-i\sqrt{ \frac{\omega_\n}{2}} \left(\hat{a}_\n-\hat{a}_{{\n}}^\dagger\right)\, 
\end{equation}
 satisfying the commutation relation $[\hat{\phi}_\n,\hat{\pi}^\dagger_{\n'}] =i \delta_{\n\n'}$.\footnote{Notice that since $\omega_\n$ is dimensionless, also the normal mode field operators and conjugate momenta are dimensionless.} 

The Hamiltonian can be expressed in terms of the normal mode field operators as
\begin{equation}
\hat{H} = \sum_\n \frac{1}{2} \left( \hat{\pi}_\n^2 +\omega_\n^2   \hat{\phi}_\n ^2 -\omega_\n \right) \,,
\end{equation}
where the extra constant term is the zero point energy difference between choosing $\hat{a}_\n$ on the right as the definition of normal ordering, instead of $\hat{\pi}_\n$ and $\hat{\phi}_\n$. From this form it is easy to find the ground state wavefunctional
\begin{equation}
\Psi_0[\phi] \equiv \langle \phi |0\rangle \propto {\rm exp}\left[- \frac12\sum_\n  \omega_\n\phi_\n^2 \right]\,. \label{eq:GS}
\end{equation}

With this background out of the way, we can now summarize the circuit complexity construction in the context of free quantum field theory in $\text{AdS}_{d+1}$.

\subsection{Circuit complexity of a free scalar}
\label{sec:wavefunction}
 
We here set up the ingredients that enter in the computation of circuit complexity: the reference and target states, the set of elementary gates and the choice of cost function. 

\paragraph{Coherent Gaussian states.} Nielsen's geometric approach to quantum circuit complexity was applied to free QFTs in \cite{Jeff}. This formalism was later developed  in~\cite{cohere} for bosonic coherent Gaussian states  of the form
\begin{equation}
  \Psi_T[\phi] \equiv \langle \phi| \alpha \rangle \propto {\rm exp}\left[-\frac12 \sum_\n \omega_\n \(\phi_\n-\sqrt{\frac{2}{\omega_\n}}\alpha_\n\)^2\right] \, , 
\end{equation}
with vanishing conjugate momenta $\langle \hat{\pi}_\n\rangle$. 
Here, to study the first law, and make contact with the previous sections,  we are interested in the general case. Hence we must extend the formalism in~\cite{cohere} to include target states with non-vanishing momentum $\langle \hat{\pi}_\n\rangle$.  These correspond to complex $\alpha_\n=|\alpha_\n|e^{i\theta_\n}$ with wave functions
\begin{equation}
	\label{eq:target}
	\Psi_T[\phi] \equiv \langle \phi| \alpha_\n \rangle \propto {\rm exp}\left\{ \sum_\n \left[-\frac{\omega_\n}{2} \left(\phi_\n-\sqrt{\frac{2}{\omega_\n}}|\alpha_\n|\cos \theta_\n \right)^2 + i \sqrt{2\, \omega_\n} |\alpha_\n| \sin \theta_\n \, \phi_\n\ \right] \right\}
\end{equation}
having non-zero first moments
\begin{equation} \label{eq:firstmoments}
	\langle \alpha_\n | \hat{\phi}_\n |\alpha_\n \rangle = \sqrt{\frac{2}{\omega_\n}}|\alpha_\n| \cos \theta_\n \,,\qquad \langle \alpha_\n | \hat{\pi}_\n |\alpha_\n \rangle = \sqrt{2\,\omega_\n} \, |\alpha_\n|\sin \theta_\n\,.  
\end{equation}	
As described in section \ref{sec:bulk-coh}, these states can be generated from the vacuum by the action of the displacement operator \eqref{eq:Da}
\begin{equation}
|\alpha_\n \rangle = e^{D(\alpha_\n)}|0\rangle\,, \quad \text{where} \quad D(\alpha_\n) =  \sum_\n \left(\alpha_\n \hat{a}_\n^\dagger  - \alpha_\n^* \hat{a}_\n \right)\,.
\end{equation}
As in previous literature, we shall use as the reference state wave function
\begin{equation}
\label{eq:reference}
\Psi_R[\phi] \equiv \langle \phi |\Psi_R\rangle \propto {\rm exp}\left[-\frac \mu 2 \sum_\n \phi_\n^2 \right] 
\end{equation}
where $\mu$ is the intrinsic frequency of the chosen reference state. This corresponds to the product state with no entanglement between the modes. 

An equivalent way of describing bosonic Gaussian states is through the expectation value of field operators and conjugate momenta, and their second momenta.  
Higher point correlation functions are simply related to these two by Wick's theorem. This formalism turns out to be convenient when dealing with computations of quantum circuit complexity for determining the unitary $U(s=1)$.

For that, we collect into a vector $\hat{\xi}$ the field and conjugate momentum operators
\begin{equation} \label{eq:canonicalvar}
\hat{\xi}_A =
 \begin{pmatrix}
\hat{\phi}_\n \\
\hat{\pi}_\n
\end{pmatrix} \,.
\end{equation} 
The displacement vector $z$  and covariance matrix $G$ are then defined as
\beq\label{eq:Gz}
z_A \equiv \langle \hat{\xi}_A \rangle \,, \quad \quad  G_{AB} \equiv \langle \hat{\xi}_A \hat{\xi}_B+\hat{\xi}_B  \hat{\xi}_A \rangle -2 z_A z_B\,,
\eeq
and a general coherent state of the form \eqref{eq:target} is then  fully specified by 
\beq \label{eq:zandG}
z = \frac{\sqrt{2} |\alpha_\n|}{\sqrt{\omega_n}}
\begin{pmatrix} 
\cos \theta_\n \\
\omega_\n \, \sin \theta_\n 
\end{pmatrix} \, ,
\qquad  G = \begin{pmatrix}
	\frac{1}{\omega_\n} & 0 \\ 0 & \omega_\n
\end{pmatrix}\, .
\eeq

\paragraph{Gate set.} Next, we discuss the gates generating states~\eqref{eq:target}. Before we begin, we note that for fields in flat space, a lattice regularization was introduced in~\cite{Jeff} to simplify the discussion. In the AdS background, we can instead use the countable mode decomposition in eq.~\eqref{eq:Phi-1} and truncate the modes with very large quantum numbers. This naturally gives a cutoff in which only $N$ modes are left. 

The natural set of Hermitian generators $\hat{\cal O}_I$ for Gaussian states with $\langle \hat{\phi}_\n\rangle = \langle \hat{\pi}_\n\rangle = 0$ are the generators of $\mathrm{Sp}(2N,\mathbb{R})$
\begin{equation}
\label{eq:gates1}
\hat{\cal O}_{AB} = \frac{\hat{\xi}_A \hat{\xi}_{B}+ \hat{\xi}_{B} \hat{\xi}_A}{2 \chi_{AB}}\,,
\end{equation}
where 
$\chi_{AB}$ are yet undetermined dimensionless coefficients.\footnote{In the previous literature~\cite{Jeff,cohere,Chapman:2018hou}, the implicit choice $\frac{1}{\chi_{AB}} = 1$ was taken for the $GL(N,\mathbb{R})$ subgroup generated by the off-diagonal block~\eqref{squeezz}, while in~\cite{Chapman:2018hou}, for the diagonal blocks the coefficients depended on a gate scale $\omega_g$ (see eqs. (37) and (59) in~\cite{Chapman:2018hou}). } For real Gaussian states~\cite{Jeff,cohere}, the set of gates studied was the $\mathrm{GL}(N,\mathbb{R})$ subgroup of $\mathrm{Sp}(2N,\mathbb{R})$ generated by the ``off-diagonal'' block
\beq\label{squeezz}
\hat{{\cal O}}_{\pi_\n\phi_\m}= \frac{\hat{\pi}_\n \hat{\phi}_\m+ \hat{\phi}_\m \hat{\pi}_\n}{2\chi_{\pi_\n\phi_\m}}\,,
\eeq
which satisfy
\beq
\label{eq:com-1}
\left[ \hat{\cal O}_{\pi_\n\phi_\m},\hat{\cal O}_{\pi_\o\phi_\p}\right]= \frac{i}{\chi_{\pi_\n\phi_\m} \chi_{\pi_\o\phi_\p}} \left(\delta_{\m\o} \chi_{\pi_\n\phi_\p} \hat{\cal O}_{\pi_\n \phi_\p}-\delta_{\n\p} \chi_{\pi_\o\phi_\m}\hat{\cal O}_{\pi_\o \phi_\m}\right)\,.
\eeq
Requiring that these commutators have canonical normalization for all $\n$, $\m$, $\o$ and $\p$ fixes $\chi_{\pi_\n\phi_\m} = \frac{f_\n}{f_\m}$ for some coefficients $f_\n$. For the more general case of $\mathrm{Sp}(2N,\mathbb{R})$, we also have generators which are quadratic in $\hat{\phi}_\n$ and in $\hat{\pi}_\n$:
\beq
\label{eq:newgens}
\hat{{\cal O}}_{\phi_\n\phi_\m}= \frac{\hat{\phi}_\n \hat{\phi}_\m+ \hat{\phi}_\m \hat{\phi}_\n}{2\chi_{\phi_\n\phi_\m}}\,, \quad \hat{{\cal O}}_{\pi_\n\pi_\m}= \frac{\hat{\pi}_\n \hat{\pi}_\m+ \hat{\pi}_\m \hat{\pi}_\n}{2\chi_{\pi_\n\pi_\m}}\,.
\eeq
These generate two abelian subgroups
\beq
\label{eq:com-2}
\left[ \hat{\cal O}_{\phi_\n\phi_\m},\hat{\cal O}_{\phi_\o\phi_\p}\right]=0\,, \quad \left[ \hat{\cal O}_{\pi_\n\pi_\m},\hat{\cal O}_{\pi_\o\pi_\p}\right]=0\,,
\eeq
that are invariant under conjugations by the $\mathrm{GL}(N,\mathbb{R})$ group discussed above
\beq
\label{eq:com-3}
\begin{aligned}
	\left[ \hat{\cal O}_{\phi_\n\phi_\m},\hat{\cal O}_{\pi_\o\phi_\p}\right] &= \frac{i}{\chi_{\phi_\n\phi_\m}\chi_{\pi_\o\phi_\p}} \left(\delta_{\n\p}\chi_{\phi_\o\phi_\m} \hat{\cal O}_{\phi_\o \phi_\m}+\delta_{\m\p} \chi_{\phi_\o\phi_\n} \hat{\cal O}_{\phi_\o \phi_\n}\right)\,,\\	
\left[ \hat{\cal O}_{\pi_\n\pi_\m},\hat{\cal O}_{\pi_\o\phi_\p}\right] &= \frac{-i}{\chi_{\pi_\n\pi_\m}\chi_{\pi_\o\phi_\p}} \left(\delta_{\n\p}\chi_{\pi_\o\pi_\m} \hat{\cal O}_{\pi_\o \pi_\m}+\delta_{\m\p} \chi_{\pi_\o\pi_\n} \hat{\cal O}_{\pi_\o \pi_\n}\right)\,.
\end{aligned}
\eeq
These subgroups fail to be normal subgroups because they are not invariant under conjugation by one another
\beq
\label{eq:com-4}
	\left[ \hat{\cal O}_{\phi_\n\phi_\m},\hat{\cal O}_{\pi_\o\pi_\p}\right] = \frac{i}{\chi_{\phi_\n\phi_\m} \chi_{\pi_\o\pi_\p}} \left(\delta_{\n\o} \hat{\cal O}_{\pi_\p \phi_\m}+\delta_{\n\p} \hat{\cal O}_{\pi_\o \phi_\m}+\delta_{\m\o} \hat{\cal O}_{\pi_\p \phi_\n}+\delta_{\m\p} \hat{\cal O}_{\pi_\o \phi_\n}\right)\,.
\eeq
The commutation relations for the full $\mathrm{Sp}(2N,\mathbb{R})$ generators are given by eqs.~\eqref{eq:com-1}, \eqref{eq:com-2}, \eqref{eq:com-3} and \eqref{eq:com-4}. For these to have a canonical normalization for all $\n$, $\m$, $\o$ and $\p$ fixes all $\chi$'s up to a sequence $\{f_\n\}$
\beq
\label{eq:chis}
\chi_{\pi_\n\phi_\m} = \frac{f_\n}{f_\m}\,, \quad 
\chi_{\phi_\n\phi_\m} = \left(f_\n f_\m\right)^{-1}\,, \quad \chi_{\pi_\n\pi_\m} =  f_\n f_\m\,.
\eeq
There are two natural choices for the sequence $\{f_\n\}$. One is to set all the $f_\n = 1$, for which the $\mathrm{Sp}(2N,\mathbb{R})$ generators have unit normalization in terms of the normal modes operators $\hat{\phi}_\n,\hat{\pi}_\n$. This normalization was adopted in \cite{Jeff,cohere}. The second natural choice is to  set  $f_\n = \sqrt{\omega_\n}$, so that $\mathrm{Sp}(2N,\mathbb{R})$ generators have unit normalization when written in terms of the vacuum creation and annihilation operators $\hat{a}^\dagger_\n,\,\hat{a}_\n$. 

For coherent Gaussian states, with non-vanishing first moments, we also need to include the displacement operators in space and momentum as part of the set of elementary gates
\begin{equation}
\label{eq:gates2}
\hat{\cal O}_{0\pi_\n} =\bar{\phi}_\n \hat{\pi}_{\n}  \,, \quad \hat{\cal O}_{0 \phi_\n} = \bar{\pi}_{\n} \hat{\phi}_\n\,,
\end{equation}
where  $\bar{\phi}_\n$ and $\bar{\pi}_\n$ are dimensionless parameters fixing a gate scale. They specify how much the fields are shifted by applying one of the displacement operators. For example
\begin{equation}
\Psi'[\phi] = \langle \phi|\hat{Q}_{0\pi_\k }| \Psi \rangle = \langle \phi|e^{i\epsilon\hat{\cal O}_{0\pi_\k }}|\Psi\rangle =\langle \phi'|\Psi \rangle =  \Psi[\phi']
\end{equation}
where 
$\phi_\n = \phi_\n$, except for $\n=\k$ where $\phi'_\k = \phi_\k+\epsilon\bar{\phi}_\k$, and $\eps$ is an infinitesimal parameter. When the momentum displacement operators are added, the algebra is not closed since
\begin{equation}
\label{eq:com-5}
[\hat{\cal O}_{0\pi_\n},\hat{\cal O}_{0\phi_{n'}}] = -i \bar{\phi}_\n \bar{\pi}_{n'} \delta_{nn'}\,.
\end{equation}
However, the commutator of these gates is simply the generator of an overall phase rotation
\begin{equation}
\langle \phi | e^{[\hat{\cal O}_{0\pi_\n},\hat{\cal O}_{0\phi_{n'}}]}|\Psi\rangle = e^{-i \bar{\phi}_\n \bar{\pi}_{n'} \delta_{nn'}} \Psi[\phi]\,,
\end{equation}
which is trivial since quantum states live in a projective Hilbert space where $e^{i\theta}|\Psi \rangle \sim |\Psi \rangle$. We can therefore proceed to quotient the phase gate subgroup. After this quotient, the displacement gates form an abelian $\mathbb{R}^{2N}$ subgroup since the right hand side of eq.~\eqref{eq:com-5} vanishes and
\beq
\label{eq:com-6}
[\hat{\cal O}_{0\phi_\n},\hat{\cal O}_{0\phi_{n'}}] = [\hat{\cal O}_{0\pi_\n},\hat{\cal O}_{0\pi_{n'}}] =0\,.
\eeq 
More precisely, the displacement gates form a normal $\mathbb{R}^{2N}$ subgroup
\beq
\begin{aligned}
\label{eq:com-7}
[ \hat{\cal O}_{0 \phi_\n}, \hat{\cal O}_{\phi_\m \phi_\p} ] &= 0 \quad & [\hat{\cal O}_{0\pi_\n},\hat{\cal O}_{\pi_{m}\pi_\p}]&= 0  \\
[ \hat{\cal O}_{0 \phi_\n}, \hat{\cal O}_{\pi_\m \phi_\p} ] &= \frac{i \bar{\pi}_\n}{\chi_{\pi_\m\phi_\p}} \frac{\delta_{\n\m}}{\bar{\pi}_\p} \hat{\cal O}_{0\phi_\p} \,, \quad &
[ \hat{\cal O}_{0 \phi_\n}, \hat{\cal O}_{\pi_\m \pi_\p} ] &= \frac{i \bar{\pi}_\n}{\chi_{\pi_\m\pi_\p}} \left(\frac{\delta_{\n\m}}{\bar{\phi}_\p} \hat{\cal O}_{0\pi_\p} + \frac{\delta_{\n\p}}{\bar{\phi}_\m} \hat{\cal O}_{0\pi_\m} \right)\,,\\
[ \hat{\cal O}_{0 \pi_\n}, \hat{\cal O}_{\pi_\m \phi_\p} ]& = \frac{-i \bar{\phi}_\n}{\chi_{\pi_\m\phi_\p}} \frac{\delta_{\n\p}}{\bar{\phi}_\m} \hat{\cal O}_{0\pi_\m} \,, \quad &
 [ \hat{\cal O}_{0 \pi_\n}, \hat{\cal O}_{\phi_\m \phi_\p} ]& = \frac{-i \bar{\phi}_\n}{\chi_{\phi_\m\pi_\p}} \left(\frac{\delta_{\n\m}}{\bar{\pi}_\p} \hat{\cal O}_{0\phi_\p} + \frac{\delta_{\n\p}}{\bar{\pi}_\m} \hat{\cal O}_{0\phi_\m} \right)\,.
\end{aligned}
\eeq
Once again, demanding that this algebra be canonically normalized fixes the coefficients of the displacement gates 
\be \label{eq:chisbis}
\bar{\phi}_\n \equiv \frac{ \lambda}{\chi_{0\pi_\n}} = \lambda  f_\n^{-1} \quad \quad     \bar{\pi}_\n  \equiv \frac{ \lambda}{\chi_{0\phi_\n}} =  \lambda f_\n \ .
\ee 
The dimensionless parameter $\lambda$ arises due to symmetry of eq.~\eqref{eq:com-7} under  rescaling of the translation gates. We will see below that it is notationally convenient to set it to $\lambda=\sqrt{2}$.

The group structure of the elementary gates is therefore affine symplectic transformation, \ie $\mathbb{R}^{2N} \rtimes \mathrm{Sp}(2N,\mathbb{R})$, and the algebra is given by eqs.~\eqref{eq:com-1}, \eqref{eq:com-2},\eqref{eq:com-3}, \eqref{eq:com-4} and \eqref{eq:com-7}. The action of the elementary gates can be illustrated by the following examples (for $f_\n = 1$)\footnote{The action of the Gaussian integral case is illustrated in the simplest case $\vec k = \vec k'$ by  
	\begin{equation}
	\begin{aligned}
	e^{\epsilon \, \partial_{x}^{2}} f(x) &=\frac{1}{\sqrt{4 \pi \epsilon}} \int_{-\infty}^{\infty} \exp \left(-\frac{(x-y)^{2}}{4 \epsilon}\right) f(y) d y \,.
	\end{aligned}
	\end{equation}
}
\begin{equation}\label{eq:gatesall}
\begin{aligned}
	&\langle \phi |\hat{Q}_{\pi_\k \phi_\k }|\Psi\rangle = e^{\epsilon/2}\Psi[\phi'] &   \quad   {\rm scale}\quad&     \phi_\k \to \phi_\k '= e^\epsilon \phi_\k \, , \\
	&\langle \phi |\hat{Q}_{\pi_\k \phi_{\k'}}|\Psi\rangle = \Psi[\phi']\ &   \quad    {\rm shift} \quad &     \phi_\k \to \phi_\k ' = \phi_\k + \epsilon \phi_{\k'} \quad ({\rm entangling~gates})\,,\\
	&\langle \phi |\hat{Q}_{0\pi_\k }|\Psi\rangle = \Psi[\phi'] &    \quad      {\rm shift} \quad  &      \phi_\k \to \phi_\k ' = \phi_\k + \epsilon \bar{\phi}_{\k}\,, \\
	&\langle \phi |\hat{Q}_{0\phi_\k }|\Psi\rangle = e^{i\epsilon \bar{\pi}_\k \phi_\k }\Psi[\phi] &     \quad    {\rm shift}   \quad &  \pi_\k \to \pi_\k ' = \pi_\k + \epsilon \bar{\pi}_{\k}\,, \\
	&\langle \phi |\hat{Q}_{\phi_\k \phi_{\k'}}|\Psi\rangle = e^{i\epsilon \phi_\k \phi_{\k'}}\Psi[\phi] &\quad    {\rm phase ~shift}\quad&  \theta \to \theta' = \theta + \epsilon \phi_\k \phi_{\k'}\,, \\
	&\langle \phi |\hat{Q}_{\pi_\k \pi_{\k'}}|\Psi\rangle = e^{-i\epsilon \partial_{\phi_\k }  \partial_{\phi_{\k'}}}\Psi[\phi] 	&\quad    \text{Gaussian integral.}  \quad & 
\end{aligned}
\end{equation}

\paragraph{Cost functions.} In the following, we focus on two classes of cost functions because of the similarity of the corresponding complexities with results in holographic complexity, \eg see \cite{Jeff,Chapman:2017rqy,Chapman:2018hou}. One class, introduced in \cite{Jeff}, takes the form 
\begin{equation}\label{eq:kappa}
	F_\kappa(U,Y)=\sum_I \left|Y^I\right|^\kappa~.
\end{equation}
These $\kappa$ cost functions  can be thought of as a generalization of the $F_1$ cost function in eq.~\reef{function_F}. The corresponding vacuum complexity compares well with the results from holographic complexity \cite{Jeff}, but these cost functions do not satisfy the homogeneity property, \ie the cost \reef{costD} is not invariant under reparametrization of $s$. We also note that the $\kappa=2$ cost function will yield exactly the same extremal trajectories or optimal circuits as the $F_2$ cost function in eq.~\reef{function_F}. Another interesting suggestion in \cite{Hackl:2018ptj} was to construct a family of new cost functions using the Schatten norm (\eg see \cite{bhatia2013matrix,watrous2018theory,gil2003operator}) 
\beq
F_p(U,Y)=\Vert V \Vert_p =\Big[
{\rm Tr}\!\(\( V^\dagger\,V\)^{p/2}\)\Big]^{1/p}\,,
\label{Schatten}
\eeq 
where $V=Y^I(s)\mO_I$ is the tangent vector defined as an operator which transforms the states -- see further discussion in \cite{cohere}. These cost functions satisfy all of the desired properties and further are independent of the particular choice of basis for the $\mO_I$ -- another issue for the $F_1$ measure and the general $\kappa$ cost functions (for $\kappa\ne2$) \cite{Jeff}.
The geometry on the space of unitaries is smooth for the $\kappa=2$ cost functions, while for the $\kappa=1$ and the Schatten $p=1$ cost functions, the resulting spaces have a generalized ``Manhattan metric''. In particular, within these two broad classes, we specialize in the $\kappa = 1,2$ and $p=1,2$ costs, which are the ones that have been mostly studied in the literature \cite{Jeff,Hackl:2018ptj,cohere}.

\subsection{Circuit complexity for coherent states}

We here set up the formalism in the general case $\mR^{2N}\rtimes \mathrm{Sp}(2N,\mathbb{R})$, review the results of \cite{Jeff,cohere} for coherent states with vanishing conjugate momenta and extend their analysis to the general coherent target states of the form \eqref{eq:target}.

 To build a  representation of the quantum circuit and its action on  coherent Gaussian states,  we extend the definition of the canonical linear variable operator  \eqref{eq:canonicalvar}  to the $2N + 1$ vector \footnote{With a notation similar to the one used for the generators we will indicate the components as  $A = \{0,\hat{\phi}_\n,\hat{\pi}_\n\}$.}
\begin{equation} \label{eq:canvarext}
\hat{\xi}_A =
 \begin{pmatrix}
 1\\
\hat{\phi}_\n \\
\hat{\pi}_\n
\end{pmatrix}\, .
\end{equation} 
The definition of covariance matrix and displacement operators can be extended accordingly in a straightforward manner, giving for the general coherent state
\be
z= \begin{pmatrix} 
1\\
\sqrt{ \frac{ 2 }{ \omega_n} } |\alpha_\n| \cos \theta_\n  \\
\sqrt{2 \omega_\n} |\alpha_\n| \,\sin \theta_\n 
\end{pmatrix} \, ,
\qquad   G = \begin{pmatrix}
	0\,  &0 &0\\
	0 & \frac{1}{\omega_\n} & 0  \\
	0 & 0 & \omega_\n
\end{pmatrix}\, . 
 \ee
The action of the circuit  is then simply represented as 
\beq\label{eq:transGz}
 G(s) = U(s)^T G_R \, U(s)  \,, \quad \quad  z(s) = U(s)^T z_R \,,
\eeq
 where
 \beq\label{eq:Umat}
U = \begin{pmatrix}
	1 & \mathbf{u}^T\\
	\textbf{0} & \tilde{U} 
\end{pmatrix}\,,
\eeq
with $\tilde{U} \in  \mathrm{Sp}(2N,\mathbb{R})$ and $\mathbf{u} \in \mathbb{R}^{2N}$. The subscript $R$ here indicates the reference state \eqref{eq:reference}, reflecting  the boundary conditions
\be
\begin{aligned}
G(0) &= G_R\, ,   \qquad &z(0) &= z_R  \,  \\
G(1) &= G_T \, ,  \qquad  &z(1) &= z_T \, .
\end{aligned}
\ee

In order to geometrize the problem we rewrite the circuit as in \eqref{unitaries} in terms  of  instantaneous control functions $Y^I(s)$ and gate generators $M_I$
\begin{equation}
U(\sigma) = \cev{\cal P}{\rm exp}  \int_0^\sigma ds  \sum_{I} Y^{I}(s) M_{I}  \, .
\end{equation}
Here $I = \{0 \phi_\n, 0 \phi_\n,\phi_\n \phi_\m,  \pi_\n \phi_\m, \pi_\n \pi_\m \}$  labels the different gates discussed in the previous section. The  explicit  representation of each $M_I$ can be found  evaluating the action of the gate generators  on  \eqref{eq:canvarext}  
\beq
\[\hat{\cal O}_I,\hat{\xi}\] =  i M^T_I \hat{\xi}\,.
\eeq
 The generators $M_{\pi_\n \phi_\m}$ of the $\mathrm{GL}(N,\mathbb{R})$ subgroup are diagonal blocks 
\beq
\[M_{\pi_\n \phi_\m}\]_{\alpha \beta} =\frac{1}{\chi_{\pi_\n\phi_\m}} \(  \delta_{\pi_\n \alpha}  \delta_{\pi_\m \beta}-\delta_{\phi_\m \alpha}  \delta_{\phi_\n \beta}\)   \, ,
\eeq
while the  $M_{\phi_\n \phi_\m}$ and  $M_{\pi_\n \pi_\m}$ generators are off-diagonal blocks
\beq
\begin{aligned}
\[M_{\phi_\n\phi_\m}\]_{\alpha \beta}&= \frac{1}{\chi_{\phi_\n\phi_\m}} \( \delta_{\phi_\n \alpha}\delta_{\pi_\m\beta}+  \delta_{\phi_\m \alpha}\delta_{\pi_\n \beta}\) \,,\\
\[M_{\pi_\n\pi_\m}\]_{\alpha \beta}&= -\frac{1}{\chi_{\pi_\n\pi_\m} }\( \delta_{\pi_\n \alpha}\delta_{\phi_\m\beta}+  \delta_{\pi_\m \alpha}\delta_{\phi_\n \beta}\) \,,
\end{aligned}
\eeq
and the remaining $\mathbb{R}^{2N}$  generators form a vector 
\begin{equation}
\begin{aligned}%
\[M_{0 \phi_\n}\]_{\alpha \beta} &= \frac{\lambda}{\chi_{0\phi_\n}}\delta_{0 \alpha}  \delta_{\pi_\n \beta}     \, , \\
 \[M_{0 \pi_\n}\]_{\alpha \beta} &= -\frac{\lambda}{\chi_{0\pi_\n}} \delta_{0 \alpha}  \delta_{\phi_\n \beta}   \, .
\end{aligned}
\end{equation}
Schematically, the different $M_I$'s appear in the following block form
 \be  \label{eq:blockformM}
M_I = \begin{pmatrix}
	0 & M_{0 \pi } & M_{0 \phi }\\
	\textbf{0} & M_{\pi  \phi } &  M_{\phi \phi } \\
	\textbf{0} & M_{\pi  \pi } &  M_{\pi \phi } 
\end{pmatrix}\,. 
\eeq
Picking for convenience $\lambda=\sqrt{2}$ as anticipated
\beq
\label{eq:normalization}
{\rm Tr}\, M_I M^T_J = \frac{2 \delta_{IJ}}{\chi^2_I} \,,
\eeq
and the corresponding control functions $Y^{I}(s)$  are then given by  
\beq
\label{eq:Y}
Y^I(s) = \frac{1}{2} \chi_{I}^{2}\, {\rm Tr} \left(\partial_sU(s) U^{-1}(s) M^T_I \right)\, .
\eeq

With these results, it is straightforward to derive the ``geometry'' defined by the cost
 \eqref{costD} for a given choice of cost function \eqref{eq:kappa} or \eqref{Schatten}. For example the $\kappa=2$ measure is
\begin{align}
{\cal D}_{\kappa =2}(U) &= \frac{1}{4}\int_0^1 ds  \sum_I\left( \chi_{I}^{2}\, {\rm Tr} \left(\partial_sU(s) U^{-1}(s) M^T_I \right)\right)^2 \, , \label{eq:costk2}
\end{align}
and the other cost functions we consider, \ie $\kappa = 1$ and $p=1,2$, have analogous expressions.
From these, one derives the geodesic equation in the space of unitaries and solves for the optimal trajectory that computes the corresponding complexity measure \eqref{compD}. This procedure was carried out in detail in \cite{cohere} for coherent states with vanishing conjugate momentum. We review those steps in appendix~\ref{sec:simple} for the $\kappa =2$ cost function, and extend the derivation to arbitrary coherent states of the form \eqref{eq:target}. We now here summarize the main findings. 
 
\paragraph{Single mode coherent states with $\langle \alpha_\n | \hat{\pi}_\n |\alpha_\n \rangle=0$.} The analysis of coherent states with vanishing conjugate momentum in \cite{cohere} found that for states with a single coherent mode $|\alpha_\k \rangle$, the geodesic remains in a $\mathbb{H}^2 \times \mathbb{R}^{N-1}$ subspace, where the hyperbolic factor is spanned by the $\hat{Q}_{0 \pi_{\k}}$ and  $\hat{Q}_{\pi_\k\phi_{\k}}$  gates and the entangling gates do not enter the optimal circuit. 
The complexity of such a coherent state, with a single real $\alpha_\k$, was computed in \cite{cohere} and found to be
\begin{align}
\mC_{\kappa=1} &= \Lambda_{\k} + \sum_{\n\ne \k}   \left|\log \sqrt{\frac{\omega_\n}{\mu}}\right|\,, \label{distant9} \\ 
\mC_{\kappa=2} &=  \Delta_\k^2 + \sum_{\n \ne \k} \(\log \sqrt{\frac{\omega_\n}{\mu}}\)^2\,,  \label{lattice_distancek} \\ 
\mC_{p=1} &= |\Delta_\k | + \sum_{\n\ne \k}  \left|\log \sqrt{\frac{\omega_\n}{\mu}}\right|\,, \\
\mC_{p=2} &= \sqrt{ \mC_{\kappa=2} }\,,
\end{align}
where
\begin{equation}
\begin{split}
\Lambda_{\k}= 
\begin{cases}
 \left|\log \sqrt{\frac{\omega_\n}{\mu}}\right|+\frac{\alpha_\k \, f_\k}{\sqrt{\omega_\k}} \, {\rm min}\(1,\sqrt{\frac{\mu}{\omega_\k}} \)\,, &{\rm for}\quad {\rm min}\(1,\sqrt{\frac{\mu}{\omega_\k}} \) \le \frac{2 \sqrt{\omega_\k}}{\alpha_\k \, f_\k} \\
 \log \sqrt{\omega_\n\, \mu} + 2\,\log \frac{\alpha_\k \, f_\k}{2\,\omega_\k} +2\,, &{\rm for}\quad {\rm min}\(1,\sqrt{\frac{\mu}{\omega_\k}} \) > \frac{2 \sqrt{\omega_\k}}{\alpha_\k \, f_\k}
\end{cases}
\end{split}\,,
\end{equation}
and
\beq\label{light8}
\Delta_\k = \log \frac{\mu+\frac{\mu}{\omega_\k}\,(\alpha_\k \, f_\k)^2 +\omega_\k + \sqrt{\left(\mu+\frac{\mu}{\omega_\k} \, (\alpha_\k \, f_\k)^2 +\omega_\k \right)^2-4\,\omega_\k\,\mu }}{2\sqrt{\omega_\k\,\mu }} \,.
\eeq
 
For $\langle \hat{\phi}_\k \rangle = 0$, that is $\alpha_\k = 0$, we recover the ground state results of ~\cite{Jeff}
\begin{align}
{\cal C}_{\kappa=1}^{\rm GS}&= {\cal C}_{p=1}^{\rm GS} = \sum_{\n} \left|\log \sqrt{\frac{\omega_\n}{\mu}}\right| \,,\\
{\cal C}_{\kappa=2}^{\rm GS} &= \({\cal C}_{p=2}^{\rm GS}\)^2=  \sum_\n  \(\log \sqrt{\frac{\omega_\n}{\mu}}\)^2 \,. \label{eq:GSk2}
\end{align}
Notice that only  the scaling gates $\hat{Q}_{\pi_\n\phi_{\n}}$ appear in the optimal circuit preparing the ground state \eqref{eq:GS} and the geodesic thus lies in a flat $\mathbb{R}^{N}$ subspace.\footnote{The diagonal coefficients $\chi_{\pi_\n\phi_\n} = 1$ and the complexity of the ground state is therefore independent from the choice of  $f_\n$. The expressions above thus directly match the result of~\cite{Jeff}.}

In terms of the mode cutoff $N$, these complexities diverge as ${\cal C}_{\kappa=1} \sim {\cal C}_{\kappa=2} \sim  {\cal C}_{p=1} \sim  {\cal C}_{p=2}^2 \sim N$. 

\paragraph{Small amplitude multi-mode coherent states with $\langle \alpha_\n | \hat{\pi}_\n |\alpha_\n \rangle=0$.} When more than one coherent mode is excited, the geodesics \emph{do not} remain in the subspace of unentangled normal modes. Despite the fact that both reference and target states have no entanglement between normal modes, the optimal circuit introduces and removes entanglement in the preparation of the state \cite{cohere}.
 
However, when only a set  of $K$ modes  $\{\vec k\}$ is excited with a small amplitude $\varepsilon \alpha_\k$,  the optimal circuits turn out to remain perturbatively close to a $\left(\mathbb{H}^2\right)^K\times \mathbb{R}^{N-K}$ submanifold with no entanglement. More precisely these circuits live in this submanifold up to corrections of  $\mathcal{O}(\varepsilon^2)$.  The variation in complexity with respect to the ground state    
\beq
\delta \mC_{\mt{QFT}} \equiv \mC - \mC^{\rm GS}
\eeq
 can thus be estimated at the leading order for each cost function  by studying geodesics in the simpler $\left(\mathbb{H}^2\right)^K\times \mathbb{R}^{N-K}$ manifold. This yields  
\begin{align}\label{collections0}
\delta \mC_{\kappa=1} & = \, \varepsilon  \sum_{\omega_\k \le  \mu} \frac{\alpha_\k \, f_\k}{\sqrt{ \omega_\k}} +  \,\varepsilon \sum_{\omega_\k >  \mu} \frac{\alpha_\k \, f_\k\,\sqrt{\mu}}{\omega_\k} +  { O}(\varepsilon^3) \\
\delta\mC_{\kappa=2} &=   \varepsilon^2 \sum_\k  \frac{\log\frac{\omega_\k}{\mu} }{\omega_\k -\mu}\, \frac{\mu}{\omega_\k}\,(\alpha_\k \, f_\k)^2  +  { O}(\varepsilon^4) \,,\\
\delta\mC_{p=1} &=  \varepsilon^2 \sum_\k \frac{\mu}{\omega_\k}\,\frac{ (\alpha_\k \, f_\k)^2}{|\omega_\k -\mu|} +  { O}(\varepsilon^4) \,,
\end{align}
where the sums run over the excited modes and the subindex labels the cost function. The variation of $p=2$ complexity is simply associated with that of the $\kappa=2$ cost function 
\begin{equation}
\delta {\cal C}_{p=2} = \frac{\delta{\cal C}_{\kappa=2}}{2 {\cal C}_{p=2}}\,,
\end{equation}
due to the simple relation $\mC_{\kappa=2} = \mC_{p=2}^2$. Because for a free QFT, $\mC_{\kappa=2}\sim N$, the variation $\delta \mC_{p=2} \sim N^{-1/2}$ approaches zero when taking the cutoff to infinity. For this reason, we will not focus on the $p=2$ complexity for more general states. In the following we will also omit the $\kappa=1$ complexity because this is linear in $\alpha_\k$, unlike the holographic complexity results, which are quadratic in the amplitude of the excitation.

\paragraph{Single mode coherent states with $\langle \alpha_\n | \hat{\phi}_\n |\alpha_\n \rangle=0$.}  For target states where only one mode $\pi_\k $ is excited, $\alpha_\k$ is purely imaginary, and a straightforward extension of the results of~\cite{cohere} leads to the complexities
\begin{align}\label{lattice_distancek2}
\mC_{\kappa=2} &=  \tilde \Delta_\k^2 + \sum_{\n \ne \k} \(\log \sqrt{\frac{\omega_\n}{\mu}}\)^2\,, \\ 
\mC_{p=1} &= |\tilde \Delta_\k | + \sum_{\n\ne \k}  \left|\log \sqrt{\frac{\omega_\n}{\mu}}\right|\,, 
\end{align}
where now
\beq\label{light8tilde}
\tilde \Delta_\k = \log \frac{\mu+ \omega_\k^2(|\alpha_\k| / f_\k)^2 +\omega_\k + \sqrt{\left(\mu+ \omega_\k^2(|\alpha_\k|/ \, f_\k)^2 +\omega_\k \right)^2-4\,\omega_\k\,\mu }}{2\sqrt{\omega_\k\,\mu }} \,.
\eeq
%

\paragraph{Small amplitude multi-mode coherent states.}  For target states with small amplitude excitations $\varepsilon \alpha_\k$ for both first moments \eqref{eq:firstmoments}, the geodesics of~\eqref{eq:costk2} can be solved perturbatively to find the complexity of this state to ${O}(\varepsilon^2)$ -- see appendix \ref{sec:simple} for details of the derivation. In particular, the increase in complexity is
\begin{align}\label{small-many}
	\delta\mC_{\kappa=2} &=\varepsilon^2 \sum_\k  \frac{\log\frac{\omega_\k}{\mu} }{\omega_\k -\mu} |\alpha_\k|^2 \,\left( \frac{\mu}{\omega_\k} \,f_\k^2 \cos^2 \theta_\k +\frac{\omega_\k^2 }{f_\k^2} \sin^2 \theta_\k \right) + { O}(\varepsilon^4) \,,\\
	\delta \mC_{p=1}&=  \varepsilon^2 \sum_\k \frac{1}{|\omega_\k-\mu|}\, |\alpha_\k|^2  \left(\frac{\mu}{\omega_\k} f_\k^2 \cos^2 \theta_\k +\frac{\omega_\k^2 }{f_\k^2} \sin^2 \theta_\k \right) + { O}(\varepsilon^4) \,,
\end{align}
where the sum runs over excited modes.

\paragraph{Time evolution.} So far we focused on the $t=0$ slice, but it is immediate to extend these results to arbitrary times, as to study the complexity time dependence. For that, let us consider the time evolution of a state where at $t=0$ only one mode is excited with real $ \varepsilon \alpha_\k$, that is $\langle \hat{\phi}_\k \rangle = \sqrt{\frac{2}{\omega_k}} \, \varepsilon\alpha_\k$ and $\langle \hat{\pi}_\k \rangle =0$.  Going back to sec.~\ref{sec:quantizedscalar}-\ref{sec:wavefunction}, we see the time dependence simply reflects in the definition of the normal modes and in their expectaction values as: $\langle \hat{\phi}_\k \rangle = \sqrt{\frac{2}{\omega_\k}}\,\varepsilon\alpha_\k\, {\rm cos}(\omega_\k t)$ and  $\langle \hat{\pi}_\k \rangle = -\sqrt{2\,\omega_\k} \, \varepsilon\alpha_\k\, {\rm sin}(\omega_\k t)$. 

The variation in complexity  with respect to the ground state at any time $t$ is then given by a simple generalization of the above results:
\beq
\begin{aligned}
\label{eq:time}
\delta\mC_{\kappa=2}(t)& = \varepsilon^2 \frac{\log\frac{\omega_\k}{\mu} }{\omega_\k -\mu} |\alpha_\k|^2 \left(\frac{\mu}{\omega_\k} f_\k^2 \, \cos^2 (\omega_\k t) + \frac{\omega_\k^2 }{f_\k^2}  \sin^2 (\omega_\k t) \right)+{\cal O}(\varepsilon^4)\,,\\
\delta\mC_{p=1}(t) &= \varepsilon^2 \frac{1 }{|\omega_\k -\mu|}\,|\alpha_\k|^2 \left(\frac{\mu}{\omega_\k} f_\k^2 \, \cos^2 (\omega_\k t) + \frac{\omega_\k^2 }{f_\k^2}  \sin^2 (\omega_\k t) \right)+{\cal O}(\varepsilon^4)\,.
\end{aligned}
\eeq
Notice these complexities would be time independent if we were to fix $f_\n^4 =\omega_\n^3/\mu \,. $

Finally, for several excited modes with complex amplitudes $\varepsilon \alpha_\k= \varepsilon |\alpha_\k|e^{i\theta_\k}$, and in a notation that matches the one we used in the bulk for $\delta {\cal C}_{\rm A}$ in \eqref{eq:CAvar-final} and  $\delta {\cal C}_{\rm V}$ in \eqref{howse}, we have
\beq
\begin{aligned}
\delta {\cal C}_{\kappa=2} =\varepsilon^2 \sum_{\vec{k}} |\alpha_\k|^2 \(\mC^{\kappa=2}_\k \cos^2\(\omega_\k t-\theta_\k\)+{\cal S}^{\kappa=2}_\k \sin^2\(\omega_\k t-\theta_\k\)\)\,,\\
\label{eq:kappa2-var}
\delta {\cal C}_{p=1} =\varepsilon^2\sum_{\vec{k}} |\alpha_\k|^2 \(\mC^{p=1}_\k \cos^2\(\omega_\k t-\theta_\k\)+{\cal S}^{p=1}_\k \sin^2\(\omega_\k t-\theta_\k\)\)\,,\\
\end{aligned}
\eeq
with 
\beq
\begin{aligned}
&\mC^{\kappa=2}_\k = \frac{\log\frac{\omega_\k}{\mu} }{\omega_\k -\mu}\,\frac{\mu}{\omega_\k}\,f_\k^2 \,, \qquad &{\cal S}^{\kappa=2}_\k& = \frac{\log\frac{\omega_\k}{\mu} }{\omega_\k -\mu}\,\frac{\omega_\k^2}{f_\k^2}\,,\\
\label{eq:kappa2-amp}
&\mC^{p=1}_\k = \frac{1 }{|\omega_\k -\mu|}\,\frac{\mu}{\omega_\k}\,f_\k^2 \,, \qquad &{\cal S}^{p=1}_\k &= \frac{1 }{|\omega_\k -\mu|}\,\frac{\omega_\k^2}{f_\k^2}\,.
\end{aligned}
\eeq

%% file: sections/discuss.tex
In this paper, we made a detailed examination of the first law of complexity proposed in \cite{Bernamonti:2019zyy}. In particular, as an application of the first law, we considered variations of holographic complexity, using both the complexity=volume \reef{defineCV} and complexity=action \reef{defineCA}  conjectures, for (spherically symmetric) perturbations of the $\text{AdS}_{d+1}$ vacuum by a free scalar field. To compare with the circuit complexity techniques developed for quantum field theories, we also explored the complexity of the same coherent states for the scalar field in a fixed AdS background.

A preliminary comparison of our results using the CA and CV approaches was given in  section~\ref{sec:comparison}. At a qualitative level, the first law variations of the holographic complexity had a number of common features in both approaches. For example, comparing the form of the results in eq.~\reef{deltaCACV}, we see that the variations are second order in the amplitudes $\varepsilon|\alpha_j|$ of the coherent states; the functional form of time dependence is the same; and, for perturbations \eqref{eq:s-mode} involving more than a single mode, they include interference terms, \ie off-diagonal contributions with $j\neq k$ coming from modes with different frequencies.  Further, both our analytic calculations and numerical analysis gave evidence that the dominant contributions to both $\delta \mC_{\mt A}$ and $\delta \mC_{\mt V}$ generically come from the diagonal terms, \ie with $j=k$. 

However, it is striking how differently $\delta \mC_{\mt A}$ and $\delta \mC_{\mt V}$ behave upon closer examination. If we consider coherent states where a single mode (characterized by the radial quantum number $j$) is excited, we found that $\delta \mC_{\mt A}$ decays as $\frac{\log j}{j}$ for large $j \gg 1$, whereas $\delta \mC_{\mt V}$ increases linearly with $\omega_j = \Delta + 2j$ in the same regime. The behaviour of the off-diagonal contributions is also very different. In particular, we found that $\mC^\mt{V}_{j,k}$ and $\mS^\mt{V}_{j,k}$ appearing in $\delta \mC_{\mt V}$ decay much more rapidly with $|j-k|$ than the corresponding coefficients $\mC^\mt{A}_{j,k}$ and $\mS^\mt{A}_{j,k}$ in $\delta \mC_{\mt A}$. In fact, $\mC^\mt{V}_{j,k}$ and $\mS^\mt{V}_{j,k}$ are only significant for $k=j\pm 1$, whereas  $\mC^\mt{A}_{j,k}$ and $\mS^\mt{A}_{j,k}$ have a richer structure, especially when both $j,k\sim\mathcal{O}(1)$. The different behaviour of these coefficients then has a major impact on the time evolution for the two approaches. In particular, the time variations of $\delta \mC_{\mt V}$ are a subleading contribution, whereas the analogous  time dependence appears at the leading order for $\delta \mC_{\mt A}$. Moreover, $\delta \mC_{\mt V}$ can be negative in the window of relevant operators, while $\delta \mC_{\mt A}$ is always manifestly positive. 

These qualitative and quantitative differences must certainly be emphasized, as they definitely distinguish the complexity=action and complexity=volume approaches. In most previous studies, holographic complexity was found to behave in essentially the same way when evaluated using either of the two approaches. Of course, differences were found between these approaches but these took a more subtle form or appeared in rather exceptional situations. For example, extra logarithmic factors were found to appear in the UV divergences for the CA approach \cite{Carmi,Chapman:2018lsv}.

Interestingly, the difference found in \cite{Chapman:2018bqj,Braccia:2019xxi,Sato:2019kik} might be interpreted in terms of the response of the complexity to a perturbation. In those cases, a conformal defect or conformal boundary was inserted in the vacuum of a $d=2$ holographic CFT. This produced a new logarithmic divergence in the holographic complexity evaluated using the CV approach, while the result was unaffected for the CA approach  \cite{Chapman:2018bqj}, or only modified by finite terms \cite{Braccia:2019xxi}. Hence in analogy to our results presented here, the CV approach was more sensitive to the perturbation, \ie the defect, than the CA approach.

\subsection{Comparison of holographic and QFT results}

To test the first law in holography most stringently, we needed target states which are well understood as quantum states, \ie we need to be able to develop a good understanding of the variation $\delta x^a$ in eq.~\reef{eq:first-law}. Hence we chose the coherent states since, as discussed in section~\ref{sec:coherent}, they can be understood in the context of the Hilbert space \reef{Fock} of a free field. Of course, this is a remarkable result of the large-N limit, \ie despite the boundary CFT being a strongly coupled theory, the corresponding excitations are described by generalized free fields to leading order in $1/N$, \eg see \cite{ElShowk:2011ag,Fitzpatrick:2011jn,kaplan2013lectures,Terashima:2017gmc}. The dual description is simply given by a free scalar $\hat{\Phi}$ propagating in the bulk AdS spacetime, and the AdS/CFT correspondence dictates that both the boundary and bulk descriptions are describing the same free Hilbert space.

Further in our test of the first law, both the initial and perturbed target states, \ie the vacuum and coherent states, respectively, are Gaussian states. This observation reminds us of the techniques developed  to evaluate the circuit complexity of Gaussian states, \ie the vacuum in \cite{Jeff} and coherent states in \cite{cohere}, in a free scalar field theory using Nielsen's geometric approach \cite{nielsen2006quantum,nielsen2008,Nielsen:2006}. Hence in section \ref{sec:q-circuit}, we applied the latter to make an analogous examination of the first law of complexity with variations from the vacuum to a coherent state (with a small amplitude) for free scalar QFT in a fixed AdS$_{d+1}$ background. In this framework, the corresponding circuits are exposed, being constructed with explicit realizations of the gates and cost functions. Hence this exercise should allow us to formulate some new intuition and insights for the holographic results, where the circuits, gates and cost functions are all left very mysterious.

While we are considering more or less the same free Hilbert space in studying the variation of the complexity using the QFT and holographic techniques, we must keep in mind that this is only an approximation valid to describe certain states near the vacuum for the holographic CFT. In the latter case, the circuits of interest are actually preparing \eg the vacuum state of a strongly coupled large-N quantum field theory. From the perspective of the bulk description of the holographic framework, we imagine that the circuit begins acting on some unentangled reference state of geometric or quantum gravity degrees of freedom, which suggests that there is nothing resembling a spacetime geometry at the outset. The corresponding complexity then includes the effort needed to build up the background spacetime, as well as preparing the ground state (or coherent state) of all of  the quantum fields in this background. In contrast, the QFT calculations are all carried out with a fixed AdS$_{d+1}$ spacetime, and the corresponding circuits prepare the vacuum or coherent states of the scalar field propagating in this fixed background, \ie the circuit does not create the spacetime geometry itself. Hence while we can match the variation of the target states (\ie $\delta x^a$) in the QFT and holographic frameworks, the full circuits are certainly different but further, the behaviour at the end of the circuits may also be different. To be precise, we may find that even after projecting into the free Hilbert space, the velocity $\dot x^a$ is different in the two frameworks. Then, even if we had the same cost function in both cases, the $p^a$ would be different. Without further knowledge of the holographic circuits in the two cases, this limits our ability to make precise quantitative comparisons between the free QFT and holographic calculations.

Despite these comments, we can look for some qualitative intuition by comparing the first law results for the free QFT and holography. Our first observation is that $\delta \mC_{\mt A}$ and $\delta \mC_{\mt V}$ are \emph{second order} in the small amplitudes $\varepsilon\alpha_i$ of the coherent states. Of course, the significance of this lies in the fact that the first law \eqref{eq:first-law} includes \emph{first order} contributions in general. Hence we are learning that $\delta x^a$ must be orthogonal to the momentum $p_a$ carried by the vacuum circuit from the holographic complexity. Of course, the leading variations of the complexity are also quadratic in the amplitudes for the free scalar in section \ref{sec:q-circuit}. In the holographic calculations, the quadratic dependence can be traced to the matter field perturbation $\Phi_{\mt{cl}}$ sourcing the metric perturbation and contributing in the matter action at order $\mathcal{O}(\varepsilon^2)$. In the explicit circuit calculations for the free field, the quadratic dependence arises because whereas preparing the vacuum only makes use of the $\mathrm{GL}(N,\mathbb{R})$ subgroup of squeezing gates \reef{squeezz}, a completely new set of gates, \ie the shift gates \reef{eq:gates2}, are needed in preparing the coherent states. This makes clear the orthogonality of $\delta x^a$ to the direction of the vacuum circuit for any reasonable cost function \cite{cohere}. Hence it is reasonable to interpret the holographic results in this way, \ie the holographic circuits invoke a new set of gates in preparing coherent states.\footnote{Of course, one can produce a first-order variation by applying the first law to a coherent state. That is, we begin with a coherent state with a small but finite amplitude and then make a small increase or decrease in this amplitude. For example, the variation of circuit complexity with $\kappa=2$ cost function for this case can be directly derived from \eqref{lattice_distancek}. This situation was also considered in the discussions of \cite{Belin:2018fxe,Belin:2018bpg}.}

It is noteworthy that the $\kappa=1$ cost function \eqref{eq:kappa} is an exception to the above property. That is,  $\delta \mC_{\kappa=1}$ is first order in the small amplitudes of the coherent state, as shown in eq.~\eqref{collections0}. This cost function is positive and homogeneous, as desired \cite{Nielsen:2006}, due to the linearity on all tangent vectors $Y^I(s)$ in eq.~\eqref{eq:kappa}. On the other hand, it is not smooth and in particular, it is not smooth at zero amplitude. This prevents our derivation of the first law of complexity \eqref{eq:first-law} from applying to this case. Hence  $\delta \mC_{\kappa=1}$ can be first-order even when $p_a$ is orthogonal to the variation $\delta x^a$ at the endpoint of the geodesic. In \cite{Hackl:2018ptj,cohere}, the $p=1$ Schatten norm \eqref{Schatten} was proposed as an alternative to the $\kappa=1$ cost function, which had similar properties. However, as well as being positive and homogeneous, the $p=1$ Schatten norm provides a smooth cost function and so eq.~\eqref{eq:first-law} applies in this case. Hence, as can be seen in eq.~\eqref{collections0}, the resulting $\delta \mC_{p=1}$ is second order.

One striking difference that is evident in comparing $\delta \mC_{\mt A}$ and $\delta \mC_{\mt V}$ with $\delta \mC_\mt{QFT}$, \ie comparing eqs.~\reef{eq:CAvar-final} and \reef{howse} with eq.~\reef{eq:kappa2-var}, is that the holographic results contain off-diagonal contributions. That is, the coefficient $\mC_{j,k}$ and $\mS_{j,k}$ are generally nonvanishing for $j\ne k$ in the holographic calculations, while they are all zero in the QFT calculations unless $j=k$. Of course, we can add that for holography, the largest coefficients are still the diagonal ones. This is most evident of the CV approach where the coefficients decay extremely rapidly away from $j=k$ (see figures~\ref{off-dia}, \ref{CV_DP} and \ref{CV_DP02}, as well as the discussion around eq.~\eqref{eq:vol-explain}). With the CA approach, the off-diagonal coefficients decay but more slowly as can be seen from figures~\ref{Cplusminus}, \ref{deltaC_Delta} and \ref{deltaC_Delta2} --- see also the discussion around eq.~\eqref{eq:action-explain}.

As a result, for $\delta \mC_\mt{QFT}$, when several modes are excited in a coherent state, the variation of the complexity is simply the sum of the variations produced by the individual modes. However, this is not the case in the holographic calculations, although to a lesser extent in the CV calculations. This seems to indicate that the holographic complexity uses a much more complex cost function, at least from the perspective of the mode functions \reef{eigenwaves} of the scalar field. One might (partially) ameliorate this disparity by working with another basis to describe the excitations of the coherent states. In particular, the mode functions of the scalar are naturally orthogonal in the QFT framework but of course, this orthogonality does not extend to the holographic calculations. However, if instead, we thought of exciting localized `wave packets' of the scalar field on a given time slice, these states should be orthogonal with both the QFT and CV approaches to complexity. However, such wave packets would have a complicated time evolution as the scalar propagates through the WDW patch and so we would still not expect this basis to provide an orthogonal basis of excitations for the CA complexity. To construct an orthogonal basis for the CA approach, one might be led to consider localized wave packets on the null boundaries of the WDW patch. It would be interesting to understand if such wave packets can be consistently defined and if so, how they propagate through the WDW patch and \eg how they would appear on the extremal constant time slice at the center of this spacetime region. It may also be interesting to reformulate the quantization of the free scalar field on such null surfaces and to consider coherent states in this context, in order to compare to the CA calculations.

A feature common to the variations in eqs.~\reef{eq:CAvar-final}, \reef{howse} and \reef{eq:kappa2-var} is the oscillatory nature of the results as a given coherent state evolves in time. Of course, the details of the oscillations are very different within the two holographic approaches, as well as the QFT construction, as the magnitude of the coefficients is very different for the various terms. At first sight, the appearance of these oscillations may seem surprising for holographic complexity, though they are compatible with operator size considerations \cite{Susskind:2019ddc}. Recall that a distinguishing feature of holographic complexity was the linear growth found when considering AdS black holes. However, there is no reason that the dynamics of a system can not produce a decreasing or oscillating complexity. An essential ingredient for the linear growth exhibited by the AdS black holes is that the dual thermofield double states were probing the chaotic spectrum of high energy states in the boundary CFT. As a result, the time evolution was exploring states further and further out in the full Hilbert space of the CFT. The coherent states in our present investigation are all very close to the vacuum and so the time evolution does not take us beyond the free Hilbert space discussed above. Hence it should not be surprising that the corresponding complexity exhibits oscillations.\footnote{The time dependence of the complexity of the thermofield double state of a free scalar was studied in \cite{Chapman:2018hou}. Recall that in this case, the complexity was constant at late times (in contrast to the linear growth seen in holography) because the time evolution only explored a particular submanifold of Gaussian states within the full Hilbert state. Further, let us add that in an initial transient phase, the complexity typically exhibited damped oscillations and was seen to decrease for certain parameter choices.}

Any comparison of the holographic results to the variation of the complexity for a free massive scalar field in a fixed AdS geometry using the circuit complexity formalism developed for QFT in \cite{Jeff}, in particular for gaussian coherent states \cite{cohere}, will depend on the choice of cost functions. Consider the result for the $\kappa=2$ measure in eqs.~\eqref{eq:kappa2-var}-\eqref{eq:kappa2-amp}. The dependence on $\log \omega_{\vec{k}}$, with $\omega_{\vec{k}} \sim j$ for large radial quantum number (see eq.~\eqref{eigenenergy}) prevents any matching with $\delta \mC_{\mt V}$, since the dominant diagonal contribution of the latter scales linearly in $j$, but could be compatible with $\delta \mC_{\mt A}$, since both dominant diagonal amplitudes $\mC^\mt{A}_{j,j} \sim \mS^\mt{A}_{j,j} \sim \log j/j$ in this regime. Unfortunately, this observation is not enough to completely match the time dependence in both complexity variations. Indeed, $\mC^{\kappa =2}_j\sim\log j/j$ only if $f_{\vec{k}}^2\sim\omega_\k \sim j$, but then $\mS^{\kappa =2}_j\sim \log j$ with no $1/j$ suppression. Alternatively, $\mS^{\kappa =2}_j\sim\log j/j$ if $f_{\vec{k}}\sim j$, but then $\mC^{\kappa =2}_j\sim \log j$ with no $1/j$ suppression. Hence, although it is intriguing that a $\log j/j$ behaviour appears in both $\delta \mC_\mt{QFT}$ (with the $\kappa=2$ measure) and $\delta \mC_{\mt A}$, we do not find a complete match between the two. This conclusion differs from our earlier results in \cite{Bernamonti:2019zyy}, which were only valid for $\langle \hat{\pi}_{\k} \rangle = 0$ for all of the modes $\k$, \eg they only considered states at a moment of time symmetry. Alternatively, we could consider the variation $\delta \mC_\mt{QFT}$ for the Schatten $p=1$ measure in \eqref{eq:kappa2-var}-\eqref{eq:kappa2-amp}. Absence of logarithmic behaviour, prevents any match with $\delta \mC_{\mt A}$ for large radial quantum number $j$, but we can compare with the linear behaviour shown in $\delta \mC_{\mt V}$ in this same regime. Choosing $f^2_{\vec{k}}\sim j^3$ or $f_{\vec{k}}\sim \text{constant}$, one could reproduce the linear dependence in $j$ observed in the holographic complexity $\delta \mC_{\mt V}$, but none of these choices matches the right time dependence, since the latter is subdominant in holography.

We should also comment on the dimensionful quantities, which are left implicit with the notation adopted here. Let us consider the result in eq.~(30) of \cite{Bernamonti:2019zyy} for the variation of the QFT complexity with the $\kappa=2$ measure,
\begin{equation}\label{smalla}
\delta \mC_{\kappa=2} = \sum \frac{ 2\, \varepsilon^2 \alpha^2_n}{\hat\mu^2x_0^2\,(\omega_n/R\hat\mu-1)}\,\log\!\Big(\frac{\omega_n}{R\hat\mu}
\Big)   \, , 
\end{equation} 
where $\omega_n$ are the dimensionless eigenfrequencies in eq.~\reef{eigenenergy}, $\hat\mu$ is the dimensionful frequency characterizing the reference state, $x_0$ is a (dimensionful) scale characterizing the shift gates needed to prepare the coherent state \cite{cohere}, and $R$ is the radius of curvature characterizing the boundary geometry \reef{cylinder_metric2} (\ie which gives dimension to the frequencies as in eq.~\reef{PPP}). This expression should be compared to the same result in our current presentation of eqs.~\reef{eq:kappa2-var} and \reef{eq:kappa2-amp} with $t=0$ and $\theta_\k=0$. In our new notation, the gate scale $x_0$ is hidden in the dimensionless parameters $f_{\vec{k}}$, \eg choosing  $x_0 \hat{\mu} \sim 1$ corresponds to the choice $f_{\vec{k}} \sim \sqrt{2\omega_\k}$ in our current notation.\footnote{More generally, we have $\hat{\mu}x_0 \sim \sqrt{2\omega_\k}/f_{\vec{k}}$. Note that in \cite{Bernamonti:2019zyy}, we assumed a single gate scale for all of the modes which corresponds to fixing $\sqrt{2\omega_\k}/f_{\vec{k}}$ to a single value for all $\k$.} Similarly, the dimensionful reference frequency $\hat\mu$ is given by $\hat\mu=\mu/R$ where $\mu$ is the dimensionless frequency introduced in eq.~\reef{eq:reference}.

Now while we found the above choices provide a convenient notation, we should mention that this may seem to be an unnatural approach. By this we mean that ordinarily one would not expect the parameters defining the complexity model, \eg the gate scale(s) or the reference frequency, are related to a scale appearing in the infrared and in the definition of the target state, \eg the size $R$ or the mass $m_\Phi$.
In particular, we might want to compare the complexity of different states where these infrared parameters are varied. With this perspective for such comparisons, we should keep in mind that the parameters $f_{\vec{k}}$ and $\mu$ should be adjusted to fix the gate scale $x_0$ and the reference frequency $\hat \mu$. 

Furthermore, it is striking that the various expressions for $\delta \mC_\mt{QFT}$  implicitly involve a number of different scales, in particular, in the coefficients $\mC_\k$ and $\mS_\k$ in eq.~\reef{eq:kappa2-amp}. In contrast, in the holographic results for 
$\delta \mC_{\mt A}$ and $\delta \mC_{\mt V}$, the corresponding coefficients only depend on the dimensionless quantum numbers  which characterize the corresponding modes of the scalar field, as well as the conformal weight $\Delta$ and the spacetime dimension $d$ in the dual boundary theory,  \eg see eqs.~\reef{33}, \reef{440} or \reef{deltaCV_d3}. This does suggest that the corresponding scales in the complexity models underlying the holographic proposals should be related. That is, the holographic complexity models would set $\hat\mu x_0\sim 1$ and $\hat\mu R\sim 1$ in eq.~\reef{smalla}. While the first relation seems reasonable, as described above, the second does not, \ie we would be relating a scale in the complexity model to an infrared scale in the target state.
Hence this observation raises a curious question for our understanding of holographic complexity.

This discussion also brings to mind the proposal that the counterterm scale $\ell_\mt{ct}$ appearing in the gravitational action \reef{eq:faction} should be connected to the scale $\hat\mu$ used in defining the reference state in the corresponding circuit model \cite{Jeff,Chapman:2017rqy,Chapman:2018lsv}. However, our holographic results for CA are independent of $\ell_\mt{ct}$,  while the circuit model results for the scalar QFT depend on $\hat\mu$ (implicitly through the appearance of $\mu$). Again, the resolution of this apparent tension would be to set $\mu=\hat \mu R\sim1$, which seems an unnatural choice (as explained above).

\subsection{Interesting lessons}
\label{sec:lessons}

We would like to comment here on some aspects of our results, which may provide a broader perspective on the interpretation of the first law of complexity \eqref{eq:first-law}.

As already stressed in \cite{Bernamonti:2019zyy}, the contribution of the counterterm $I_{\mt{ct}}$ in the full gravitational action \eqref{eq:faction} is essential to achieve the cancellation among the different gravitational contributions to $\delta\mC_{\mt A}$ for the spherically symmetric matter perturbations considered in this work. This is an interesting observation on its own, highlighting the relevance of this term from another different perspective.\footnote{This term was introduced in \cite{Lehner:2016vdi} to ensure that the WDW action was invariant under reparameterizations of the null boundaries. The importance of this term in properly defining the WDW action was further elucidated and emphasized in \cite{Reynolds:2017lwq,Chapman:2018dem,Chapman:2018lsv}. } However, this cancellation of the gravitational contributions is \emph{not} true in general, as recently reported in \cite{Hashemi:2019aop}. It fails when introducing perturbations of a black hole background, or when considering less symmetric perturbations of the vacuum AdS. It would be important to understand the relevance of these statements in the broader picture of using holographic complexity to learn about spacetime reconstruction.

Independently of the cancellation, all the gravitational contributions to $\delta \mC_{\mt A}$ can be written as an integral over the \emph{boundary} of the original, unperturbed, WDW patch,
\begin{equation}
  \delta \mC_{\mt A} = \int_{\del\mt{WDW}} ds\,d^{d-1}\Omega\,\sqrt{\gamma}\,\mathcal{T}(s)
\label{eq:gvar}
\end{equation}
for some computable response $\mathcal{T}(s)$ determined by the perturbation. This is interesting for several reasons. First, notice the same situation occurs in our derivation of the first law of complexity \eqref{eq:first-law}. In the quantum circuit discussion, the variation of the complexity is a  boundary contribution coming from the (target state) end of the circuit, \eg see figure \ref{variation}. Hence one may speculate that the boundary of the WDW patch may correspond to the `end of the circuit' in the CA conjecture.\footnote{We might contrast this feature of complexity variation in the CA approach with the results for the CV approach. The latter involves an integral over the entire extremal surface and so this does not obviously lend itself to a similar interpretation.} This suggests a picture where the AdS spacetime is built up by adding layers of null cones. This interpretation may have connections with the surface/state correspondence of \cite{Miyaji:2015yva}. Second, $\delta \mC_{\mt A}$ can still be written as in \eqref{eq:gvar} when considering more general on-shell backgrounds $g_0$ and perturbations $\delta g$. This reinforces the first point since the quantum circuit variation result is also general. Furthermore, as stressed in \cite{Hashemi:2019aop}, this statement can have interesting purely gravitational consequences, since $\mathcal{T}(s)$ may be interpretable in terms of some quasi-local stress tensor defined on $\del\mt{WDW}$. 

As noted above, it is interesting that the variation of holographic complexity (in both the CA and CV approaches) is independent of any scales, \ie independent of any dimensionful parameters appearing in the problem (up to the frequencies controlling the time dependence). In contrast, the full holographic complexity contains a variety of scales, \eg in the CA approach, the leading UV divergence has the form $\mC_A \sim \log\!\(2\ell_\mt{ct} /L\)  {\rm Vol}(\Sigma) / \delta^{d-1}$ with  $\delta$ being the short-distance cutoff \cite{Carmi:2016wjl,Chapman:2018lsv}.  An interesting question to ask is how general this statement is, \ie the present observation applies for a limited family of excitations above the vacuum, but does it still hold for more general perturbations, such as those without spherical symmetry? Of course, these comments are closely related to our previous discussion below eq.~\reef{smalla} where we saw that a variety of dimensionful parameters defining the complexity model appear in $\delta \mC_\mt{QFT}$.

All our holographic calculations involved spherically symmetric matter perturbations $\varepsilon\,\Phi_{\mt{cl}}$ with a small amplitude $\varepsilon$ and their second-order backreaction on the metric $\delta g\sim \varepsilon^2$. We found that the linear term in eq.~\reef{eq:first-law} vanished, which has the interpretation that the directions associated with introducing these excitations are orthogonal to the underlying quantum circuit which prepares the vacuum state. One expects that the same result applies for general excitations of matter fields because the matter action will only contain terms which are quadratic (and higher-order) in the fields. Hence the directions associated with exciting the corresponding single trace operators will be orthogonal to the circuit preparing the vacuum. 

In the absence of matter perturbations, we could have considered linear \emph{gravitational} excitations of the global AdS vacuum. These were studied in \cite{Ishibashi:2004wx}. Since $\delta \mC_{\rm V}$ is only sensitive to \emph{scalar} perturbations and it involves an integral over the ($d$--1)-sphere in global AdS$_{d+1}$, the only linear order contribution to $\delta \mC_{\rm V}$ comes from the spherically symmetric scalar excitations. By Birkhoff's theorem, these perturbations are time-independent and their nonlinear resummation would give rise to a spherically symmetric AdS black hole. Hence, we conclude the only linear contributions to $\delta \mC_{\rm V}$ are those corresponding to introducing a spherically symmetric black hole.\footnote{Note that this assumes $\delta {\cal V}_{\delta X} +\delta   {\cal V}_{\delta \mt{cutoff}}$ do not contribute.} Note that from the bulk perspective, these excitations are completely changing the topology of spacetime geometry. Interpreting this result from the boundary perspective, it indicates that for the CV approach, the directions associated with almost all single trace operators are orthogonal to the underlying circuit which prepares the vacuum state. The only exception would be a spherically symmetric mode of the stress tensor.

We may expect a similar result will apply for $\delta\mC_{\mt A}$, but in fact, this is not the case. Here we note the detailed calculations of \cite{Flory:2018akz,Flory:2019kah}. In particular, $\delta\mC_{\mt A}$ was evaluated in \cite{Flory:2019kah} for variations of the vacuum of a two-dimensional CFT under small local conformal transformations, which produce small excitations of the stress tensor 
due to the Schwarzian derivative. In the bulk, this involves a careful evaluation of variations of the WDW action for excitations corresponding to Banados geometries \cite{Banados:1998gg}. Surprisingly, the results showed that there were not just linear contributions, \ie $\delta\mC_{\mt A}\sim\veps$, but also contributions proportional to $\veps \log\veps$. The interpretation of the latter terms in terms of a complexity model of some underlying quantum circuits is particularly challenging. However, it would also be interesting to better understand to what extent these results for $\delta\mC_{\mt A}$ extend to metric excitations in  higher dimensions. The first steps in this direction can be found in \cite{Hashemi:2019aop}.

Recall that in section \ref{sec:connection}, we commented on the relation of our results for $\delta\mC_\mt{V}$ with similar variations studied in \cite{Belin:2018bpg,Jacobson:2018ahi}. Here, we would like to consider the possible connection of our first law of complexity \eqref{eq:first-law} with the second law in \cite{Brown:2017jil}. There the increase in the complexity is interpreted in terms of the increase in the entropy of an auxiliary system. Of course, our nomenclature suggests a similar thermodynamic interpretation, however, the latter is \emph{not} immediately apparent. In particular, eq.~\eqref{eq:first-law} refers to general small variations in quantum circuit complexity and our explicit holographic calculations did not involve any black holes, although they could have, as in \cite{Hashemi:2019aop}. However, the relation between $\delta \mC_{\rm V}$ and the first law of AdS WDW patches, as  discussed in \cite{Jacobson:2018ahi} as a limiting case of the first law of causal diamonds, and reviewed here in section~\ref{sec:connection}, can offer a fresh, and technically precise, perspective on this matter. Indeed, the gravitational tools used to derive this result are analogous to the ones leading to other gravitational first laws whose thermodynamic interpretation is well established. 

Let us add that the first law \eqref{eq:first-law} can be thought of as describing a balance equation in which $\delta \mC$ attempts to quantify a \emph{resource}, like free energy is in standard thermodynamics or relative entropy is in entanglement theory. In \cite{Brown:2017jil}, it was suggested that \emph{uncomplexity}, the difference in complexity from the one in the density matrix $\rho\propto \mathbb{I}$ is a resource related to the available volume in the interior of the black hole. From the definition of uncomplexity, \ie $\Delta \mC \equiv\mC_{\rm max}- \mC$, one naively finds that $\delta \Delta\mC= - \delta \mC$ for the variations that we have been studying. Hence, applying the first law may be an interesting approach to better understand the properties of uncomplexity and sharpen the idea that it provides a resource, as defined in quantum information theory.\footnote{See, for example \cite{gour2015resource}, and references therein, for an accurate definition and presentation of this topic.}  However, to make the equality $\delta \Delta\mC= - \delta \mC$ rigorous, one would have to understand how the Hilbert space of the holographic boundary theory should be regulated, \ie how is $\mC_{\rm max}$ defined for a quantum field theory, in particular, one with bosonic degrees of freedom.\footnote{We recall that regulating $\mC_{\rm max}$ is distinct from introducing a UV regulator in the theory -- see discussion in \cite{Jeff}. In the context of holographic complexity, the complexity was regulated with a simple late time cutoff in the interesting discussion in \cite{Zhao:2017isy}.} 
Of course, this would in itself be a useful step towards making precise the notion that uncomplexity as the basis of a proper resource theory. Further, there is interesting recent work in \cite{Brandao:2019sgy} where precise definitions on state and unitary complexity were given allowing to derive rigorous mathematical results on the number of highly complex states and the rate of complexity growth.

\subsection{Future directions}

The first law of complexity provides a new approach to investigate holographic complexity and in particular, to build a concrete bridge to standard approaches to circuit complexity. While we focussed on the complexity=volume \reef{defineCV} and complexity=action \reef{defineCA} proposals, the same approach could also be used to investigate the complexity=spacetime volume  conjecture \cite{Couch:2016exn}. Further, our derivation in section \ref{sec:firstcm} assumed that complexity is defined with Nielsen geometric approach, however, the complexity is similarly defined in terms of an extremization procedure for the Fubini-Study approach of \cite{Chapman:2017rqy} and for the path integral optimization procedure of \cite{Bhattacharyya:2018wym, Takayanagi:2018pml, Caputa:2017urj, Caputa:2017yrh}. Hence our approach should be useful to investigate these directions as well. In this respect it would also be interesting to explore other state-dependent measures, such as the $F_{\langle H^2 \rangle} \equiv \sqrt{\langle\psi(s) |H(s)^2|\psi(s)\rangle}$, which was argued to provide a tighter bound on circuit complexity with respect to the cost functions considered in this work \cite{Magan:2018nmu,Bueno:2019ajd}. 
In the present paper, we considered coherent state excitations of a real massive scalar field on the (global) AdS vacuum, but the same ideas can be applied for arbitrary matter fields and different quantum states allowing a good classical bulk description.

Our holographic calculations focused on conformal dimensions $\Delta > \frac{d}{2}$, however, this leaves the window $\frac{d}2-1\le\Delta\le\frac{d}2$, which is still compatible with unitarity in the boundary theory. The present analysis needs to be extended for this range because we found that new divergences appear in both $\delta \mC_{\mt A}$ and $\delta \mC_{\mt V}$. At present, it is not clear if these divergences are simply a technical challenge requiring a new treatment,\footnote{For example, the alternate quantization for the corresponding bulk scalars, \eg see  \cite{Klebanov:1999tb,Casini:2016rwj}, might suggest that the WDW action requires additional boundary terms involving the bulk scalar.} or if the first law produces qualitatively new behaviour in this regime. Of course, this presents interesting questions for further study.

In section~\ref{sec:lessons}, the effect of linear gravitational perturbations on global AdS for $\delta \mC_{\rm V}$ was already briefly discussed. Consider the same problem for perturbations around spherically symmetric AdS black holes. In the absence of matter, the gauge-invariant analysis of the required metric perturbations was performed in \cite{Kodama:2003jz}. As before, only scalar zero mode spherical harmonic perturbations give rise to a non-vanishing first order $\delta \mC_{\rm V}$. By Birkhoff's theorem, these perturbations change the mass of the black hole. Hence, it follows
\begin{equation}
  \delta \mC_{\rm V}\propto \left(\int^\infty_{r_h^\prime} \frac{r^{d-2}}{\sqrt{f}}\,dr - \int^\infty_{r_h} \frac{r^{d-2}}{\sqrt{f_0}}\,dr\right)
\end{equation}
where 
\begin{equation}
  f(r)=f_0(r) - \frac{2\delta M}{r^{d-3}} = 1+ \frac{r^2}{L^2} - \frac{2(M+\delta M)}{r^{d-3}}\,,
\end{equation}
and $r_h$ and $r_h^\prime$ refer to the black hole event horizon of the initial and perturbed black holes, respectively. Working with large black holes, \eg $r_h\gg L$, and using the complexity=volume results in  \cite{Chapman:2016hwi}
\begin{equation}
   \delta \mC_{\rm V} \approx \tilde{k}_d\, \delta S_{\mt{BH}}\,,
\end{equation}
where we used the same notation as in \cite{Chapman:2016hwi}. That is,   $\tilde{k}_d$ is an order one coefficient depending on the spacetime dimension, $S_{\mt{BH}}$ is the black hole entropy and we neglected a subleading piece due to  $r_h\gg L$. The standard first law of black hole mechanics allows to equivalently write this as\footnote{As above, this assumes $\delta {\cal V}_{\delta X} + \delta  {\cal V}_{\delta \mt{cutoff}}$ do not contribute.}
\begin{equation}
   \delta \mC_{\rm V}  \approx \tilde{k}_d\, \frac{\delta M}{T_{\mt{BH}}}\,.
\label{eq:cv-gravity}
\end{equation}
Hence these calculations may be useful in developing a thermodynamic understanding of the first law of complexity. Preliminary calculations using the results in \cite{Hashemi:2019aop} suggest a similar result can be derived for $\delta \mC_{\rm A}$. It would be interesting to fully develop this line of investigation.

In the context of black holes, the connection between complexity and holography was originally made through the time evolution of chaotic systems. Hence it is natural to ask if the first law, \ie complexity variations, might be a useful probe of the latter time evolution. To be more precise, consider a target state $|\Psi_{\mt{T}}\rangle$ and some perturbed state $|\Psi_{\mt{T}} + \delta\Psi\rangle = \mathcal{O}\,|\Psi_{\mt{T}}\rangle$ obtained by the action of some local operator $\mathcal{O}$. If $H$ is the hamiltonian of the system, we could examine the time evolution of the complexity variation between the two states, \ie
\begin{equation}
  \Delta\mathcal{C}(\mathcal{O}) \equiv \mathcal{C}\left[e^{-iHt}\mathcal{O}\,|\Psi_{\mt{T}}\rangle\right] - \mathcal{C}\left[e^{-iHt}\,|\Psi_{\mt{T}}\rangle\right]
\end{equation}
and ask how this encodes information on the \emph{operator growth} due to the time evolution $\mathcal{O}(-t)= e^{-iHt}\,\mathcal{O}\,e^{iHt}$. This line of reasoning was discussed for small perturbations in \cite{Bueno:2019ajd,Magan:2020iac}, based on earlier work \cite{Magan:2018nmu}, and more recently in \cite{Barbon:2019tuq} in connection to the momentum/complexity duality using the conjecture=volume \cite{Susskind:2018tei,Magan:2018nmu,Susskind:2019ddc,Lin:2019kpf}.

Finally, it would also be interesting to study the first law of complexity for mixed states.  In particular, the purification complexity, defined in \cite{Agon:2018zso,Caceres:2019pgf}, is the minimal complexity of all purifications of the mixed target state. Hence, one possibility is to study the effect on this minimization procedure due to a small perturbation in the mixed state. In \cite{future}, an alternate approach was proposed extending the Fubini-Study method to compute mixed state complexity. It should be possible to apply our methods to examine the first law of complexity in both situations.

%% file: sections/append1.tex
We show the caustics at the tip of the WDW patch do not contribute any additional term to the action \eqref{eq:faction}.

\begin{figure}[htbp]
	\centering 	\subfigure{\includegraphics[width=0.3 \textwidth]{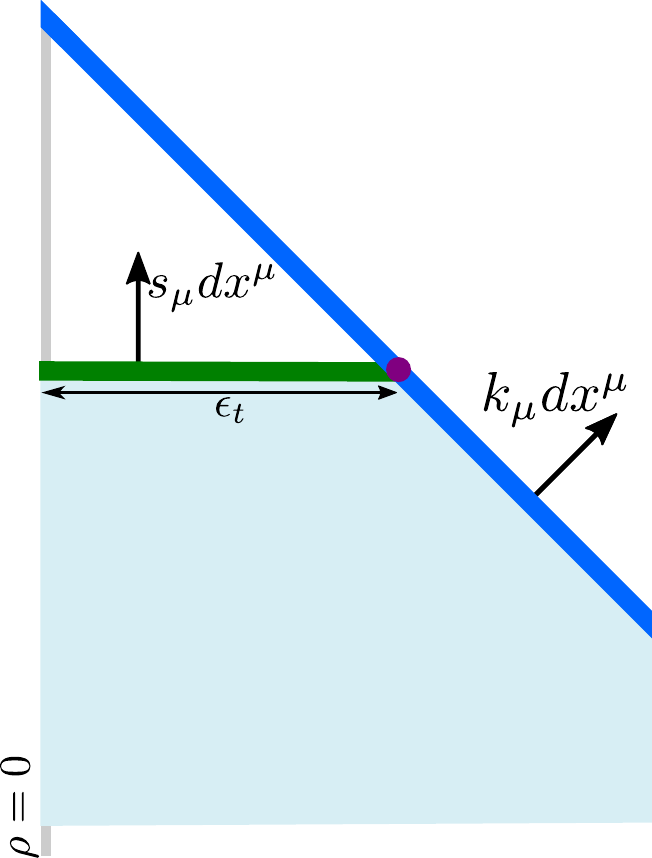}}
	\caption{The caustic of the WDW and its regularization obtained with a spacelike hypersurface. The location of the hypersurface is such that the corresponding spacelike boundary of the WDW extends for a length $\eps_t$ in the radial direction.} \label{fig:caustic}
\end{figure}

This question was studied in \cite{Chapman:2016hwi} for vacuum AdS solutions by regularizing this tip cutting it with a spacelike surface, as schematically depicted in figure.~\ref{fig:caustic}. 
After this regularization,  the boundary of the WDW patch includes this new spacelike boundary together with a joint piece, where the null boundary of the WDW patch and the new spacelike hypersurface meet. It is the corresponding GHY and joint action terms that must be added to retain a good variational principle  \cite{Lehner:2016vdi} that we compute below, following the same procedure described in \cite{Chapman:2016hwi}. For simplicity we only consider the future tip of the WDW patch, but the same analysis and conclusion goes through for the past tip of the WDW patch.
Working in our perturbative set-up, the normalized outward directed normal to the hypersurface at constant $t$ is
\be
 s_{\mu}dx^{\mu} = \frac{L}{\cos \rho}\left(1+\frac{1}{2}\varepsilon^2 (a_2-2b_2)\right) dt\,.
\label{eq:ds}
\ee
The corresponding GHY term yields 
\be 
     I_{\mt{GHY}} =  \frac{1}{8\pi \GN} \int_{t=\rm{const}}\!\!\!\!\!  \!\!\!\!  d^{d}x\,\sqrt{|h|}\,K\, =   \frac{\varepsilon^2}{16\pi \GN} \int_{0}^{\eps_t} d\rho \, d\Omega_{d-1}L^{d-1}\tan^{d-1} \rho   \, \del_t a_2   \, .
 \ee
This is $\mathcal{O}(\varepsilon^2)$, in agreement with \cite{Chapman:2016hwi}, since the extrinsic curvature vanishes for vacuum AdS. We introduced the parameter $\eps_t$ to indicate the  radial size of the spacelike region arising from the regularization procedure.   The regularity conditions \eqref{eq:reg} imply  $\del_t a_2  \sim \mO(\rho^2)$ close to the origin. It follows
$I_{\mt{GHY}} \to 0 $ when the regulator of the caustic is removed, \ie for $\eps_t \to 0$. 

The additional joint piece equals  
\be
      I_{\mt{jt}} = \frac{1}{8\pi \GN} \int_{\rm joints}\!\!\!\!\!  d\Omega_{d-1} \,\sqrt{\sigma}\,a_{\mt{jt}} \quad \text{with} \quad a_{\mt{jt}}  = \zeta  \log |k_{\mu} s^{\mu}|
\ee
$\zeta$ is a sign that will turn out to be irrelevant for the present discussion, $k_\mu$ is the null normal vector to the null WDW boundary given in \eqref{dk} and $s^\mu$ is the vector associated to the normal \eqref{eq:ds}. It follows 
\begin{equation}
  k_{\mu} s^{\mu} = \cos\rho  \( -1 + \frac{ \varepsilon^2 }{2}     \( a_2 - 2 b_2  \) \)\,,
\end{equation} 
and the joint piece yields
\be
I_{\mt{jt}} = \frac{\zeta}{8\pi \GN} \int_{\rm joint}\!\!\!\!\!  d\Omega_{d-1}\, L^{d-1}\tan^{d-1} \rho\left[ \log \cos\rho -  \frac{ \varepsilon^2 }{2}     \( a_2 - 2 b_2  \)  \right] 
\ee
The first term corresponds to the vacuum AdS value. As in \cite{Chapman:2016hwi}, this term goes to zero when evaluated at the joint , \ie for $\rho=\eps_t \to 0$. The second conclusion holds for the $\varepsilon^2$ term since the regularity conditions \eqref{eq:reg} determine $a_2  \sim \mO(\rho^2)$ and $b_2 \sim  \mO(1)$ at the origin.

%% file: sections/append2.tex
The evaluation of the holographic complexity $\delta\mathcal{C}_{\text{A}}(\Sigma)$ in \eqref{varC} includes the term $\delta I_{\delta \mt{cutoff}}$ due to the change of the radial location of the AdS boundary regulator surface. In this appendix, the relation between the global AdS cutoff $\epsilon_\rho$ and the perturbed cutoff $\epsilon_{\mt{pert}}$ is derived. Then, the contribution $\delta I_{\delta \mt{cutoff}}$ is evaluated, explicitly showing that it vanishes when the cutoff is removed.

\subsection{Matching of cutoffs}

The matching of the cutoffs requires an (asymptotic) change of coordinates in the perturbed metric
\begin{equation}
\begin{split} \label{eq:metricpert}
ds^2 &=  \frac{L^2}{\cos^2 \rho}\bigg[ - \left(1+\varepsilon^2 (a_2-2b_2)\right) dt^2 + \left(1-\varepsilon^2a_2\right) d\rho^2 + \sin^2\rho\, d\Omega_{d-1}^2\bigg] 
\end{split}
\end{equation}
to match the radial structure of the metric with the standard Fefferman-Graham expansion. For vacuum AdS, this just amounts to a redefinition of the radial variable in \eqref{eq:metricpert}. For our current purpose, it will suffice to bring the perturbed metric \eqref{eq:metricpert} to the ``almost  Fefferman-Graham'' form
\begin{equation}
\begin{split} 
ds^2 &=  \frac{L^2}{\cos^2 R}\bigg[ g_{TT}(T,R)dT^2 + dR^2 + g_{\Omega\Omega}(T ,R) d\Omega_{d-1}^2\bigg] 
\end{split}
\end{equation}
and to match the radial cutoff in the $R$ coordinate with the vacuum AdS one.

It is natural to look for such diffeomorphism perturbatively in $\varepsilon$
\begin{eqs}
  t= T +  \varepsilon^2 t_2(T,R)  + \dots \\
  \rho = R +  \varepsilon^2 \rho_2(T,R) + \dots 
\end{eqs}
where the ellipsis indicate higher order terms in the $\varepsilon$ expansion. Plugging these into \eqref{eq:metricpert}, the leading order terms are
\begin{eqs}
ds^2 &=  \frac{L^2}{\cos^2 R }\Bigg\{  - dT^2 + dR^2 + \sin^2R \, d\Omega_{d-1}^2  \\
&~~~~ + \varepsilon^2 \Big[ - \( a_2 - 2b_2 + 2 \del_{T}t_2  + \tan R ~ \rho_2 \)dT^2   + \tan R ~\rho_2  ~d\Omega_{d-1}^2  \\
 &~~~~+ 2 \( \del_{T} \rho_2 - \del_R t_2 \) dT dR  + \( - a_2 + 2 \del_{R} \rho_2  +  2 \tan R ~ \rho_2 \)dR^2 \Big] \Bigg\}\,.
\end{eqs}
Requiring the last two terms to vanish, determines
\begin{eqs}
\rho_2(T,R)  &= \frac{1}{2} \cos R \int^{R}_{\pi/2}  dr \frac{a_2(T,r)}{\cos r}\,, \\
t_2(T,R)  &=   \int^{R}_{\pi/2}  dr ~\del_{T} \rho_2(T,r) =  \frac{1}{2} \int^{R}_{\pi/2}  dr ~ \cos r  ~  \int^{r}_{\pi/2}  d\tilde r ~\del_{T} \frac{a_2(T, \tilde r)}{\cos \tilde r}\,.
\end{eqs}
Notice integration constants were conveniently fixed to match the AdS boundary. 

Matching the vacuum AdS and perturbed metric cutoffs corresponds to impose
\be
\pi/2- \eps_{\mt{pert}}  = \pi/2- \eps_{\rho}  + \varepsilon^2 \rho_2(t,\pi/2- \eps_\rho)
\ee
or,  equivalently, the relation between both cutoffs $\eps_\rho$ and $ \eps_{\mt{pert}}$ is given by 
\be
 \eps_{\mt{pert}}= \eps_\rho  \( 1+ \frac{1}{2} \varepsilon^2  a_2(t, \pi/2-\eps_\rho )\)\, .
 \ee

\subsection{Vacuum CA}

We review the calculation of the CA for global AdS originally performed in \cite{Chapman:2016hwi}, but including the counterterm $I_{\mt{ct}}$, so that the full CA consists of
\begin{equation}
  I_{\mt{vac}} = I_{\mt{EH}} + I_{\mt{GHY}} + I_{\mt{jt}} + I_{\kappa} + I_{\mt{ct}}\,,
\end{equation}
evaluated on the WDW patch anchored at the boundary time $t_\Sigma$ and bounded by the null geodesics $t_\pm(\rho)=t_\Sigma \pm (\pi/2-\rho)$ in \eqref{eq:nullsurf}.

Using the on-shell relation ${\cal R}_0=-d(d+1)/ L^2$, the EH term equals
\begin{eqs}
  I_{\mt{EH}} & = \frac{1}{16\pi\GN} \int_{\mt{WDW}}  d^{d+1} y \sqrt{|g_0|} \bigg[{\cal R}_0+ \frac{d(d-1)}{L^2}\bigg] \\
     &= - \frac{d ~ \VO L^{d-1}}{8 \pi\GN} \int_{0}^{\pi/2 - \eps_\r} dr \int^{t_\Sigma+ (\pi/2 - \rho)}_{t_\Sigma- (\pi/2 - \rho)} dt \frac{\tan^{d-1}\rho}{\cos^2 \rho} \\
 &= - \frac{d ~ \VO L^{d-1}}{8 \pi\GN} \int_{0}^{\pi/2 - \eps_\r} d\r ~2  ~(\pi/2 - \rho) ~ \frac{\tan^{d-1}\rho}{\cos^2 \rho} ~ \\
  &= -\frac{d}{d-1} \frac{ \VO L^{d-1}}{4 \pi\GN} \eps_{\r}^{1-d} + \dots
\end{eqs}
where we only kept the dominant contribution in the cutoff $\eps_\r$, which is enough for our purpose. 

Using the extrinsic curvature $K_0=\frac{d-1 +  \sin^2\rho}{L \sin \rho}$ of the AdS boundary regulator surface, the dominant contribution to the GHY term equals
\begin{eqs}
     I_{\mt{GHY}} &=  \frac{1}{8\pi \GN} \int_{\mt{regulator}}\!\!\!\!\!  \!\!\!\!  d^{d}x\,\sqrt{|h_0|}\,K_0\, \\
     &=  \frac{ \VO L^{d-1}}{8\pi \GN}  \int^{t_\Sigma+ (\pi/2 - \rho)}_{t_\Sigma- (\pi/2 - \rho)} dt  \frac{\tan^{d-1}\rho}{\cos\rho}\frac{d-1 +  \sin^2\rho}{\sin \rho} \bigg|_{\rho = \pi/2- \eps_{\r}}\\
     &=  d \frac{ \VO L^{d-1}}{4\pi \GN}   \eps^{1-d} + \dots
\end{eqs}

The counterterm $I_{\kappa}$ vanishes since $\kappa=0$ for the affine parameterization used to describe the null boundaries of global AdS. Using $a_{0,\mt{jt}}=-\log|n_{0,\mu} k_0^\mu|$ for the normals \eqref{dk} and \eqref{dn}, the dominant contribution to the joint term equals
\begin{eqs}
  I_{\mt{jt}} &=  \frac{1}{8\pi \GN} \int_{\rm joints}\!\!\!\!\!  d^{d-1}x\,\sqrt{\sigma}\,a_{0,\mt{jt}}  \\
  &= - \frac{\VO L^{d-1} }{8\pi \GN} \tan^{d-1}\rho \log \cos\rho  \bigg|_{\rho = \pi/2- \eps_{\r}} \\
   &= - \frac{\VO L^{d-1} }{8\pi \GN} \eps_{\rho}^{d-1}  \log\eps_\rho  +\dots \\
\end{eqs}
Finally, using $\Theta_0 =\frac{ (d-1)}{L}\frac{\cos\rho}{\sin\rho}$, the dominant contribution to the gravitational counterterm equals
\begin{eqs}
 I_{\mt{ct}} &= \frac{1}{8\pi \GN} \int_{\del{\rm WDW}}\!\!\!\!\! ds \,d^{d-1}\Omega\,\sqrt{\gamma}\, \Theta_0 \log (\ell_{\mt{ct}} \Theta_0)\\
 &= \frac{\VO L^{d-1} }{8\pi \GN}\eps_\rho^{1-d}  \( \frac{ 1}{d-1}   + \log \frac{  \ell_{\mt{ct}}(d-1)}{L } +  \log \eps_\rho  +\dots  \)
\end{eqs}
Summing all contributions
\begin{equation}
 I_{\mt{vac}} = \frac{\VO L^{d-1} }{8\pi \GN}\eps_\rho^{1-d} \( 2(d-1) - \frac{1}{d-1} + \log \frac{  \ell_{\mt{ct}}(d-1)}{L }  +\dots  \)
\end{equation}
reproduces the result in \cite{Chapman:2016hwi} together with the dependence on the arbitrary scale $\ell_{\mt{ct}}$ introduced by the gravitational counterterm. This reproduces the result
\eqref{eq:deltaIcutoff} in the main text.

%% file: sections/Riemannian.tex
An essential assumption in our derivation of the first law of complexity \reef{eq:first-law} was that the optimal trajectories form a smooth continuous family $x^a(s,z)$ as we vary the parameters (\ie $z$) characterizing the target state. In particular, we assumed that with a small perturbation of the target state, the optimal circuit preparing the new state remains close to the original optimal circuit. In this appendix, we first investigate this assumption for cost functions described by Riemannian metrics. Afterwards we construct a simple model to illustrate how with Nielsen's geometric approach, which effectively applies fractional gates, the circuit space is smoothed relative to that found by applying on discrete gates -- this point was discussed at the end of section \ref{sec:intro2}.

\subsection{Conjugate points and globally minimizing geodesics}
\label{sec:conjugate}

Consider the subset of circuit complexities \eqref{eq:qc-comp} with cost function described by a Riemannian metric $g_{ab}(x)$, \eg the $F_2$ and the $\kappa=2$ cost functions in eqs.~\eqref{function_F} and \eqref{eq:kappa}, or also the Fubini-Study method to define complexity \cite{Chapman:2017rqy,cohere}. This restriction allows us to borrow standard results on geodesic variations in Riemannian geometry, \eg see the textbooks \cite{frankel2011geometry,jost2008riemannian}. 
  
The deviation between geodesics in Riemannian geometry is described by a vector $V$ satisfying the \emph{geodesic deviation equation}
\begin{equation}\label{geodesic_deviation}
\frac{D^2 V^\mu}{\partial s^2} =\tensor{R}{^\mu_{\nu\rho\sigma}}T^\nu T^\rho V^\sigma  \,.
\end{equation} 
Here, $s$ is an affine parameter, $T$ is the tangent vector to the original geodesic and $D/ds = T^\mu \nabla_\mu$ denotes the directional covariant derivative. This equation is equivalent to \eqref{Jacobi_equations} in the main text, also known as Jacobi equation. We refer to its solutions as Jacobi fields.

Consider now manifolds with \emph{constant sectional curvature} $K$. Using the property
\be
\tensor{R}{^\mu_{\nu\rho\sigma}}T^\nu T^\rho J^\sigma = K J^\mu
\ee
where $J=V^{\bot}$ corresponds to the perpendicular component of $V$ along the tangent vector $T$, the projection of the geodesic deviation \eqref{geodesic_deviation} along this perpendicular direction gives rise to
\begin{equation}\label{Jacobi_example}
\frac{D^2 J}{\partial s^2} + KJ =0 \qquad \Longrightarrow \qquad J(s)=\left\{
\begin{array}{lr}
\frac{w(0)}{\sqrt{K}}\sin (\sqrt{K}s), & K>0\,,\\
\\
w(0)s\,, & K=0\,,\\
\\
\frac{w(0)}{\sqrt{-K}}\sinh (\sqrt{-K}s), & K<0\,,\\
\end{array}
\right.
\end{equation} 
with boundary conditions $J(0)=0,\dot{J}(0)=w(0)$.

The mathematical analysis of the geodesic deviation equation \eqref{geodesic_deviation} allows to reach a first important conclusion in the discussion of locally length extremizing geodesics vs globally minimizing ones: a \emph{necessary condition} for a geodesic to have \emph{globally minimizing} length is the absence of conjugate points along it.\footnote{If a nontrivial Jacobi field $J$ along a geodesic $PQ$ vanishes at point $P$ and another point $P'$ at the interior of geodesic, we call $P'$ a conjugate point to $P$.} The relevance of Jacobi fields and conjugate points for quantum circuit complexity was originally discussed by Dowling and Nielsen \cite{nielsen2008}. 

\begin{figure}[htbp]
	\centering
	\subfigure{\includegraphics[width=0.40 \textwidth]{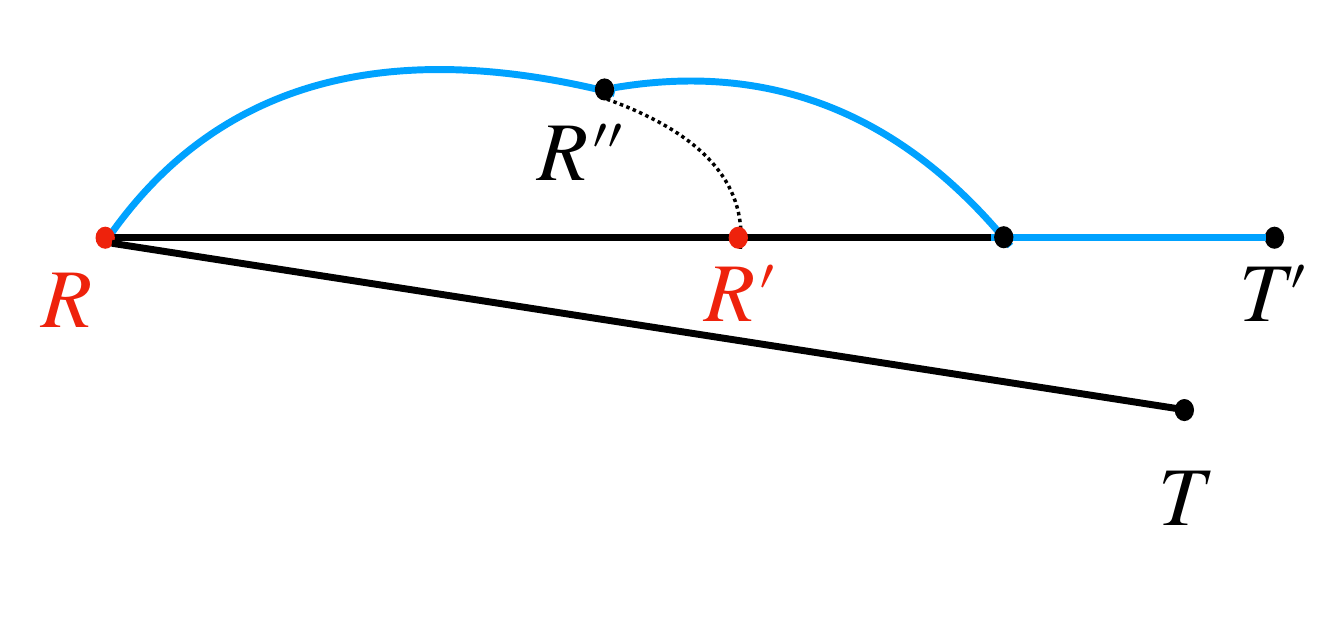}}
	\subfigure{
		\begin{tikzpicture} 
		
		
		\def\R{2.5} 
		\def\angEl{25} 
		\def\angAz{-100} 
		\def\angPhiOne{-50} 
		\def\angPhiTwo{-35} 
		\def\angBeta{33} 
		
		
		\pgfmathsetmacro\H{\R*cos(\angEl)} 
		\LatitudePlane[equator]{\angEl}{0}
		
		
		\fill[ball color=white] (0,0) circle (\R); 
		\draw (0,0) circle (\R);
		
		\coordinate (O) at (0,0);
		\coordinate[mark coordinate] (T) at (0+.32,\H-.2);
		\coordinate[mark coordinate] (T') at (0-.32,\H+.125);
		\coordinate[mark coordinate,red] (R') at (0,\H);
		\coordinate[mark coordinate,red] (R) at (0,-\H);
		\DrawLongitudeCircle[\R,thick,blue]{\angPhiOne} 
		\DrawLatitudeCircle[\R]{0} 
		
		\node[above=4pt,red] at (R') {$\mathbf{R}'$};
		\node[below=8pt,red] at (R) {$\mathbf{R}$};
		\node[below=4pt] at (T) {$\mathbf{T}$};
		\node[above=4pt] at (T') {$\mathbf{T}'$};
		
		\end{tikzpicture}
	}
	\caption{The geodesic $RT'$ contains the point $R'$, which is conjugate to $R$. One can find the blue curve has a shorter length than the geodesic $RR'T'$ because we can make the curve $RR''R'$ have the same length as that of $RR'$. One can apply this to the perturbed geodesic $ RT'$ from an original geodesic $RT$ and then the perturbed geodesic is not even locally length minimizing. The right figure is an example of the theorem in the case of a sphere. The north pole $R'$ is the conjugate point to the south pole $R$.} \label{conjugate_points}
\end{figure}
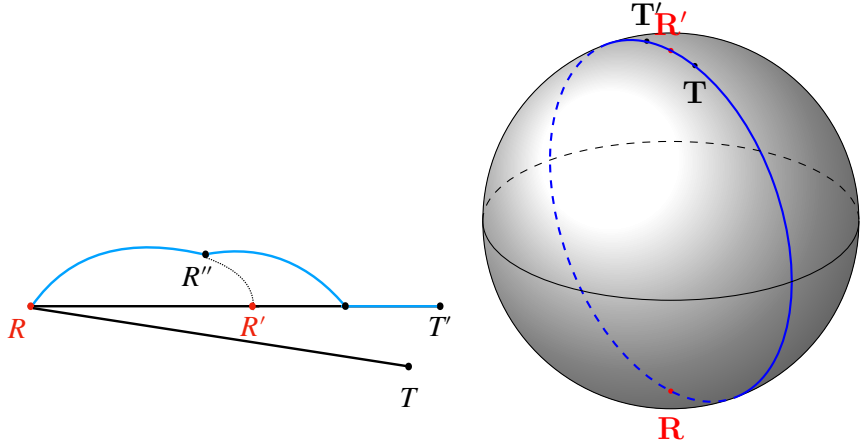

As a well-known example, consider geodesics on a $n$-sphere, as shown in figure \ref{conjugate_points}. 
Take the south pole $R$ as the initial point of the geodesic and $T$ as its endpoint, representing respectively the reference and target state, $|\Psi_{\mt{R}} \rangle$ and $|\Psi_{\mt{T}} \rangle$ in section~\ref{firstL}. The geodesic connecting these points is a portion of a great circle. Identify the perturbed target state $|\Psi_{\mt{T}} + \delta\Psi \rangle$ with the point $T^\prime$. Assuming the shortest geodesic lies near the original $RT$, one would identify the new optimal trajectory as $RTT^\prime$. However, there exists a shorter path, the globally minimizing one, corresponding to $RT^\prime$ in figure \ref{conjugate_points}.

The $n$-sphere example is a particular case of the theorem in Riemannian geometry  (see Theorem 12.11 in \cite{frankel2011geometry} and also \cite{Witten:2019qhl,jost2008riemannian} for more details)
\begin{theorem}
  If a geodesic contains the conjugate point to its initial point, then it is not a length minimizing one. 
\end{theorem}

This theorem implies that a geodesic is \emph{locally length minimizing} iff it has \emph{no} conjugate points along it. Hence, given any geodesic, the first task is to determine whether it contains conjugate points. For example, in the $n$-sphere discussion, the curve $RTT'$ passes through the north pole, which is conjugate to the south pole, and indeed there exists a shorter $RT^\prime$ geodesic in such situation.

The appearance of conjugate points is strictly related to the sectional curvature of the manifold. Indeed, from the Jacobi eq.~\eqref{Jacobi_example}, we see that manifolds with positive sectional curvature -such as the $n$-sphere- do have conjugate points, that is two zeroes of $J(s)$. On the contrary, the geodesics in manifolds with only non-positive sectional curvature do not have conjugate points and thus are always locally length minimizing. 

The existence of conjugate points is fairly generic in Nielsen's geometric approach to circuit complexity. It is in fact proven by Milnor \cite{milnor1976curvatures} that any unimodular Lie group\footnote{A group with both left-invariant and right -invariant Haar measure is called unimodular. For example: Abelian groups, finite groups, compact Lie groups and semi-simple Lie groups are all unimodular.} with left or right invariant metric must contain strictly positive sectional curvature, if it is not completely flat.  It thus follows that these geometries generically have conjugate points. This situation arises for instance in the studies of qubits or fermions associated with the special unitary group. See \cite{Balasubramanian:2019wgd} for a recent discussion on circuit complexity and conjugate points in manifolds associated with $\mathrm{SU}(2^N)$. 

Let us now consider the geodesics of this work. These are defined on $\mathbb{R}^{2N} \rtimes \mathrm{Sp}(2N,\mathbb{R})$ group manifolds, and given the semi-product of a semi-simple Lie group and abelian group is also unimodular, for generic perturbations they will have conjugate points. In other words, we can not make all the geodesics in this manifold to be locally length minimizing. However, it was found in \cite{cohere} that the simple hyperbolic geometry 
\begin{equation}\label{hyperbolic}
ds^2 = dy^2 + e^{-2y} du^2\,,
\end{equation}
with $K =-1$, effectively captures the geometry of circuit complexity for perturbations from the vacuum to coherent states with vanishing conjugate momentum expectation value.  This two-dimensional hyperbolic geometry indeed originates from affine transformations, which is an example of non-unimodular group.  
Given that all relevant geodesics for coherent states lie in such hyperbolic submanifold, it follows from \eqref{Jacobi_example} that there are no conjugate points, and thus all geodesics on this special surface are locally length minimizing. 
Of course, we stress we can not claim the absence of conjugate points in the full manifold of Lie group $\mR^{2N}\rtimes \mathrm{Sp}(2N,\mathbb{R})$. 
It is in fact only proven that on a complete Riemannian manifold with a non-positive sectional curvature, there are no conjugate points (Cartan-Hadamard theorem \cite{jost2008riemannian}).  

The previous argument explains the absence of conjugate points in the specific geodesic perturbations considered in this work. However, this is not sufficient for them to be globally minimizing since the topology of the manifold can also play a role. For example, a torus ($\mathbb{S}^1\times \mathbb{S}^1$) with completely flat metric has no conjugate points, but any pair of points can be connected by infinite geodesics which have different lengths and belong to different homotopy classes. Obviously most of them are not global length minimizing. In the next subsection~\ref{globall}, we take a simpler example on a circle ($\mathbb{S}^1$) and discuss the effect of taking a continuum limit, like the one used in Nielsen's geometry, in the presence of non-trivial topology.

%% file: sections/append_smoothness.tex

As discussed at the end of section \ref{sec:intro2}, if we consider discrete gates as in standard complexity models discussed in quantum information, our assumption on the existence of a smooth continuous family of optimal circuits $x^a(s,z)$ typically fails. That is, the discrete nature of such complexity models may produce wildly different complexities for nearby states, and correspondingly these states are prepared by dissimilar circuits. However, within Nielsen's geometric approach, the unitary circuits \reef{unitaries} are effectively constructed with arbitrary fractional gates. This approach generally gives rise to smooth complexity functions over the space of target states. We would now like to illustrate this point with a simple model. 

Let us begin with quantum mechanics on a circle and focus on $\delta$-function localized states at particular angles, \ie $\ket{\theta}$. Choose the reference state $\ket{\psi_\mt{R}}=\ket{0}$ and ask for the complexity of the target state $\ket{\psi_\mt{T}}=\ket{\theta_\mt{T}}$ using a single gate which rotates by an angle $\Delta \theta$, \ie $g=\exp[i\hat\ell\,\Delta\theta]$ with
$\hat\ell=-i\partial_\theta$. To parallel the Nielsen approach more closely, we allow for the application of the inverse gate $g^{-1}$ as part of the circuits. 

Now there may not be any such circuit that yields precisely the desired target state. However, in order to approximate a generic $\theta_\mt{T}$, we first choose $\Delta \theta$ to be an irrational fraction of $2\pi$, making the gate set universal. Second, we introduce a tolerance $\eps$ declaring the circuit $g^m$ achieves the preparation of the requested target state $\ket{\theta_\mt{T}}$ whenever the state $\ket{m\,\Delta\theta}= g^m\ket{0}$ satisfies
\beq
|\theta_\mt{T}+2\pi n-m\,\Delta\theta|\le \eps\,, 
\label{chump}
\eeq
where $m$ and $n$ may be either positive or negative integers (or zero).\footnote{In principle, one might also consider general  circuits composed of $g$ and $g^{-1}$ separately, \eg $g^{m_1}(g^{-1})^{m_2}g^{m_3}(g^{-1})^{m_4}g^{m_5}(g^{-1})^{m_6}\cdots$, however, it is clear that they will never be optimal.}  By definition, the complexity is simply  
\begin{equation} \label{eq:Ctheta}
\mC (\ket{\theta_\mt{T}})=\text{Min}\,|m| = \text{Min}  \frac{|\theta_\mt{T}+2\pi n|}{\Delta\theta}\,,
\end{equation}
the minimal (integer) number of times $g$ must be applied to produce the desired target state. 

\begin{figure}[htbp]
	\subfigure{\includegraphics[width=0.49 \textwidth]{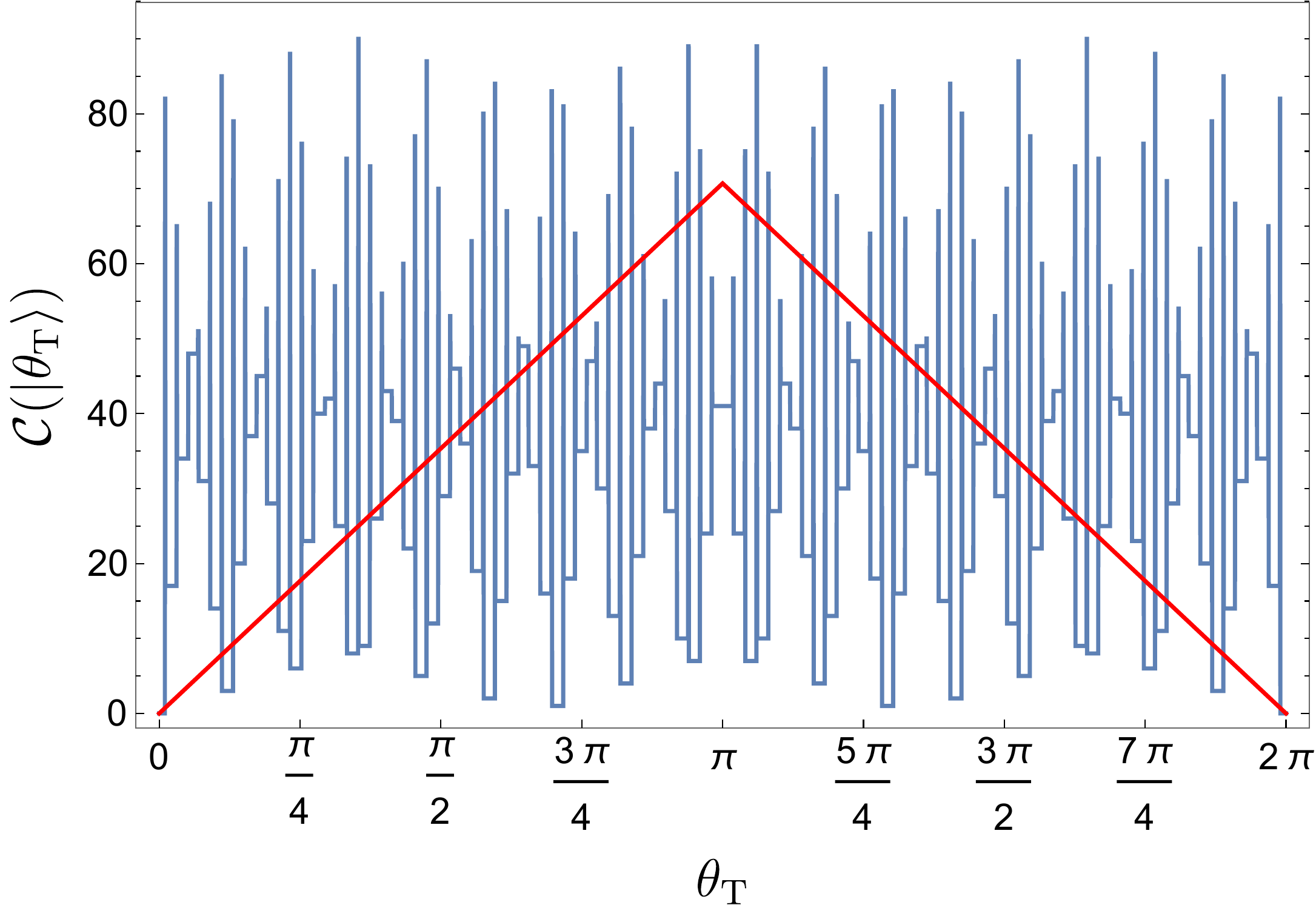}}
	\subfigure{\includegraphics[width=0.49\textwidth]{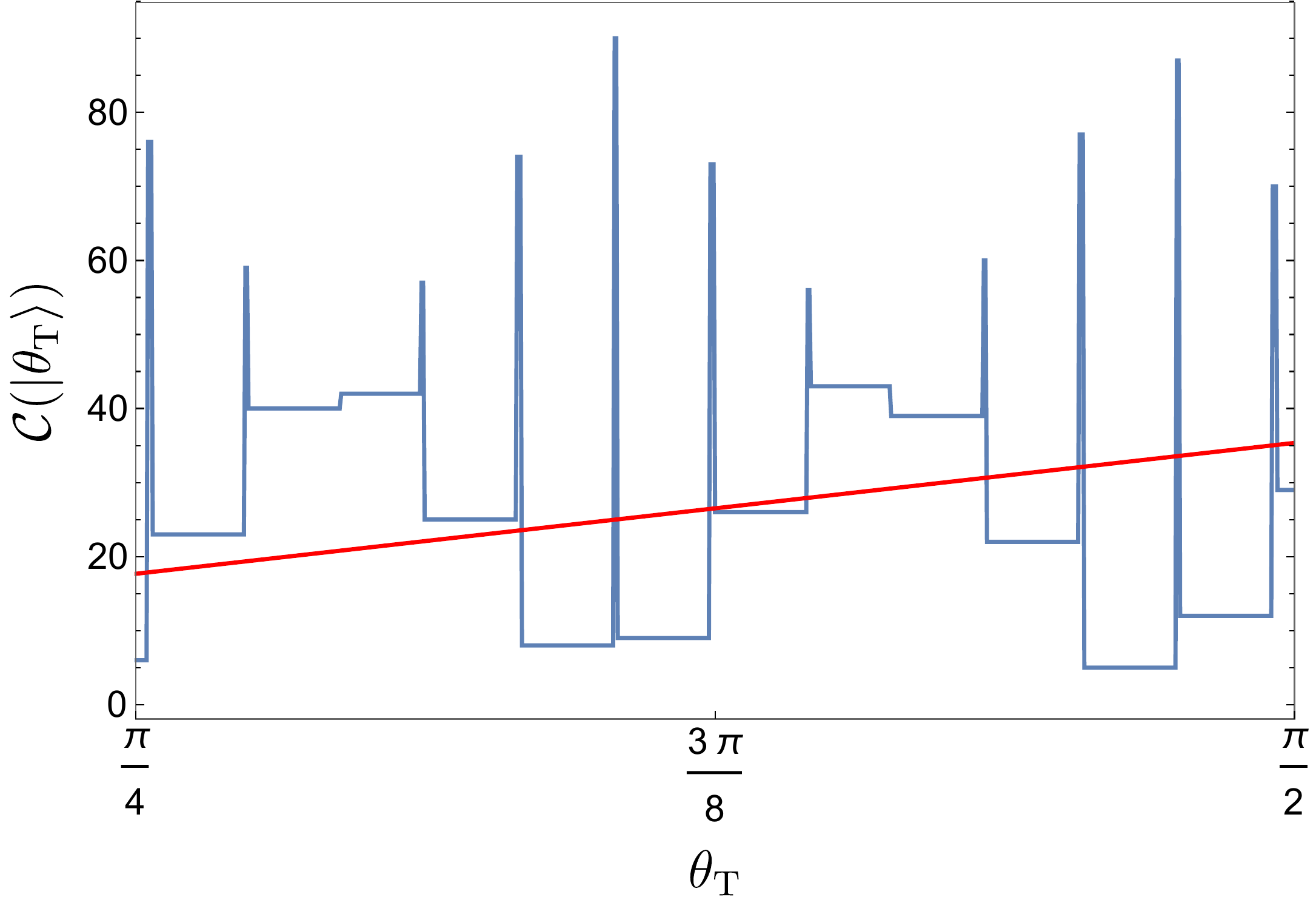}}
	\caption{The complexity (blue) for $\delta$-function states on a circle evaluated with the following parameters: gate angle $\Delta \theta = \frac{\pi}{\sqrt{2}}$ and tolerance $\epsilon =\frac{\pi}{100}$.  
The right panel shows more detail for the region $\frac{\pi}{4} \le \theta_{\mt T} \le \frac{\pi}{2}$. The complexity landscape is characterized by plateaus of width roughly $2\eps$ separated by sharp spikes. For contrast, the red line represents the  complexity \eqref{chump2} ({\it times a factor of 50!}) evaluated using a continuous circuit model.}
	\label{circle}
\end{figure}

Figure~\ref{circle} illustrates the complexity for a specific choice of the parameters, \ie $\Delta \theta = \frac{\pi}{\sqrt{2}}$ and $\epsilon =\frac{\pi}{100}$. One can see that the complexity landscape is very rough, characterized by plateaus of width roughly $2\eps$ separated by sharp spikes. Of course, the circuits associated with these plateaus and spikes are all very different. Hence if we consider a small perturbation of some target state $\ket{\theta_\mt{T}}\to \ket{\theta_\mt{T} + \delta \theta}$, we may find that the complexity of the perturbed state, and the corresponding circuit, jumps enormously.\footnote{Of course, in the cases where the complexity does not jump, it will instead not change at all! That is, if $\ket{\theta_\mt{T}}$ and $\ket{\theta_\mt{T} + \delta \theta}$ sit on the same plateau, then $\delta\mC=0$. This simply emphasizes that the first law is really only a concept that should be considered in the context of Nielsen's geometric approach.}

The origin of the sharp transitions above is the discrete nature of the underlying circuits, \ie we only ever apply $g$ an integer number of times. Now we want to show that continuous circuits, analogous to those \reef{unitaries} constructed in the main text, will smooth out this rugged complexity landscape. The continuous Hamiltonian in eq.~\reef{unitaries} for the present problem would take the form 
\begin{equation}
 {\cal H}(s)\equiv Y(s)\,\mathcal{O} \qquad{\rm with}\ \ 
\mathcal{O}=-\hat\ell\,\Delta\theta\,,
\label{unitariesB}
\end{equation}
where the control function $Y(s)$ takes real values. Choosing the $F_1$ cost function \reef{function_F}, the corresponding complexity is
\beq
\mC (\ket{\theta_\mt{T}})=
\begin{cases}
 \ \, \frac{\theta_\mt{T}}{\Delta\theta}\qquad\,&{\rm for}\ \ {0\le\theta_\mt{T}\le\pi}\,,\\
 \frac{2\pi-\theta_\mt{T}}{\Delta\theta}\quad&{\rm for}\ \ {\pi\le\theta_\mt{T}\le2\pi}\,.
 \end{cases}
\label{chump2}
\eeq
This function (multiplied by a factor of fifty) is plotted with the red lines in figure~\ref{circle}. Of course, the distinguishing feature is that with real $Y(s)$, we are effectively inserting arbitrary fractional versions of the gate $g$ in the continuous circuits. Hence we can always prepare the desired target state, without any need for tolerance. Further, this approach greatly reduces the complexity of a typical state. In particular, as the state evolves along the circuit, it never winds around the circle many times as in the discrete case.\footnote{In general, for the discrete complexity model, we expect the `average' complexity will decrease as $\Delta\theta$ decreases, but it will increase when $\eps$ is decreased.} We expect this smoothing, due to the use of continuous unitaries, to be a generic feature when applying Nielsen's geometric approach for states in quantum field theory or holography. 

We finish this discussion pointing out that even for the continuous circuits, the complexity is not `smooth' at  $\theta_\mt{T}=\pi$ (as well as at $\theta_\mt{T}=0$). That is, the complexity is continuous but the first derivative jumps sharply here. This sharp feature arises because the space of unitaries has nontrivial topology.\footnote{It also reflects the choice of the $F_1$ cost function. For example, the complexity would be smooth at $\theta_\mt{T}=0,\,\pi$ if one chose the $\kappa=2$ cost function \reef{eq:kappa}.} That is, we are considering rotations on a circle where the two points separated by $2\pi$ are identified, \eg $\theta_{\mt T}=0, 2\pi$ are identified. Indeed, the optimal circuit for both perturbed states $\ket{\pi\pm\delta\theta}$ is \emph{not} a small variation, since $\ket{\pi-\delta\theta}$ uses the control function $Y(s)=(\pi-\delta\theta)/\Delta\theta$ to build the optimal circuit, whereas the state $\ket{\pi+\delta\theta}$ requires $Y(s)=-(\pi-\delta\theta)/\Delta\theta$. Notice, the second state could have been prepared using the cost function $Y(s)=(\pi+\delta\theta)/\Delta\theta$, but this is \emph{not} optimal. Hence, a small perturbation of the target state in the vicinity of  $\theta_\mt{T}=\pi$ produces a small variation in the complexity, but the change in the minimal circuit due to the variation $Y(s)$ is large. This example shows that assuming the perturbed circuit remains close to the original one, even within Nielsen's geometric formulation, may fail at special points when the space of unitaries has nontrivial topology.

%% file: sections/append3.tex
In this section we show how to find the distance using the cost functional~\eqref{eq:costk2} for some simple target states. In particular, we start by focusing on target states with perturbatively small excitation for one mode only. That is, $\langle \hat{\phi}_\k \rangle =\sqrt{\frac{2}{\omega_\k}} \, \varepsilon a_\k$ and $\langle \hat{\pi}_\k \rangle = \sqrt{2\omega_\k}\, \varepsilon b_\k$ with all other first moments vanishing. Concretely, we want to find the distance  between  
\beq
U(s=0)=\mathbbm{I}
\qquad {\rm and} \qquad 
U(s=1) = \begin{pmatrix}
	1 & \mathbf{u}^T_{\rm T} \\
	0 & \tilde{U}_{\rm RT}
\end{pmatrix}\, ,
\eeq
where 
\beq 
\tilde{U}_{\rm RT} = {\rm diag}\(\sqrt{\omega_\n/\mu},\sqrt{\mu/\omega_\n}\)\,, \quad \mathbf{u}_\k^T = \(0,\cdots,\sqrt{\frac{2}{\omega_\k}} \, \varepsilon \, a_\k,\cdots,\sqrt{2\omega_\k}\,\varepsilon \, b_\k,\cdots,0\)\,,
\eeq
as given by the distance functional~\eqref{eq:costk2}
\be
\begin{aligned}
\label{eq:costk2-app}
 {\cal D}_{\kappa =2}(U) &= \frac{1}{4}\int_0^1 ds  \sum_I\left( \chi_{I}^{2}\, {\rm Tr} \left(\partial_sU(s) U^{-1}(s) M^T_I \right)\right)^2\,.
\end{aligned}
\ee

By the arguments of section 4.1 of~\cite{cohere}, it is possible to show that the optimal geodesic remains perturbatively close to the submanifold
\beq\label{eq:Us}
U(s) = \begin{pmatrix}
	1 & \mathbf{u}^T(s) \\
	0 & \tilde{U}_{\rm diag}(s) 
\end{pmatrix}\,,
\eeq
where 
\beq
\tilde{U}_{\rm diag}(s) = {\rm diag} \(e^{y_\n(s)},e^{-y_\n(s)}\)\,, \quad \mathbf{u}^T(s)= \(0,\cdots,u(s)_\k,\cdots,v(s)_\k,\cdots,0\)\,.
\eeq
To do this, we consider small perturbations from these types of trajectories
\beq
\hat{U} = U + \eta\, \delta U\,, \qquad {\rm with}\qquad \delta U = \begin{pmatrix}
	0 & \delta\mathbf{u}^T\\
	0 & Z
\end{pmatrix}\,,
\eeq
where $\delta \mathbf{u}$ has zero $\k$-th components and Z is off-diagonal. For very small $\eta$, the order $\eta$ terms in the cost function~\eqref{eq:costk2-app} vanish, leaving out the possibility of having source terms for the $\delta U$ components, allowing to consistently set $\delta U=0$ in the equations of motion (for more details, refer to section 4.1 of~\cite{cohere}). 
We therefore look at trajectories of the form~\eqref{eq:Us}. 

For these trajectories, the cost function~\eqref{eq:costk2-app} reduces to 
\beq
\label{eq:metricgen}
{\cal D}_{\kappa =2}(U) = \int_0^1 ds\left( \sum_\n \dot{y}_\n^2 \chi_{\phi_\n\phi_\n}^2+ e^{-2y_\k} \dot{u}_\k^2 \,   \frac{\chi^2_{0\pi_\k}}{2} + e^{2y_\k} \dot{v}_\k^2 \, \frac{\chi^2_{0\phi_\k}}{2}  \right)\, ,
\eeq
The normalization constants come from the fact that the coordinates $y_\n$ are associated with $M_{\pi_\n \phi_\n}$, $u_\k$      with $M_{0\phi_\k}$, and   $v_\k$  with $M_{0\pi_\k}$, defined in  \eqref{eq:chis} and  \eqref{eq:chisbis}
\begin{equation}
\chi_{\phi_\n\phi_\n} =1\,, \qquad   \chi_{0\phi_\k}=  f^{-1}_{\k} \, ,  \qquad   \chi_{0\pi_\k}=  f_{\k} \,  .
\end{equation}

This metric corresponds to the square of the metric of $\mathbb{R}^{N-1}$ together with a three dimensional geometry given by the $\k$ terms. 
One can treat the two parts of the metric separately. Minimizing the $\mathbb{R}^{N-1}$ part corresponds to finding the ground state circuit for the set of modes  $\n\neq \k$, with complexity $  \sum_{\n \neq \k}  \(\log \sqrt{ \omega_\n / \mu} \)^2$  (see eq.~\eqref{eq:GSk2}). We will therefore focus on solving the three dimensional part of the metric associated with the $\k$ terms. Dropping for compactness the subscript $\vk$, we are interested in the distance 
\beq\label{eq:metric}
{\cal D}_{\kappa =2}(U) =\int_0^1 ds \left(\dot{y}^2 + \frac{ f^2}{2}  e^{-2y} \, \dot{u}^2 +   \frac{ 1 }{2 f^2} e^{2y} \dot{v}^2 \right)
\eeq
with initial conditions
\beq
	y(0)=u(0)=v(0)=0\,,
\eeq
and final conditions
\beq
	 y(1) = {\rm log}\sqrt{\frac{\omega_\k}{\mu}}\,\qquad u(1)=\sqrt{\frac{2}{\omega_\k}} \, \varepsilon \, a_\k \,, \qquad v(1) =\sqrt{2\,\omega_\k} \, \varepsilon \, b_\k \,.
\eeq

The first integrals of the equations of motions derived from~\eqref{eq:metric} read 
\beq
\begin{aligned} \label{eq:eomapp}
e^{-2y}\dot{u} &=  c_1\,, \\
e^{2y}\dot{v} &=  c_2	\, ,\\
\dot{y}&=   \frac{ 1}{2f^2}  \,  c_2  \,v -    \frac{ f^2 }{2} c_1\,u + c_3\,,
\end{aligned}
\eeq
where $c_1$, $c_2$ and $c_3$ are integration constants.  

These equations can be solved perturbatively in $\varepsilon$. By inspection, one can check that up to order $\varepsilon^3$ the following perturbative expansion  is compatible with the equations of motion  
\beq
\begin{aligned}
y(s) &= y_{(0)}(s)   + \varepsilon^2 y_{(2)}(s) +\cdots\,, \\
u(s) &= \varepsilon u_{(1)}(s) + \cdots\,, \\
v(s) &= \varepsilon v_{(1)}(s) + \cdots\,.
\end{aligned}
\eeq
and 
\beq
\begin{aligned}
c_1 &=\varepsilon \Delta u + \dots \, ,\\
c_2&= \varepsilon  \Delta v + \dots \, ,\\
c_3 &= \Delta y  + \varepsilon^2 \delta y+ \dots  \, .
\end{aligned}
\eeq
At leading order, the only equation is 
\beq
\dot{y}_{(0)} = \Delta y\, , 
\eeq
which imposing the boundary conditions integrates to 
\beq
y_{(0)} =s\, {\rm log}\sqrt{\frac{\omega_\k}{\mu}}\,, \qquad {\rm with} \quad \Delta y= {\rm log}\sqrt{\frac{\omega_\k}{\mu}} \, .
\eeq
The first order equations of motion for $u$ and $v$  are
\beq
\begin{aligned}
\dot{u}_{(1)} &= e^{2y_{(0)}}\Delta u  = e^{2\Delta y \,s} \Delta u \,, \\
 \dot{v}_{(1)} &= e^{-2y_{(0)}}\Delta v = e^{-2\Delta y\, s} \Delta v \,. 
\end{aligned}
\eeq
Integrating and imposing the boundary conditions, we find
\beq
\begin{aligned}
u_{(1)} &= \frac{\Delta u}{2\Delta y} \left(e^{2\Delta y\,s}-1\right)= \frac{(\omega_\k/\mu)^s-1}{\omega_\k/\mu-1} \sqrt{\frac{2}{\omega_\k}} a_\k \,,\\
v_{(1)} &=\frac{\Delta v}{2\Delta y} \left(1-e^{-2\Delta y\,s}\right) = \frac{1-(\mu/\omega_\k)^{s}}{1-(\mu/\omega_\k)} \sqrt{2\,\omega_\k} b_\k\, ,
\end{aligned}
\eeq
which corresponds to having fixed the integration constants to
\beq
\begin{aligned}
\Delta u &= \frac{ 2 \Delta y }{ e^{2 \Delta y }-1} \sqrt{\frac{2}{\omega_\k}} \, a_\k= \frac{\log \frac{\omega_\k}{\mu}}{\frac{\omega_\k}{\mu}-1} \sqrt{\frac{2}{\omega_\k}} \, a_\k \,, \\
 \Delta v  &= \frac{ 2 \Delta y }{ 1- e^{-2 \Delta y }} \sqrt{2\,\omega_\k}\, b_\k= \frac{\log \frac{\omega_\k}{\mu}}{1-\frac{\mu}{\omega_\k}} \sqrt{2\,\omega_\k}\, b_\k\,.
\end{aligned}
\eeq
Having solved $u_{(1)}$ and $v_{(1)}$, we can proceed to integrate the equation for $y_{(2)}$ 
\beq
 \dot{y}_{(2)}=   \frac{ 1 }{2f^2}  \,   v_{(1)} \Delta v \,  -  \frac{ f^2 }{2 }  \,    u_{(1)} \Delta u +\delta y\,  
\eeq
to find  
\beq
y_{(2)} = \frac{\Delta u^2 f^2}{8  \, \Delta y^2} \left( 1-e^{2\Delta y\,s}+(e^{2\Delta y}-1)s \right) + \frac{\Delta v^2  }{8 f^2 \, \Delta y^2 } \left( e^{-2\Delta y\,s}-1+(1-e^{-2\Delta y})s \right)\,,
\eeq
where the integration constant $\delta y$ has been fixed to
\beq
\delta y = \frac{\Delta u^2 f^2   }{8  \, \Delta y^2} \left( e^{2\Delta y}-1\right)+\frac{\Delta v^2 }{8\,f^2 \Delta y^2}\left( 1-e^{-2\Delta y} \right) - \frac{\Delta u^2 f^{2}+\Delta v^2 f^{-2}}{4\Delta y}\,.
\eeq
Given the perturbative solution, we  expand the distance to ${\cal O}(\varepsilon^2)$  
\beq
\begin{aligned}
{\cal D}_{\kappa =2}(U) &= \int_0^1 ds\left[ \dot{y}_{(0)}^2 + \varepsilon^2 \left(2\, \dot{y}_{(0)} \dot{y}_{(2)} +   \frac{ f^2}{2} e^{-2y_{(0)}}\dot{u}_{(1)}^2 +    \frac{ 1 }{2 f^2} e^{2y_{(0)}}\dot{v}_{(1)}^2\right) + \cdots\right]\\
&= \Delta y^2 + \varepsilon^2 \(2\,\Delta y\, \delta y + \frac{\Delta u^2 f^2 }{2}  +\frac{\Delta v^2 }{2 f^2}\) + \cdots\\
\end{aligned}
\eeq
and noticing 
\beq
2\,\Delta y\, \delta y + \frac{\Delta u^2 f^2 }{2}  +\frac{\Delta v^2 }{2 f^2}=  \,  \frac{\log\frac{\omega_\k}{\mu} }{\omega_\k -\mu}\,\left( \frac{\mu}{\omega_\k} f^2 a_\k^2 +\omega^2_\k \, \frac{b_\k^2}{f^2} \right)\,
\eeq
we have 
\beq
\begin{aligned}
{\cal D}_{\kappa =2}(U)  = \( \log \,\sqrt{ \frac{\omega_\k}{\mu}} \)^2 + \, \varepsilon^2  \frac{\log\frac{\omega_\k}{\mu} }{\omega_\k -\mu}\,\left( \frac{\mu}{\omega_\k} f^2 a_\k^2 +\omega^2_\k \, \frac{b_\k^2}{f^2} \right) + \cdots
\end{aligned}
\eeq

All in all, adding back the contribution from the $\mathbb{R}^{N-1}$ part,  one gets
\be
\begin{aligned}
{\cal C}_{\kappa =2}  &= \sum_{\n}   \( \log \,\sqrt{ \frac{\omega_\k}{\mu}} \)^2 + \varepsilon^2  \frac{\log\frac{\omega_\k}{\mu} }{\omega_\k -\mu}\,\left( \frac{\mu}{\omega_\k} f_\k^2 a_\k^2 +\omega^2_\k \, \frac{b_\k^2}{f_\k^2} \right)+ \cdots \, .
\end{aligned}
\eeq
which is consistent with the result in eq.~\eqref{small-many} for $a_\k = |\alpha_\k| \cos \theta_\k $, $b_\k = |\alpha_\k| \sin \theta_\k $.

When more than one mode is excited, the ${\cal O}(\varepsilon^2)$ contributions to the distance remain unentangled and the generalization of eq.~\eqref{eq:metricgen} is
\beq
{\cal D}_{\kappa =2}(U) = \int_0^1 ds\left( \sum_{\n} \dot{y}_\n^2 \chi_{\phi_\n\phi_\n}^2 +\sum_\k   e^{-2y_\k} \dot{u}_\k^2 \,   \frac{\chi^2_{0\pi_\k}}{2} + e^{2y_\k} \dot{v}_\k^2 \, \frac{\chi^2_{0\phi_\k}}{2}  \right)\,,
\eeq
where $\n$ runs over all modes and $\k$ runs only over excited modes. The distance of these geodesics is then obtained with a straightforward generalization of the single mode case
\beq
\begin{aligned}
	{\cal C}_{\kappa =2} = \sum_{\n}  \( \log \,\sqrt{ \frac{\omega_\k}{\mu}} \)^2  + \, \varepsilon^2  \sum_\k  \frac{\log\frac{\omega_\k}{\mu} }{\omega_\k -\mu}\,\left(  \frac{\mu}{\omega_\k} f_\k^2 a_\k^2 +\omega^2_\k \, \frac{b_\k^2}{f_\k^2} \right)  + \cdots\,.
\end{aligned}
\eeq

%% file: AspectsFirstLaw.bbl
\providecommand{\href}[2]{#2}\begingroup\raggedright\begin{thebibliography}{100}

\bibitem{Bernamonti:2019zyy}
A.~Bernamonti, F.~Galli, J.~Hernandez, R.~C. Myers, S.-M. Ruan, and J.~Simon,
  {\it {First Law of Holographic Complexity}},  {\em Phys. Rev. Lett.} {\bf
  123} (2019), no.~8 081601, [\href{http://arxiv.org/abs/1903.04511}{{\tt
  arXiv:1903.04511}}].

\bibitem{johnw}
J.~Watrous, {\it Quantum computational complexity},  in {\em Encyclopedia of
  complexity and systems science}, pp.~7174--7201.
\newblock Springer, 2009.

\bibitem{AaronsonRev}
S.~Aaronson, {\it {The Complexity of Quantum States and Transformations: From
  Quantum Money to Black Holes}},  2016.
\newblock \href{http://arxiv.org/abs/1607.05256}{{\tt arXiv:1607.05256}}.

\bibitem{Susskind:2014rva}
L.~Susskind, {\it {Computational Complexity and Black Hole Horizons}},  {\em
  Fortsch. Phys.} {\bf 64} (2016) 44--48,
  [\href{http://arxiv.org/abs/1403.5695}{{\tt arXiv:1403.5695}}]. [Fortsch.
  Phys.64,24(2016)].

\bibitem{Stanford:2014jda}
D.~Stanford and L.~Susskind, {\it {Complexity and Shock Wave Geometries}},
  {\em Phys. Rev.} {\bf D90} (2014), no.~12 126007,
  [\href{http://arxiv.org/abs/1406.2678}{{\tt arXiv:1406.2678}}].

\bibitem{Brown:2015bva}
A.~R. Brown, D.~A. Roberts, L.~Susskind, B.~Swingle, and Y.~Zhao, {\it
  {Holographic Complexity Equals Bulk Action?}},  {\em Phys. Rev. Lett.} {\bf
  116} (2016), no.~19 191301, [\href{http://arxiv.org/abs/1509.07876}{{\tt
  arXiv:1509.07876}}].

\bibitem{Brown:2015lvg}
A.~R. Brown, D.~A. Roberts, L.~Susskind, B.~Swingle, and Y.~Zhao, {\it
  {Complexity, action, and black holes}},  {\em Phys. Rev.} {\bf D93} (2016),
  no.~8 086006, [\href{http://arxiv.org/abs/1512.04993}{{\tt
  arXiv:1512.04993}}].

\bibitem{Couch:2016exn}
J.~Couch, W.~Fischler, and P.~H. Nguyen, {\it {Noether charge, black hole
  volume, and complexity}},  {\em JHEP} {\bf 03} (2017) 119,
  [\href{http://arxiv.org/abs/1610.02038}{{\tt arXiv:1610.02038}}].

\bibitem{nielsen2006quantum}
M.~A. Nielsen, M.~R. Dowling, M.~Gu, and A.~C. Doherty, {\it Quantum
  computation as geometry},  {\em Science} {\bf 311} (2006), no.~5764
  1133--1135, [\href{http://arxiv.org/abs/quant-ph/0603161}{{\tt
  quant-ph/0603161}}].

\bibitem{nielsen2008}
M.~R. Dowling and M.~A. Nielsen, {\it The geometry of quantum computation},
  {\em Quantum Info. Comput.} {\bf 8} (Nov., 2008) 861--899,
  [\href{http://arxiv.org/abs/quant-ph/0701004}{{\tt quant-ph/0701004}}].

\bibitem{Nielsen:2006}
M.~A. Nielsen, {\it A geometric approach to quantum circuit lower bounds},
  {\em Quantum Info. Comput.} {\bf 6} (May, 2006) 213--262,
  [\href{http://arxiv.org/abs/quant-ph/0502070}{{\tt quant-ph/0502070}}].

\bibitem{Susskind:2014moa}
L.~Susskind, {\it {Entanglement is not enough}},  {\em Fortsch. Phys.} {\bf 64}
  (2016) 49--71, [\href{http://arxiv.org/abs/1411.0690}{{\tt
  arXiv:1411.0690}}].

\bibitem{Couch:2018phr}
J.~Couch, S.~Eccles, T.~Jacobson, and P.~Nguyen, {\it {Holographic Complexity
  and Volume}},  {\em JHEP} {\bf 11} (2018) 044,
  [\href{http://arxiv.org/abs/1807.02186}{{\tt arXiv:1807.02186}}].

\bibitem{Susskind:2014jwa}
L.~Susskind and Y.~Zhao, {\it {Switchbacks and the Bridge to Nowhere}},
  \href{http://arxiv.org/abs/1408.2823}{{\tt arXiv:1408.2823}}.

\bibitem{Susskind:2015toa}
L.~Susskind, {\it {The Typical-State Paradox: Diagnosing Horizons with
  Complexity}},  {\em Fortsch. Phys.} {\bf 64} (2016) 84--91,
  [\href{http://arxiv.org/abs/1507.02287}{{\tt arXiv:1507.02287}}].

\bibitem{Roberts:2014isa}
D.~A. Roberts, D.~Stanford, and L.~Susskind, {\it {Localized shocks}},  {\em
  JHEP} {\bf 03} (2015) 051, [\href{http://arxiv.org/abs/1409.8180}{{\tt
  arXiv:1409.8180}}].

\bibitem{Lehner:2016vdi}
L.~Lehner, R.~C. Myers, E.~Poisson, and R.~D. Sorkin, {\it {Gravitational
  action with null boundaries}},  {\em Phys. Rev.} {\bf D94} (2016), no.~8
  084046, [\href{http://arxiv.org/abs/1609.00207}{{\tt arXiv:1609.00207}}].

\bibitem{Cai:2016xho}
R.-G. Cai, S.-M. Ruan, S.-J. Wang, R.-Q. Yang, and R.-H. Peng, {\it {Action
  growth for AdS black holes}},  {\em JHEP} {\bf 09} (2016) 161,
  [\href{http://arxiv.org/abs/1606.08307}{{\tt arXiv:1606.08307}}].

\bibitem{Reynolds:2016rvl}
A.~Reynolds and S.~F. Ross, {\it {Divergences in Holographic Complexity}},
  {\em Class. Quant. Grav.} {\bf 34} (2017), no.~10 105004,
  [\href{http://arxiv.org/abs/1612.05439}{{\tt arXiv:1612.05439}}].

\bibitem{Chapman:2016hwi}
S.~Chapman, H.~Marrochio, and R.~C. Myers, {\it {Complexity of Formation in
  Holography}},  {\em JHEP} {\bf 01} (2017) 062,
  [\href{http://arxiv.org/abs/1610.08063}{{\tt arXiv:1610.08063}}].

\bibitem{Carmi:2016wjl}
D.~Carmi, R.~C. Myers, and P.~Rath, {\it {Comments on Holographic Complexity}},
   {\em JHEP} {\bf 03} (2017) 118, [\href{http://arxiv.org/abs/1612.00433}{{\tt
  arXiv:1612.00433}}].

\bibitem{Moosa:2017yvt}
M.~Moosa, {\it {Evolution of Complexity Following a Global Quench}},  {\em
  JHEP} {\bf 03} (2018) 031, [\href{http://arxiv.org/abs/1711.02668}{{\tt
  arXiv:1711.02668}}].

\bibitem{Couch:2017yil}
J.~Couch, S.~Eccles, W.~Fischler, and M.-L. Xiao, {\it {Holographic complexity
  and noncommutative gauge theory}},  {\em JHEP} {\bf 03} (2018) 108,
  [\href{http://arxiv.org/abs/1710.07833}{{\tt arXiv:1710.07833}}].

\bibitem{Cai:2017sjv}
R.-G. Cai, M.~Sasaki, and S.-J. Wang, {\it {Action growth of charged black
  holes with a single horizon}},  {\em Phys. Rev.} {\bf D95} (2017), no.~12
  124002, [\href{http://arxiv.org/abs/1702.06766}{{\tt arXiv:1702.06766}}].

\bibitem{Brown:2017jil}
A.~R. Brown and L.~Susskind, {\it {Second law of quantum complexity}},  {\em
  Phys. Rev.} {\bf D97} (2018), no.~8 086015,
  [\href{http://arxiv.org/abs/1701.01107}{{\tt arXiv:1701.01107}}].

\bibitem{Carmi:2017jqz}
D.~Carmi, S.~Chapman, H.~Marrochio, R.~C. Myers, and S.~Sugishita, {\it {On the
  Time Dependence of Holographic Complexity}},  {\em JHEP} {\bf 11} (2017) 188,
  [\href{http://arxiv.org/abs/1709.10184}{{\tt arXiv:1709.10184}}].

\bibitem{Swingle:2017zcd}
B.~Swingle and Y.~Wang, {\it {Holographic Complexity of
  Einstein-Maxwell-Dilaton Gravity}},  {\em JHEP} {\bf 09} (2018) 106,
  [\href{http://arxiv.org/abs/1712.09826}{{\tt arXiv:1712.09826}}].

\bibitem{Flory:2017ftd}
M.~Flory, {\it {A complexity/fidelity susceptibility $g$-theorem for
  AdS$_{3}$/BCFT$_{2}$}},  {\em JHEP} {\bf 06} (2017) 131,
  [\href{http://arxiv.org/abs/1702.06386}{{\tt arXiv:1702.06386}}].

\bibitem{Zhao:2017isy}
Y.~Zhao, {\it {Uncomplexity and Black Hole Geometry}},  {\em Phys. Rev.} {\bf
  D97} (2018), no.~12 126007, [\href{http://arxiv.org/abs/1711.03125}{{\tt
  arXiv:1711.03125}}].

\bibitem{Abt:2017pmf}
R.~Abt, J.~Erdmenger, H.~Hinrichsen, C.~M. Melby-Thompson, R.~Meyer, C.~Northe,
  and I.~A. Reyes, {\it {Topological Complexity in AdS$_3$/CFT$_2$}},  {\em
  Fortsch. Phys.} {\bf 66} (2018), no.~6 1800034,
  [\href{http://arxiv.org/abs/1710.01327}{{\tt arXiv:1710.01327}}].

\bibitem{Abt:2018ywl}
R.~Abt, J.~Erdmenger, M.~Gerbershagen, C.~M. Melby-Thompson, and C.~Northe,
  {\it {Holographic Subregion Complexity from Kinematic Space}},  {\em JHEP}
  {\bf 01} (2019) 012, [\href{http://arxiv.org/abs/1805.10298}{{\tt
  arXiv:1805.10298}}].

\bibitem{Alishahiha:2018tep}
M.~Alishahiha, A.~Faraji~Astaneh, M.~R. Mohammadi~Mozaffar, and A.~Mollabashi,
  {\it {Complexity Growth with Lifshitz Scaling and Hyperscaling Violation}},
  {\em JHEP} {\bf 07} (2018) 042, [\href{http://arxiv.org/abs/1802.06740}{{\tt
  arXiv:1802.06740}}].

\bibitem{An:2018xhv}
Y.-S. An and R.-H. Peng, {\it {Effect of the dilaton on holographic complexity
  growth}},  {\em Phys. Rev.} {\bf D97} (2018), no.~6 066022,
  [\href{http://arxiv.org/abs/1801.03638}{{\tt arXiv:1801.03638}}].

\bibitem{Fu:2018kcp}
Z.~Fu, A.~Maloney, D.~Marolf, H.~Maxfield, and Z.~Wang, {\it {Holographic
  complexity is nonlocal}},  {\em JHEP} {\bf 02} (2018) 072,
  [\href{http://arxiv.org/abs/1801.01137}{{\tt arXiv:1801.01137}}].

\bibitem{Mahapatra:2018gig}
S.~Mahapatra and P.~Roy, {\it {On the time dependence of holographic complexity
  in a dynamical Einstein-dilaton model}},  {\em JHEP} {\bf 11} (2018) 138,
  [\href{http://arxiv.org/abs/1808.09917}{{\tt arXiv:1808.09917}}].

\bibitem{Chapman:2018dem}
S.~Chapman, H.~Marrochio, and R.~C. Myers, {\it {Holographic complexity in
  Vaidya spacetimes. Part I}},  {\em JHEP} {\bf 06} (2018) 046,
  [\href{http://arxiv.org/abs/1804.07410}{{\tt arXiv:1804.07410}}].

\bibitem{Chapman:2018lsv}
S.~Chapman, H.~Marrochio, and R.~C. Myers, {\it {Holographic complexity in
  Vaidya spacetimes. Part II}},  {\em JHEP} {\bf 06} (2018) 114,
  [\href{http://arxiv.org/abs/1805.07262}{{\tt arXiv:1805.07262}}].

\bibitem{Cano:2018aqi}
P.~A. Cano, R.~A. Hennigar, and H.~Marrochio, {\it {Complexity Growth Rate in
  Lovelock Gravity}},  {\em Phys. Rev. Lett.} {\bf 121} (2018), no.~12 121602,
  [\href{http://arxiv.org/abs/1803.02795}{{\tt arXiv:1803.02795}}].

\bibitem{Barbon:2018mxk}
J.~L.~F. Barb\'on and J.~Mart\'in-Garc\'ia, {\it {Terminal Holographic
  Complexity}},  {\em JHEP} {\bf 06} (2018) 132,
  [\href{http://arxiv.org/abs/1805.05291}{{\tt arXiv:1805.05291}}].

\bibitem{Susskind:2018fmx}
L.~Susskind, {\it {Black Holes and Complexity Classes}},
  \href{http://arxiv.org/abs/1802.02175}{{\tt arXiv:1802.02175}}.

\bibitem{Susskind:2018tei}
L.~Susskind, {\it {Why do Things Fall?}},
  \href{http://arxiv.org/abs/1802.01198}{{\tt arXiv:1802.01198}}.

\bibitem{Cooper:2018cmb}
S.~Cooper, M.~Rozali, B.~Swingle, M.~Van~Raamsdonk, C.~Waddell, and D.~Wakeham,
  {\it {Black Hole Microstate Cosmology}},  {\em JHEP} {\bf 07} (2019) 065,
  [\href{http://arxiv.org/abs/1810.10601}{{\tt arXiv:1810.10601}}].

\bibitem{Numasawa:2018grg}
T.~Numasawa, {\it {Holographic Complexity for disentangled states}},
  \href{http://arxiv.org/abs/1811.03597}{{\tt arXiv:1811.03597}}.

\bibitem{Brown:2018kvn}
A.~R. Brown, H.~Gharibyan, A.~Streicher, L.~Susskind, L.~Thorlacius, and
  Y.~Zhao, {\it {Falling Toward Charged Black Holes}},  {\em Phys. Rev.} {\bf
  D98} (2018), no.~12 126016, [\href{http://arxiv.org/abs/1804.04156}{{\tt
  arXiv:1804.04156}}].

\bibitem{Goto:2018iay}
K.~Goto, H.~Marrochio, R.~C. Myers, L.~Queimada, and B.~Yoshida, {\it
  {Holographic Complexity Equals Which Action?}},  {\em JHEP} {\bf 02} (2019)
  160, [\href{http://arxiv.org/abs/1901.00014}{{\tt arXiv:1901.00014}}].

\bibitem{Agon:2018zso}
C.~A. Ag\'on, M.~Headrick, and B.~Swingle, {\it {Subsystem Complexity and
  Holography}},  {\em JHEP} {\bf 02} (2019) 145,
  [\href{http://arxiv.org/abs/1804.01561}{{\tt arXiv:1804.01561}}].

\bibitem{Chapman:2018bqj}
S.~Chapman, D.~Ge, and G.~Policastro, {\it {Holographic Complexity for Defects
  Distinguishes Action from Volume}},  {\em JHEP} {\bf 05} (2019) 049,
  [\href{http://arxiv.org/abs/1811.12549}{{\tt arXiv:1811.12549}}].

\bibitem{Flory:2018akz}
M.~Flory and N.~Miekley, {\it {Complexity change under conformal
  transformations in AdS$_{3}$/CFT$_{2}$}},  {\em JHEP} {\bf 05} (2019) 003,
  [\href{http://arxiv.org/abs/1806.08376}{{\tt arXiv:1806.08376}}].

\bibitem{Flory:2019kah}
M.~Flory, {\it {WdW-patches in AdS$_{3}$ and complexity change under conformal
  transformations II}},  {\em JHEP} {\bf 05} (2019) 086,
  [\href{http://arxiv.org/abs/1902.06499}{{\tt arXiv:1902.06499}}].

\bibitem{Braccia:2019xxi}
P.~Braccia, A.~L. Cotrone, and E.~Tonni, {\it {Complexity in the presence of a
  boundary}},  \href{http://arxiv.org/abs/1910.03489}{{\tt arXiv:1910.03489}}.

\bibitem{Sato:2019kik}
Y.~Sato and K.~Watanabe, {\it Does boundary distinguish complexities?},  {\em
  JHEP} {\bf 11} (2019) 132, [\href{http://arxiv.org/abs/1908.11094}{{\tt
  arXiv:1908.11094}}].

\bibitem{Barbon:2015ria}
J.~L.~F. Barbon and E.~Rabinovici, {\it {Holographic complexity and spacetime
  singularities}},  {\em JHEP} {\bf 01} (2016) 084,
  [\href{http://arxiv.org/abs/1509.09291}{{\tt arXiv:1509.09291}}].

\bibitem{Barbon:2015soa}
J.~L.~F. Barbon and J.~Martin-Garcia, {\it {Holographic Complexity Of Cold
  Hyperbolic Black Holes}},  {\em JHEP} {\bf 11} (2015) 181,
  [\href{http://arxiv.org/abs/1510.00349}{{\tt arXiv:1510.00349}}].

\bibitem{Auzzi:2018pbc}
R.~Auzzi, S.~Baiguera, M.~Grassi, G.~Nardelli, and N.~Zenoni, {\it {Complexity
  and action for warped AdS black holes}},  {\em JHEP} {\bf 09} (2018) 013,
  [\href{http://arxiv.org/abs/1806.06216}{{\tt arXiv:1806.06216}}].

\bibitem{Auzzi:2018zdu}
R.~Auzzi, S.~Baiguera, and G.~Nardelli, {\it {Volume and complexity for warped
  AdS black holes}},  {\em JHEP} {\bf 06} (2018) 063,
  [\href{http://arxiv.org/abs/1804.07521}{{\tt arXiv:1804.07521}}].

\bibitem{Bhattacharya:2019zkb}
A.~Bhattacharya, K.~T. Grosvenor, and S.~Roy, {\it Entanglement entropy and
  subregion complexity in thermal perturbations around pure-ads spacetime},
  {\em Phys.Rev.D} {\bf 100} (2019), no.~12 126004,
  [\href{http://arxiv.org/abs/1905.02220}{{\tt arXiv:1905.02220}}].

\bibitem{Ghosh:2019jgd}
A.~Ghosh and R.~Mishra, {\it Inhomogeneous jacobi equation and holographic
  subregion complexity},  \href{http://arxiv.org/abs/1907.11757}{{\tt
  arXiv:1907.11757}}.

\bibitem{Chapman:2017rqy}
S.~Chapman, M.~P. Heller, H.~Marrochio, and F.~Pastawski, {\it {Towards
  Complexity for Quantum Field Theory States}},  {\em Phys. Rev. Lett.} {\bf
  120} (2018), no.~12 121602, [\href{http://arxiv.org/abs/1707.08582}{{\tt
  arXiv:1707.08582}}].

\bibitem{Brown:2016wib}
A.~R. Brown, L.~Susskind, and Y.~Zhao, {\it {Quantum Complexity and Negative
  Curvature}},  {\em Phys. Rev.} {\bf D95} (2017), no.~4 045010,
  [\href{http://arxiv.org/abs/1608.02612}{{\tt arXiv:1608.02612}}].

\bibitem{Jeff}
R.~A. Jefferson and R.~C. Myers, {\it {Circuit complexity in quantum field
  theory}},  {\em JHEP} {\bf 10} (2017) 107,
  [\href{http://arxiv.org/abs/1707.08570}{{\tt arXiv:1707.08570}}].

\bibitem{Khan:2018rzm}
R.~Khan, C.~Krishnan, and S.~Sharma, {\it {Circuit Complexity in Fermionic
  Field Theory}},  {\em Phys. Rev.} {\bf D98} (2018), no.~12 126001,
  [\href{http://arxiv.org/abs/1801.07620}{{\tt arXiv:1801.07620}}].

\bibitem{Molina-Vilaplana:2018sfn}
J.~Molina-Vilaplana and A.~Del~Campo, {\it {Complexity Functionals and
  Complexity Growth Limits in Continuous MERA Circuits}},  {\em JHEP} {\bf 08}
  (2018) 012, [\href{http://arxiv.org/abs/1803.02356}{{\tt arXiv:1803.02356}}].

\bibitem{Hackl:2018ptj}
L.~Hackl and R.~C. Myers, {\it {Circuit complexity for free fermions}},  {\em
  JHEP} {\bf 07} (2018) 139, [\href{http://arxiv.org/abs/1803.10638}{{\tt
  arXiv:1803.10638}}].

\bibitem{Alves:2018qfv}
D.~W.~F. Alves and G.~Camilo, {\it {Evolution of complexity following a quantum
  quench in free field theory}},  {\em JHEP} {\bf 06} (2018) 029,
  [\href{http://arxiv.org/abs/1804.00107}{{\tt arXiv:1804.00107}}].

\bibitem{Magan:2018nmu}
J.~M. Mag\'an, {\it {Black holes, complexity and quantum chaos}},  {\em JHEP}
  {\bf 09} (2018) 043, [\href{http://arxiv.org/abs/1805.05839}{{\tt
  arXiv:1805.05839}}].

\bibitem{Camargo:2018eof}
H.~A. Camargo, P.~Caputa, D.~Das, M.~P. Heller, and R.~Jefferson, {\it
  {Complexity as a novel probe of quantum quenches: universal scalings and
  purifications}},  {\em Phys. Rev. Lett.} {\bf 122} (2019), no.~8 081601,
  [\href{http://arxiv.org/abs/1807.07075}{{\tt arXiv:1807.07075}}].

\bibitem{Chapman:2018hou}
S.~Chapman, J.~Eisert, L.~Hackl, M.~P. Heller, R.~Jefferson, H.~Marrochio, and
  R.~C. Myers, {\it {Complexity and entanglement for thermofield double
  states}},  {\em SciPost Phys.} {\bf 6} (2019), no.~3 034,
  [\href{http://arxiv.org/abs/1810.05151}{{\tt arXiv:1810.05151}}].

\bibitem{cohere}
M.~Guo, J.~Hernandez, R.~C. Myers, and S.-M. Ruan, {\it {Circuit Complexity for
  Coherent States}},  {\em JHEP} {\bf 10} (2018) 011,
  [\href{http://arxiv.org/abs/1807.07677}{{\tt arXiv:1807.07677}}].

\bibitem{Caceres:2019pgf}
E.~Caceres, S.~Chapman, J.~D. Couch, J.~P. Hernandez, R.~C. Myers, and S.-M.
  Ruan, {\it {Complexity of Mixed States in QFT and Holography}},
  \href{http://arxiv.org/abs/1909.10557}{{\tt arXiv:1909.10557}}.

\bibitem{Ali:2018fcz}
T.~Ali, A.~Bhattacharyya, S.~Shajidul~Haque, E.~H. Kim, and N.~Moynihan, {\it
  {Time Evolution of Complexity: A Critique of Three Methods}},  {\em JHEP}
  {\bf 04} (2019) 087, [\href{http://arxiv.org/abs/1810.02734}{{\tt
  arXiv:1810.02734}}].

\bibitem{Bhattacharyya:2018bbv}
A.~Bhattacharyya, A.~Shekar, and A.~Sinha, {\it {Circuit complexity in
  interacting QFTs and RG flows}},  {\em JHEP} {\bf 10} (2018) 140,
  [\href{http://arxiv.org/abs/1808.03105}{{\tt arXiv:1808.03105}}].

\bibitem{Jiang:2018nzg}
J.~Jiang and X.~Liu, {\it {Circuit Complexity for Fermionic Thermofield Double
  states}},  {\em Phys. Rev.} {\bf D99} (2019), no.~2 026011,
  [\href{http://arxiv.org/abs/1812.00193}{{\tt arXiv:1812.00193}}].

\bibitem{Chapman:2019clq}
S.~Chapman and H.~Z. Chen, {\it {Complexity for Charged Thermofield Double
  States}},  \href{http://arxiv.org/abs/1910.07508}{{\tt arXiv:1910.07508}}.

\bibitem{Camargo:2019isp}
H.~A. Camargo, M.~P. Heller, R.~Jefferson, and J.~Knaute, {\it {Path integral
  optimization as circuit complexity}},  {\em Phys. Rev. Lett.} {\bf 123}
  (2019), no.~1 011601, [\href{http://arxiv.org/abs/1904.02713}{{\tt
  arXiv:1904.02713}}].

\bibitem{Doroudiani:2019llj}
M.~Doroudiani, A.~Naseh, and R.~Pirmoradian, {\it {Complexity for Charged
  Thermofield Double States}},  {\em JHEP} {\bf 01} (2020) 120,
  [\href{http://arxiv.org/abs/1910.08806}{{\tt arXiv:1910.08806}}].

\bibitem{Bueno:2019ajd}
P.~Bueno, J.~M. Mag\'an, and C.~S. Shahbazi, {\it {Complexity measures in QFT
  and constrained geometric actions}},
  \href{http://arxiv.org/abs/1908.03577}{{\tt arXiv:1908.03577}}.

\bibitem{Ali:2019zcj}
T.~Ali, A.~Bhattacharyya, S.~S. Haque, E.~H. Kim, N.~Moynihan, and J.~Murugan,
  {\it {Chaos and Complexity in Quantum Mechanics}},  {\em Phys. Rev.} {\bf
  D101} (2020), no.~2 026021, [\href{http://arxiv.org/abs/1905.13534}{{\tt
  arXiv:1905.13534}}].

\bibitem{Ge:2019mjt}
D.~Ge and G.~Policastro, {\it {Circuit Complexity and 2D Bosonisation}},  {\em
  JHEP} {\bf 10} (2019) 276, [\href{http://arxiv.org/abs/1904.03003}{{\tt
  arXiv:1904.03003}}].

\bibitem{Caputa:2018kdj}
P.~Caputa and J.~M. Mag\'an, {\it {Quantum Computation as Gravity}},  {\em
  Phys. Rev. Lett.} {\bf 122} (2019), no.~23 231302,
  [\href{http://arxiv.org/abs/1807.04422}{{\tt arXiv:1807.04422}}].

\bibitem{Bekenstein:1973ur}
J.~D. Bekenstein, {\it {Black holes and entropy}},  {\em Phys. Rev.} {\bf D7}
  (1973) 2333--2346.

\bibitem{Bardeen:1973gs}
J.~M. Bardeen, B.~Carter, and S.~W. Hawking, {\it {The Four laws of black hole
  mechanics}},  {\em Commun. Math. Phys.} {\bf 31} (1973) 161--170.

\bibitem{Hawking:1974sw}
S.~W. Hawking, {\it {Particle Creation by Black Holes}},  {\em Commun. Math.
  Phys.} {\bf 43} (1975) 199--220. [,167(1975)].

\bibitem{Jacobson:1995ab}
T.~Jacobson, {\it {Thermodynamics of space-time: The Einstein equation of
  state}},  {\em Phys. Rev. Lett.} {\bf 75} (1995) 1260--1263,
  [\href{http://arxiv.org/abs/gr-qc/9504004}{{\tt gr-qc/9504004}}].

\bibitem{Nozaki:2013vta}
M.~Nozaki, T.~Numasawa, A.~Prudenziati, and T.~Takayanagi, {\it {Dynamics of
  Entanglement Entropy from Einstein Equation}},  {\em Phys. Rev.} {\bf D88}
  (2013), no.~2 026012, [\href{http://arxiv.org/abs/1304.7100}{{\tt
  arXiv:1304.7100}}].

\bibitem{Lashkari:2013koa}
N.~Lashkari, M.~B. McDermott, and M.~Van~Raamsdonk, {\it {Gravitational
  dynamics from entanglement 'thermodynamics'}},  {\em JHEP} {\bf 04} (2014)
  195, [\href{http://arxiv.org/abs/1308.3716}{{\tt arXiv:1308.3716}}].

\bibitem{Faulkner:2013ica}
T.~Faulkner, M.~Guica, T.~Hartman, R.~C. Myers, and M.~Van~Raamsdonk, {\it
  {Gravitation from Entanglement in Holographic CFTs}},  {\em JHEP} {\bf 03}
  (2014) 051, [\href{http://arxiv.org/abs/1312.7856}{{\tt arXiv:1312.7856}}].

\bibitem{VanRaamsdonk:2010pw}
M.~Van~Raamsdonk, {\it {Building up spacetime with quantum entanglement}},
  {\em Gen. Rel. Grav.} {\bf 42} (2010) 2323--2329,
  [\href{http://arxiv.org/abs/1005.3035}{{\tt arXiv:1005.3035}}]. [Int. J. Mod.
  Phys.D19,2429(2010)].

\bibitem{Bianchi:2012ev}
E.~Bianchi and R.~C. Myers, {\it {On the Architecture of Spacetime Geometry}},
  {\em Class. Quant. Grav.} {\bf 31} (2014) 214002,
  [\href{http://arxiv.org/abs/1212.5183}{{\tt arXiv:1212.5183}}].

\bibitem{Maldacena:2013xja}
J.~Maldacena and L.~Susskind, {\it {Cool horizons for entangled black holes}},
  {\em Fortsch. Phys.} {\bf 61} (2013) 781--811,
  [\href{http://arxiv.org/abs/1306.0533}{{\tt arXiv:1306.0533}}].

\bibitem{Yang:2017nfn}
R.-Q. Yang, {\it {Complexity for quantum field theory states and applications
  to thermofield double states}},  {\em Phys. Rev.} {\bf D97} (2018), no.~6
  066004, [\href{http://arxiv.org/abs/1709.00921}{{\tt arXiv:1709.00921}}].

\bibitem{Yang:2018nda}
R.-Q. Yang, Y.-S. An, C.~Niu, C.-Y. Zhang, and K.-Y. Kim, {\it {Principles and
  symmetries of complexity in quantum field theory}},  {\em Eur. Phys. J.} {\bf
  C79} (2019), no.~2 109, [\href{http://arxiv.org/abs/1803.01797}{{\tt
  arXiv:1803.01797}}].

\bibitem{Lloyd_2000}
S.~Lloyd, {\it Ultimate physical limits to computation},  {\em Nature} {\bf
  406} (Aug, 2000) 1047--1054.

\bibitem{kirillov2004lectures}
A.~A. Kirillov, {\em Lectures on the orbit method}, vol.~64.
\newblock American Mathematical Soc., 2004.

\bibitem{Susskind:2018pmk}
L.~Susskind, {\it {Three Lectures on Complexity and Black Holes}},  2018.
\newblock \href{http://arxiv.org/abs/1810.11563}{{\tt arXiv:1810.11563}}.

\bibitem{Brown:2019whu}
A.~R. Brown and L.~Susskind, {\it {Complexity geometry of a single qubit}},
  {\em Phys. Rev.} {\bf D100} (2019), no.~4 046020,
  [\href{http://arxiv.org/abs/1903.12621}{{\tt arXiv:1903.12621}}].

\bibitem{Balasubramanian:2019wgd}
V.~Balasubramanian, M.~Decross, A.~Kar, and O.~Parrikar, {\it {Quantum
  Complexity of Time Evolution with Chaotic Hamiltonians}},  {\em JHEP} {\bf
  01} (2020) 134, [\href{http://arxiv.org/abs/1905.05765}{{\tt
  arXiv:1905.05765}}].

\bibitem{Witten:1998qj}
E.~Witten, {\it {Anti-de Sitter space and holography}},  {\em Adv. Theor. Math.
  Phys.} {\bf 2} (1998) 253--291,
  [\href{http://arxiv.org/abs/hep-th/9802150}{{\tt hep-th/9802150}}].

\bibitem{Papadodimas:2013jku}
K.~Papadodimas and S.~Raju, {\it {State-Dependent Bulk-Boundary Maps and Black
  Hole Complementarity}},  {\em Phys. Rev.} {\bf D89} (2014), no.~8 086010,
  [\href{http://arxiv.org/abs/1310.6335}{{\tt arXiv:1310.6335}}].

\bibitem{Almheiri:2014lwa}
A.~Almheiri, X.~Dong, and D.~Harlow, {\it {Bulk Locality and Quantum Error
  Correction in AdS/CFT}},  {\em JHEP} {\bf 04} (2015) 163,
  [\href{http://arxiv.org/abs/1411.7041}{{\tt arXiv:1411.7041}}].

\bibitem{Harlow:2018fse}
D.~Harlow, {\it {TASI Lectures on the Emergence of Bulk Physics in AdS/CFT}},
  {\em PoS} {\bf TASI2017} (2018) 002,
  [\href{http://arxiv.org/abs/1802.01040}{{\tt arXiv:1802.01040}}].

\bibitem{ElShowk:2011ag}
S.~El-Showk and K.~Papadodimas, {\it {Emergent Spacetime and Holographic
  CFTs}},  {\em JHEP} {\bf 10} (2012) 106,
  [\href{http://arxiv.org/abs/1101.4163}{{\tt arXiv:1101.4163}}].

\bibitem{Fitzpatrick:2011jn}
A.~L. Fitzpatrick and J.~Kaplan, {\it {Scattering States in AdS/CFT}},
  \href{http://arxiv.org/abs/1104.2597}{{\tt arXiv:1104.2597}}.

\bibitem{kaplan2013lectures}
J.~Kaplan, {\it Lectures on ads/cft from the bottom up},  2013.

\bibitem{Terashima:2017gmc}
S.~Terashima, {\it {AdS/CFT Correspondence in Operator Formalism}},  {\em JHEP}
  {\bf 02} (2018) 019, [\href{http://arxiv.org/abs/1710.07298}{{\tt
  arXiv:1710.07298}}].

\bibitem{Berenstein:2019tcs}
D.~Berenstein and J.~Sim\'on, {\it {Localized states in global AdS}},
  \href{http://arxiv.org/abs/1910.10227}{{\tt arXiv:1910.10227}}.

\bibitem{Avis:1977yn}
S.~J. Avis, C.~J. Isham, and D.~Storey, {\it {Quantum Field Theory in anti-De
  Sitter Space-Time}},  {\em Phys. Rev.} {\bf D18} (1978) 3565.

\bibitem{Burgess:1984ti}
C.~P. Burgess and C.~A. Lutken, {\it {Propagators and Effective Potentials in
  Anti-de Sitter Space}},  {\em Phys. Lett.} {\bf 153B} (1985) 137--141.

\bibitem{Cotabreveescu:1999em}
I.~I. Cot\ifmmode~\u{a}\else \u{a}\fi{}escu, {\it Remarks on the quantum modes
  of the scalar field on ${\mathrm{ads}}_{d+1}$ spacetime},  {\em Phys. Rev. D}
  {\bf 60} (Oct, 1999) 107504.

\bibitem{IZ}
C.~Itzykson and J.~B. Zuber, {\em {Quantum Field Theory}}.
\newblock International Series In Pure and Applied Physics. McGraw-Hill, New
  York, 1980.

\bibitem{BottaCantcheff:2015sav}
M.~Botta-Cantcheff, P.~Mart\'inez, and G.~A. Silva, {\it {On excited states in
  real-time AdS/CFT}},  {\em JHEP} {\bf 02} (2016) 171,
  [\href{http://arxiv.org/abs/1512.07850}{{\tt arXiv:1512.07850}}].

\bibitem{Marolf:2017kvq}
D.~Marolf, O.~Parrikar, C.~Rabideau, A.~Izadi~Rad, and M.~Van~Raamsdonk, {\it
  {From Euclidean Sources to Lorentzian Spacetimes in Holographic Conformal
  Field Theories}},  {\em JHEP} {\bf 06} (2018) 077,
  [\href{http://arxiv.org/abs/1709.10101}{{\tt arXiv:1709.10101}}].

\bibitem{BottaCantcheff:2019apr}
M.~Botta-Cantcheff, P.~J. Mart\'inez, and G.~A. Silva, {\it {Holographic
  excited states in AdS Black Holes}},  {\em JHEP} {\bf 04} (2019) 028,
  [\href{http://arxiv.org/abs/1901.00505}{{\tt arXiv:1901.00505}}].

\bibitem{Bizon:2011gg}
P.~Bizon and A.~Rostworowski, {\it {On weakly turbulent instability of anti-de
  Sitter space}},  {\em Phys. Rev. Lett.} {\bf 107} (2011) 031102,
  [\href{http://arxiv.org/abs/1104.3702}{{\tt arXiv:1104.3702}}].

\bibitem{Buchel:2012uh}
A.~Buchel, L.~Lehner, and S.~L. Liebling, {\it {Scalar Collapse in AdS}},  {\em
  Phys. Rev.} {\bf D86} (2012) 123011,
  [\href{http://arxiv.org/abs/1210.0890}{{\tt arXiv:1210.0890}}].

\bibitem{Buchel:2013uba}
A.~Buchel, S.~L. Liebling, and L.~Lehner, {\it {Boson stars in AdS spacetime}},
   {\em Phys. Rev.} {\bf D87} (2013), no.~12 123006,
  [\href{http://arxiv.org/abs/1304.4166}{{\tt arXiv:1304.4166}}].

\bibitem{Kim:2014ida}
N.~Kim, {\it {Time-periodic solutions of massive scalar fields in dynamical AdS
  background: Perturbative constructions}},  {\em Phys. Lett.} {\bf B742}
  (2015) 274--278, [\href{http://arxiv.org/abs/1411.1633}{{\tt
  arXiv:1411.1633}}].

\bibitem{PhysRevLett.28.1082}
J.~W. York, {\it Role of conformal three-geometry in the dynamics of
  gravitation},  {\em Phys. Rev. Lett.} {\bf 28} (Apr, 1972) 1082--1085.

\bibitem{PhysRevD.15.2752}
G.~W. Gibbons and S.~W. Hawking, {\it Action integrals and partition functions
  in quantum gravity},  {\em Phys. Rev. D} {\bf 15} (May, 1977) 2752--2756.

\bibitem{Klebanov:1999tb}
I.~R. Klebanov and E.~Witten, {\it {AdS / CFT correspondence and symmetry
  breaking}},  {\em Nucl. Phys.} {\bf B556} (1999) 89--114,
  [\href{http://arxiv.org/abs/hep-th/9905104}{{\tt hep-th/9905104}}].

\bibitem{Casini:2016rwj}
H.~Casini, D.~A. Galante, and R.~C. Myers, {\it {Comments on Jacobson's
  "entanglement equilibrium and the Einstein equation"}},  {\em JHEP} {\bf 03}
  (2016) 194, [\href{http://arxiv.org/abs/1601.00528}{{\tt arXiv:1601.00528}}].

\bibitem{fefferman1985elie}
C.~Fefferman and C.~R. Graham, {\it {Conformal invariants}},  {\em \'Elie
  Cartan et les math\'ematiques d'aujourd'hui - Lyon, 25-29 juin 1984,
  Ast\'erisque} {\bf S131} (1985) 95--116.

\bibitem{fefferman2007ambient}
C.~Fefferman and C.~R. Graham, {\it {The ambient metric}},  {\em Ann. Math.
  Stud.} {\bf 178} (2011) 1--128, [\href{http://arxiv.org/abs/0710.0919}{{\tt
  arXiv:0710.0919}}].

\bibitem{Emparan:1999pm}
R.~Emparan, C.~V. Johnson, and R.~C. Myers, {\it {Surface terms as counterterms
  in the AdS / CFT correspondence}},  {\em Phys. Rev.} {\bf D60} (1999) 104001,
  [\href{http://arxiv.org/abs/hep-th/9903238}{{\tt hep-th/9903238}}].

\bibitem{deHaro:2000vlm}
S.~de~Haro, S.~N. Solodukhin, and K.~Skenderis, {\it {Holographic
  reconstruction of space-time and renormalization in the AdS / CFT
  correspondence}},  {\em Commun. Math. Phys.} {\bf 217} (2001) 595--622,
  [\href{http://arxiv.org/abs/hep-th/0002230}{{\tt hep-th/0002230}}].

\bibitem{Skenderis:2002wp}
K.~Skenderis, {\it {Lecture notes on holographic renormalization}},  {\em
  Class. Quant. Grav.} {\bf 19} (2002) 5849--5876,
  [\href{http://arxiv.org/abs/hep-th/0209067}{{\tt hep-th/0209067}}].

\bibitem{usefulformulas}
M.~Robert, {\em Conventions, Definitions, Identities, and Formulas}.
\newblock Available at
  \href{http://jacobi.luc.edu/Useful.html}{http://jacobi.luc.edu/Useful.html}.
\newblock [Accessed 8-September-2019].

\bibitem{gradshteyn2007}
I.~S. Gradshteyn and I.~M. Ryzhik, {\em Table of integrals, series, and
  products}.
\newblock Elsevier/Academic Press, Amsterdam, seventh~ed., 2007.
\newblock Translated from the Russian, Translation edited and with a preface by
  Alan Jeffrey and Daniel Zwillinger, With one CD-ROM (Windows, Macintosh and
  UNIX).

\bibitem{Belin:2018bpg}
A.~Belin, A.~Lewkowycz, and G.~S\'arosi, {\it {Complexity and the bulk volume,
  a new York time story}},  {\em JHEP} {\bf 03} (2019) 044,
  [\href{http://arxiv.org/abs/1811.03097}{{\tt arXiv:1811.03097}}].

\bibitem{Jacobson:2018ahi}
T.~Jacobson and M.~Visser, {\it {Gravitational Thermodynamics of Causal
  Diamonds in (A)dS}},  {\em SciPost Phys.} {\bf 7} (2019), no.~6 079,
  [\href{http://arxiv.org/abs/1812.01596}{{\tt arXiv:1812.01596}}].

\bibitem{Belin:2018fxe}
A.~Belin, A.~Lewkowycz, and G.~S\'arosi, {\it {The boundary dual of the bulk
  symplectic form}},  {\em Phys. Lett.} {\bf B789} (2019) 71--75,
  [\href{http://arxiv.org/abs/1806.10144}{{\tt arXiv:1806.10144}}].

\bibitem{bhatia2013matrix}
R.~Bhatia, {\em Matrix analysis}, vol.~169.
\newblock Springer Science \& Business Media, 2013.

\bibitem{watrous2018theory}
J.~Watrous, {\em Theory of Quantum Information}.
\newblock Cambridge University Press, 2018.
\newblock (See section 1.1).

\bibitem{gil2003operator}
M.~I. Gil, {\em Operator functions and localization of spectra}.
\newblock Springer, 2003.

\bibitem{Carmi}
D.~Carmi, R.~C. Myers, and P.~Rath, {\it {Comments on Holographic Complexity}},
   {\em JHEP} {\bf 03} (2017) 118, [\href{http://arxiv.org/abs/1612.00433}{{\tt
  arXiv:1612.00433}}].

\bibitem{Susskind:2019ddc}
L.~Susskind, {\it {Complexity and Newton's Laws}},
  \href{http://arxiv.org/abs/1904.12819}{{\tt arXiv:1904.12819}}.

\bibitem{Reynolds:2017lwq}
A.~Reynolds and S.~F. Ross, {\it {Complexity in de Sitter Space}},  {\em Class.
  Quant. Grav.} {\bf 34} (2017), no.~17 175013,
  [\href{http://arxiv.org/abs/1706.03788}{{\tt arXiv:1706.03788}}].

\bibitem{Hashemi:2019aop}
S.~S. Hashemi, G.~Jafari, and A.~Naseh, {\it {On the first law of holographic
  complexity}},  \href{http://arxiv.org/abs/1912.10436}{{\tt
  arXiv:1912.10436}}.

\bibitem{Miyaji:2015yva}
M.~Miyaji and T.~Takayanagi, {\it {Surface/State Correspondence as a
  Generalized Holography}},  {\em PTEP} {\bf 2015} (2015), no.~7 073B03,
  [\href{http://arxiv.org/abs/1503.03542}{{\tt arXiv:1503.03542}}].

\bibitem{Ishibashi:2004wx}
A.~Ishibashi and R.~M. Wald, {\it {Dynamics in nonglobally hyperbolic static
  space-times. 3. Anti-de Sitter space-time}},  {\em Class. Quant. Grav.} {\bf
  21} (2004) 2981--3014, [\href{http://arxiv.org/abs/hep-th/0402184}{{\tt
  hep-th/0402184}}].

\bibitem{Banados:1998gg}
M.~Banados, {\it {Three-dimensional quantum geometry and black holes}},  {\em
  AIP Conf. Proc.} {\bf 484} (1999), no.~1 147--169,
  [\href{http://arxiv.org/abs/hep-th/9901148}{{\tt hep-th/9901148}}].

\bibitem{gour2015resource}
G.~Gour, M.~P. M{\"u}ller, V.~Narasimhachar, R.~W. Spekkens, and N.~Y. Halpern,
  {\it The resource theory of informational nonequilibrium in thermodynamics},
  {\em Physics Reports} {\bf 583} (2015) 1--58,
  [\href{http://arxiv.org/abs/1309.6586}{{\tt arXiv:1309.6586}}].

\bibitem{Brandao:2019sgy}
F.~G. S.~L. Brand\~ao, W.~Chemissany, N.~Hunter-Jones, R.~Kueng, and
  J.~Preskill, {\it {Models of quantum complexity growth}},
  \href{http://arxiv.org/abs/1912.04297}{{\tt arXiv:1912.04297}}.

\bibitem{Bhattacharyya:2018wym}
A.~Bhattacharyya, P.~Caputa, S.~R. Das, N.~Kundu, M.~Miyaji, and T.~Takayanagi,
  {\it {Path-Integral Complexity for Perturbed CFTs}},  {\em JHEP} {\bf 07}
  (2018) 086, [\href{http://arxiv.org/abs/1804.01999}{{\tt arXiv:1804.01999}}].

\bibitem{Takayanagi:2018pml}
T.~Takayanagi, {\it {Holographic Spacetimes as Quantum Circuits of
  Path-Integrations}},  {\em JHEP} {\bf 12} (2018) 048,
  [\href{http://arxiv.org/abs/1808.09072}{{\tt arXiv:1808.09072}}].

\bibitem{Caputa:2017urj}
P.~Caputa, N.~Kundu, M.~Miyaji, T.~Takayanagi, and K.~Watanabe, {\it {Anti-de
  Sitter Space from Optimization of Path Integrals in Conformal Field
  Theories}},  {\em Phys. Rev. Lett.} {\bf 119} (2017), no.~7 071602,
  [\href{http://arxiv.org/abs/1703.00456}{{\tt arXiv:1703.00456}}].

\bibitem{Caputa:2017yrh}
P.~Caputa, N.~Kundu, M.~Miyaji, T.~Takayanagi, and K.~Watanabe, {\it {Liouville
  Action as Path-Integral Complexity: From Continuous Tensor Networks to
  AdS/CFT}},  {\em JHEP} {\bf 11} (2017) 097,
  [\href{http://arxiv.org/abs/1706.07056}{{\tt arXiv:1706.07056}}].

\bibitem{Kodama:2003jz}
H.~Kodama and A.~Ishibashi, {\it {A Master equation for gravitational
  perturbations of maximally symmetric black holes in higher dimensions}},
  {\em Prog. Theor. Phys.} {\bf 110} (2003) 701--722,
  [\href{http://arxiv.org/abs/hep-th/0305147}{{\tt hep-th/0305147}}].

\bibitem{Magan:2020iac}
J.~M. Magan and J.~Simon, {\it {On operator growth and emergent Poincar\'e
  symmetries}},  \href{http://arxiv.org/abs/2002.03865}{{\tt
  arXiv:2002.03865}}.

\bibitem{Barbon:2019tuq}
J.~L.~F. Barb\'on, J.~Mart\'in-Garc\'ia, and M.~Sasieta, {\it
  {Momentum/Complexity Duality and the Black Hole Interior}},
  \href{http://arxiv.org/abs/1912.05996}{{\tt arXiv:1912.05996}}.

\bibitem{Lin:2019kpf}
H.~W. Lin and L.~Susskind, {\it {Complexity Geometry and Schwarzian Dynamics}},
   {\em JHEP} {\bf 01} (2020) 087, [\href{http://arxiv.org/abs/1911.02603}{{\tt
  arXiv:1911.02603}}].

\bibitem{future}
S.-M. Ruan {\em to appear}.

\bibitem{frankel2011geometry}
T.~Frankel, {\em The geometry of physics: an introduction}.
\newblock Cambridge university press, 2011.

\bibitem{jost2008riemannian}
J.~Jost, {\em Riemannian geometry and geometric analysis}, vol.~42005.
\newblock Springer, 2008.

\bibitem{Witten:2019qhl}
E.~Witten, {\it {Light Rays, Singularities, and All That}},
  \href{http://arxiv.org/abs/1901.03928}{{\tt arXiv:1901.03928}}.

\bibitem{milnor1976curvatures}
J.~Milnor, {\it Curvatures of left invariant metrics on lie groups},  1976.

\end{thebibliography}\endgroup
